\documentclass{article}

\usepackage[pdftex]{graphicx}
\usepackage{float}
\def\SdS{Schwarzschild--de~Sitter}
\def \dS{de~Sitter}
\def\KdS{Kerr--de~Sitter}
\def\be{\begin{equation}} \def\ee{\end{equation}}
\def\bea{\begin{eqnarray}} \def\eea{\end{eqnarray}}
\def\din{\,\mathrm{d}} \def\dbe{\mathrm{d}}
\def\bet{\begin{tabular}} \def\ent{\end{tabular}}
\def\oder#1#2{\frac{\dbe#1}{\dbe#2}}
\def\cale{{\cal E}} \def\calk{{\cal K}}

\begin{document}
	\title{Light escape cones in local reference frames of \KdS\, black hole spacetimes and related black hole shadows}
	
	\author{Zden\v{e}k Stuchl\'{i}k, Daniel Charbul\'{a}k and Jan Schee \\
		\emph{Institute of Physics} \\
		\emph{and Research Centre of Theoretical Physics and Astrophysics,} \\
		\emph{Faculty of Philosophy and Science,}\\ 
		\emph{Silesian university in Opava,}\\ 
		\emph{Bezru\v{c}ovo n\'{a}m. 13,}\\
		\emph{CZ-746 01 Opava, Czech Republic}\\
		\small{zdenek.stuchlik@fpf.slu.cz, daniel.charbulak@fpf.slu.cz, jan.schee@fpf.slu.cz}}
	\date{}

	\maketitle
	
	\begin{abstract}
We construct the light escape cones of isotropic spot sources of radiation residing in special classes of reference frames in the \KdS\, (KdS) black hole spacetimes, namely in the fundamental class of 'non-geodesic' locally non-rotating reference frames (LNRFs), and two classes of 'geodesic' frames, the radial geodesic frames (RGFs), both falling and escaping, and the frames related to the circular geodesic orbits (CGFs). We compare the cones constructed in a given position for the LNRFs, RGFs, and CGFs. We have shown that the photons locally counter-rotating relative to LNRFs with positive impact parameter and negative covariant energy are confined to the ergosphere region. Finally, we demonstrate that the light escaping cones govern the shadows of black holes located in front of a radiating screen, as seen by the observers in the considered frames. For shadows related to distant static observers the LNRFs are relevant.    
	\end{abstract}

\section{Introduction}

Cosmological observations indicate that recent accelerated expansion of the Universe is governed by a dark energy that can be represented by a very small relict vacuum energy (repulsive cosmological constant $\Lambda > 0$), or by a quintessential field~\cite{Ada-etal:2013:ASTRA:DifLiYoClG,ArP-Muk-Ste:2000:PHYRL:,Bah-etal:1999:SCIEN:,Cal-Dav-Ste:1998:PHYRL:,Far:2000:PHYSR4:,Far-Jens:2006:CQG:,Ost-Ste:1995:NATURE:,Ade-etal:2014:ASTRA:,Rie-etal:2004:ASTRJ2:,Wan-etal:2000:ASTRJ2:}. The dark energy represents about $70\%$ of the energy content of the observable universe \cite{Cal-Kam:2009:NATURE:CosDarkMat, Spe-etal:2007:ApJSuppl:} and the dark energy equation of state is very close to those corresponding to the vacuum energy~\cite{Cal-Kam:2009:NATURE:CosDarkMat}. Therefore, it is worth to study the astrophysical effects of the recently observed relict cosmological constant implied by the cosmological tests to be $\Lambda \approx 1.3\times 10^{-56}\,\mathrm{cm^{-2}}$, with the related vacuum energy $\rho_{\mathrm{vac}} \sim 10^{-29}\,\mathrm{g/cm^{3}}$ that is close to the critical density of the Universe. We can consider even much larger values of $\Lambda$, which could be related to very early stages of the evolution of the Universe. 

The repulsive cosmological constant alters significantly the asymptotic spacetime structure of black holes, naked singularities and superspinars, or any gravitationally bounded configurations, as such backgrounds become asymptotically \dS{} spacetimes, rather than flat, and an event horizon (cosmological horizon) always exists, behind which the geometry is dynamic, if the vacuum region around such an object is extended enough \cite{Stu:1984:BULAI:}. 

Significant influence of the repulsive cosmological constant has been demonstrated for both Keplerian and toroidal fluid accretion disks related to active galactic nuclei and their central supermassive black holes~\cite{Kol-Stu:2010:PHYSR4:CurCarStrLoops,Kuc-Sla-Stu:2011:JCAP:ToroPerFlRNadS:,Pug-Stu:2015:ApJS:,Rez-Zan-Fon:2003:ASTRA:,Sla-Stu:2005:CLAQG:,Stu:2005:MODPLA:,Stu-Kol:2012:JCAP:StringLoops:,Stu-Kov:2008:INTJMD:PsNewtSdS:,Stu-Sla-Hle:2000:ASTRA:,Stu-etal:2005:PHYSR4:AschenUnexpTopo:}. The spherically symmetric black hole spacetimes with the $\Lambda$ term are described by the vacuum \SdS{} (SdS) geometry \cite{Kot:1918:ANNPH2:PhyBasEinsGr,Stu-Hle:1999:PHYSR4:}. Their relevance has been discussed in \cite{Arra:2014:PHYSR4:, Arraut:2013bqa, Arra:2017:Universe:,  Far:2016:PDU:,Far-Lap-Pra:2015:JCAP:,Stu:1984:BULAI:}. The internal, uniform density SdS spacetimes are studied in~\cite{Boh:2004:GENRG2:, Stu:2000:ACTPS2:} and polytropic configurations in \cite{Stu-Hle-Nov:2016:PHYSR4:,Stu-etal:2017:JCAP:}. The axially symmetric, rotating black holes are governed by the Kerr-de Sitter (KdS) geometry \cite{Car:1973:BlaHol:,Gib-Haw:1977:PHYSR4:}. Motion of photons in the SdS and KdS spacetimes was studied for special cases in~\cite{Bak-etal:2007:CEURJP:,Gre-Per-Lam:2014:PHYSR4:,Lak:2002:PHYSR4:BendLiCC,Mul:2008:GENRG2:FallSchBH,Sch-Zai:2008:0801.3776:CCTimeDelay,Ser:2008:PHYSR4:CCLens,Stu-Cal:1991:GENRG2:,Stu-Hle:2000:CLAQG:,Vil-etal:2013:ASTSS1:PhMoChgAdS:}, while motion of massive test particles was studied in~\cite{Che:2008:CHINPB:DkEnGeoMorSchw,Cru-Oli-Vil:2005:CLAQG:GeoSdSBH,Hac-etal:2010:PHYSR4:KerrBHCoStr:,Ior:2009:NEWASTR:CCDGPGrav,Kag-Kun-Lam:2006:PHYLB:SolarSdS,Kra:2005:DARK:CCPerPrec,Kra:2004:CLAQG:,Kra:2007:CLAQG:Periapsis:,Kra:2011:CLAQG:,Kra:2014:CLAQG:,Lak-Zan:2016:PHYSR4:,Oli-etal:2011:MODPLA:ChaParRNadS:,Cha-Har:2012:PHYSR4:BEConGRStar,Zhou-Chen:2011:AstrSpSc:,Stu:1983:BULAI:,Stu-Hle:1999:PHYSR4:,Stu-Hle:2002:ACTPS2:,Stu-Sch:2011:JCAP:CCMagOnCloud:,Stu-Sla:2004:PHYSR4:}. Note that the KdS geometry can be relevant also for the so called Kerr superspinars representing an alternative to black holes inspired by String theory~\cite{Gim-Hor:2009:PHYLB:AstVioSignStr,Gim-Hor:2004:hep-th0405019:GodHolo, 0264-9381-29-6-065002}, breaking the black hole bound on the dimensionless spin, and exhibiting extraordinary physical phenomena~\cite{Bla-Stu:2016:PHYSR4:,deFel:1974:ASTRA:,deFel:1978:NATURE:InstabNS, Hio-Mae:2009:PHYSR4:KerrSpinMeas,Stu:1980:BULAI:,Stu-Bla-Sch:2017:PHYSR4:,Stu-Sche:2010:CLAQG:, 0264-9381-29-6-065002, Stu-Sch:2013:CLAQG:UHEKerrGeo}. Moreover, the SdS and KdS spacetimes are equivalent to some solutions of the f(R) gravity representing black holes and naked singularities \cite{Per-Rom-PeB:2013:ASTRA:AccDiBHModGra:, Stu-Sla-Kov:2009:CLAQG:PseNewSdS}. 

Recently, we have studied in detail the general photon motion in the \KdS black hole and naked singularity spacetimes, giving the effective potentials of the latitudinal and radial photon motion and discussing the properties of the spherical photon orbits \cite{Char-Stu:2017:EPJC:}, generalizing thus the previous work concentrated on the properties of the equatorial photon motion \cite{Stu-Hle:2000:CLAQG:}. \footnote{Recall that the spherical photon orbits can be relevant also in the eikonal limit of the equations governing perturbative fields in the black hole backgrounds \cite{Car-etal:2009:PHYSR4:,Kon-Stu:2017:PLB:,Stu-etal:2017:JCAP:}.} Here we use the properties of the effective potentials of the general photon motion, and construct local escape cones of photons radiated by point sources located at rest in the most significant local reference frames, namely, the locally non-rotating frames, radial geodesic frames, and circular geodesic frames. We concentrate here on the three classes of the KdS black hole spacetimes according to classification introduced in \cite{Char-Stu:2017:EPJC:}, in the case of circular geodesic systems we restrict our attention to the marginally stable geodesics of both corotating and counter-rotating type. Note that the construction of the escape light cones is very important, being a crucial ingredient of calculations of wide range of the optical phenomena arising in vicinity of black holes, as appearance of Keplerian or toroidal disks and related profiled spectral lines, as viewed by distant observers; note however that also the self-occult effect could be significant is such situations, as shown in \cite{Bao-Stu:1992:ApJ:}. 

The light escape cones also directly govern the shadow (silhouette) of a black hole located between an observer and a radiating screen. Of special observational interest are the effects expected in vicinity of the supermassive black hole in the Galaxy centre object Sgr A* that were extensively discussed in a large series of papers - see, e. g., \cite{Doeleman:2008qh, PhysRevLett.116.031101, Zakh:2014:PRD:, Zakh:2015:JApA:, Zakh-etal:2005:A&A,  Zakh:2012:NewAR:, ZAKHAROV2005479}. The shadow can be defined for any orbiting or falling (escaping) observer, but from the astrophysical and observational point of view the most relevant are shadows related to distant static observers that could be represented by distant locally non-rotating observers \cite{Sche-Stu:2009:IJMPD:}. However, in the KdS black hole spacetimes the free static observers are expected near the so called static radius \cite{Stu:1983:BULAI:,Stu-Sla:2004:PHYSR4:} representing the outermost region of gravitationally bound systems in Universe with accelerated expansion \cite{Far:2016:PDU:,Stu:2005:MODPLA:,Stu-Hle-Nov:2016:PHYSR4:}, while behind the static radius the radially escaping geodesic observer is a more appropriate choice. 

	\section{Motion of test particles and photons in \KdS\ spacetimes}
	In the standard Boyer-Lindquist coordinates with geometrical units ($c = G = 1$) the line element in the \KdS\ geometry is given by
	\bea
	\dbe s^2 =
	&-&\frac{\Delta_r}{I^2 \rho^2}\left(\dbe t - a \sin^2\theta \din \phi\right)^2 + \frac{\Delta_\theta \sin^2\theta}{I^2\rho^2}\left[a \din t - \left(r^2 +   a^2\right)\din \phi \right]^2 \nonumber \\
	&+&\frac{\rho^2}{\Delta_r}\din r^2 + \frac{\rho^2}{\Delta_\theta}\din \theta^2. \label{line}
	\eea
	The source body has mass $M$ and angular momentum per unit mass $a,$ the cosmological constant $\Lambda$ is present in the parameter $y = \frac{1}{3}\Lambda  M^2$ in terms of the line elements. If we use dimensionless quantities by transition $s/M \rightarrow s$, $t/M \rightarrow t, r/M \rightarrow r, a/M \rightarrow a,$ then
	\bea
	\Delta_r&=&\left(1 - y r^2\right)\left(r^2 + a^2\right) - 2r,\\
	\Delta_\theta&=&1 + a^2 y \cos^2\theta,\\
	I&=&1 + a^2 y,\\
	\rho^2&=&r^2 + a^2\cos^2\theta. 
	\eea
	
	The event horizons of the \KdS\ geometry are given by
	\be
	y_{h}(r;\:a^2)\equiv \frac{r^2-2r+a^2}{r^2(r^2+a^2)}.\label{yh}
	\ee
	The geodetic motion of test particles and photons is described by the Carter equations \cite{Car:1973:BlaHol:}
		\bea
		\rho^2 \oder{r}{\lambda}&=&\pm \sqrt{R}, \label{CarterR}\\
		\rho^2 \oder{\theta}{\lambda}&=&\pm \sqrt{W}, \label{CarterW}\\
		\rho^2 \oder{\varphi}{\lambda}&=& \frac{aIP_{r}}{\Delta_r}-\frac{IP_{\theta}}{\Delta_\theta \sin^2\theta},\label{CarterPhi}\\
		\rho^2 \oder{t}{\lambda} &=&\frac{I(r^2+a^2)P_{r}}{\Delta_r}-\frac{aIP_{\theta}}{\Delta_\theta },\label{CarterT}
		\eea
		where
		\be
		R(r) = P_{r}^2 - \Delta_r(m^2r^2+\calk), \label{R} 
		\ee
		\be
		W(\theta) = (\calk-a^2m^2\cos^2\theta)\Delta_{\theta}-\left( \frac{P_{\theta}}{\sin\theta}\right) ^2, \label{W}
		\ee
		\be P_{r}=I\cale (r^2+a^2)-Ia\Phi,\ee
		\be P_{\theta}=Ia\cale \sin^2\theta-I\Phi. \ee
		
		Here $\frac{\din x^{\mu}}{\dbe \lambda}=p^{\mu}$ is the particle four-momentum, $m$ is its rest mass, energy 
		\be \cale=-p_{t}=-g_{t\mu}p^{\mu},\label{cale}
		\ee
		and axial angular momentum 
		\be 
		\Phi=p_{\phi}=g_{\phi \mu} p^{\mu} \label{phi}
		\ee
		are constants of motion following, respectively, from the time and axial symmetry of present geometry. The quantity
		$\calk$ is fourth Carter's constant, connected with a hidden symmetry of the \KdS\ spacetime. The affine parameter $\lambda$ is normalized such that 
		\be 
		p^\mu p_{\mu}=-m^2. \label{norm_p}
		\ee The four-momentum of photons ($m=0$) we shall denote $k^{\mu}$. 
		The discussion of general off-equatorial photon motion in the \KdS\ spacetimes has been performed in \cite{Char-Stu:2017:EPJC:} by method of so called `effective potentials`, where a new constants of motion $X,q$ were introduced by the relations
		\be X=\ell-a,\quad   \ell=\Phi/\cale \label{def_X,ell}\ee
		 and 
		\be q=\calk/I^2\cale^2-X^2.\label{def_q}\ee
		The motion constant $\ell$ can be considered as impact parameter of the photon, and the motion constant $q$ is constructed in such a way that for photons confined to the equatorial plane takes the value $q=0.$
		The functions in (\ref{CarterR}), (\ref{CarterW}) then read	
		 \bea
		 R(r)&\equiv& I^2\cale^2\left[\left(r^2 - aX\right)^2 -\Delta_r\left(X^2 + q\right)\right],\label{RXq} \\
		 W(\theta)&\equiv&I^2\cale^2\left[ (X^2+q)\Delta_\theta-\frac{(a\cos^2\theta+X)^2}{\sin^2\theta}\right] .\label{Thetatheta}
		 \eea
These functions determine effective potentials governing the motion of test particles and photons, as discussed in detail in \cite{Char-Stu:2017:EPJC:}. In the following, we use these effective potentials to determine the local light escape cones constructed for most relevant families of reference frames related to astrophysically significant sources (observers). As basic, locally non-rotating frames (LNRFs) are considered that give the simplest view of physical phenomena in KdS spacetimes, as they properly balance the effect of spacetime dragging \cite{Bar:1973:BlaHol:}. Then we consider the reference frames that could be considered as most important and representative from the astrophysical point of view: radial geodesic frames (RGFs), and circular geodesic frames (CGFs). Here we restrict our attention to the reference frames moving in the vicinity of the event horizons of the KdS spacetimes. The light escape cones constructed in such reference frames play crucial role role in treating the observationally relevant optical phenomena; e.g., in the case of the CGFs they are relevant in modelling optical effects related to the Keplerian accretion disks \cite{Stu-Sche:2010:CLAQG:}. Note that for distant nearly static observers the light escape cones represent the shape of the black hole shadow \cite{Bar:1973:BlaHol:,Sche-Stu:2009:IJMPD:,Stu-Sche:2010:CLAQG:}. 
	  
		\section{Light escape cones in locally non-rotating frames}
		The locally non-rotating reference frames (LNRF) are in general regarded as the most natural platform for description of optical effects and motion of test particles in axially symmetric spacetimes. The reason is that the observer in rest relatively to the LNRF sees the $+\phi$ and $-\phi$ directions as being equivalent \cite{Bar:1973:BlaHol:}. We therefore start our investigations from the perspective of the observers sending photons in various directions in the LNRF and compare the obtained results with that received in other reference frames.\par 
		\subsection{Tetrad of basis vectors}
		The locally non-rotating observer maintains constant radial coordinate (e. g. by means of rocket engines) and let himself to be dragged by the rotation of the spacetime in azimuthal direction with angular velocity 
		\be
		\Omega_{LNRF}=-\frac{g_{t\phi}}{g_{\phi \phi}}\label{OLNRF}
		\ee
		 relative to static distant observers. Using the metric coefficients from (\ref{line}) we obtain
		\be\Omega_{LNRF}=\frac{a[\Delta_{\theta}(r^2+a^2)-\Delta_r]}{A},  \ee
		where
		\be A=(r^2+a^2)^2\Delta_\theta-a^2\Delta_r\sin^2\theta. \label{A} \ee     
		The orthonormal tetrad of basis vectors of such observers reads
		\bea
		e_{(t)}&=&\sqrt{\frac{I^2A}{\Delta_r \Delta_{\theta} \rho^2}}\left( \frac{\partial}{\partial t}+\Omega_{LNRF} \frac{\partial}{\partial \phi}\right) ,\\
		e_{(r)}&=&\sqrt{\frac{\Delta_{r}}{\rho^2}}\frac{\partial}{\partial r},\\
		e_{(\theta)}&=&\sqrt{\frac{\Delta_{\theta}}{\rho^2}}\frac{\partial}{\partial \theta},\\
		e_{(\phi)}&=&\sqrt{\frac{I^2 \rho^2}{A \sin^2\theta}}\frac{\partial}{\partial \phi}, \label{e_i_LNRF}
		\eea
		and the corresponding dual 1-forms are
		\bea
		\omega^{(t)}&=&\sqrt{\frac{\Delta_{r} \Delta_{\theta} \rho^2}{I^2 A}} \din t, \\
		\omega^{(r)}&=&\sqrt{\frac{\rho^2}{\Delta_{r}}} \din r,\\
		\omega^{(\theta)}&=&\sqrt{\frac{\rho^2}{\Delta_{\theta}}} \din \theta,\\
		\omega^{(\phi)}&=&\sqrt{\frac{A \sin ^2 \theta}{I^2 \rho^2}}(\din \phi-\Omega_{LNRF} \din t).\label{o_i_LNRF} 
		\eea
		The coordinate components of the photon 4-momentum are given by equations (\ref{CarterR})-(\ref{CarterT}), together with (\ref{RXq}), (\ref{Thetatheta}), 		   
		their projections onto the tetrads determine the locally measured components
		\be k^{(a)}= \omega^{(a)}_{\mu}k^{\mu}, \quad k_{(b)}=k_{\nu} e_{(b)}^{\nu}. \ee
		The orthonormality of the tetrads ensures that these quantities are associated by relations 
		\be k^{(t)}=-k_{(t)},\quad k^{(r)}=k_{(r)},\quad k^{(\theta)}=k_{(\theta)},\quad k^{(\phi)}=k_{(\phi)}.\ee
		
		\subsection{Definition of directional angles}
		 We use the standard definition of directional angles $\alpha, \beta, \gamma$  of a photon in a local frame of reference, which are connected with the frame components of the photon 4-momentum by expressions (for details see, e.g., \cite{Stu-Sche:2010:CLAQG:})
		\bea
		k^{(r)}&=&k^{(t)}\cos \alpha,\label{k_r}\\
		k^{(\theta)}&=&k^{(t)}\sin \alpha \cos \beta,\label{k_theta}\\
		k^{(\phi)}&=&k^{(t)}\sin \alpha \sin \beta=k^{(t)}\cos \gamma. \label{k_phi}
		\eea
		Further, there is a correspondence between these locally measured components and the constants of motion due to the relations
		\bea
		-\cale&=&k_{t}=k_{(a)}\omega^{(a)}_{t}=	-k^{(t)}\omega^{(t)}_{t}+k^{(\phi)}\omega^{(\phi)}_{t}, \label{E} \\
		\Phi&=&k_{\phi}=k_{(b)}\omega^{(b)}_{\phi}=k^{(\phi)}\omega^{(\phi)}_{\phi} \label{azimom}
		\eea
		and
		\be
		q=\frac{1}{\Delta_{\theta}}\left[ \frac{\rho^4(k^{\theta})^2}{I^2\cale^2}+\frac{(a \cos^2\theta+X)^2}{\sin^2\theta}\right] -X^2, \label{4thCart}
		\ee
		where the connection with the frame components is provided by
		\be
		k^{\theta}=k^{(a)}e_{(a)}^{\theta}=k^{(\theta)}e_{(\theta)}^{\theta}. \label{kth}
		\ee
		
		\subsection{Effective potentials of the photon motion and their relation to directional angles}
\begin{figure}
			\begin{tabular}{cc}
				\includegraphics[width=5cm]{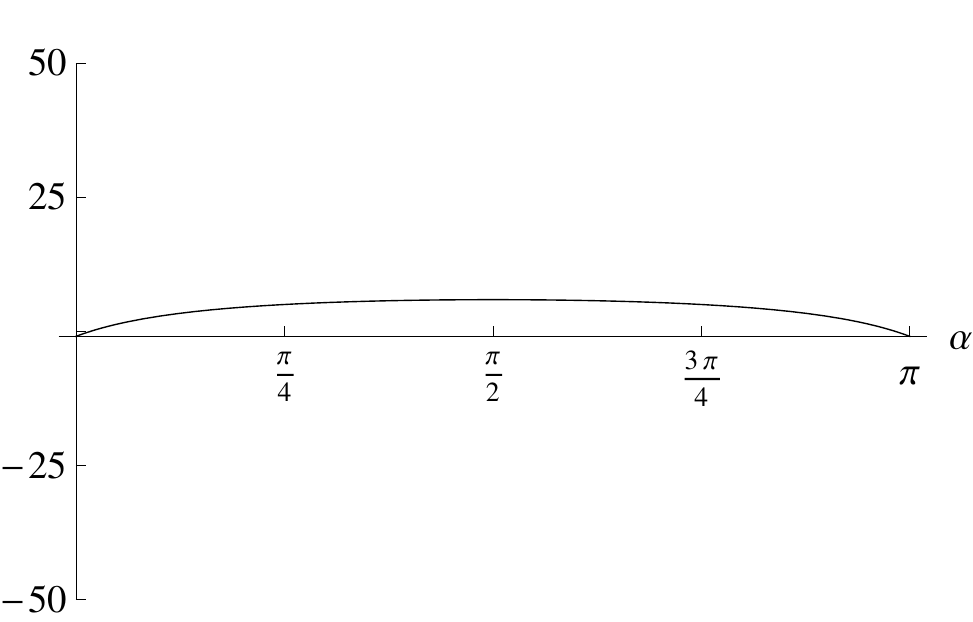}&\includegraphics[width=5cm]{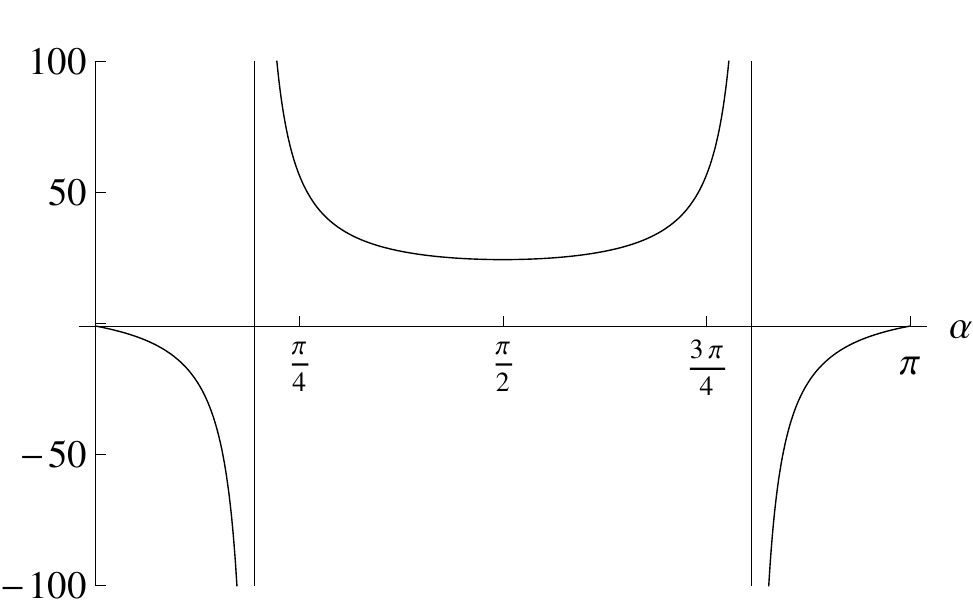}\\
				\includegraphics[width=5cm]{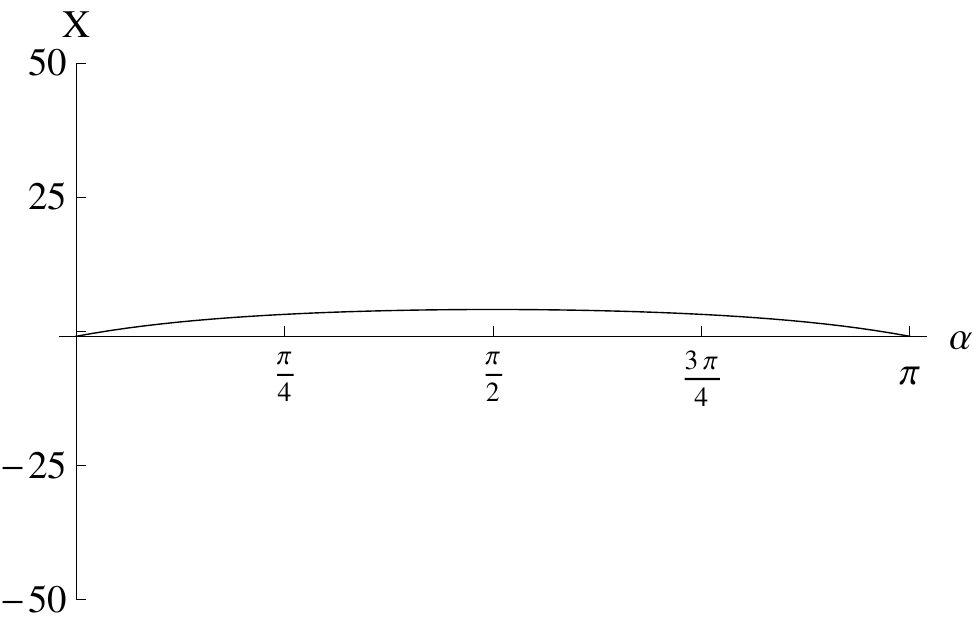}&\includegraphics[width=5cm]{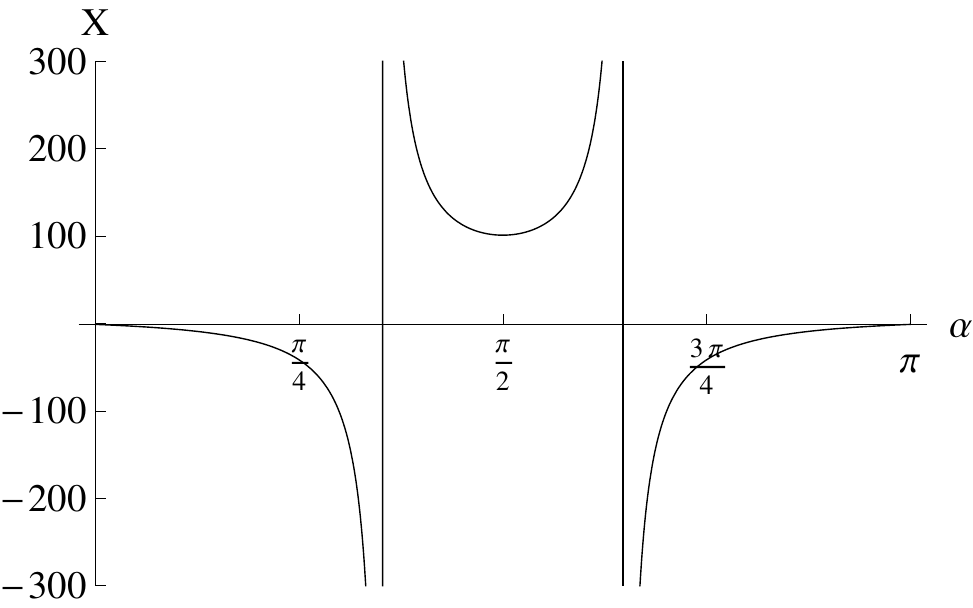}\\
				\includegraphics[width=5cm]{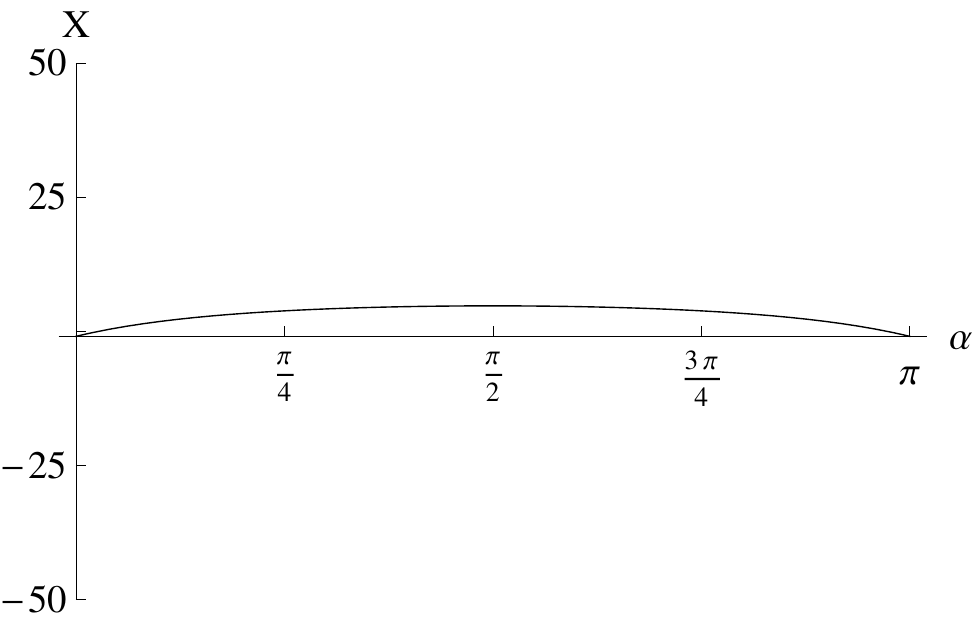}&\includegraphics[width=5cm]{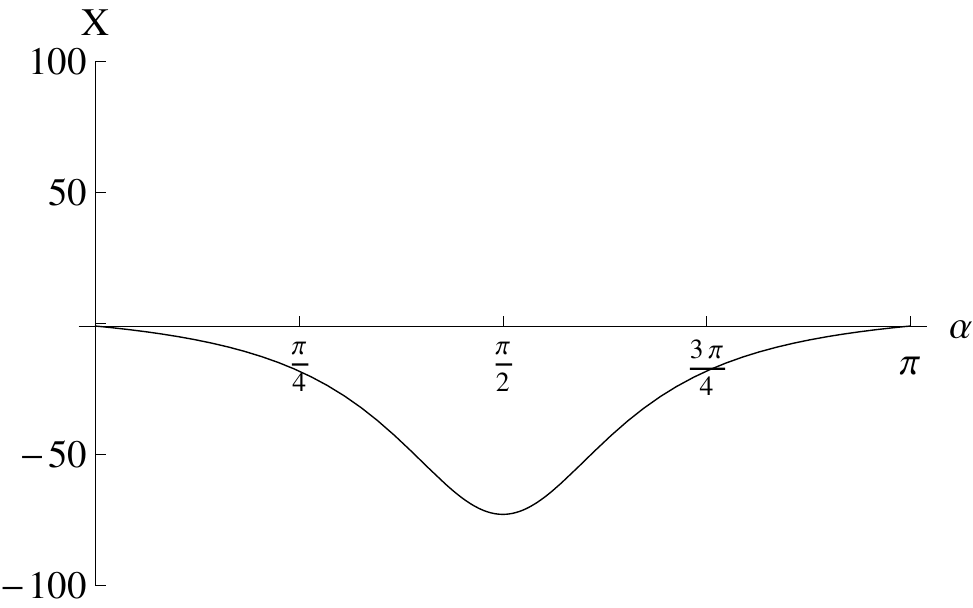}\\
			\end{tabular}	
			\caption{Graphs of dependence of parameter $X$ on directional angle $\alpha$ in the KdS spacetime with parameters $y=0.04, a^2=0.9$ are generated for $\beta=90\,^{\circ}$ (left column) and $\beta=270\,^{\circ}$ (right column) and successively for $\theta=90\,^{\circ}$ (upper row), $\theta=40\,^{\circ}$ (middle row) and $\beta=30\,^{\circ}$ (lower row). Radial position of the emitter is $r_{e}=3,$ corresponding marginal latitude reads $\theta_{m}=37.17\,^{\circ}$. Graphs in the same row are mutually complementary in the sense that they describe rotation of photon's three-momentum in one particular plane. The curves intersect the horizontal axis at $\alpha=0,$ $\alpha=\pi$ since it is lowered by $-a$ in order to demonstrate that $X(0)=X(\pi)=-a,$ i. e., the radially directed photons have $\ell=0.$   }\label{fig_X(alpha)LNRF}
		\end{figure}

\begin{figure}
			\begin{tabular}{cc}
				\includegraphics[width=5cm]{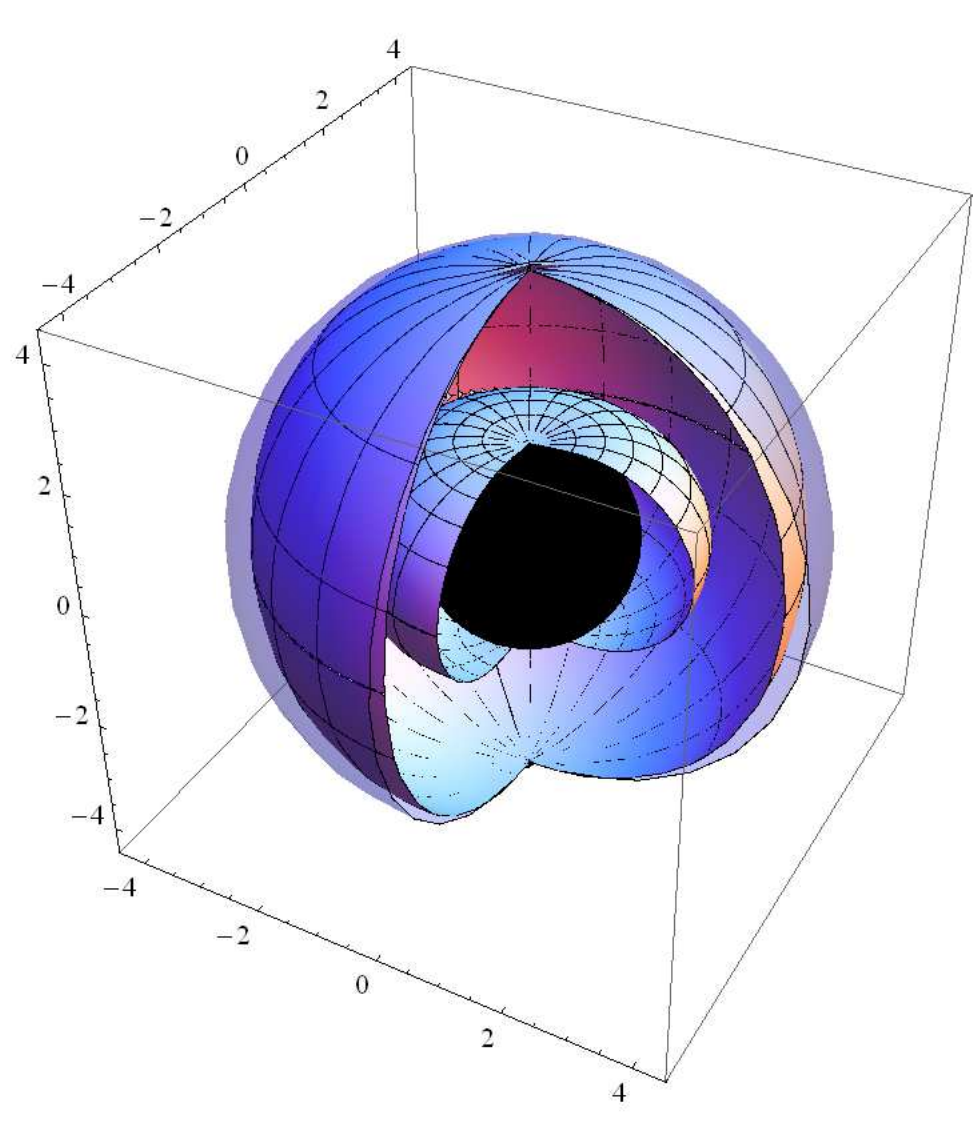}&\includegraphics[width=5cm]{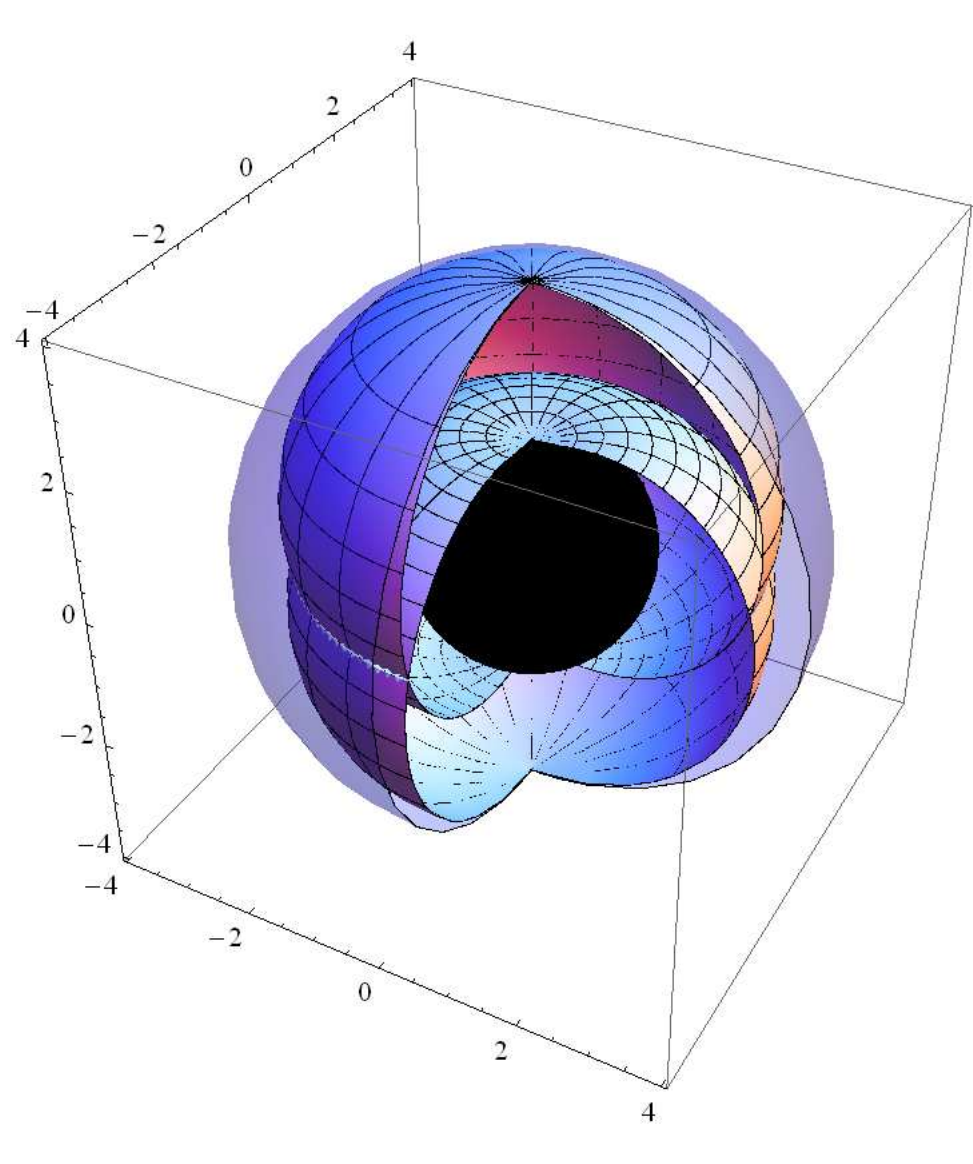}\\
				(a) $y=0.03, a^2=0.9$ & (b) $y=0.033762, a^2=0.9$\\
				\includegraphics[width=5cm]{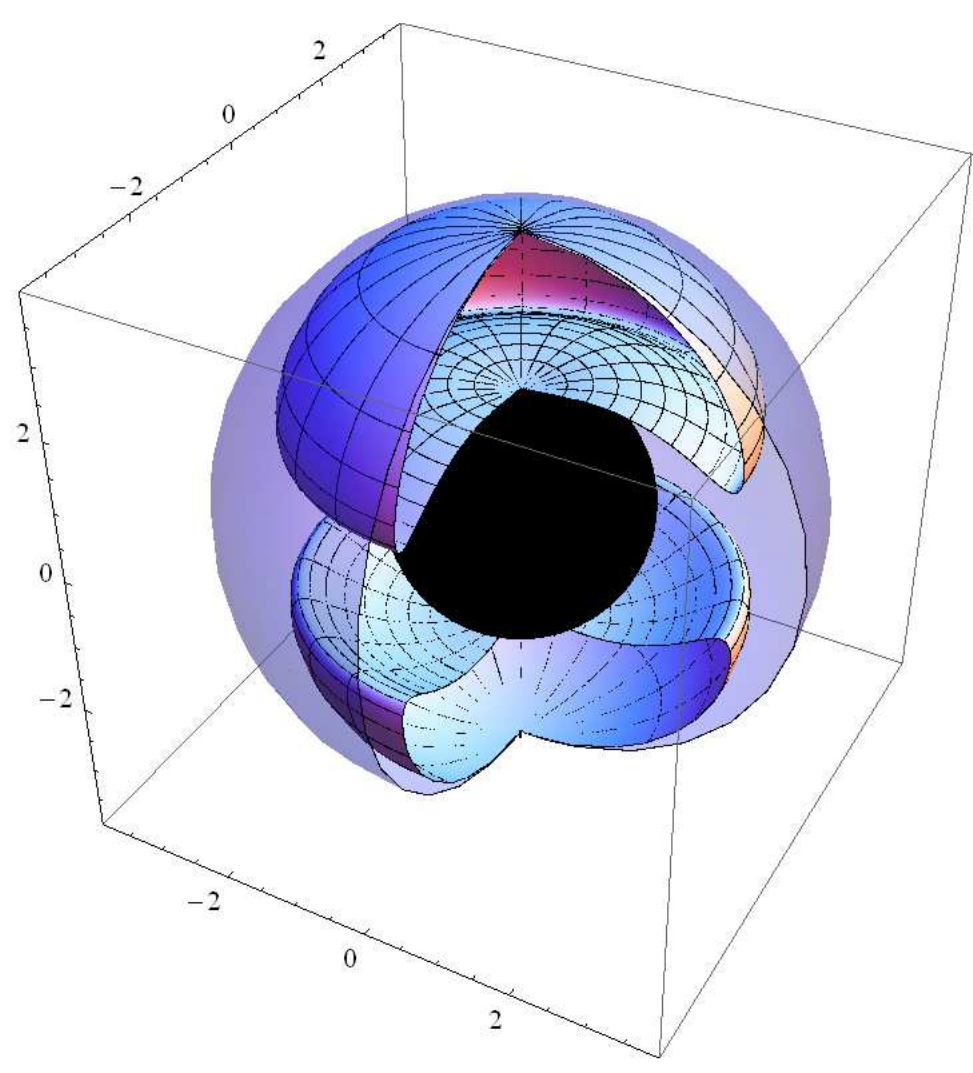}&\includegraphics[width=5cm]{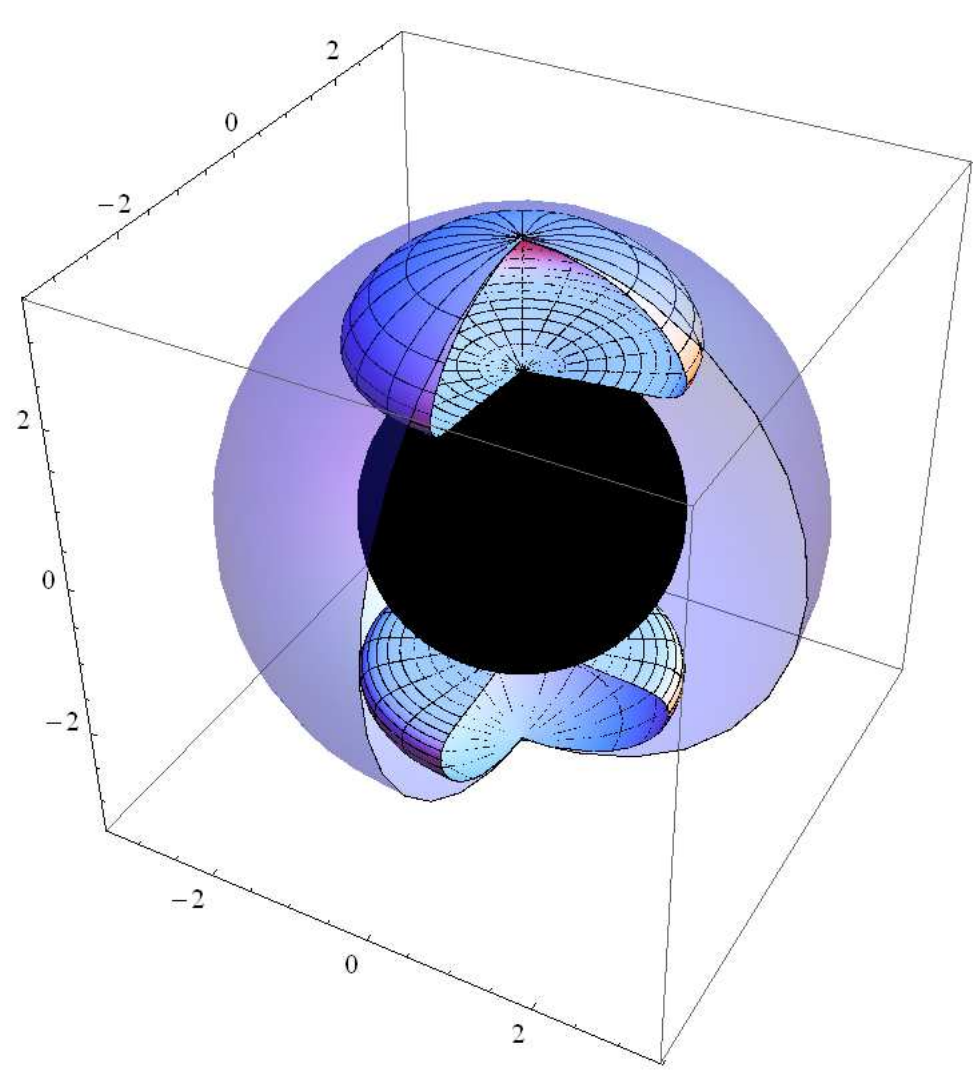}\\
				(c) $y=0.035, a^2=0.9$ & (d) $y=0.04, a^2=0.9$
			\end{tabular}
			\caption{Depiction of the ergosphere, which is found to be the region allowing the existence of retrograde photons, as seen in the LNRFs, with positive impact parameter $X,$ and therefore negative energy $E.$ In the spacetimes with DRB \textbf{(a)}, the ergosphere extends between the outer black hole horizon illustrated as black sphere and the static limit surface, depicted as the 'inner' surface, which touches this horizon at the poles and at the equatorial plane is the widest. Another part of the ergosphere, also widest in the equatorial plane, is located between the 'outer' static limit surface and the cosmological horizon, drown as the outer sphere, contacting each other at polar axis. Both inner and outer static limit surfaces mutually connect for marginal spacetime parameters $(a^2, y_{max(d)}(a^2))$ in equatorial plane \textbf{(b)}, and in the spacetimes with RRB they form two separated surfaces in each hemisphere \textbf{(c,d)}. }\label{fig_ergos}
		\end{figure}

		Through the equations (\ref{k_r})-(\ref{kth}) we can therefore relate the locally measured directional angles of the photon with its constants of motion and vice versa. \\
		The dependence of the directional angle $\alpha$, which is the rate of deviation from the radial direction on these constants, can be obtained from the expression (\ref{k_r}) by substitution
		\bea
		\cos \alpha&=&\frac{k^{(r)}}{k^{(t)}}=-\frac{k^{(r)}}{k_{(t)}}=-\frac{\omega^{(r)}_{r}k^{r}}{e_{(t)}^{t}k_{t}+e_{(t)}^{\phi}k_{\phi}}\nonumber \\ 
		&=&\pm\frac{\sqrt{\Delta_{\theta}}\sqrt{(r^2-aX)^2-\Delta_{r}(X^2+q)}}{\sqrt{A} [1-\Omega_{LNRF}(X+a)]}, \label{cosalphaLNRF}
		\eea
		where the plus (minus)sign has to be taken for the outgoing (ingoing) photon. The turning points of the radial motion are governed by the effective potentials
		\be
		X_{\pm}(r;q,\:y,\:a)\equiv \frac{ar^2\pm\sqrt{\Delta_r\left[r^4+q(a^2-\Delta_r)\right]}}{a^2-\Delta_r},\label{Xpm(r)} 
		\ee
		their properties were in detail studied in \cite{Char-Stu:2017:EPJC:}. Clearly, for $X=X_{\pm}(r;\:q,\:y,\:a)$, there is $\alpha=\pi/2.$ \\
		The inverse relation expressing at given radial coordinate $r$ the parameter $X$ in dependence on the directional angle $\alpha$ of an observer can be derived combining definitions (\ref{def_X,ell}) and equalities (\ref{E}), (\ref{azimom}), which yield 
		\be
		X=\frac{A\sin\alpha \sin\beta \sin\theta (1-a\Omega)-a\sqrt{\Delta_{r}\Delta_{\theta}}\rho^2}{\sqrt{\Delta_{r}\Delta_{\theta}}\rho^2+A\Omega_{LNRF} \sin\alpha \sin\beta \sin\theta}, \label{X(alpha)LNRF}
		\ee
		or
		\be
		\ell=\frac{A\sin\alpha \sin\beta \sin\theta}{\sqrt{\Delta_{r}\Delta_{\theta}}\rho^2+A\Omega_{LNRF} \sin\alpha \sin\beta \sin\theta} \label{ell(alpha)LNRF}
		\ee 
		alternatively. We see that the value $X=-a$ (respectively $\ell=0$) implies $\alpha=0, \pi,$ which corresponds to pure radial motion, or, in more general case, $\beta=0, \pi,$, corresponding to photons crossing the polar axis, eventually released from the polar axis ($\theta=0, \pi$). \\
		Further, in order to construct the light escape cones, we have to express the directional angle $\beta$ in dependence on the conserved quantities $X, q.$ Using the relation (\ref{k_theta}), and appropriate identities of this section, one can write 
		\bea
		\sin \alpha \cos \beta&=&-\frac{k^{(\theta)}}{k_{(t)}}=-\frac{\omega^{(\theta)}_{\theta}k^{\theta}}{e_{(t)}^{t}k_{t}+e_{(t)}^{\phi}k_{\phi}}\nonumber \\ 
		&=&\pm \sqrt{\frac{\Delta_{r}}{A}}\frac{\sqrt{(X^2+q)\Delta_{\theta}\sin^2\theta-(a\cos^2\theta+X)^2}}{ \sin\theta \left[ 1-\Omega_{LNRF}(X+a)\right]} \label{sincosbeta},\\
		\sin\alpha\sin\beta&=&-\frac{k_{(\phi)}}{k_{(t)}}=\frac{e_{(\phi)}^{\phi}k_{\phi}}{e_{(t)}^{t}k_{t}+e_{(t)}^{\phi}k_{\phi}}\nonumber \\
		&=&\frac{\sqrt{\Delta_{r}\Delta_{\theta}}\rho^2(X+a)}{A \sin\theta\left[ 1-\Omega_{LNRF}(X+a)\right] }. \label{sinsinbeta}
		\eea
		The equation (\ref{sinsinbeta}), which is equivalent to (\ref{X(alpha)LNRF}), represents the rate of motion in the azimuthal direction in such a way that its positive sign corresponds to prograde photons, while the negative sign to retrograde ones. Identical expression defines the photon azimuthal motion component of spherical orbits in \cite{Char-Stu:2017:EPJC:} (see the definition for $\sin \Psi$), where it is shown  that in the ergosphere, there exist photons with positive impact parameter $X\ge X_{+}>0$ appearing to be retrograde in LNRFs.  This is seen in the behaviour of the function $X(\alpha)$ in (\ref{X(alpha)LNRF}), when the appropriate photons have directional angle $\alpha$ in interval $\alpha \in \langle -\alpha_{d},\alpha_{d}\rangle.$ Here $\pm \alpha_{d}$ are points of divergence of $X(\alpha),$ symmetrically placed with respect to $\alpha = 3\pi/2.$ Such situation occurs also in the case of Kerr--Newman--~de Sitter spacetimes (see \cite{Stu-Hle:2000:CLAQG:}). The angle $\alpha_{d}$ occurs for latitudinal coordinate $\theta \in \langle\theta_{m}; \pi - \theta_{m} \rangle,$ where $\theta_{m}$ is the marginal latitudinal coordinate defining the boundary of the ergosphere at given radial coordinate $r_{e}$ following from the condition $g_{tt}=0$. The marginal latitude reads 
		\be
		\cos^2 \theta_{m}=\frac{a^2y-1+\sqrt{1-2[a^2+2r_{e}(r_{e}-2)]y+(2r_{e}^2+a^2)^2y^2}}{2a^2y}. \label{cos2_theta}
		\ee 
		The dependence of the function $X(\alpha)$ on the latitudinal coordinate of the source is demonstrated in Fig. \ref{fig_X(alpha)LNRF} . Because we deal with the spatial cones, the angle $\alpha$ runs from $0$ to $\pi.$ It is then appropriate to present couples of graphs constructed for $\beta$ differing by $\pi,$ which cover complete revolution of photon's three-momentum $\vec{k}$ in the particular plane. The absolute value of $X(\alpha)$ at given $\alpha$ increases with increasing absolute value of $\sin\beta,$ and the divergent points $\alpha_{d},$ where the function $X(\alpha)$ diverges, first occurs, if varying $\beta,$ for $\beta=3\pi/2$. This is also the angle for which the relation (\ref{cos2_theta}) is valid.\\
		In \cite{Char-Stu:2017:EPJC:} we have justified that the locally retrograde photons with positive impact parameter $X$ must have negative energy $E,$ hence, it is worth to devote some attention to the properties of the ergosphere, which is illustrated in Fig. \ref{fig_ergos}. In the spacetimes with divergent repulsive barrier (DRB) of the radial motion (see \cite{Stu-Ba_Ost:1998:KNdSrest.rep.bar.epm} for details), the ergosphere consists  of two surfaces (see Fig.\ref{fig_ergos}a) - one being just above to the outer black hole horizon, the latter being located in the vicinity of the cosmological horizon. It is most extended in the equatorial plane, where the inner surface reaches the radii $r_{d1}$, while the outer spreads from $r_{d2}$ to the cosmological horizon. Here $r_{d1,2}$ denote the divergent points of the effective potential $X_{+}.$  These radii coalesce for the marginal spacetime parameters $(a^2, y_{max(d)}(a^2))$ (see the definition in \cite{Char-Stu:2017:EPJC:}), for which the barrier just becomes restricted, into the value $r_{d(ex)},$ where $r_{d(ex)}$ denotes the divergent point of the function $X_{(ex)}(r;\:y,\:a),$ and both surfaces touch each other in the equatorial plane (Fig.\ref{fig_ergos}b). In the spacetimes with restricted repulsive barrier (RRB) of the radial photon motion, the ergosphere fills the whole region between the outer and cosmological horizon along the equatorial plane, while the region of possible static observers shrinks to the polar axis ( Figs.\ref{fig_ergos}c,d). We could intuitively expect such extension of the ergosphere as consequence of large value of the spin parameter, which strengthens the dragging of the spacetime by the rotation of the gravitating body, or large value of the cosmological parameter, which compensates the gravitational attraction at larger distances, while the dragging survives.\\
		
		\subsection{Construction of the light escape cones}  	
\begin{figure}[htbp]
			\centering
			\begin{tabular}{cc}
				\includegraphics[width=5.5cm]{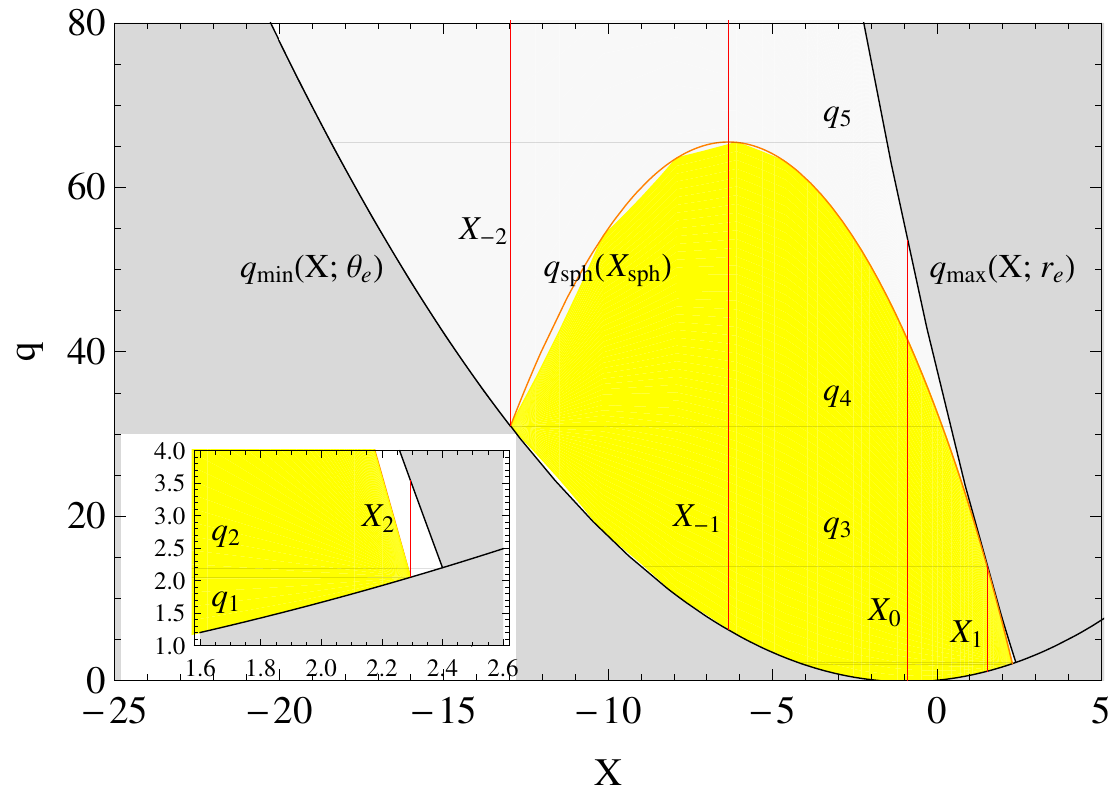}&\includegraphics[width=5.5cm]{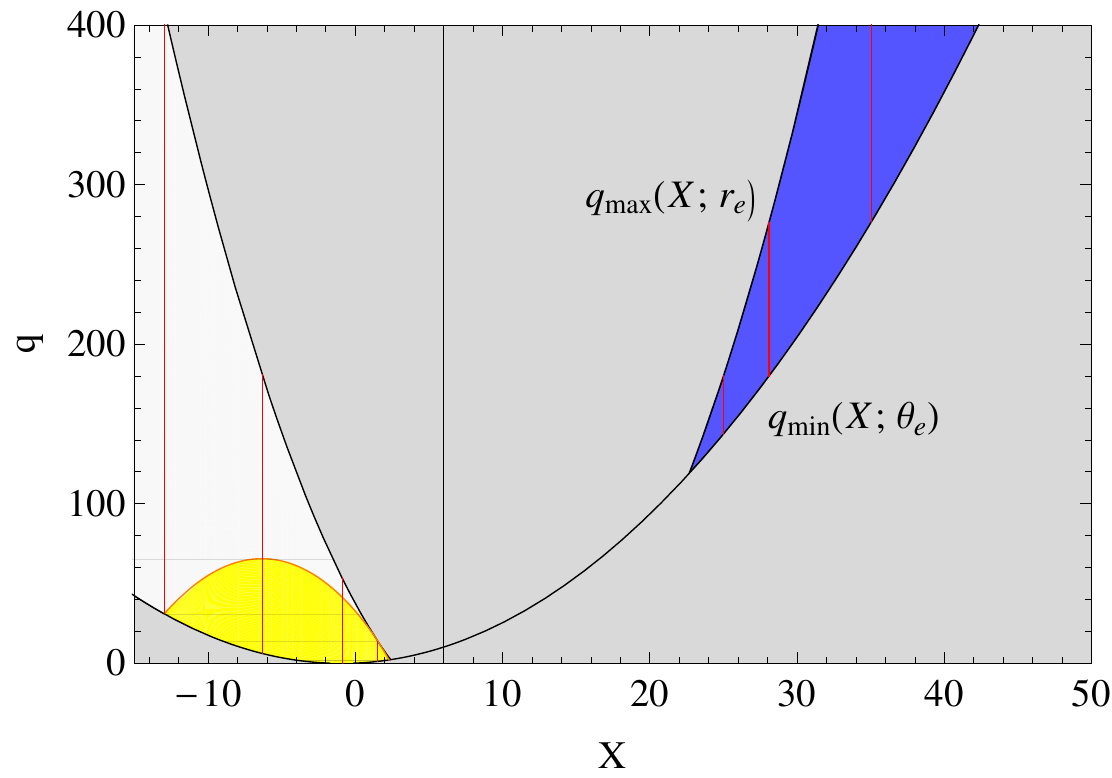}\\
				(a)&(b)

			\end{tabular}
			\caption{The motion constant plane $X-q$ of photons emitted at position $r_e=2$ and $\theta_{e}=65\,^{\circ}$ in the KdS spacetime with DRB, represented by the parameters $y=0.02,$ $a^2=0.9$, is shown in detail \textbf{(a)}, and for larger range of both parameters $X,q$  \textbf{(b)}. Photons with the motion constants from the yellow region have no turning point of the radial motion. If initially outwards directed, they escape to cosmological horizon, otherwise they plunge into the black hole. The curve denoted by $q_{sph}(X_{sph})$ is given parametrically by functions $q_{sph}(r),$ $X_{sph}(r),$ where the parameter $r$ takes values between $r_1=1.77$ and $r_2=3.36,$ where $r_1, r_2$ are limits of the spherical orbits at given latitudinal coordinate. Its touch point with the curve $q_{max}(X;\:r_e)$ is at $X_{sph}(r_e)=X_1,$ $q_{sph}(r_e)=q_3.$ The small white area on the right side of this touch point depicted in detail of the Fig. \textbf{(a)} pertains to photons which escape even if they are initially inwards directed, since they have radial turning point somewhere at $r<r_e,$ while the light grey region describes photons with no turning point if initially inwards directed, or with one turning point at $r>r_e$ if outwards directed, hence, in both cases they are captured by the black hole. The dark grey areas under the curves $q_{min}(X;\:\theta_{e})$ and above $q_{max}(X;\:r_e)$ represent regions forbidden by conditions imposed on reality of the latitudinal and the radial motion, respectively. To demonstrate in the following picture how these parameters transform into directional angles, some special $X=const.$ and $q=const.$ lines were shown. Here $X_0=-a$ ($\ell=0$), meaning of the other values arises from the picture. The considered coordinates $r_e=2,$ $\theta_{e}=65\,^{\circ}$ of the source satisfy the ergosphere condition $g_{tt}>0.$ In such a case, as shown in Fig. \textbf{(b)}, the function $q_{max}(X;\:r_e)$ is convex and the parabola $q_{min}(X;\:\theta_{e})$ is unfolded such that there is another allowed region shown in blue with high impact parameter $X,$ which however correspond to locally retrograde photons with negative energy $E$, all being captured by the black hole. The above inequality means that such situation occurs in the ergosphere. The sequence of lines of constant $X$, $q$ - value starts with $X=25.$   }	 	\label{fig_crit_loc} 	
		\end{figure}

\begin{figure}
			\centering
			\begin{tabular}{cc}
				\includegraphics[width=5.5cm]{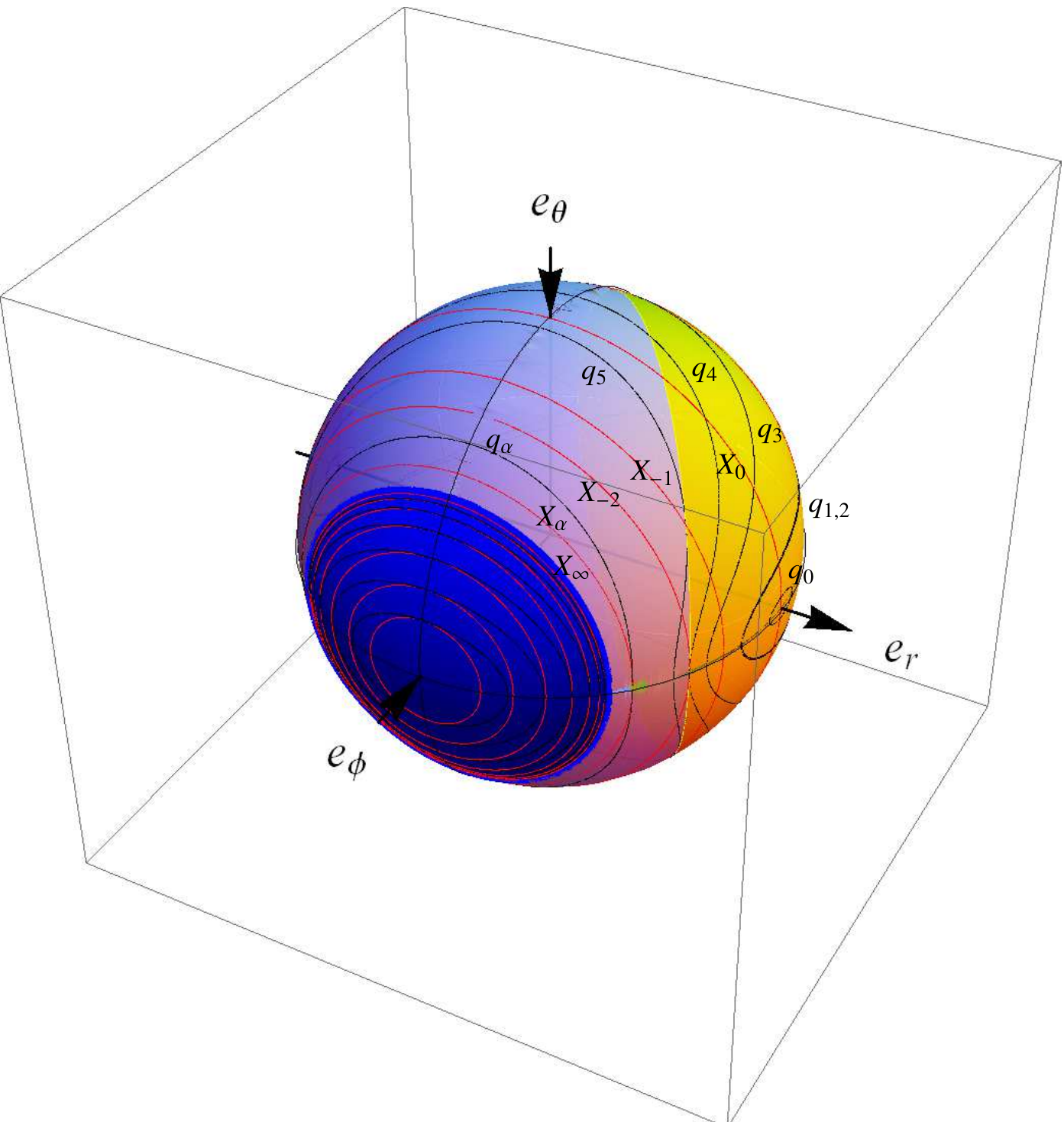}&\includegraphics[width=5.5cm]{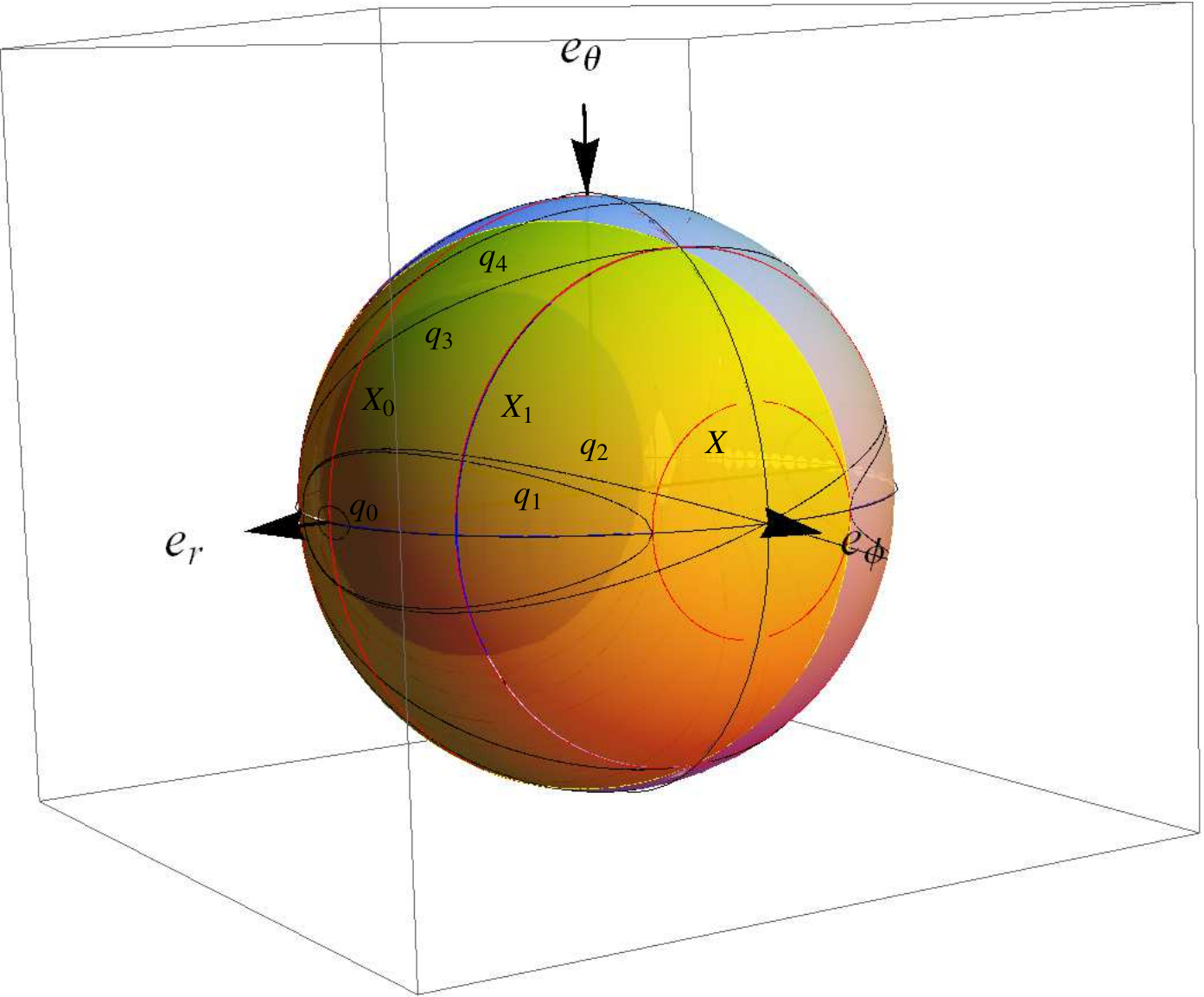}
			\end{tabular}
				\includegraphics[width=4.5cm]{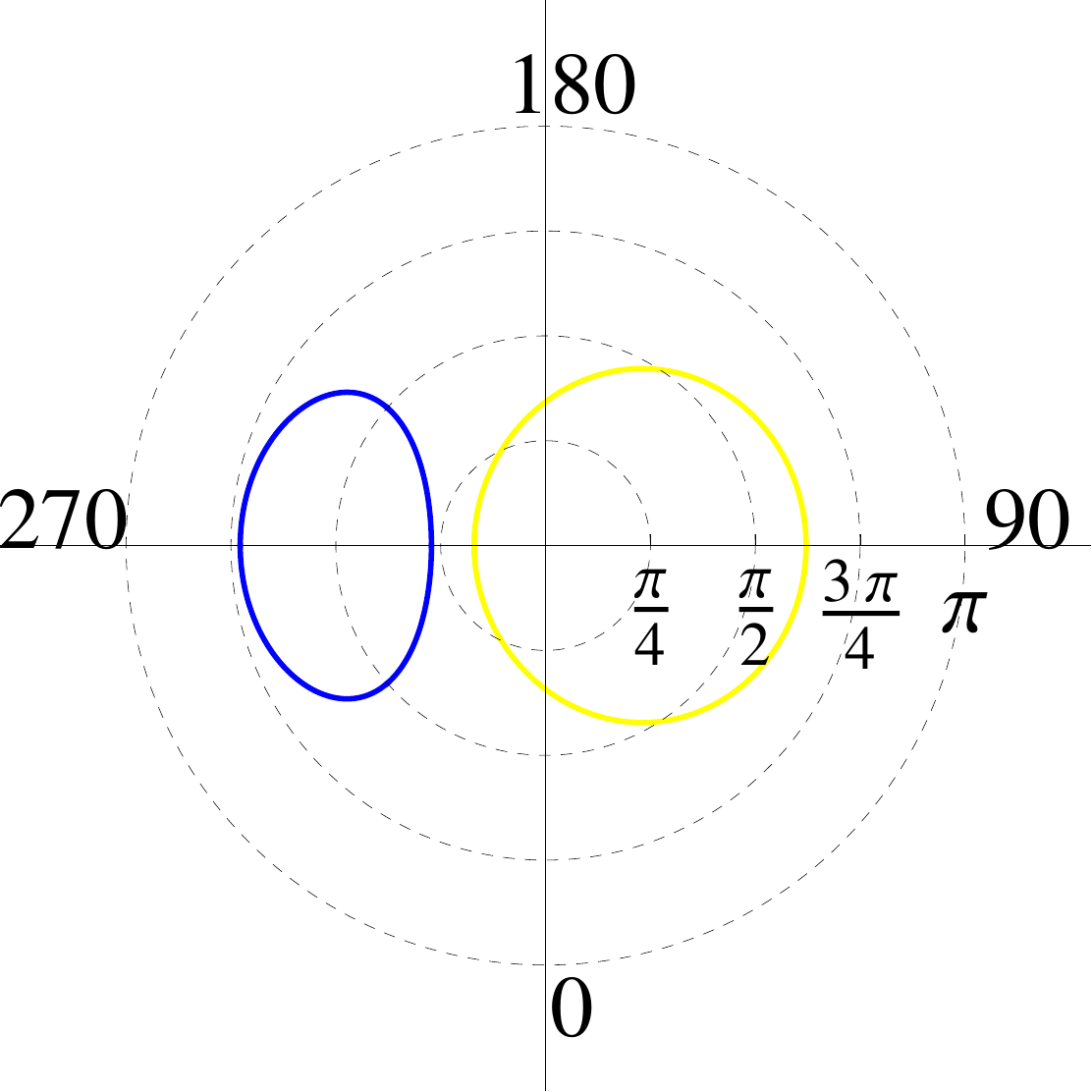}
			\caption{Light escape cone in 3D (upper row) as a result of transformation of the motion constant plane $(X-q)$ into directional angles, represented by the two-sphere consisting of end-points of corresponding photon's three-momentum vector of particular magnitude located in its centre. As indicated by the orientation of the basis vectors, the left illustration is the view towards the black hole, and partly in the direction of motion of the LNRF, the right one is the view from roughly the opposite side. Such two different angles of view we shall use also in the next illustrations, if not stated otherwise. The colouring of the surface corresponds to that in Fig. \ref{fig_crit_loc}: the yellow part represents a surface through which the light flux escape to the cosmological horizon, the rest of the sphere corresponds to flux captured by the black hole; photons passing through the boundary between the two areas have motion constants $X_{sph},$ $q_{sph};$ they enter some spherical orbit $r_1\leq r_{sph} \leq r_2,$ especially, photons with constants denoted by $X_1,$ $q_3$ maintain the radius $r_{sph}=r_e.$ The blue area demarcates the cone of locally retrograde photons with positive impact parameter $X$ and negative energy $E$. Analogous correspondence between areas of the same colour we use in further figures. The extra lines of constant $X,$ $q$ - value, which have not its counterpart in Fig. \ref{fig_crit_loc}, are $q_0=0,$ $q_\alpha=10^3=q_{min}(X_\alpha;\:\theta_{e})$ and $X_\infty \to \infty.$ The figure in the lower row is a representation of the same cone in 2D ($\alpha - \beta$) plane that we shall use in the following sections of this article. For convenience we use different notation of angles: $\alpha \in (0,360)$, $\beta \in (0, \pi).$} \label{fig_crit_loc1}
		\end{figure}

\begin{figure}[htbp]
			\centering
			\begin{tabular}{c}
				\includegraphics[width=5.5cm]{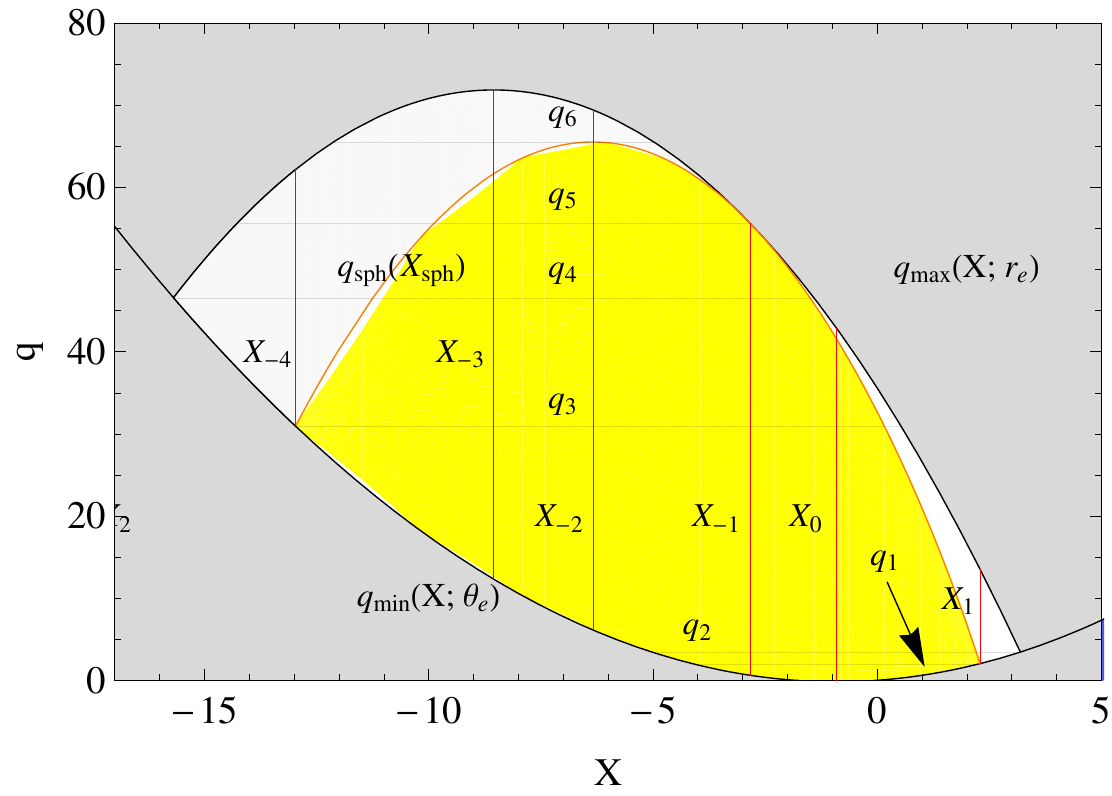}
							
			\end{tabular}

			\begin{tabular}{cc}
				\includegraphics[width=5.5cm]{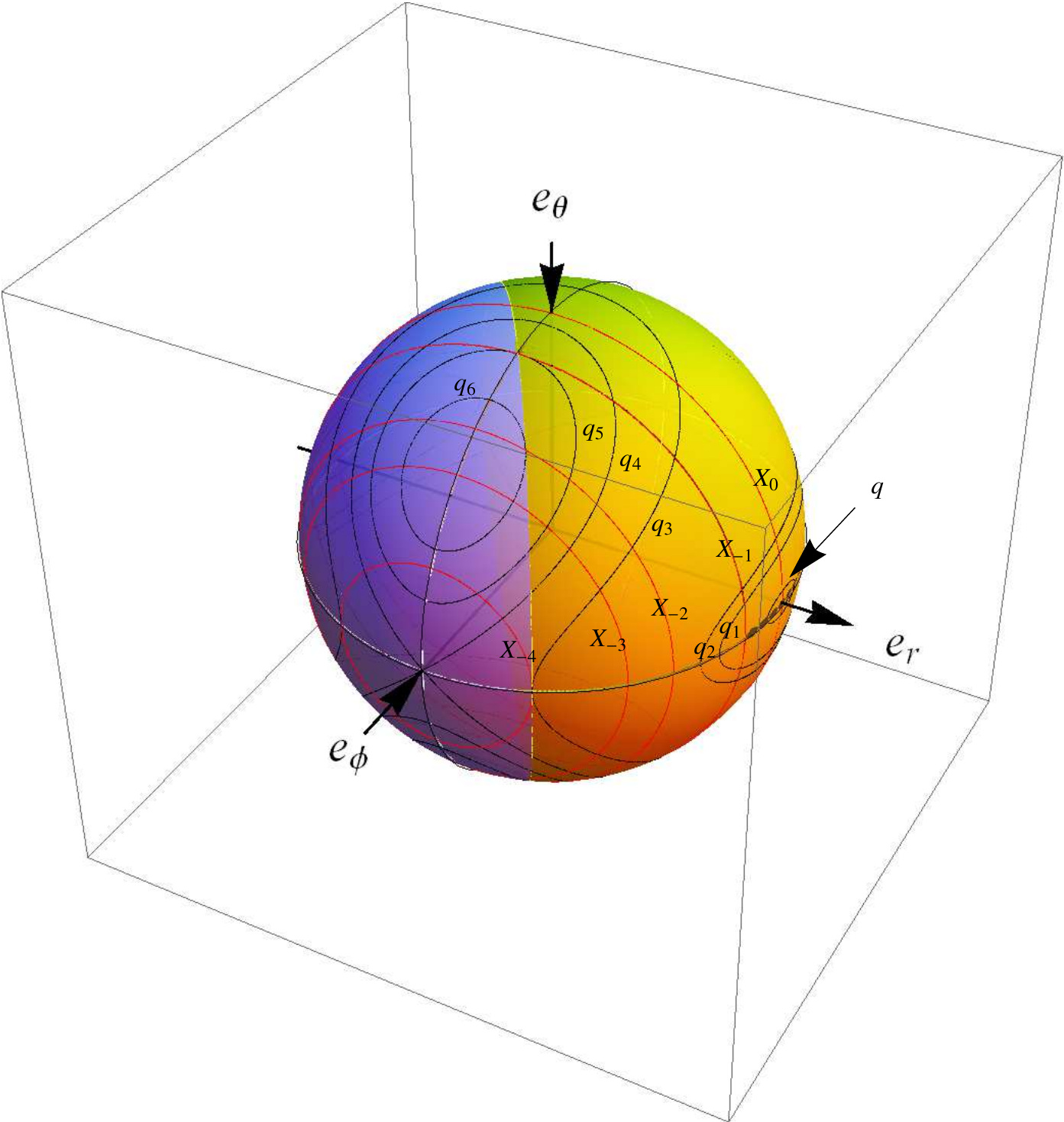}&\includegraphics[width=5.5cm]{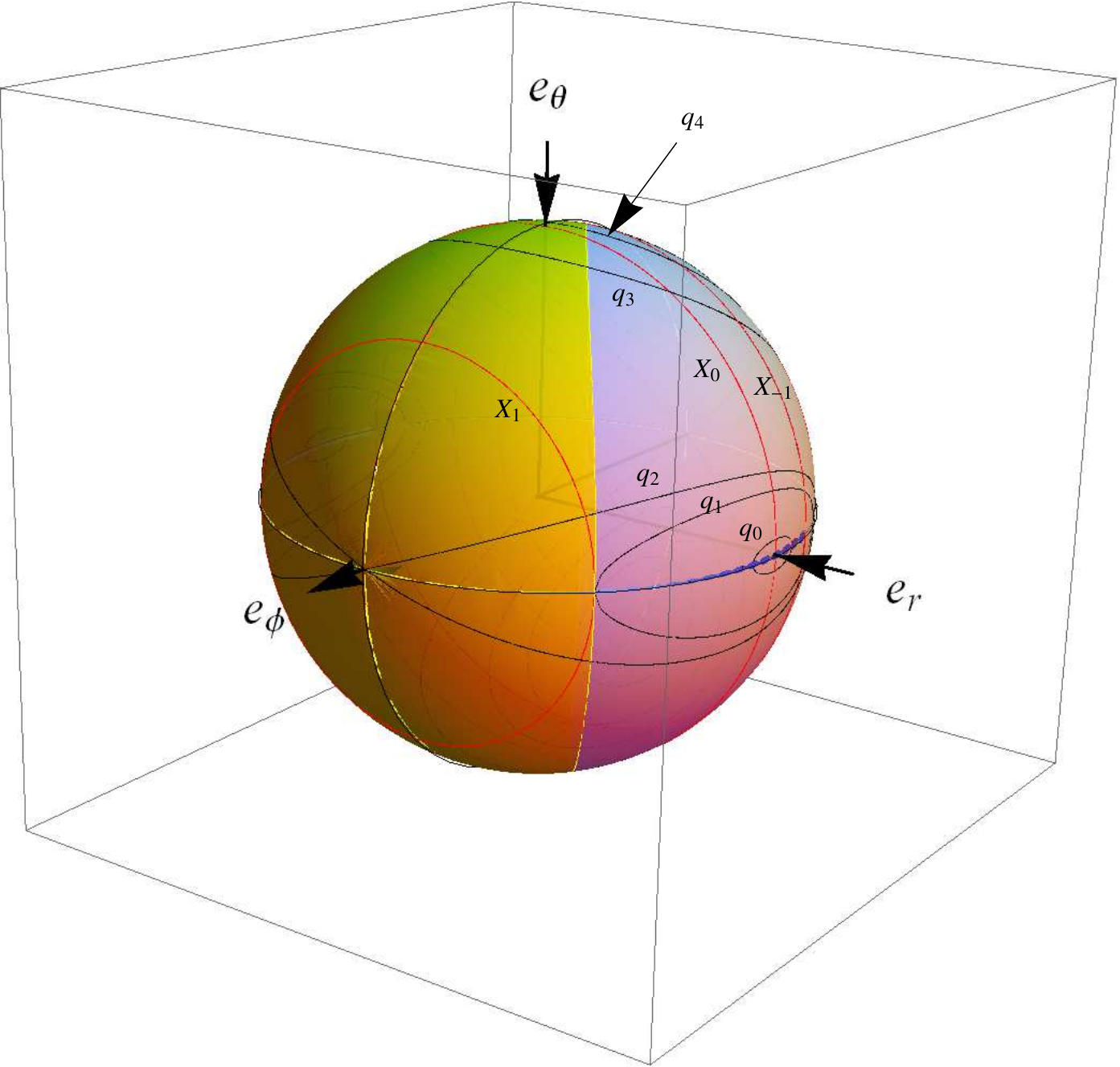}
			\end{tabular}

			\caption{The motion constant plane $X-q$ and associated light escape cone for spacetime with $a^2=0.9,$ $y=0.02$ again, but now the source is located outside the ergosphere at $r_e=2.75$ and $\theta_{e}=65\,^{\circ}.$ The function $q_{max}(X;\:r_e)$ is now concave, hence the range of allowed impact parameter $X$ is bound, and there are no locally retrograde photons with positive impact parameter $X$ and negative energy $E.$     }	 	\label{fig_crit_loc2} 
		\end{figure}

\begin{figure}[htbp]
			\centering
			\begin{tabular}{c}
				\includegraphics[width=5.5cm]{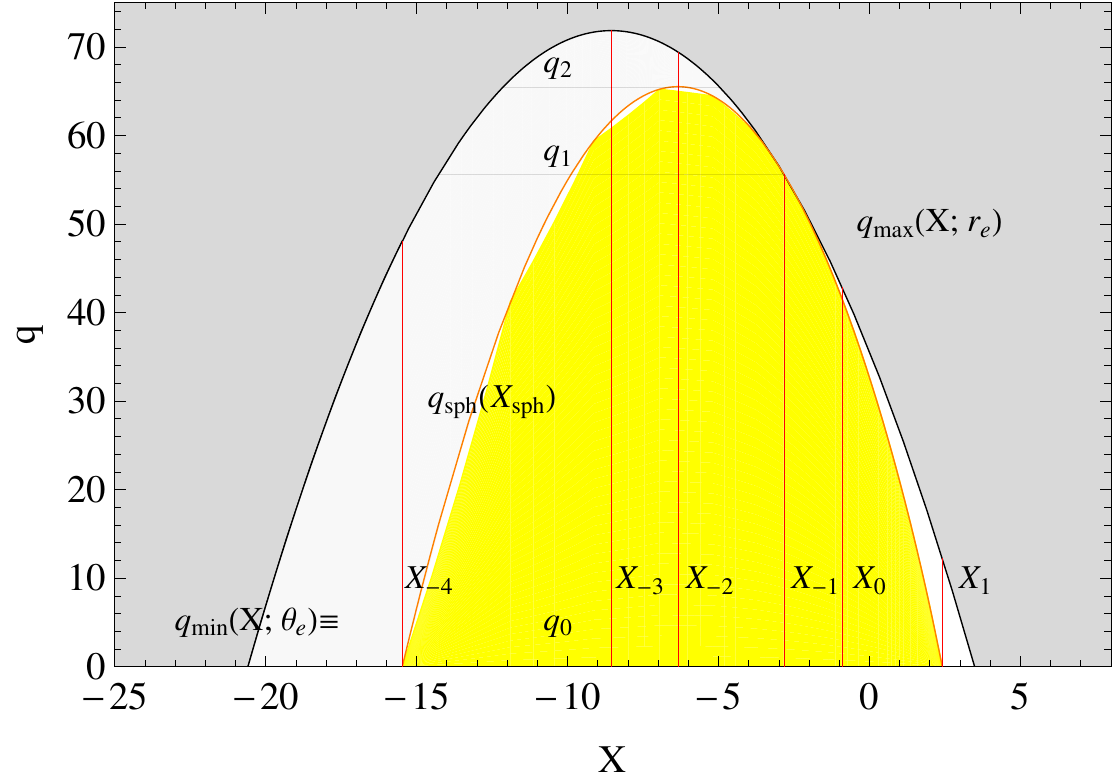}
				
			\end{tabular}

			\begin{tabular}{cc}
				\includegraphics[width=5.5cm]{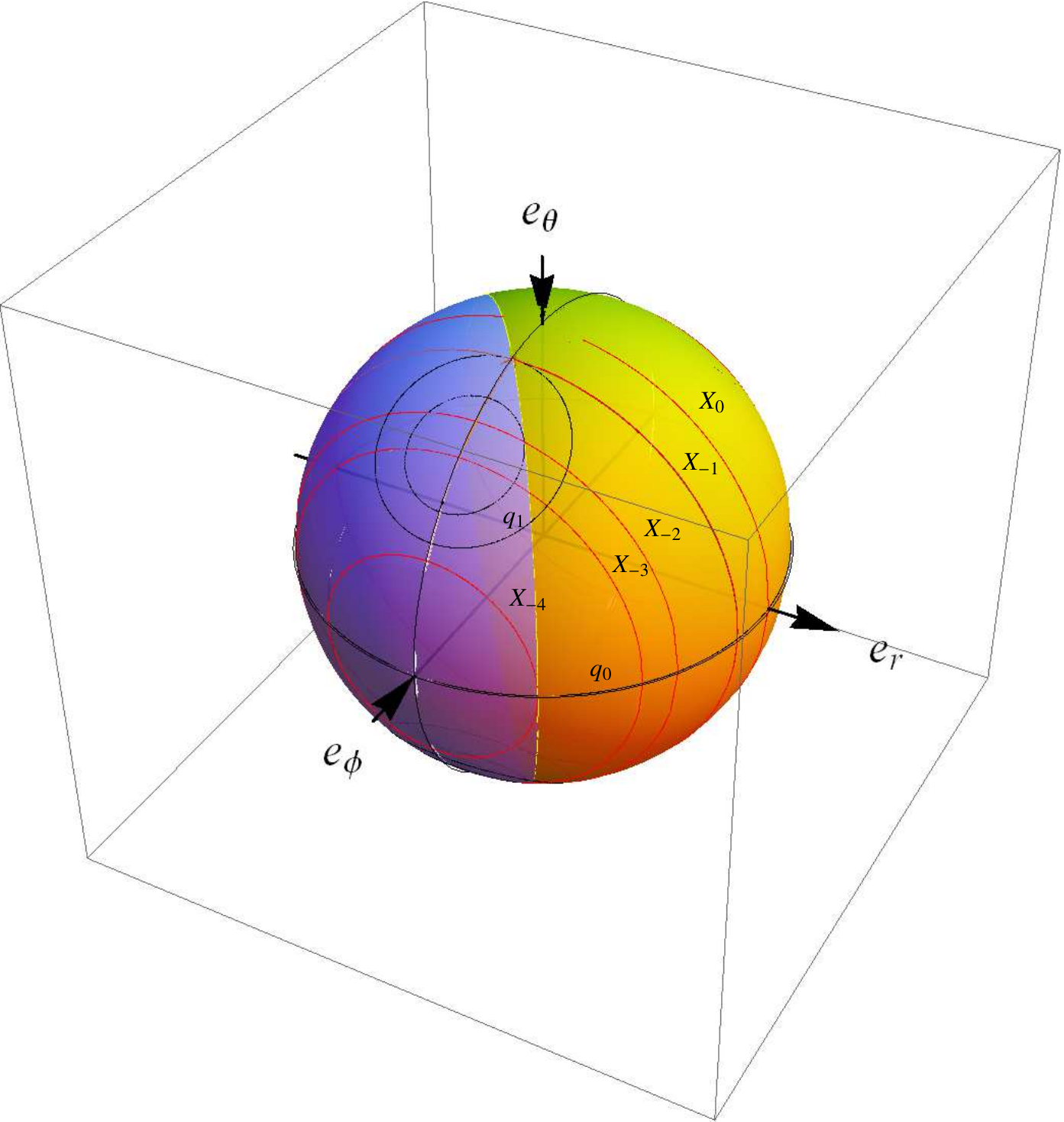}&\includegraphics[width=4.5cm]{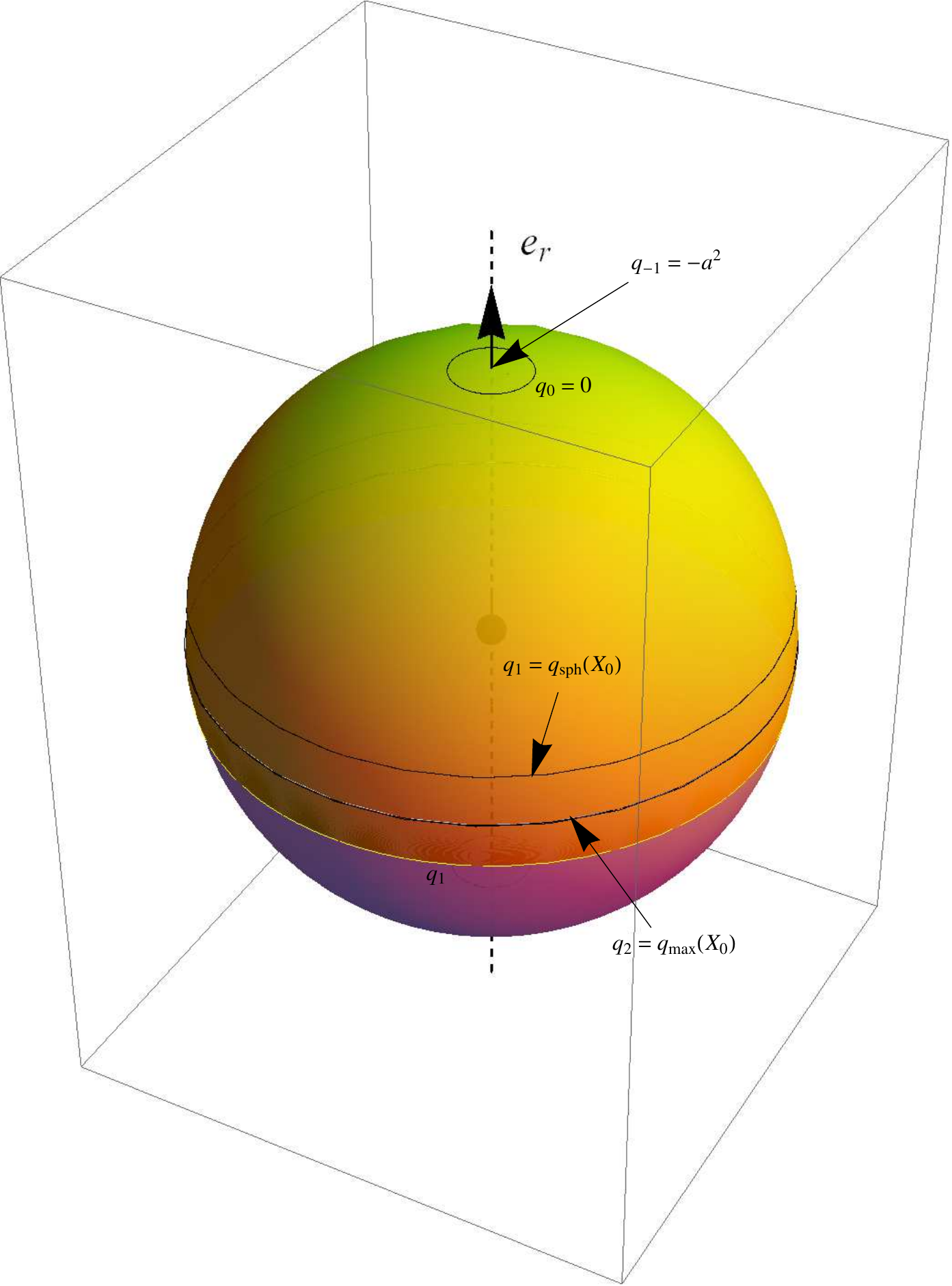}\\

			\end{tabular}
			\caption{Comparison of the the motion constant plane $X-q$ and associated light escape cones differing from the previous case by the location of the source in the equatorial plane $\theta_e=\pi/2$ (left column) and on the black hole spin axis $\theta_e=0$ (right column). Of course, in the latter case, only the value $X=-a$ ($\ell=0$) is allowed. The radial coordinate and the spacetime parameters are the same as in the previous case.     }	 	\label{fig_crit_loceq} 
		\end{figure}

\begin{figure}[htbp]
			\centering
			
			\includegraphics[width=5.5cm]{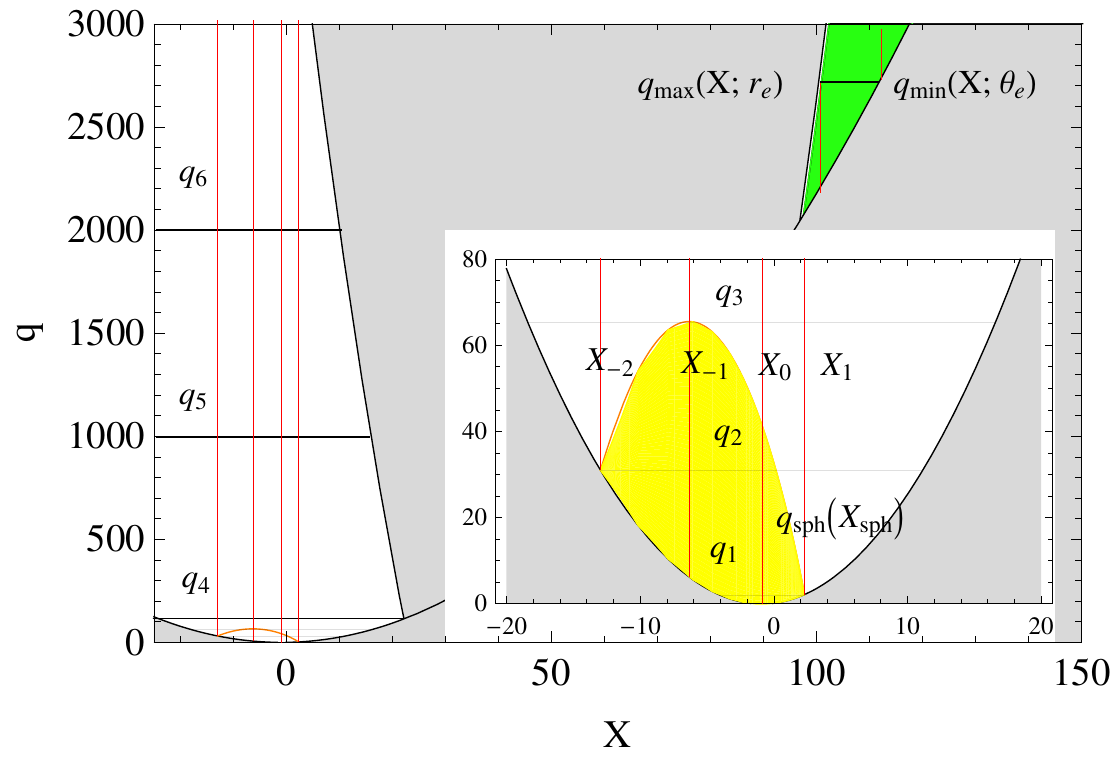}
						
			\begin{tabular}{cc}
				\includegraphics[width=5.5cm]{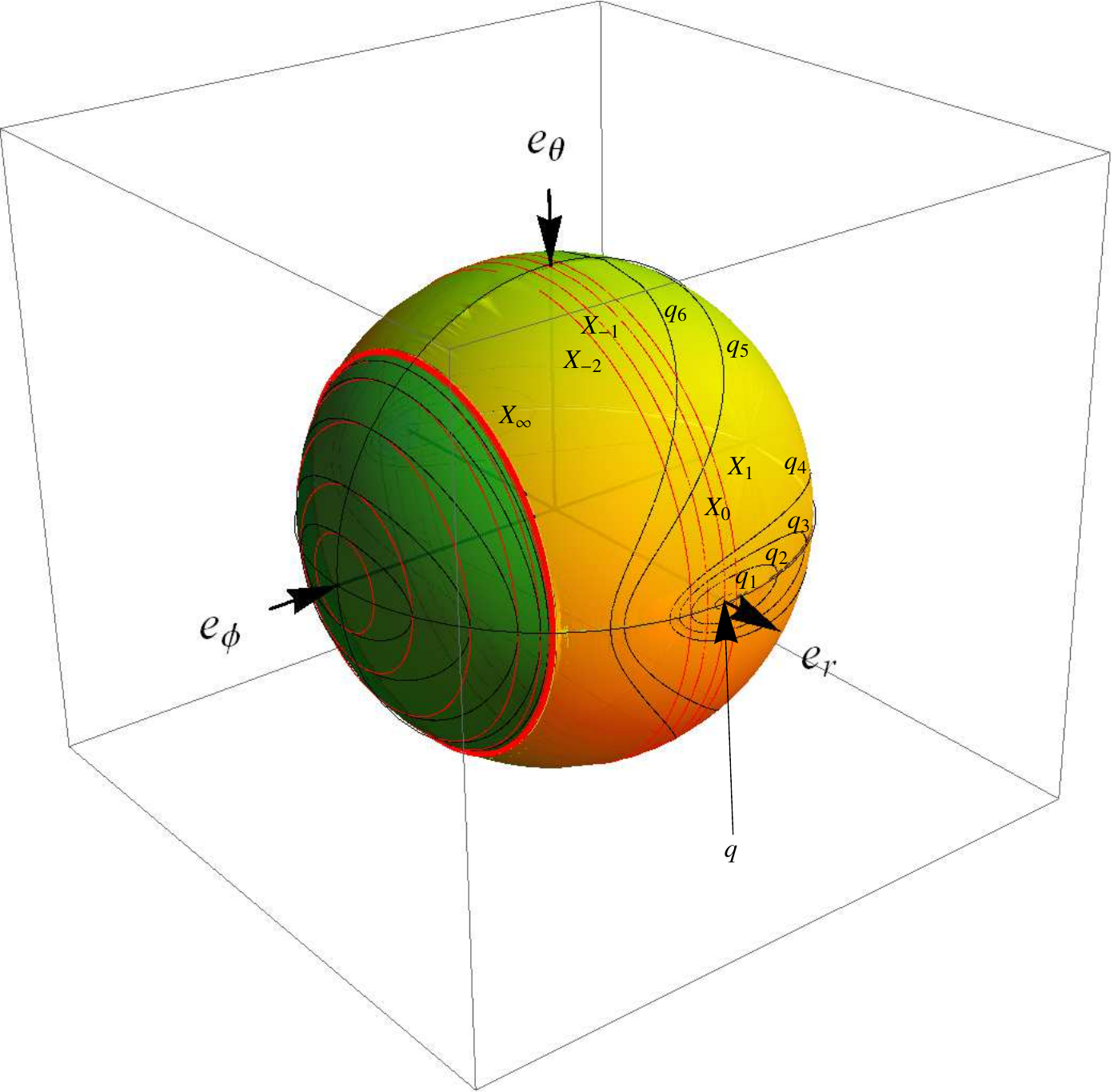}&\includegraphics[width=5.5cm]{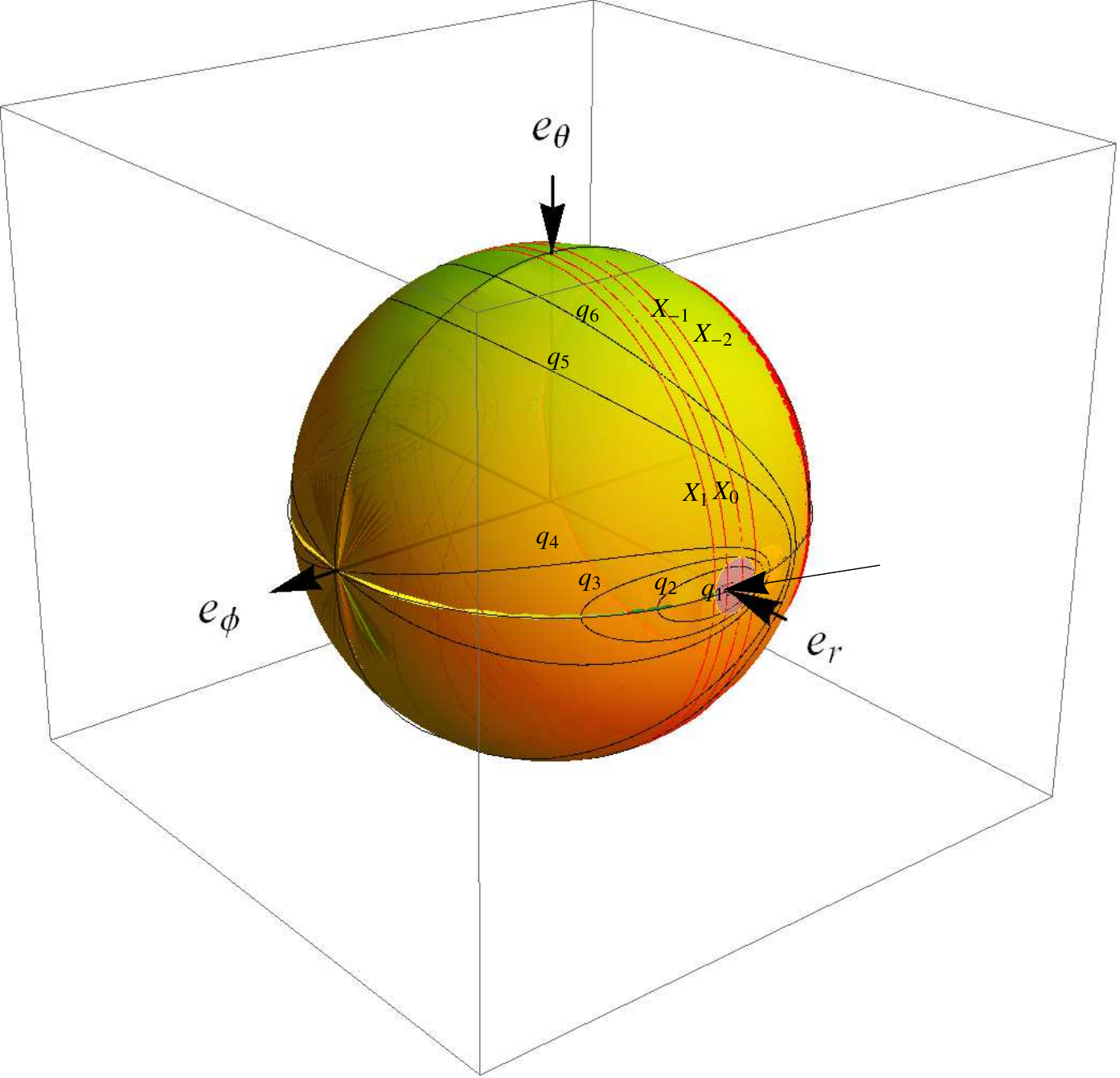}
			\end{tabular}

			\caption{The same as in the above with the source located inside the 'outer' ergosphere in the vicinity of the cosmological horizon at $r_e=5.7$ and $\theta_{e}=65\,^{\circ}.$ The function $q_{max}(X;\:r_e)$ is convex again, but the appropriate retrograde photons with high positive impact parameter $X$ and negative energy $E$ are escaping  beyond the cosmological horizon. They are distinguished by the green colour.   }	 	\label{fig_crit_loc4} 	
		\end{figure}

\begin{figure}[htbp]
				\centering
				\begin{tabular}{cc}
					\includegraphics[width=5.5cm]{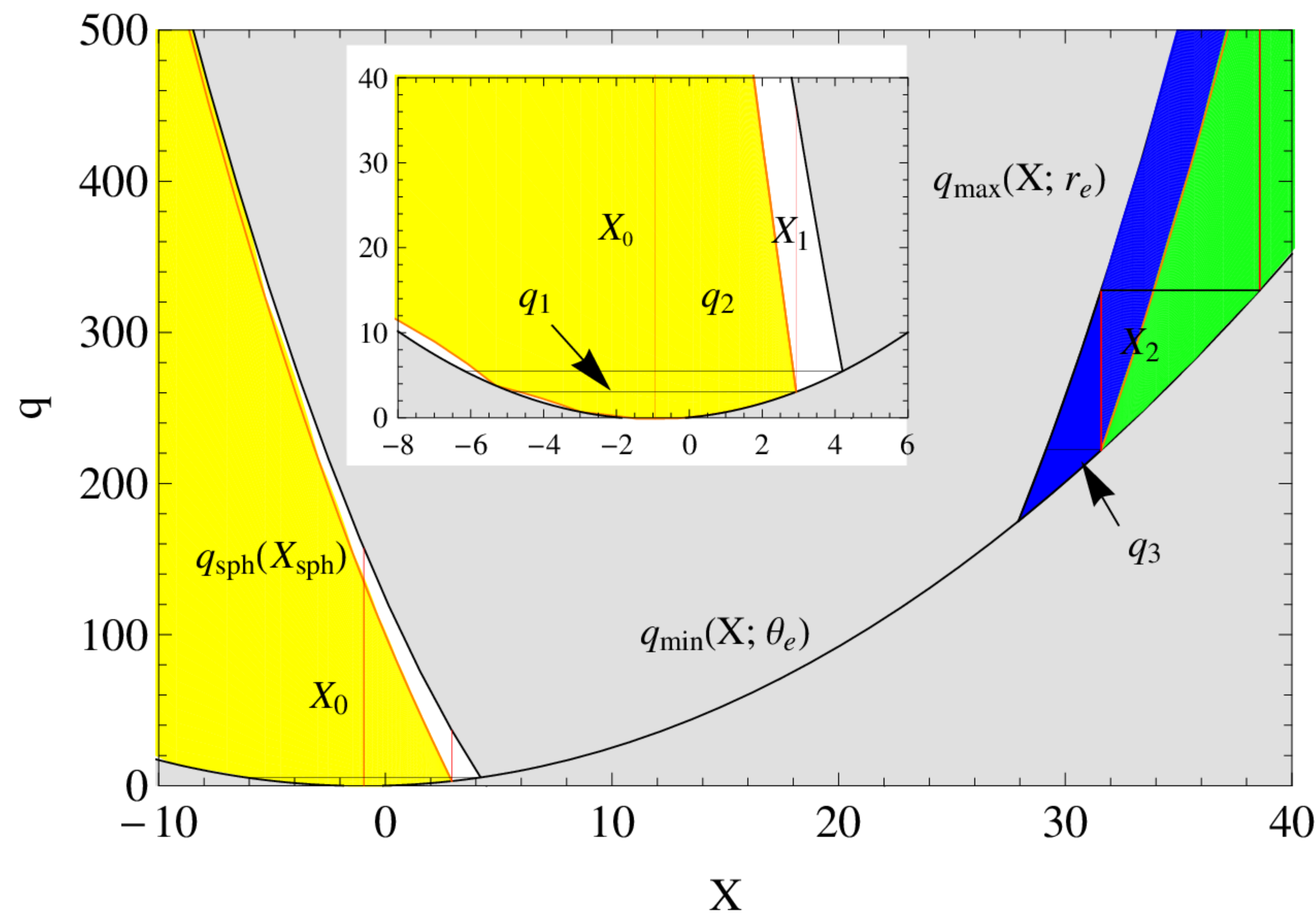}&\includegraphics[width=5.5cm]{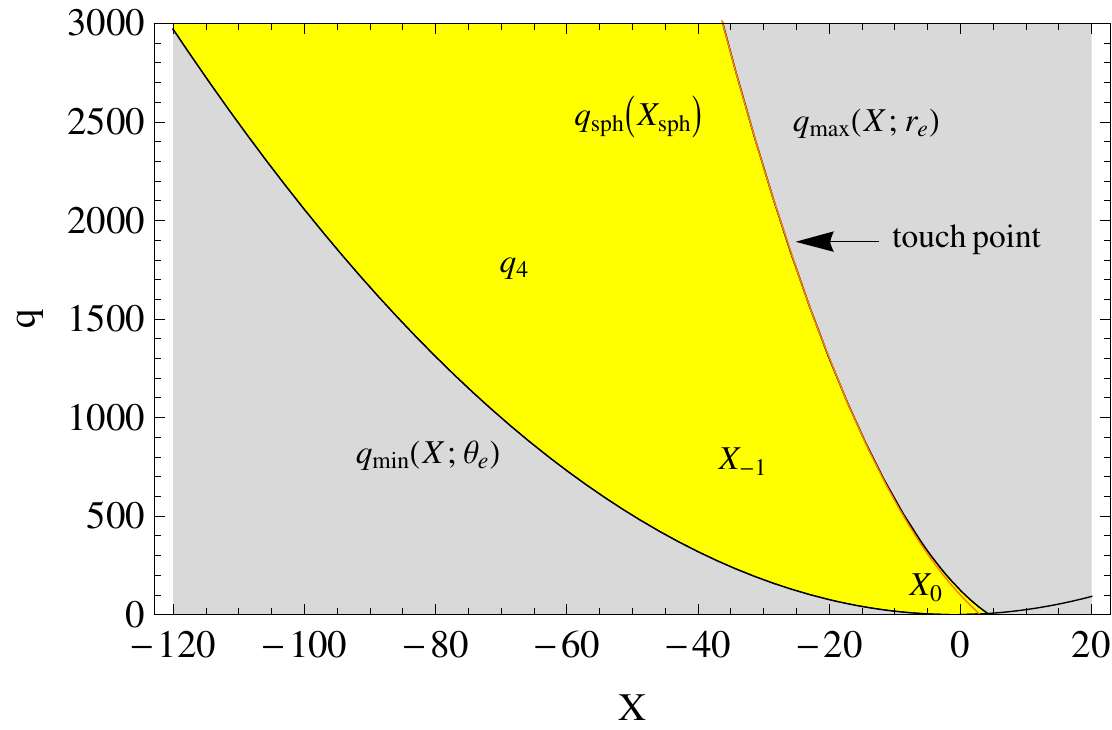}
				\end{tabular}
				
				\begin{tabular}{cc}
					\includegraphics[width=5.5cm]{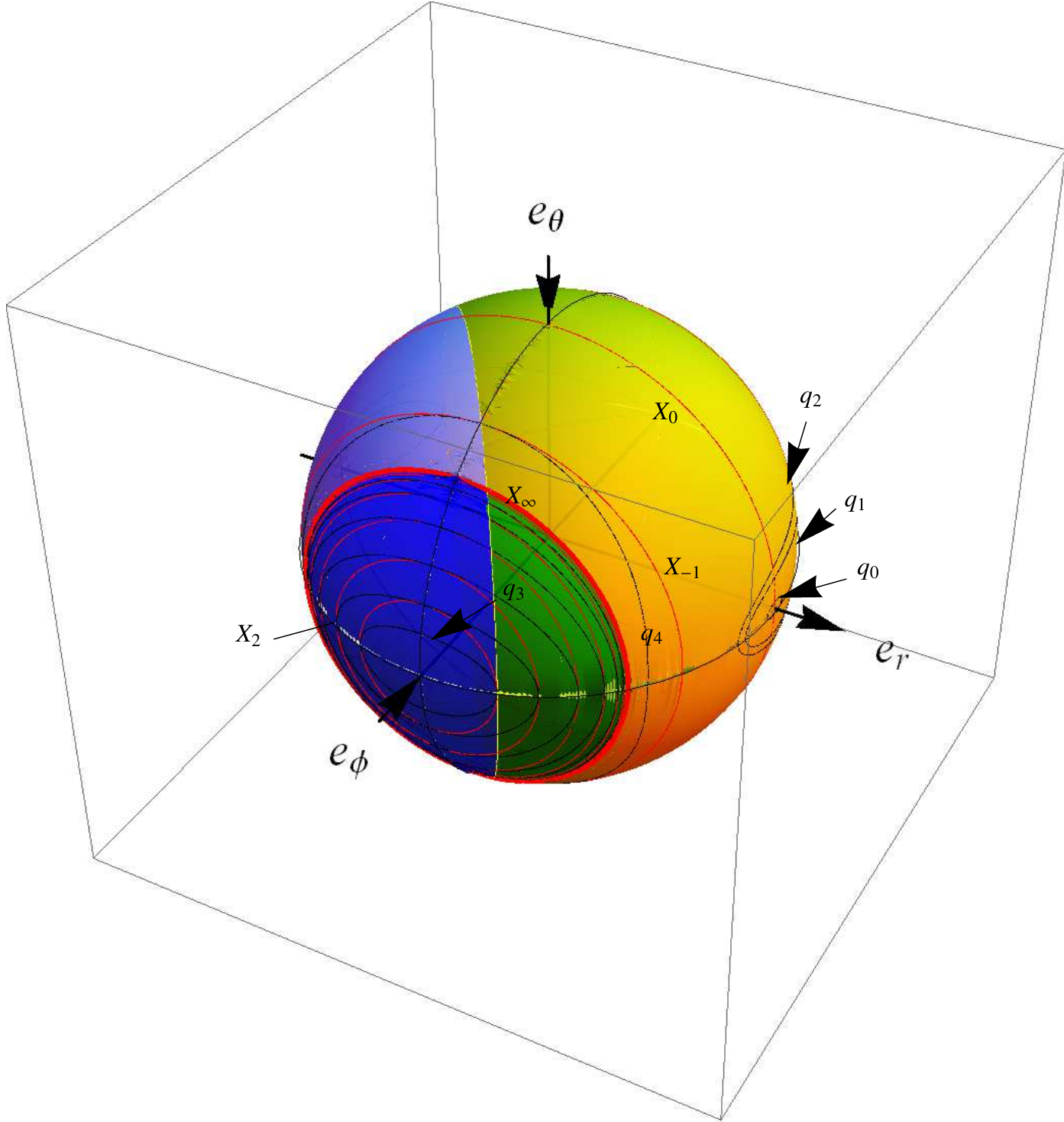}&\includegraphics[width=5.5cm]{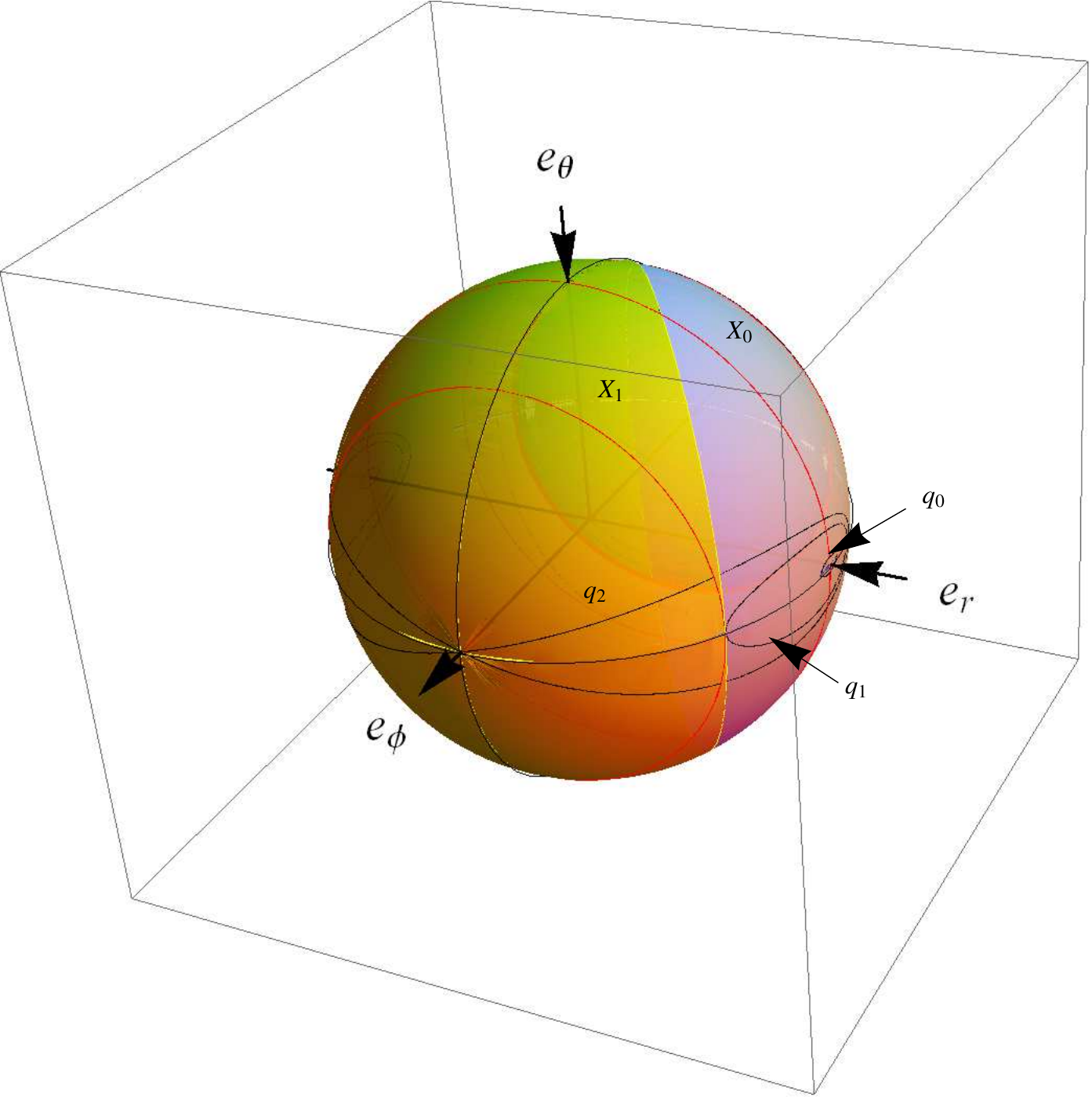}
				\end{tabular}

				\caption{The motion constant plane and light escape cones of a source in the KdS spacetime with the RRB having the spacetime parameters $a^2=0.9$ and $y=0.04,$. The source is located inside the ergosphere, at position with coordinates $r_e=2.65,$ $\theta_e=65\,^{\circ}.$ The range of the spherical orbits is limited to interval $r_1=1.90971\leq r\leq r_2=2.84748.$ The locally counter-rotating photons with high positive impact parameter $X$ occur in the whole stationary region; they can be captured by the black hole, as well as escaping through the cosmological horizon. The radial coordinate of the emitter is chosen so that $r_e<r_{d(ex)}=2.73484.$ In this case, the functions $q_{max}(X;\:r_e)$ and $q_{sph}(X_{sph}),$ which are now both convex, have touch point for $X<X_0,$ and all the above mentioned escaping photons are outwards directed.      }	 	\label{fig_crit_loc6} 	
		\end{figure}

\begin{figure}
				\centering
				\begin{tabular}{cc}
					\includegraphics[width=5.5cm]{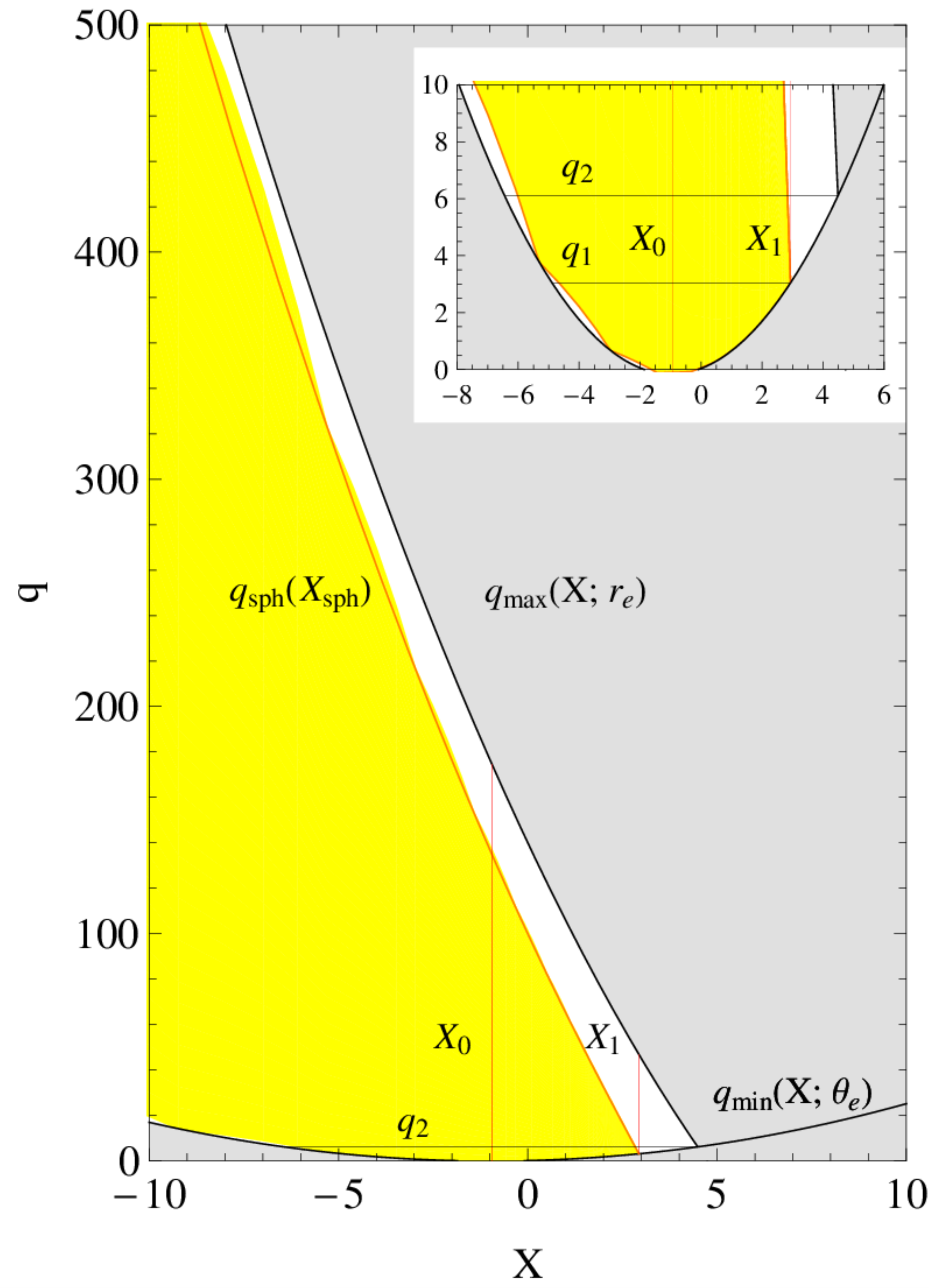}&\includegraphics[width=5.5cm]{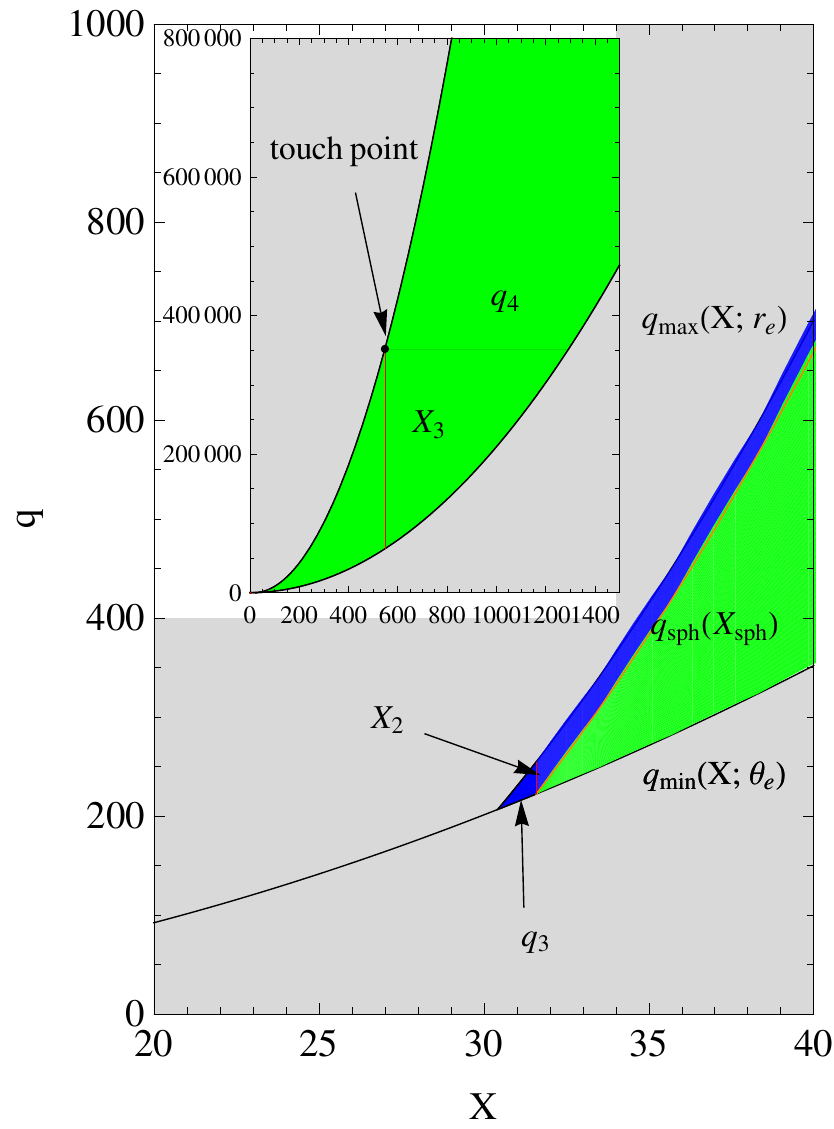}	
				\end{tabular}
						
				\begin{tabular}{cc}
					\includegraphics[width=5.5cm]{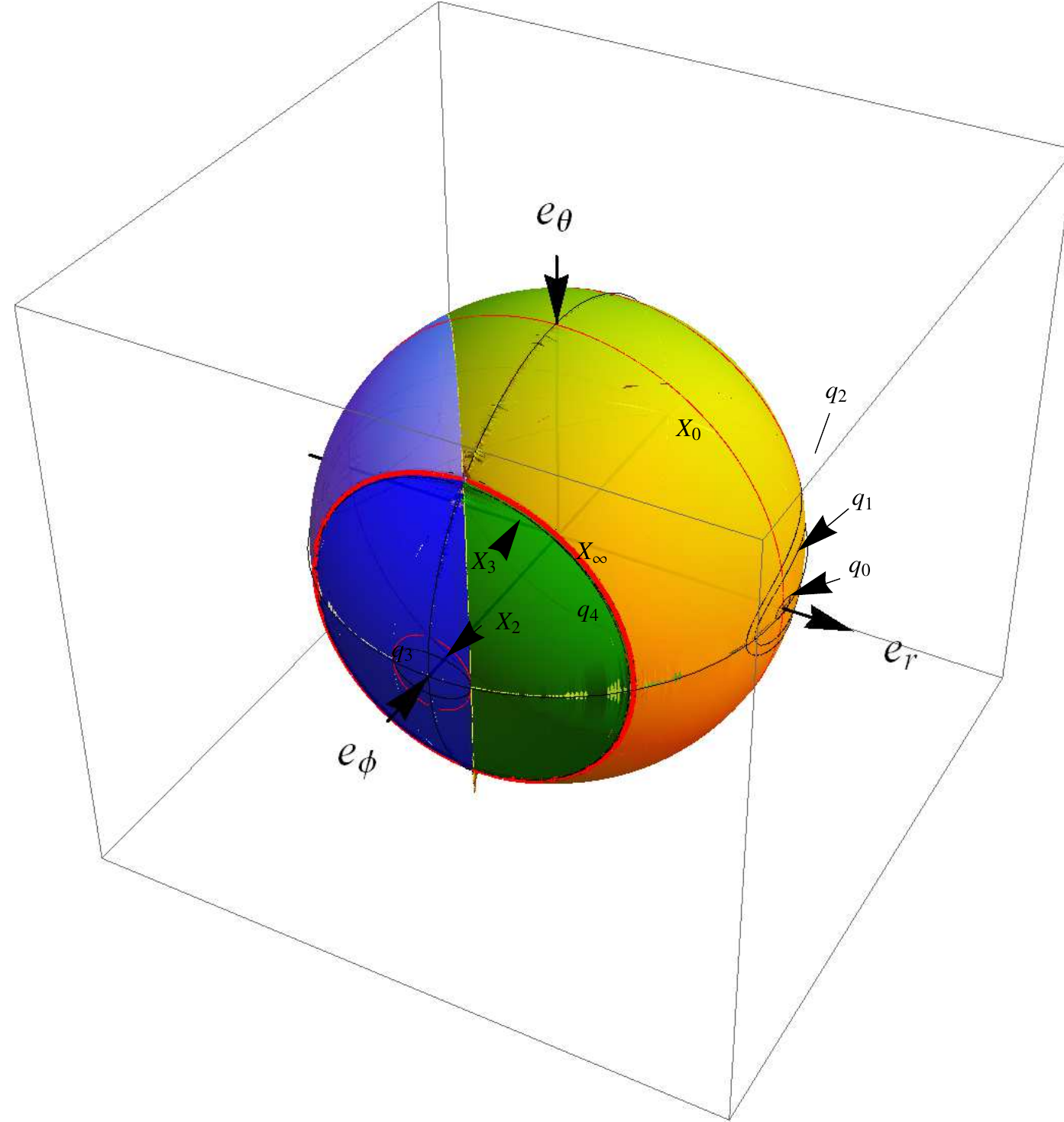}&\includegraphics[width=5.5cm]{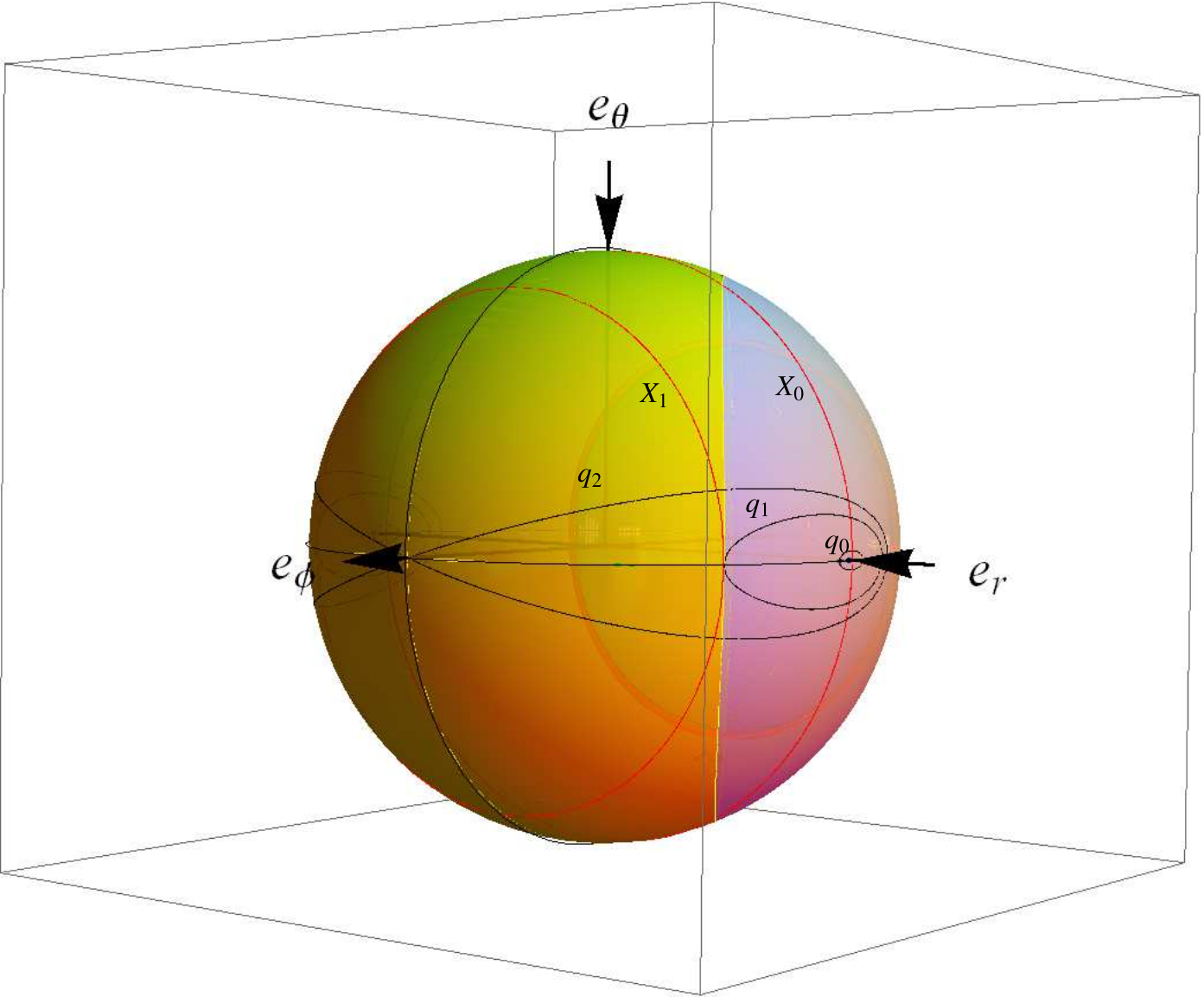}
				\end{tabular}


			\caption{ Depiction of the motion constant plane and corresponding light escape cones differing from the previous case by the radial coordinate $r_e=2.74,$ slightly exceeding the value $r_{d(ex)},$ in which case the functions $q_{max}(X;\:r_e)$ and $q_{sph}(X_{sph})$ have touch point for $X>0,$ hence, some locally counter-rotating photons with high positive impact parameter $X,$ which are escaping through the cosmological horizon, inwards directed emerge. Such situation first occurs at the radius $r_{d(ex)};$ at radii $r_e\geq r_{ph-}$, all captured locally counter-rotating photons become inwards directed (c.f. Figs. in pages \pageref{cones_rph-}, \pageref{cones_r>rph-}).       }	 	\label{fig_crit_loc7}
			\end{figure}

		The light escape cones are given by directional angles, for which the photon released by emitter (source) at position with given radial and latitudinal coordinates ($r_{e},$ $\theta_{e}$) reaches some unstable spherical orbit with radius $r_{sph}$. The relations (\ref{cosalphaLNRF}), (\ref{sincosbeta}), (\ref{sinsinbeta}), which we use for computation of these angles, thus have to be calculated for $r=r_{e},$ $\theta=\theta_{e}$ and parameters $q=q_{sph}$, $X=X_{sph},$ where 
		\be
		q_{sph}= q_{ex}(r_{sph};\:y,\:a^2) \label{qs}
		\ee
		is the value of $q,$ for which one of the potentials $X_{\pm}(r)$ has extreme at point $r_{sph},$ and
		\be
		X_{sph}= X_{(ex)}(r_{sph};\:y,\:a) \label{Xs}
		\ee
		is the value of that extreme, i. e. local maximum of $X_{+},$ or local minimum of $X_{-}.$ The function $q_{ex}(r;\:y,\:a^2),$ which determines the loci of local extrema of the potentials $X_{\pm},$ reads \cite{Char-Stu:2017:EPJC:}
		\be
		q_{ex}(r;\:y,\:a^2)\equiv -\frac{r^3}{a^2}\; \frac{y^2a^4r^3+2ya^2r^2(r+3)+r(r-3)^2-4a^2}{[2yr^3+(ya^2-1)r+1]^2}. \label{qex2} 
		\ee
		 The function in (\ref{Xs}) is a projection of local extrema of potentials $X_{\pm}(r)$ onto a hyperplane orthogonal to $q$-axis, therefore, 
		\bea
		X_{(ex)}(r;\:y,\:a)&=&X_{\pm}(r;\:q=q_{ex}(r;\:y,\:a^2),\:y,\:a)\nonumber \\
		&\equiv&\frac{r[(1-a^2y)r^2-3r+2a^2]}{a[yr(2r^2+a^2)-r+1]}. \label{Xex}
		\eea
		For fixed spacetime parameters $y, a^2$, the curve $(X_{sph},q_{sph})$ forms so called 'critical locus' (see e. g. \cite{1993A&A...272..355V}), which in the $(X,q)$-plane separates the region of the motion constants corresponding to photons for which a barrier is formed somewhere between the emitter and the black hole horizon (the cosmological horizon), from those corresponding to photons that move freely in the whole stationary region. It is parametrized by all permissible values of radius $r_{sph}$ of spherical orbits that can be reached by photon released from given position $(r_{e},\theta_{e}).$ The range of these orbits is confined by restrictions imposed on values $X,q$ of a photon at given coordinate $\theta_{e}$ that follows from the requirement on the reality of latitudinal motion. Since the reality condition $W\geq 0$ in (\ref{CarterW}) can be expressed, using linearity in $q,$ by inequality
		\be
		q\geq q_{min}(X;\: \theta,\:y,\:a)\equiv \frac{\cot^2\theta}{\Delta_{\theta}}[(1-a^2y\sin^2\theta)X^2+2aX+a^2\cos^2\theta], \label{qmin(theta)}
		\ee
		we can compute the limits of the $r_{sph}$ radii as points corresponding to intersections of the curve $q_{min}(X;\: \theta_e,\:y,\:a)$ with the curve $(X_{sph},q_{sph}).$ We therefore solve the equation
		\be
		q_{ex}(r;\:y,\:a^2)=q_{min}(X=X_{(ex)}(r;\:y,\:a);\:\theta_{e},\:y,\:a), \label{rsrange}            
		\ee
		which leads to sixth-degree polynomial in radius that has to be computed numerically. For the parameter sets ($a,$ $y,$) corresponding to the spacetimes with DRB, the curve $(X_{sph},q_{sph})$ is  concave, while for the spacetimes with RRB it is convex. However, in both cases, there are two intersections with the curve $q_{min}$ (see the figures bellow), which result in two solutions $r_1,r_2$ of the equation (\ref{rsrange}). The range of the spherical orbits are then given by $r_1\leq r_{sph}\leq r_2.$ Note that for $\theta_{e}=\pi/2$, there is $q_{min}=0,$ and the limits are the equatorial circular photon orbits $r_{1}=r_{ph+}$ and $r_{2}=r_{ph-},$ where $r_{ph+}$ denotes radius of the 'inner' co-rotating, and $r_{ph-}$ denotes the 'outer' counter-rotating photon orbit.\\
		The restrictions for the motion constants $X,$ $q,$ for which the photon motion is even possible, is given by the inequality (\ref{qmin(theta)}), together with 
		\be
		q \leq q_{max}(X;\:r,\:y,\:a)\equiv \frac{(r^2-a X)^2}{\Delta_r}-X^2 \label{qmax} 
		\ee
		following from the reality condition $R\geq 0$ in (\ref{CarterR}). The resulting parameter planes $X-q$ of permissible motion constants are presented in some representative, qualitatively different cases in Figs. \ref{fig_crit_loc}-\ref{fig_crit_loc7}, together with the corresponding light escape cones modelled in 3D. Note that in all cases the rising part of the curve $(X_{sph},q_{sph})$ matches local maxima of the potential $X_{+}(r;\:q,\:y,\:a),$ and descending part corresponds to local minima of $X_{-}(r;\:q,\:y,\:a).$\\

		The final shape of the light escape cones is decisively influenced, besides the values $y,a$ determining the geometry, by the radial position $r_e$ of the emitting body relative to both outer black hole ($r_{o}$) and cosmological horizon (${r_{c}}$), radii of the lowest ($r_{1}$) and highest ($r_{2}$) spherical orbit, loci of divergent points ($r_{d1},\,r_{d2}$) of the potential $X_+(r;\:q,\:y,\:a),$ if they exist, loci od divergent point ($r_{d(ex)}$) of the function $X_{ex}(r;\:,y,\:a),$ if it exists, and the latitudinal coordinate $\theta$ related to the critical value $\theta_{m}.$ The depiction of the light escape cones in the LNRFs for all outstanding radial positions of the source we postpone to following sections, where they are presented in comparison with the cones obtained in other reference frames in combined figures. Now we construct 3D representative illustrations of the light escape cones, later, in the case of the RGFs and CGFs we give simpler 2D schematic illustrations, i.e., two-dimensional projections of these cones. The relation of 3D and 2D schemes is presented in Fig. \ref{fig_crit_loc1}. We describe construction of the light escape cones in LNRFs in detail for some typical situations, in order to give a clear representation of the construction procedure. We give special attention also to the case of the photons that are counter-rotating relative LNRFs, while having positive impact axial parameter ($\ell > 0$) and negative covariant energy ($E<0$). 
		
		   
	   
	\section{Light escape cones in radial geodesic frames}	
	
	In this section we examine the light escape cones as they appear in frames related to observers (test particles) moving radially with respect to the locally non-rotating frames, orbiting momentarily at the same radii. 
		
	\subsection{Radially moving observers}
	First we determine conditions, under which the only non-zero locally measured component of the test particle (observer) 3-velocity relative to the corresponding LNRF is the radial one.\par
	The locally measured components of the observer (test particle) 3-velocity, related to the LNRF, $v^{(j)}, j=r,\theta,\phi$, are given by the general formula
	\be
	v^{(j)}=\frac{u^{(j)}}{u^{(t)}}=\frac{\omega^{(j)}_{\mu}u^{\mu}}{\omega^{(t)}_{\nu}u^{\nu}},\label{3-velocity}
	\ee
	where $u^\mu=\oder{x^\mu}{\tau}$ are coordinate components of particle's four-velocity and $\tau$ is its proper time related to the affine parameter by $\tau=m \lambda$ . The normalization condition (\ref{norm_p}) then takes the form
	\be
	u^{\mu}u_{\mu}=-1. \label{norm_u}
	\ee
	In our case, the latitudinal and radial components are of greatest importance. They are given by Carter's equations (\ref{CarterR})-(\ref{CarterT}).\\
	In order to simplify further discussion, it is convenient to introduce new parameters (motion constant of the test particle)
	\be E=\frac{I\cale}{m},\label{parE} \ee
	
	\be L=\frac{I\Phi}{m},  \label{parL} \ee and
	\be K=\frac{\calk}{m^2}, \ee \label{parK}
	from which we construct another constant of motion
	\be Q=K-(aE-L)^2 \label{parQ} \ee that has to be zero for the motion confined to equatorial plane.\\
	In the following, we must determine constraints that have to be imposed on the above constants of motion to allow a pure radial geodesic motion with respect to the LNRF.\par 
	Clearly, the condition $v^{(\phi)}=0$ is fulfilled if 
	\be
	\Phi=0.\label{zerophi}
	\ee
	Using (\ref{parE}) - (\ref{zerophi}), the equations (\ref{CarterR}), (\ref{CarterW}) can be rewritten in the form
	\be
	\rho^2 u^{r}=\pm \sqrt{R'}, \label{CarterR'}
	\ee
	
	\be
	\rho^2 u^{\theta}=\pm \sqrt{W'}, \label{CarterW'}
	\ee
	where
	\be
	R'(r;\:E,\:Q,\:y,\:a^2)=E^2[(r^2+a^2)^2-\Delta_{r}a^2]-\Delta_{r}(r^2+Q),\label{R'}
	\ee
	\be
	W'(\theta;\:E,\:Q,\:y,\:a^2)=Q \Delta_{\theta}+a^2\cos^2\theta(IE^2-\Delta_{\theta}). \label{W'}
	\ee

	The condition $v^{(\theta)}=0$ implies $u^{\theta}= 0,$ and finally $W'(\theta;\:E,\:Q,\:y,\:a^2) = 0,$ which can be expressed in terms of the parameter $Q$ by the equation
	\be
	Q = Q_{\theta}(\theta;\:E,\:y,\:a^2)\equiv \frac{a^2 \cos^2\theta (\Delta_{\theta}-IE^2)}{\Delta_{\theta}}. \label{Qt}
	\ee
	The reality condition of the latitudinal motion of a test particle with $\Phi=0$ then reads $Q\geq Q_{\theta}(\theta;\:E,\:y,\:a^2)$, see also \cite{Bic-Stu:1976:BULAI:}. In the following discussion we restrict ourselves to angles $0\leq \theta \leq \pi/2,$ since the situation is symmetric with respect to equatorial plane. Clearly, there is always one zero point of $Q_{\theta}(\theta;\:E,\:y,\:a^2)$ at $\theta=\pi/2.$ Another zeros are determined by
	\be
	E^2=E^2_{z(\theta)}(\theta;\; y,\;a^2)\equiv \frac{\Delta_{\theta}}{I},
	\ee 
	i. e., they are located at
	\be
	\theta_{z}=\arccos \sqrt{\frac{IE^2-1}{a^2y}} 
	\ee 
	for $I^{-1}\leq E^2\leq 1.$ \par 
	The function $Q_{\theta}(\theta;\:E,\:y,\:a^2)$ is non-negative for $E^2\le I^{-1},$ while for $E^2\ge 1$ it is non-positive. Its extrema are given by the condition $\din Q_{\theta}/\din \theta=0,$ which is fulfilled for $\theta=0,\pi/2$ and at loci given by
	\be
	E^2=E^2_{ex(\theta)}(\theta;\; y,\;a^2)\equiv \frac{\Delta_{\theta}^2}{I}.
	\ee 
	Verifying the value $\din^2Q_{\theta}/\din \theta^2\;|_{\theta=0, \pi/2},$ we find that at $\theta=0$ the function takes minimum (maximum) for $E^2>I\; (E^2<I),$ while at $\theta=\pi/2$ there is minimum (maximum) for $E^2<I^{-1}\;(E^2>I^{-1}).$ Here $$I^{-1}=E^2_{ex(\theta)min}=E^2_{ex(\theta)}(\pi/2;y,a^2)=E^2_{z(\theta)min}=E^2_{z(\theta)}(\pi/2;y,a^2),$$
	$$1=E^2_{z(\theta)max}=E^2_{z(\theta)}(0;a,y),$$
	$$I=E^2_{ex(\theta)max}=E^2_{ex(\theta)}(0;y,a^2).$$ Note that for $a^2=0$ or $y=0$ the functions $E^2_{z(\theta)}(\theta;\; y,\;a^2),$ $E^2_{ex(\theta)}(\theta;\;y,\;a^2),$ degenerate to constant value $E^2=1$ and $Q_{\theta}(\theta;\:E,\:y,\:a^2)$ then yields $Q=0.$ All these functions are depicted for some representative values of its parameters in Fig. \ref{figE2,Qt}.
	\begin{figure}[htbp]
		\centering
		\begin{tabular}{cc}
			\includegraphics[scale=0.7]{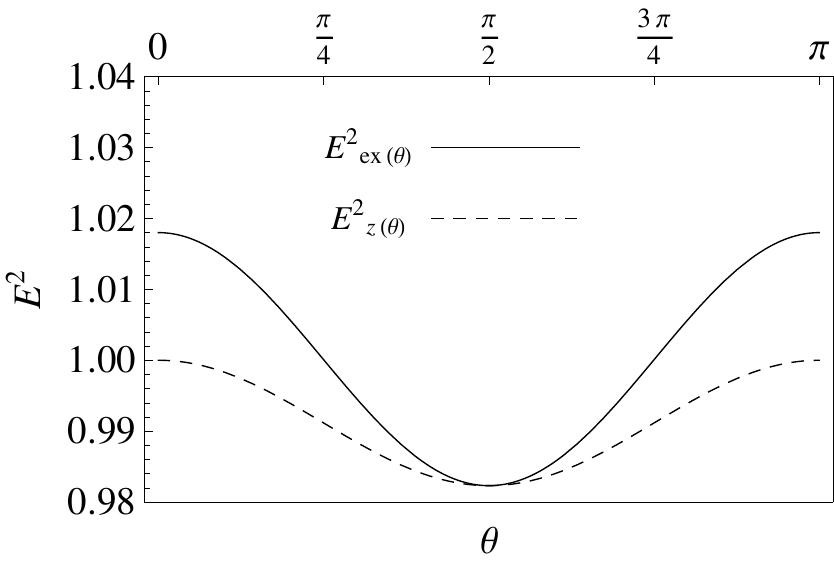}&\includegraphics[scale=0.7]{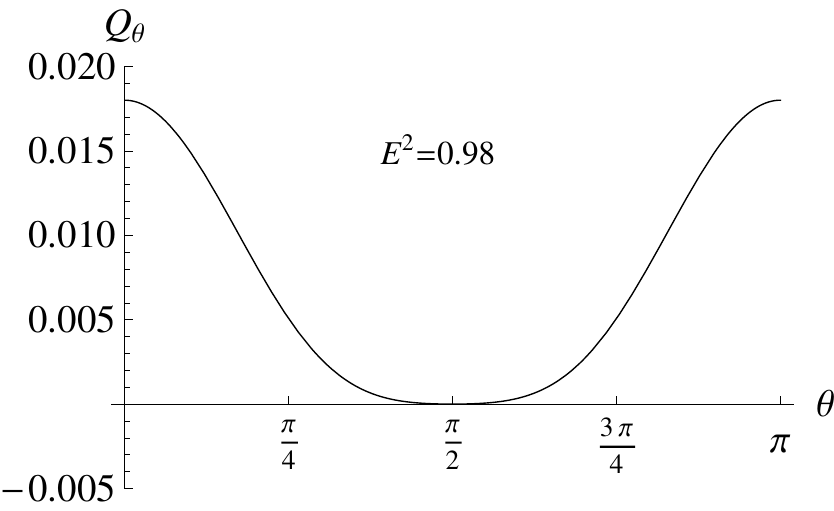}\\
			(a)&(b)\\
			\includegraphics[scale=0.7]{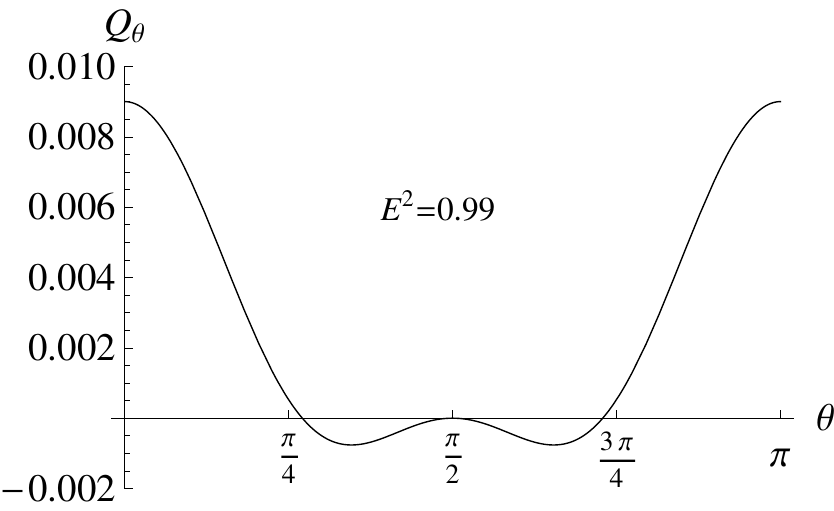}&\includegraphics[scale=0.7]{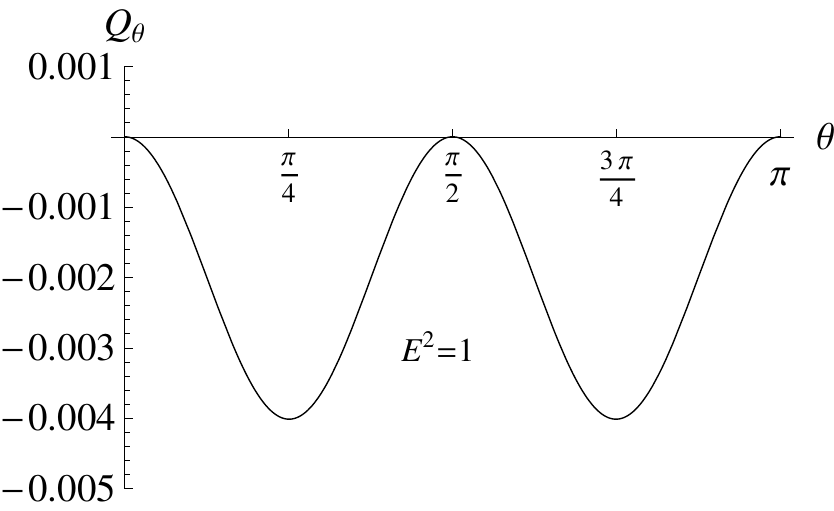}\\
			(c)&(d)\\
			\includegraphics[scale=0.7]{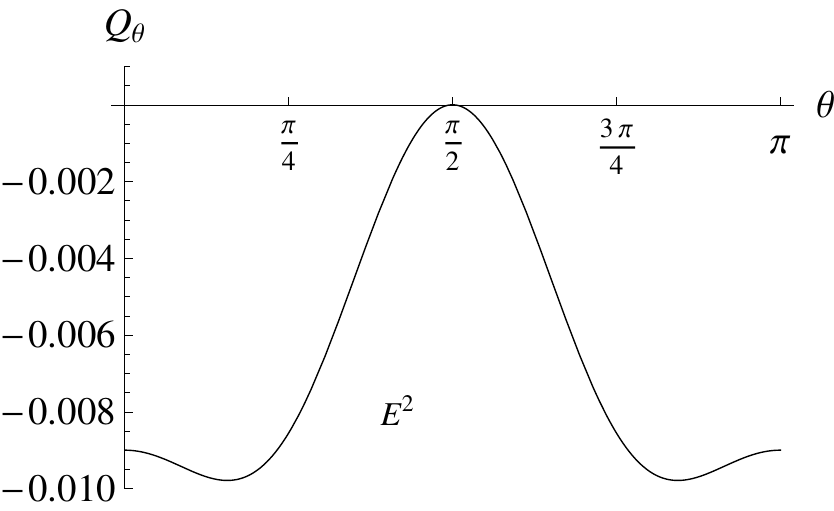}&\includegraphics[scale=0.7]{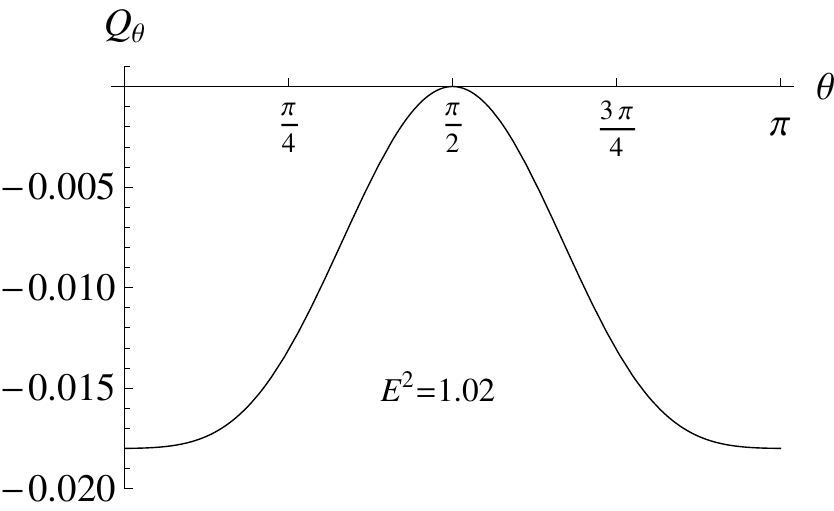}\\
			(e)&(f)
			
		\end{tabular}
		\caption{Latitudinal motion of particles with $L=0$. Functions $E^2_{z(\theta)}(\theta;\:y,\:a^2)$ and  $E^2_{ex(\theta)}(\theta;\:y,\:a^2)$ \textbf{(a)} governing the behaviour of the function $Q_{\theta}(\theta;\:E^2,\:y,\:a^2)$ displayed for $E^2$ corresponding successively to cases $E^2<I^{-1}$ \textbf{(b)}, $I^{-1}<E^2<1$ \textbf{(c)}, $E^2=1$ \textbf{(d)}, $1<E^2<I$ \textbf{(e)} and $E^2>I$ \textbf{(f)} with the spacetime parameters $y=0.02,$ $a^2=0.9,$ giving $I=1.018,$ $I^{-1}=0.9823.$}
		\label{figE2,Qt}
	\end{figure}
	\par
	It can be shown by the perturbation analysis that the local minima of the function $Q_{\theta}(\theta;\:E,\:y,\:a^2)$ correspond to constant latitudes on which the particle persists stably during its radial fall towards the outer black hole horizon or radial escape towards the cosmological horizon. The motion constants of such stable pure radial motion and the appropriate latitudes $\theta_{0}$ are therefore mutually bound by the relations
	
	\begin{equation}
		Q = \left\{
		\begin{array}{l}
			0\quad\mbox{for}\quad E^2\leq I^{-1}\quad\mbox{giving}\quad \theta_{0}=\pi/2;\\
			\\
			\frac{-(1-\sqrt{IE^2})^2}{y}\quad \mbox{for}\quad I^{-1}\leq E^2\leq I\quad\mbox{giving}\quad\theta_{0}= \arccos\sqrt[4]{\frac{-Q}{a^4y}}\\
			\\
			a^2(1-E^2)\quad \mbox{for}\quad E^2\geq I \quad \mbox{which yields}\quad \theta_{0}=0.
			\end{array}\right.\label{Q_E^2_plane}
	\end{equation}
	\par 
	The case $I^{-1}\leq E^2\leq I$ in (\ref{Q_E^2_plane}) is equivalent to relations
	\bea
	E^2&=&E^2_{ex(\theta)}(\theta_{0};\:y,\:a^2) \label{Eex} \\
	Q&=&Q_{ex(\theta)}\equiv Q_{\theta}(\theta_{0};\:E^2_{ex(\theta)}(\theta_{0};\:y,\:a^2),\:y,\:a^2)=-a^4y\cos^4\theta_{0}. \label{Qex}
	\eea
	Evidently, for spacetime parameters  $y=0,$ or $a^2=0,$ the relations (\ref{Eex}), (\ref{Qex}) yield well known results $E^2=1,$ $Q=0$ for any latitude $\theta_{0}$ \cite{1999GReGr..31...53S}. 
	In the following we shall decide, whether some of these constants allow any turning points of the radial motion, that is, the radii at which a particle, if freely released, originate its stable radial fall with respect to the LNRF, or, if equilibrium points located at stationary radii (unstable or stable) are allowed. The turning points are determined by the condition $R'(r;\:E,\:Q,\:y,\:a^2)=0,$ or, alternatively, by 
	\be
	Q=Q_{r}(r;\:E^2,\:y,\:a^2)\equiv \frac{E^2(r^2+a^2)^2}{\Delta_{r}}-r^2-E^2a^2. \label{Qr}
	\ee
	\par
\begin{figure}
		\centering
		\begin{tabular}{cc}
			\includegraphics[scale=0.7]{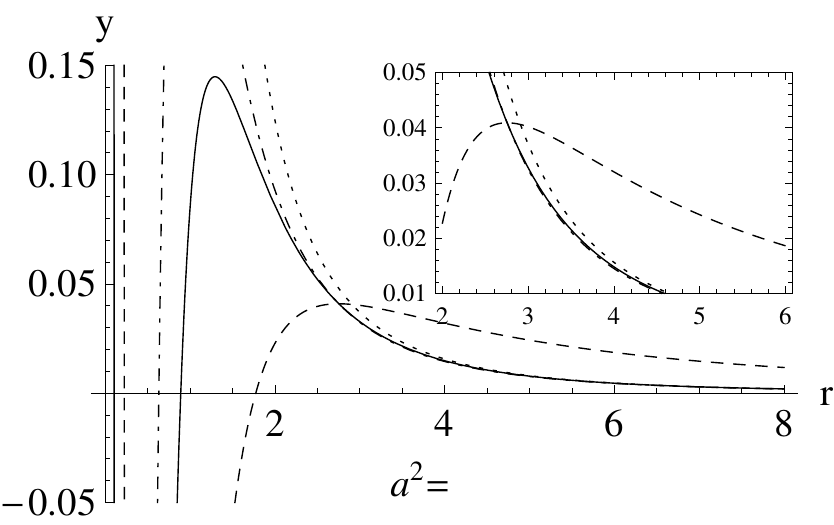}&\includegraphics[scale=0.7]{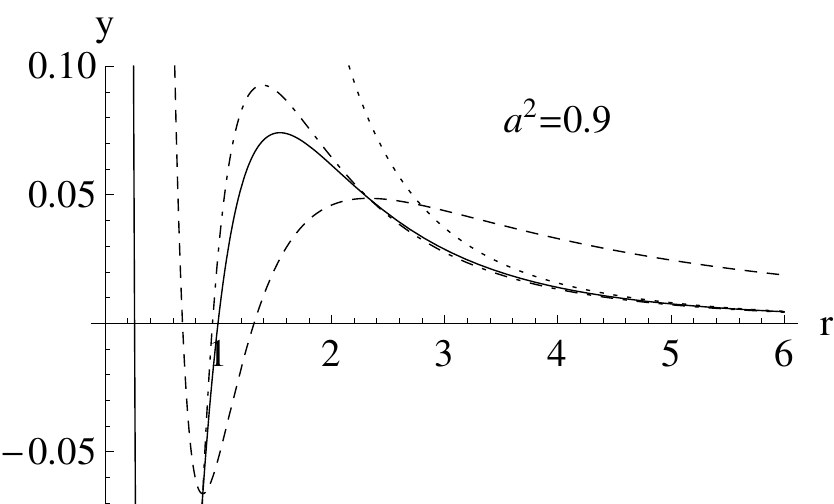}\\
			(a)&(b)\\
			\includegraphics[scale=0.7]{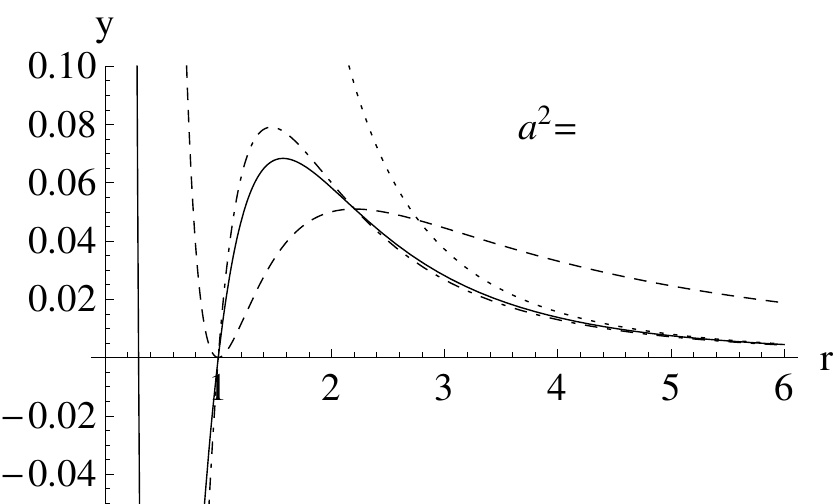}&\includegraphics[scale=0.7]{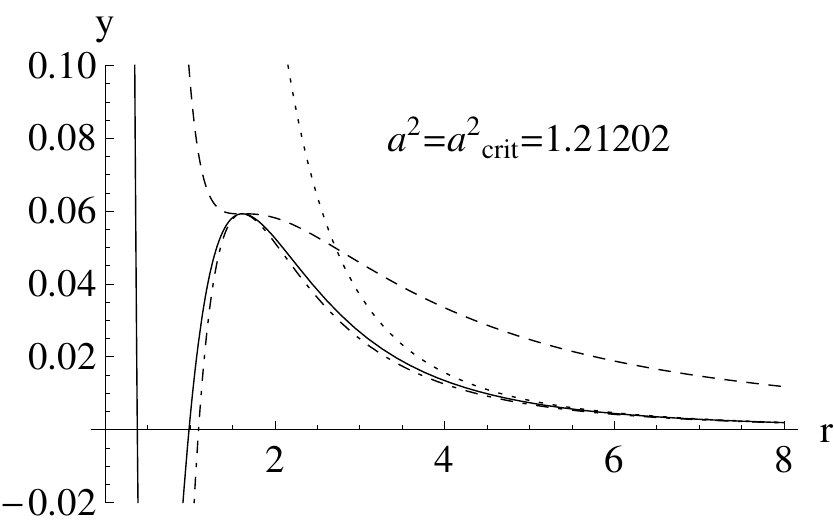}\\
			(c)&(d)
					
		\end{tabular}
		\caption{Radially moving particles. We give characteristic functions $y_{ex(z(eq))}(r;\:a^2)$ (solid curve), $y_{h}(r;\:a^2)$ (dashed curve), $y_{ex(z(ax))}(r;\:a^2)$ (dash-dotted curve) and $1/r^3$ (dotted) given for typical values of the spin parameter $a^2.$}
		\label{Fig.yex(z)}
	\end{figure}
\begin{figure}
	\centering
	\begin{tabular}{cc}
		\includegraphics[scale=0.55]{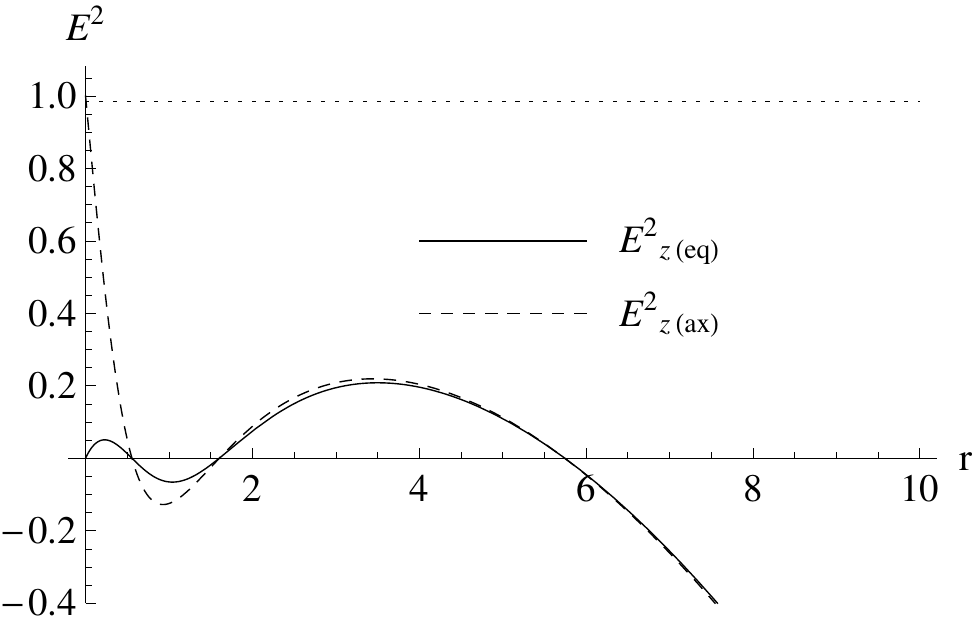}&\includegraphics[scale=0.55]{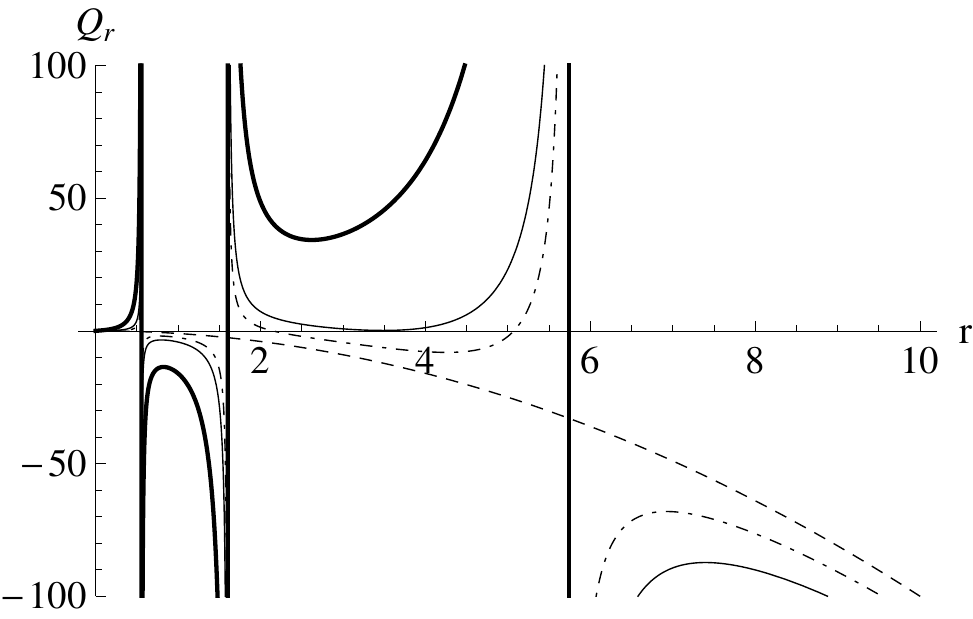}

	\end{tabular}
	\caption{Effective potential for radially moving particles. Typical behaviour of the functions $E^2_{z(eq)}(r;\:y,\:a^2),$ $E^2_{z(ax)}(r;\:y,\:a^2)$ (left) and $Q_{r}(r;\:E^2,\:y,\:a^2)$ (right)  generated for the spin parameter $a^2=0.9$ and the cosmological parameter $y=0.02.$ The black dotted horizontal line represent the values of $E^2=I^{-1}.$ Behaviour of the function $Q_{r}(r;\:E^2,\:y,\:a^2)$ is illustrated for parameters $E^2=I^{-1}$ (bold curve), $E^2= E^2_{z(eq),max}=0.2107$ (solid curve), $E^2=0.1$ (dot-dashed curve) and $E^2=0$ (dashed curve). The event horizons are given by the intersections of the functions $E^2_{z(eq)}(r;\:y,\:a^2),$ $E^2_{z(ax)}(r;\:y,\:a^2)$ with the $r$- axis, or by the asymptotes of the function $Q_{r}(r;\:E^2,\:y,\:a^2)$. }
	\label{Fig.e2z(r)}
\end{figure}

	We focus on the behaviour of this function between the outer black hole ($r_{o}$) and cosmological ($r_{c}$) horizons. Apparently, $Q_{r}(r;\:E^2,\:y,\:a^2) \to \infty$ as $r \to r_{o}$ from above, or $r \to r_{c}$ from bellow. According to (\ref{Q_E^2_plane}), the stable radial motion of constant latitude exists only for $Q \leq 0$ and $E^2 \geq I^{-1},$ we therefore have to determine zeros of the function in (\ref{Qr}) designating the turning points in the equatorial plane. They are given by
	\be
	E^2=E^2_{z(eq)}(r;\:y,\:a^2)\equiv \frac{\Delta_{r}r^2}{(r^2+a^2)^2-\Delta_{r}a^2}. \label{E2z}
	\ee
	The denominator of this function is always positive for $y>0,$  hence its zeros coincide with loci of the event horizons and in the stationary regions it is positive. It can be shown that it always holds $E^2_{z(eq)}(r;\:y,\:a^2)<I^{-1}$ for $y>0,$ therefore, there are no turning points of the radial motion, except for the equatorial plane $\theta_{o}=\pi/2.$ If $y=0,$ then $$\lim_{r\rightarrow \infty} E^2_{z(eq)}(r;\:y,\:a^2)=I^{-1}=1.$$ 
	The local extrema can be expressed from the condition 
	\be
	\partial E^2_{z(eq)}/\partial r=0 \label{dE2z}
	\ee
	 by the equation
	\bea
	y&=&y_{ex(z(eq))}(r;\:a^2)\\ \nonumber
	&\equiv& \frac{-\sqrt{r}[r(r^2+a^2)+4a^2]+\sqrt{r^2+a^2}\sqrt{r^3(r^2+a^2)+4a^2(3r^2+a^2)}}{2\sqrt{r^3}a^2(r^2+a^2)}.\label{yexz}
	\eea
	The curve $y_{ex(z(eq))}(r;\:a^2)$ determines radii $r_{eq}(a^2,y)$ of unstable equatorial circular orbits of particles with zero angular momentum. Note that in the case of pure \SdS\ spacetime ($a=0$) the solution of (\ref{dE2z}) yields $y=1/r^3,$ i. e., loci of the so called static radius $r_{s}=1/\sqrt[3]{y}.$
	The common points of the function $y_{ex(z(eq))}(r;\:a^2)$ and $y_{h}(r;\:a^2),$ which determines the loci of the spacetime horizons, are determined by
	\be
	a^2=\frac{1}{2}(r-2r^2+r\sqrt{1+8r}),
	\ee
	which is identical with the solution of the equation $\partial y_{h}(r;\:a^2)/\partial r =0.$ Hence, the intersections
	and the local extrema of the function $y_{h}(r;\:a^2)$, if they exist, coincide, and there are no other intersections of these curves. In the stationary regions it is $y_{ex(z(eq))}(r;\:a^2)<y_{h}(r;\:a^2).$ The asymptotic behaviour is given by
	$$y_{ex(z(eq))}(r;\:a^2)\to 0\quad \mbox{as}\quad r \to \infty \quad \mbox{with} \quad 0<y_{ex(z(eq))}(r;\:a^2)<y_{h}(r;\:a^2).$$ For completeness, 
	$$y_{ex(z(eq))}(r;\:a^2)\to \infty\quad \mbox{as}\quad r \to 0 \quad \mbox{and again} \quad 0<y_{ex(z(eq))}(r;\:a^2)<y_{h}(r;\:a^2).$$ The behaviour of the characteristic functions $y_{ex(z(eq))}(r;\:a^2),$ $E^2_{z(eq)}(r;\:y,\:a^2)$ and $Q_{r}(r;\:E^2,\:y,\:a^2),$ which	is shown for some representative values of parameters $a^2,$ $y$ and $E^2$ in Figs. \ref{Fig.yex(z)}-\ref{Fig.e2z(r)}, suggests that in the field of the KdS black hole spacetimes there always exists one  equatorial circular orbit of a test particles with $Q=0, L=0$ and $E^2=E^2_{z(eq),max},$ corresponding to the local maximum of (\ref{E2z}). For particles with $0<E^2<E^2_{z(eq),max}$ there exist trajectories with either one turning point above the outer black hole horizon, or one under the cosmological horizon. These turning points enclose a forbidden region, which extends to both horizons as $E^2\to 0.$ 

	We can summarize that
	there are no turning points of the stable pure radial motion for trajectories of constant latitude with $\theta_{0}\neq \pi/2,$ since they are given by the local minima $Q_{\theta(min)}<0$ of $Q_{\theta}(\theta;\:E,\:y,\:a^2),$ which occur for $E^2>I^{-1},$ but $Q_{r}(r;E^2,y,a^2)>0$ for $E^2>I^{-1}.$  In the equatorial plane, such turning points occur for $Q=Q_{\theta,min}=0$ in the vicinity of the event horizons for particles with small enough energies $E^2\leq E^2_{z(eq),max}<I^{-1}.$ 
	
	At the spin axis, there can exist some turning points of an unstable radial motion of test particles with motion constants $Q=Q_{\theta,max}>0,$ $0\leq E^2 \leq E^2_{z(ax)}.$ Here we denoted by $Q_{\theta,max}$ local maxima of the function $Q_{\theta}(\theta;\:E,\:y,\:a^2)$ at $\theta=0,\pi,$ which read $Q_{\theta,max}=a^2(1-E^2)$ and $E^2_{z(ax)}$ is a solution of the equation $Q_{r}(r;E^2,y,a^2)=a^2(1-E^2),$ which yields
	\be
	E^2=E^2_{z(ax)}\equiv \frac{\Delta_{r}}{r^2+a^2}. \label{E2r(ax)}
	\ee
	The function $E^2_{z(ax)}$ determines the turning points of the unstable radial motion along the spin axis. Its local maxima, given by the function
	\be
	y=y_{ex(z(ax))}(r;\:a^2)\equiv \frac{r^2-a^2}{r(r^2+a^2)} \label{yex(r(ax))}
	\ee
	determines the radial coordinate $r_{ax}(a^2,y)$ on the spin axis, where the test particle can stay in an unstable equilibrium. We present the behaviour of this function in comparison with the function $y_{ex(z(ax))}(r;\:a^2)$ in Fig. \ref{Fig.yex(z)}. They intersect each other at the local extrema of the function $y_h(r;\:a^2)$ and in the stationary regions of the black hole spacetimes it holds $y_{ex(z(ax))}(r;\:a^2) < y_{ex(z(eq))}(r;\:a^2) < y_h(r;\:a^2),$ i. e., it is $r_{ax}(a^2,y)<r_{eq}(a^2,y).$ 
	
    \subsection{Radial geodesic frames}	
\begin{figure}[H]
			\begin{tabular}{cc}
				\includegraphics[scale=0.7]{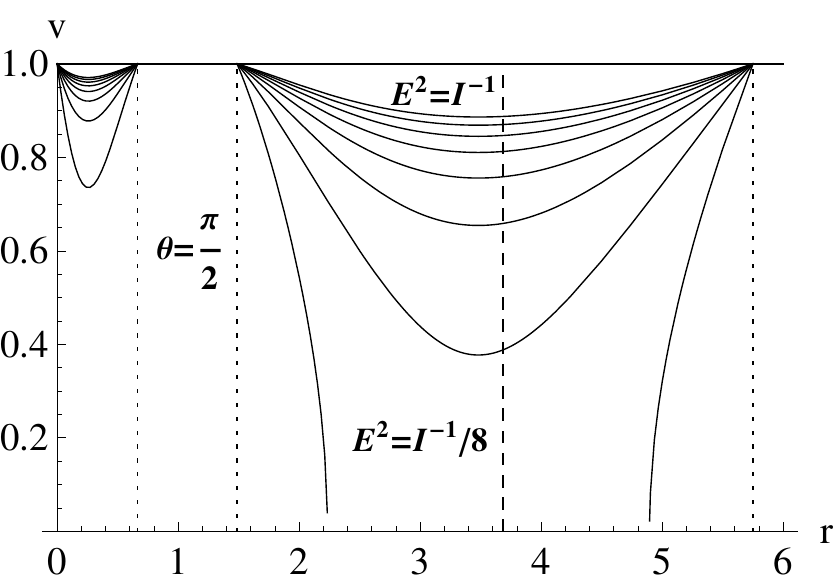}&\includegraphics[scale=0.7]{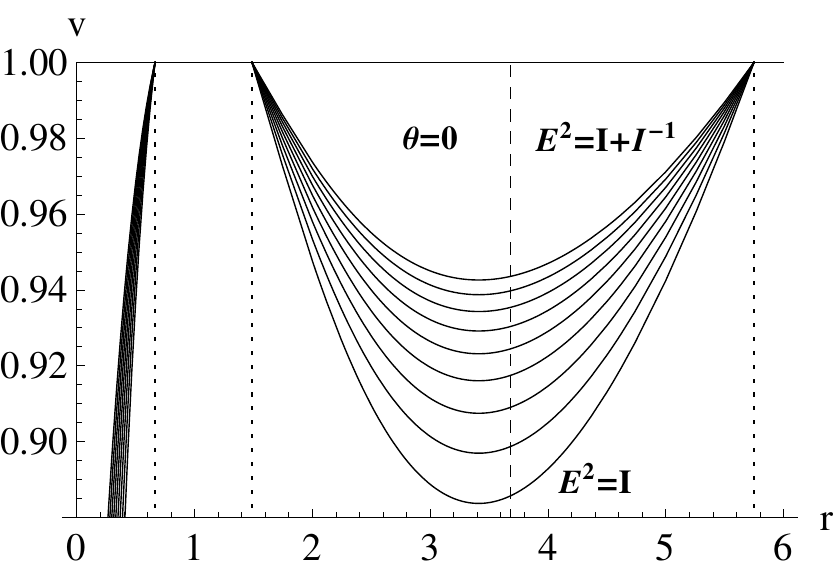}				
			\end{tabular}
			\caption{Radial velocity of purely radial motion of a massive test particle, as locally measured in the LNRF, is presented for the KdS spacetime with parameters $y=0.02,$ $a^2=0.9,$ are given for particles with energies $E^2\leq I^{-1}=0.982$ corresponding to the motion confined to the equatorial plane (left), and $E^{2}>I,$ for which the particles move along the spin axis (right). All curves are generated for $E^{2}$ heightened by $I^{-1}/8$. The radial turning points in the first case occur for energies $E^2< E^2_{z,max}=0.2107.$ Vertical lines demarcate loci of the event horizons (dotted) and the static radius (dashed).    }\label{fig_radvel}
	\end{figure}

	In the following we shall concentrate on the two extremal cases of radially falling frames, namely, the radial fall confined to the equatorial plane and along the spin axis. The general formula for the locally measured radial velocity $v^{(r)}$ we get using (\ref{3-velocity}) and associated relations. We arrive to
	\be
	v^{(r)}=\frac{I}{\Delta_{r}}\sqrt{\frac{A}{\Delta_{\theta}}}\frac{u^r}{u^t},\label{gen_radvel}
	\ee
	 where the radial component $u^r$ is given by (\ref{CarterR'}), (\ref{R'}) and the temporal one can be expressed considering (\ref{cale}) and (\ref{parE}) by means of the rescaled parameter $E$ in the form 
	\be
	u^{t}=-g^{tt}\frac{E}{I}.
	\ee
	In order to get the dependence of the radial velocity on the latitudinal coordinate $\theta_{0}$, we substitute for parameters $E,Q$ from (\ref{Eex}),(\ref{Qex}), and after some algebraic manipulations we finally obtain
	\be
	v^{(r)}=\pm \sqrt{1-\frac{I\rho^2\Delta_{r}}{A \Delta_{\theta}}}. \label{radvel}
	\ee
	In the case of pure Kerr spacetimes ($y=0$) this relation goes over to well known formula $v^{(r)}=\pm \sqrt{1-\frac{\rho^2\Delta_{r}}{A}}$ (see, e.g., \cite{2005ragt.meet..143S}). However, for energies $E^2<I^{-1}$ or $E^2>I$ the formula (\ref{radvel}) does not work and one must use the more general expression (\ref{gen_radvel}). In the special cases of our interest we can modify the expression (\ref{radvel}) inserting $\theta=0, \pi/2.$ The first case corresponds to the stable radial fall/escape along the spin axis with the minimum allowed energy $E^2=I$ and velocity
	\be
	v_{ax}^{(r)}=\pm \sqrt{1-\frac{\Delta_{r}}{I(r^2+a^2)}}, \label{radvel_ax}
	\ee
	the latter case corresponds to the pure radial fall/escape stably confined to the equatorial plane with maximum allowed energy $E^2=I^{-1}$ and velocity
	\be
	v_{eq}^{(r)}=\pm \sqrt{1-\frac{I\Delta_{r}r^2}{(r^2+a^2)^2-\Delta_{r}a^2}}. \label{radvel_eq}
	\ee
	Considering the relations (\ref{E2r(ax)}), (\ref{E2z}) we can write the formulas (\ref{radvel_ax}), (\ref{radvel_eq}) in an alternative form
	\be
	v_{ax}^{(r)}=\pm \sqrt{1-\frac{E^2_{z(ax)}}{E^2}},\label{vax(E2)}
	\ee
	\be
	v_{eq}^{(r)}=\pm \sqrt{1-\frac{E^2_{z(eq)}}{E^2}},\label{veq(E2)}
	\ee
	which, as can be verified, hold for any allowed energy. The formula (\ref{vax(E2)}) then describes the unstable radial fall/escape for energies $E^2_{z(ax)}\leq E^2 < I$ along the spin axis, and stable radial fall/escape for energies $I\leq E^2,$ while the formula (\ref{veq(E2)}) governs the stable radial fall/escape for energies $E^2_{z(eq)}\leq E^2 \leq I^{-1},$ and unstable radial fall/escape for energies $E^2>I^{-1}$ in the equatorial plane. The dependence of the radial velocities on the particle energy is illustrated in Fig. \ref{fig_radvel}.

	With the knowledge of the radial velocity, and using the standard Lorentz transformation
	\be
	k^{(\hat{a})}=\Lambda^{(\hat{a})}_{(b)}k^{(b)}, \label{LT}
	\ee
	where $k^{(\hat{a})},$ $k^{(a)}$ denote the components of a photon`s four-momentum locally measured by the radially falling observers and by the locally non-rotating observers, respectively, we can relate the directional angles between the both frames. The matrix elements in (\ref{LT}) are given by
	\be
	\Lambda^{(\tilde{a})}_{(b)}=\left( \begin{array}{cccc}
		\gamma & -\gamma v & 0 & 0\\
		-\gamma v & \gamma & 0 & 0\\
		0 & 0 & 1 & 0\\
		0 & 0 & 0 & 1
	\end{array}	\right),
	\ee
	where 
	\be
	\gamma=(1-v^2)^{-1/2}
	\ee
	is the Lorentz factor. 
	
	\subsection{Construction of light escape cones}
	Since the directional angles are defined by the same way in every frame, one can derive
	\be
	\cos\hat{\alpha}=\frac{k^{(\hat{r})}}{k^{(\hat{t})}}=\frac{\cos\alpha - v}{1-v \cos\alpha}, \label{cosalpha}
	\ee
	\be
	\sin\hat{\alpha}\cos\hat{\beta}=\frac{k^{(\hat{\theta})}}{k^{(\hat{t})}}=\frac{\sin\alpha \cos\beta}{\gamma (1-v \cos\alpha)}, \label{sinalphacosbeta}
	\ee
	\be	\sin\hat{\alpha}\sin\hat{\beta}=\frac{k^{(\hat{\theta})}}{k^{(\hat{t})}}=\frac{\sin\alpha \sin\beta}{\gamma (1-v \cos\alpha)}.\label{sinalphasinbeta}
	\ee
	In the above relations and hereafter, we denote the observer velocity by $v$ instead of $v^{(r)}$ for simplicity; $v>0$ for radially escaping observer, $v<0$ for the falling one. We shall focus on observers with the radial velocities given by the relation (\ref{radvel}).
	Since 
	$$\left[\frac{\sin\alpha}{\gamma (1-v \cos\alpha)}\right]^2+\left[\frac{\cos\alpha-v}{1-v \cos\alpha}\right]^2=1,$$
	one can put
	\be
	\sin\hat{\alpha}=\frac{\sin\alpha}{\gamma (1-v \cos\alpha)} \label{sinhatalpha}
	\ee
	and the relations (\ref{sinalphacosbeta}), (\ref{sinalphasinbeta}) imply 
	\be
	\hat{\beta}=\beta, \label{hatbeta}
	\ee
	hence, the angle $\hat{\beta}$ is determined by the same functions as $\beta$ in the case of the LNRF.\par
\begin{figure}[H]
			\begin{tabular}{cc}
				\includegraphics[scale=0.7]{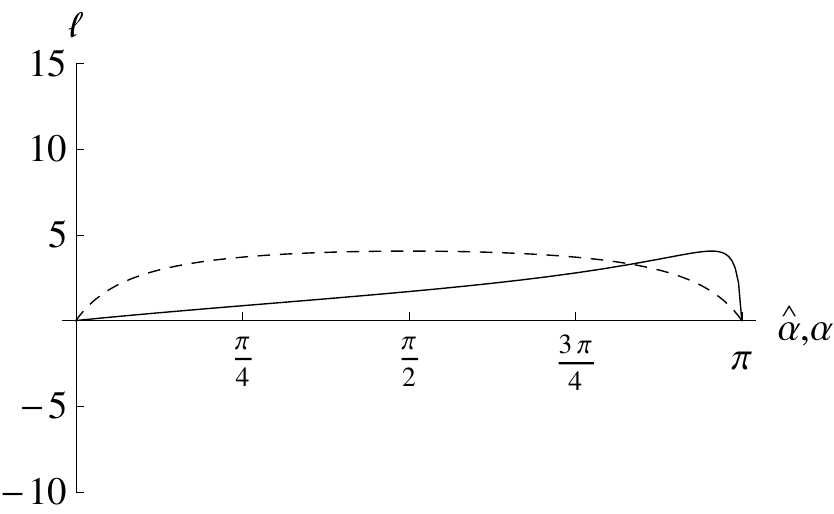}&\includegraphics[scale=0.7]{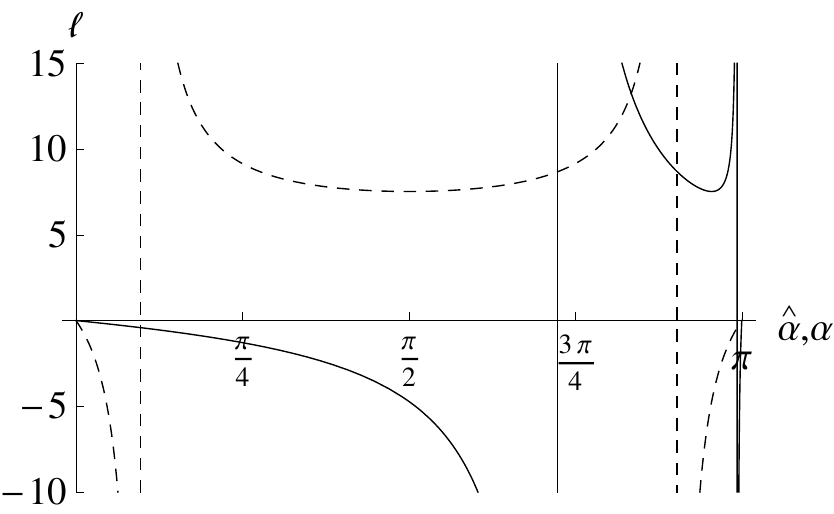}\\
				\includegraphics[scale=0.7]{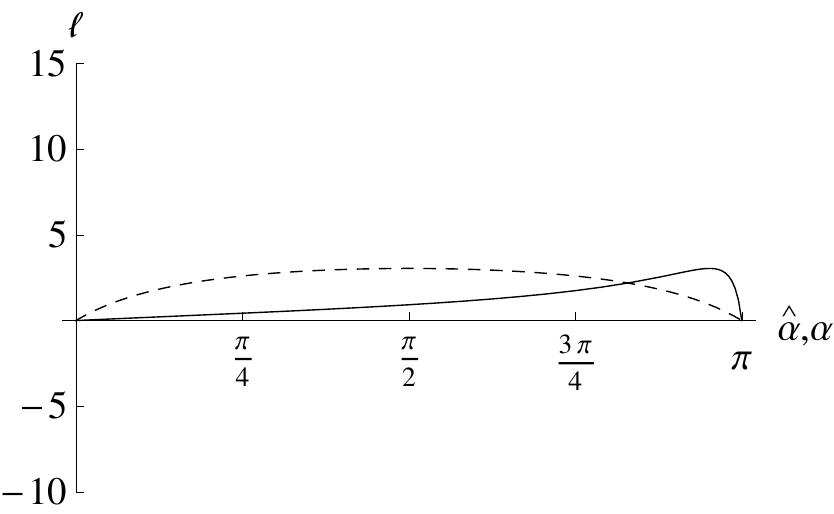}&\includegraphics[scale=0.7]{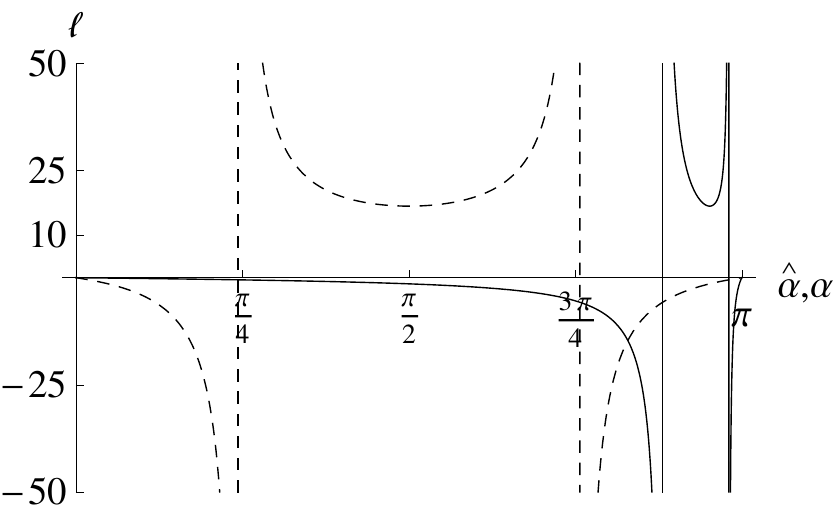}\\
				\includegraphics[scale=0.7]{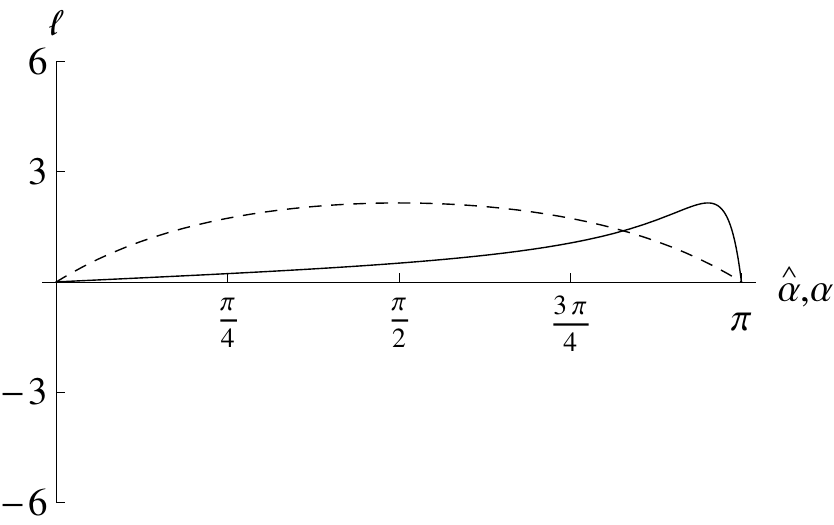}&\includegraphics[scale=0.7]{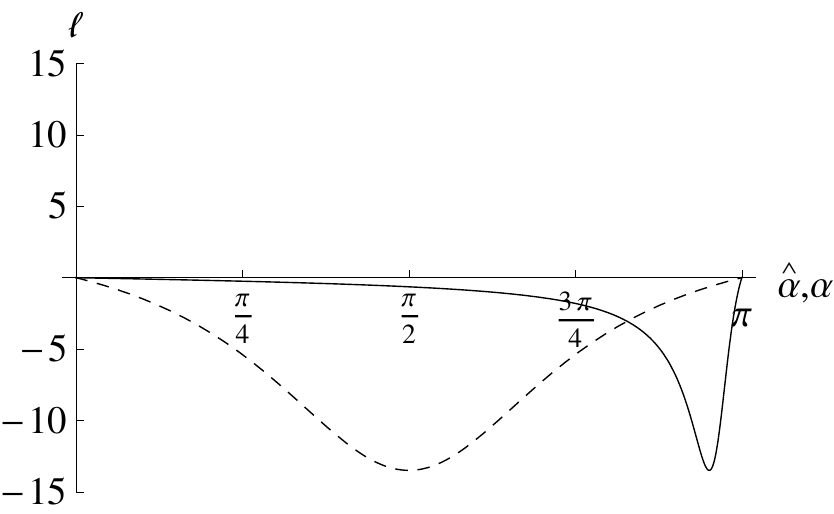}\\
				
			\end{tabular}
			\caption{Dependence of the impact parameter $\ell$ on the directional angle $(\hat{\alpha})$ (full curves) related to the observers radially escaping in the KdS spacetime with $y=0.04,$ $a^2=0.9$, at instantaneous radial position $r_{o}=2$, and at fixed latitude $\theta_o=90\,^{\circ}$-equatorial plane (upper row), $\theta_{o}=30\,^{\circ}$ (middle row) and $\theta_{o}=15\,^{\circ}$ (lower row). The left column is generated for $\hat{\beta}=90\,^{\circ}$, the right one holds for $\hat{\beta}=270\,^{\circ}.$ It is compared with the corresponding dependence of $\ell$ on the angle $\alpha$ measured by the LNRF located momentarily at the same position (dashed curves). The critical latitude $\theta_{c}=20.7\,^{\circ},$ for which the discontinuity appears, is common for both observers. In LNRFs, the curves are symmetric, having the maximum at $\alpha = \pi/2$, but in RGFs the local extrema are shifted to angles $\alpha_{ex}=\arccos (-v).$ The elementary properties of trigonometric functions in (\ref{ell(alpha)RGF}) imply $\ell(\hat{\alpha},-v)\equiv\ell(\pi-\hat{\alpha},v),$ hence, in the case of the falling observers, the graphs of $\ell(\hat{\alpha},v)$ are symmetric to those of $\ell(\pi-\hat{\alpha},-v),$ with respect to the vertical $\hat{\alpha}=\pi/2.$  }\label{fig_ell(alpha)}
		\end{figure}

 
	The expression (\ref{cosalpha}) can be regarded as an implicit function of the impact parameter $X,$ or $\ell,$ respectively, so it is worth to deal with an inverse relation determining the dependence of the impact parameter $\ell$ on the directional angles. Following the same procedure as in the LNRF case, and after performing the appropriate transformations, we obtain
	\be
	\ell=\frac{A\sin\hat{\alpha}\sin\hat{\beta}\sin\theta}{\gamma\rho^2\sqrt{\Delta_{r}\Delta_{\theta}}(1+v\cos\hat{\alpha})+\Omega A\sin\hat{\alpha}\sin\hat{\beta}\sin\theta }. \label{ell(alpha)RGF}
	\ee
	We present the behaviour of this function in Fig. \ref{fig_ell(alpha)} for the case of the radially escaping observers.\par 
	
\begin{figure}[H]
	\flushleft Class III: $y=0.04,\quad a^2=0.9$\\
	\begin{tabular}{|cc|}
		\hline
		\multicolumn{2}{|c|}{}\\	
		\multicolumn{2}{|c|}{$r_{e}=3.2$}\\
		\includegraphics[width=5cm]{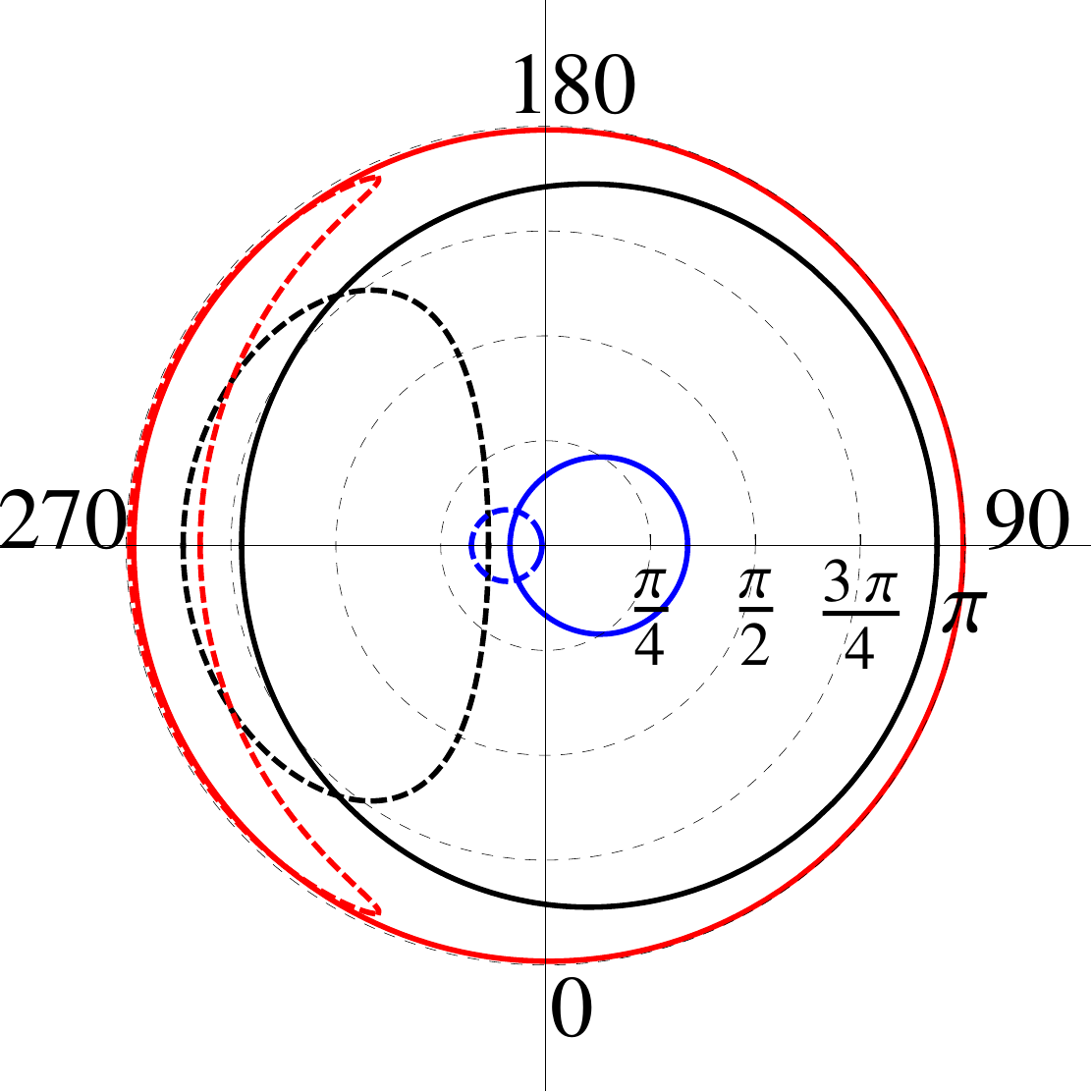}&\includegraphics[width=5cm]{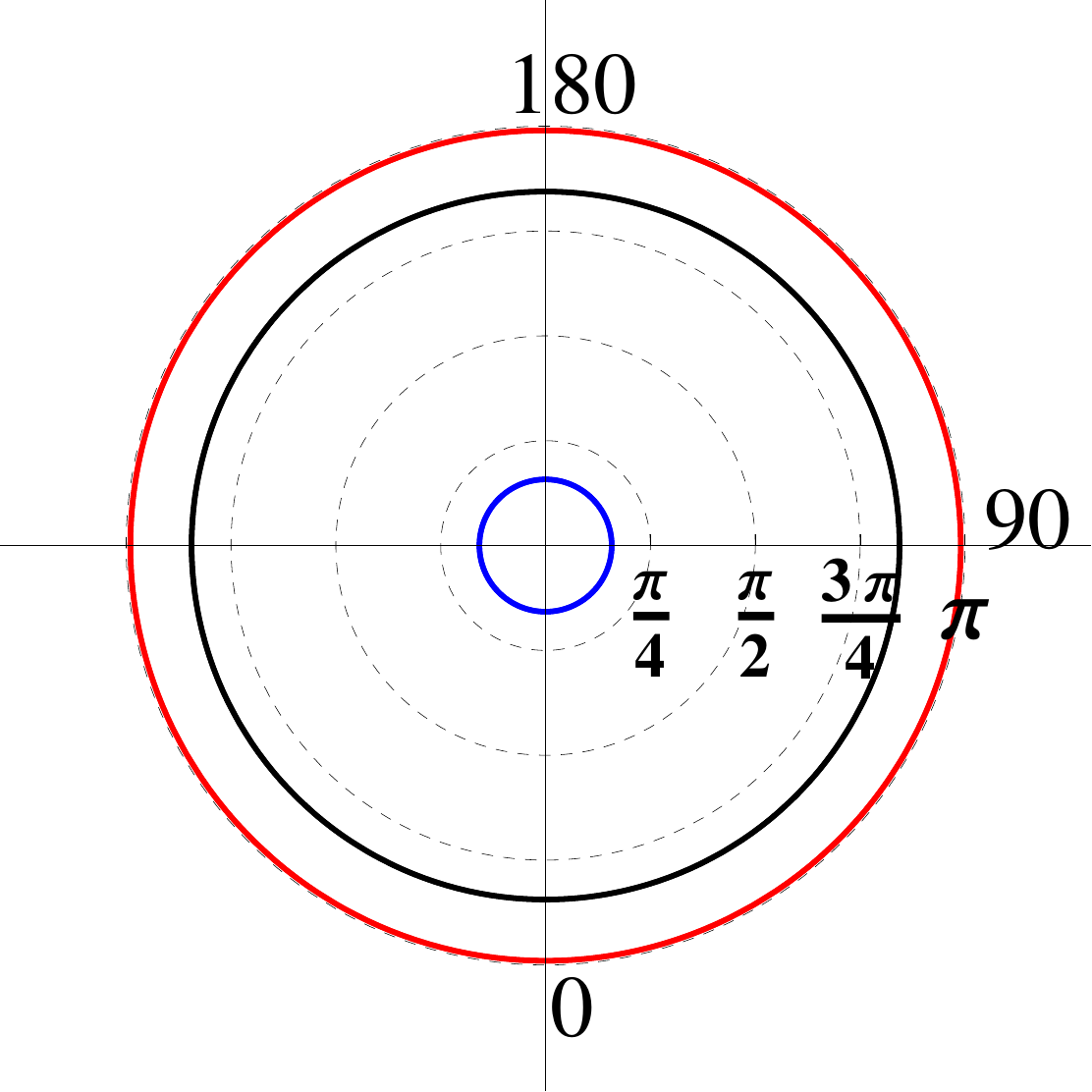}\\
		\hline
	\end{tabular}
	\caption{Light escape cones as seen by the radially falling (blue curves) and the radially escaping observers (red curves), compared with the related local escape cones of the locally non-rotating observers (black curves). Inner parts of these curves demarcate escaping photons. The dashed curves, if present, delimit flux transmitted by the retrograde photons with positive impact parameter $X$ and negative energy $E.$ Directional angle $\alpha$ is measured in radians, directional angle $\beta$ in degrees. The radial coordinate of $r_{e}$ runs significant values in attached table. Indistinguishable 'central' parts were zoomed in the right upper corner. Depiction in the left lower corner, if present, is the sight 'backwards' in negative azimuthal direction, showing the behaviour of the light escape cones in the LNRF in the vicinity of radii $r_{d(ex)}.$ Significance of directional angle $\gamma$ follows from (\ref{k_phi}).} \label{fig_RGF_cones}
\end{figure}

	The light escape cones in the radially moving frames are constructed using the same procedure as those applied and described in details in the LNRF case, and compared with the cones constructed in the LNRF orbiting momentarily at the same position. The results are illustrated for typical situations in Fig. \ref{fig_RGF_cones}. The KdS spacetime parameters are chosen so that they correspond to the three classes of the \KdS\ black hole spacetimes,  differing by the rank of outstanding radii relevant for the character of the photon motion as introduced and described in \cite{Char-Stu:2017:EPJC:}. At a given position, both falling and escaping RGFs are considered. The light escape cones are in all the considered cases represented by the 2D diagrams in the ($\alpha - \beta$) plane. The pure radial inward direction corresponds to $\alpha=0$. 
	
	\begin{figure}[H]
		\flushleft \textbf{Class I: $y=0.01,\quad a^2=0.1$}\\
		
		\bet{lccr}
		\hline
		\bet{l}$r_{o}=2.0376$\\$r_{c}=8.7908$\ent & \bet{l}$r_{ph+}=2.6524$\\$r_{ph-}=3.2838$ \ent &\bet{l} $r_{d1}=2.0939$\\$r_{d2}=8.7822$\ent & \bet{l} $r_{pol}=2.9517$ \ent
		\ent
		
		\begin{tabular}{|cc|}
			\hline
			\multicolumn{2}{|c|}{}\\
			\multicolumn{2}{|c|}{$r_{e}=2.05$}\\ \includegraphics[width=5cm]{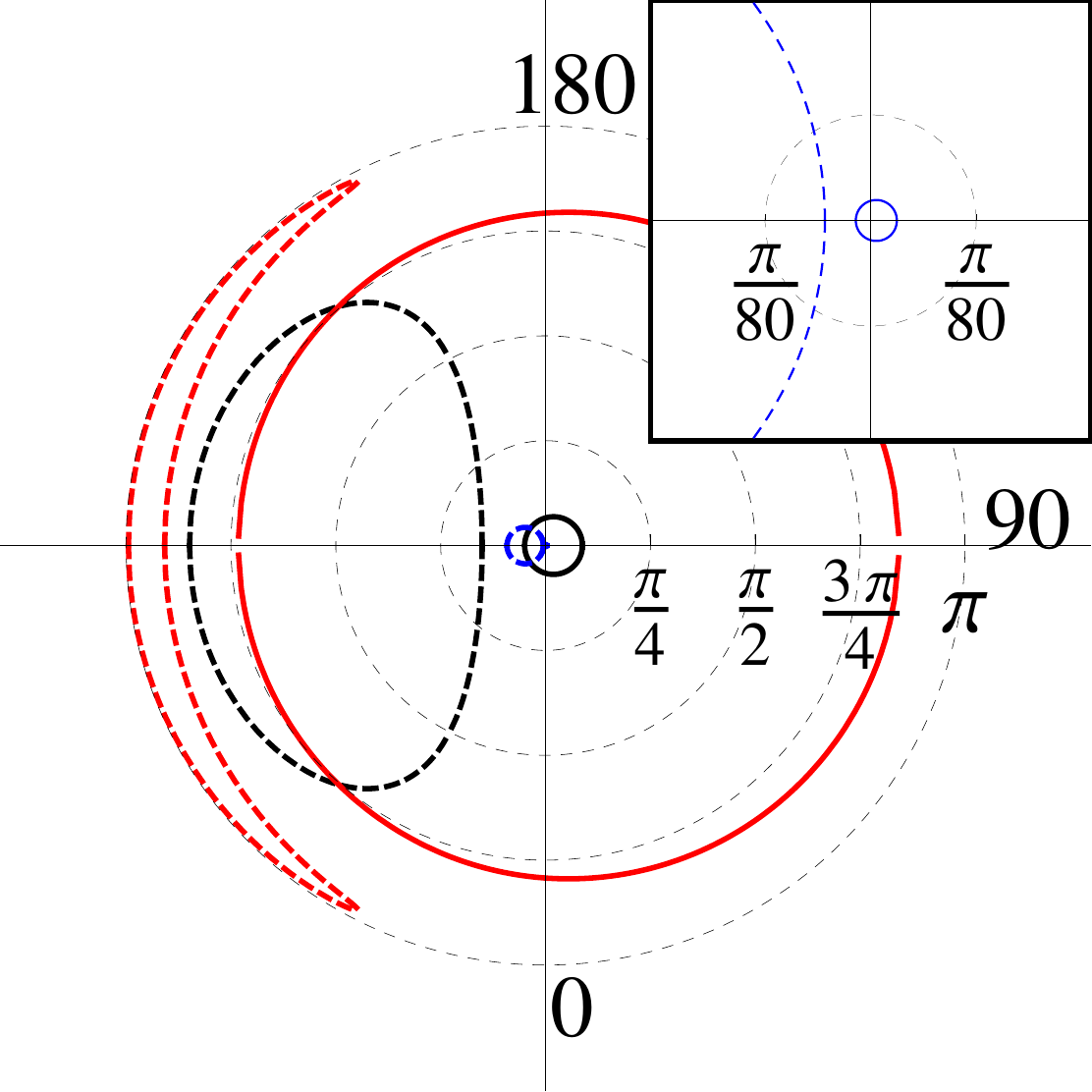}&\includegraphics[width=5cm]{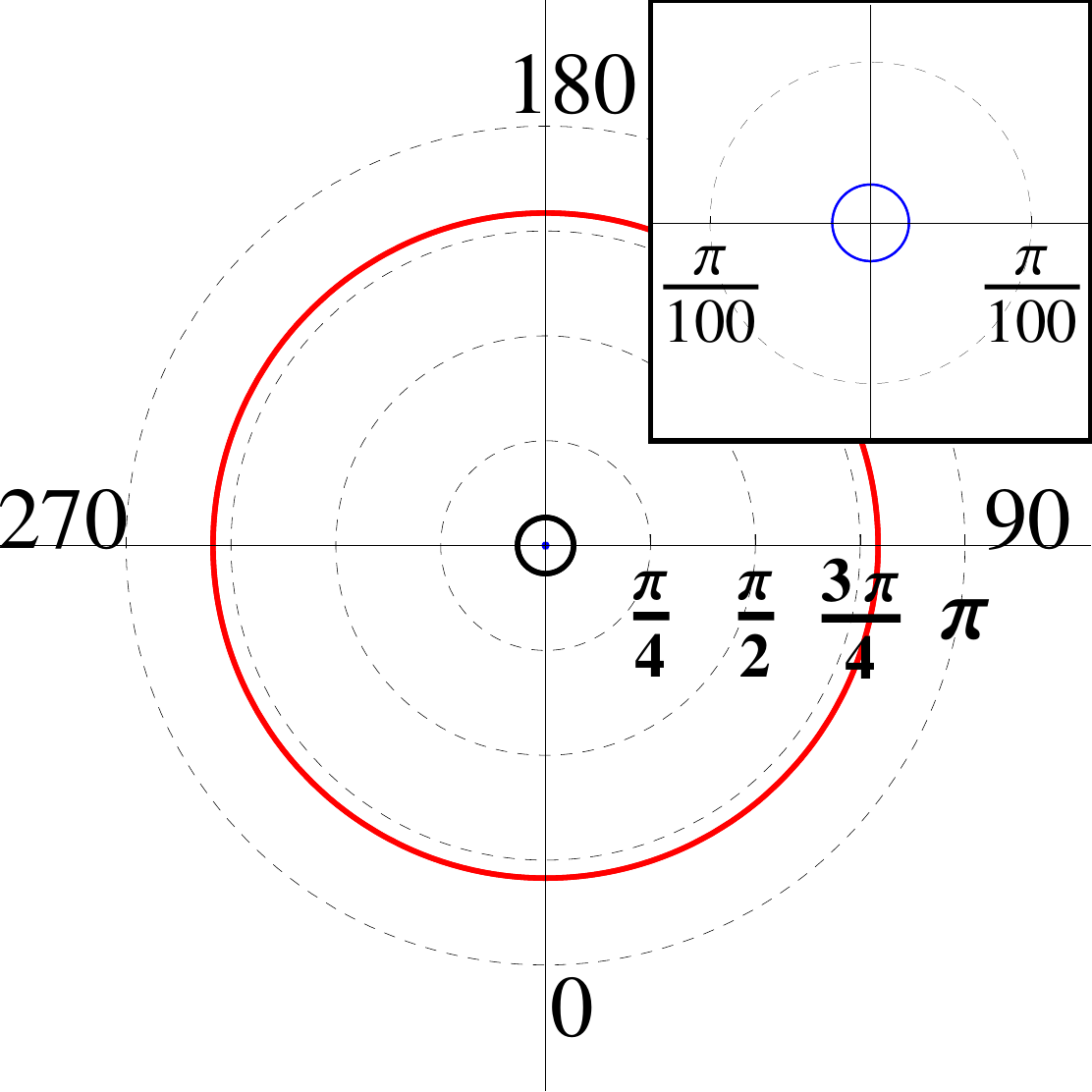}\\
			\hline
				
			\multicolumn{2}{|c|}{}\\
			\multicolumn{2}{|c|}{$r_{e}=2.5$}\\
			\includegraphics[width=5cm]{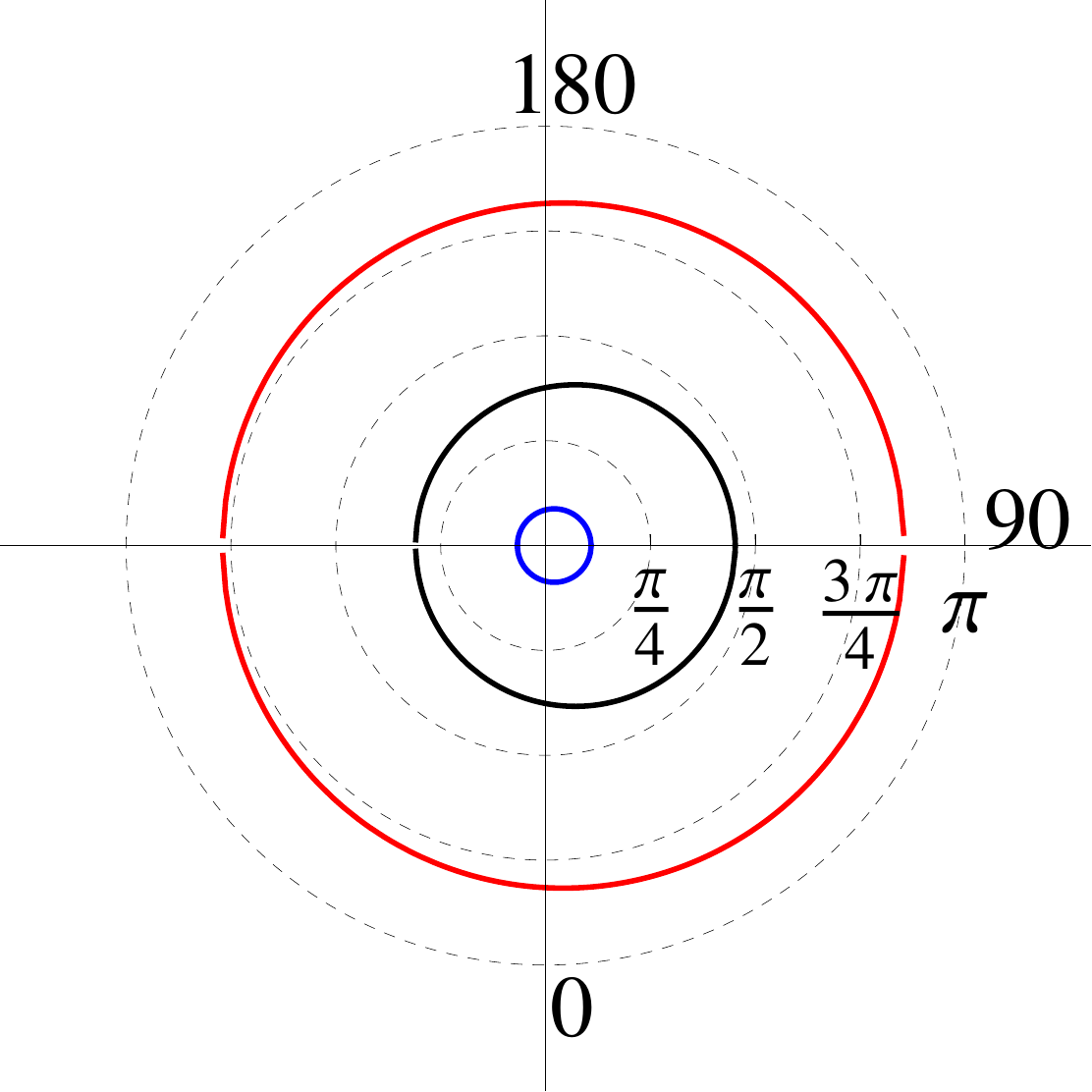}&\includegraphics[width=5cm]{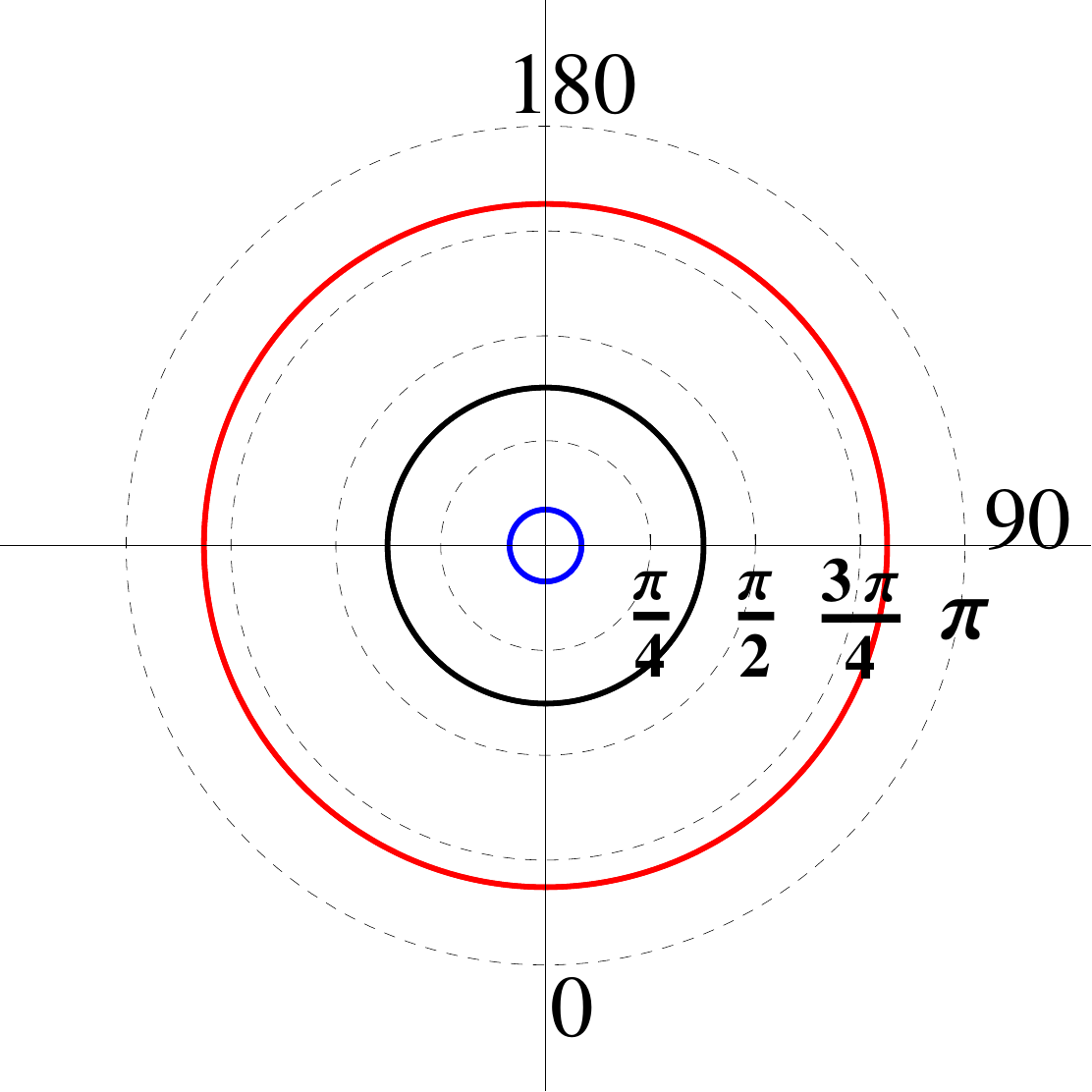}\\
				
			\hline
			\multicolumn{2}{|c|}{}\\
			\multicolumn{2}{|c|}{$r_{e}=r_{ph+}$}\\
			\includegraphics[width=5cm]{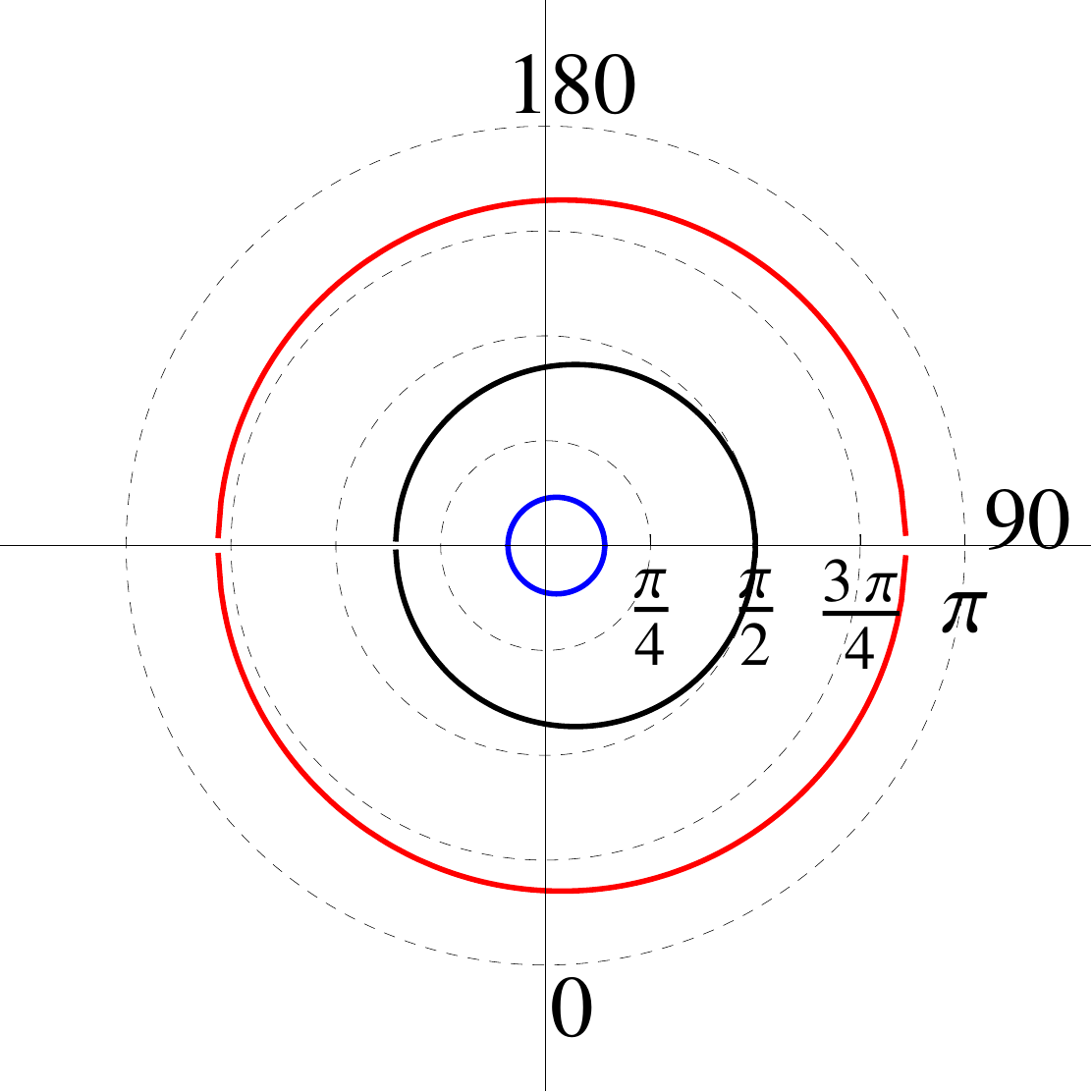}&\includegraphics[width=5cm]{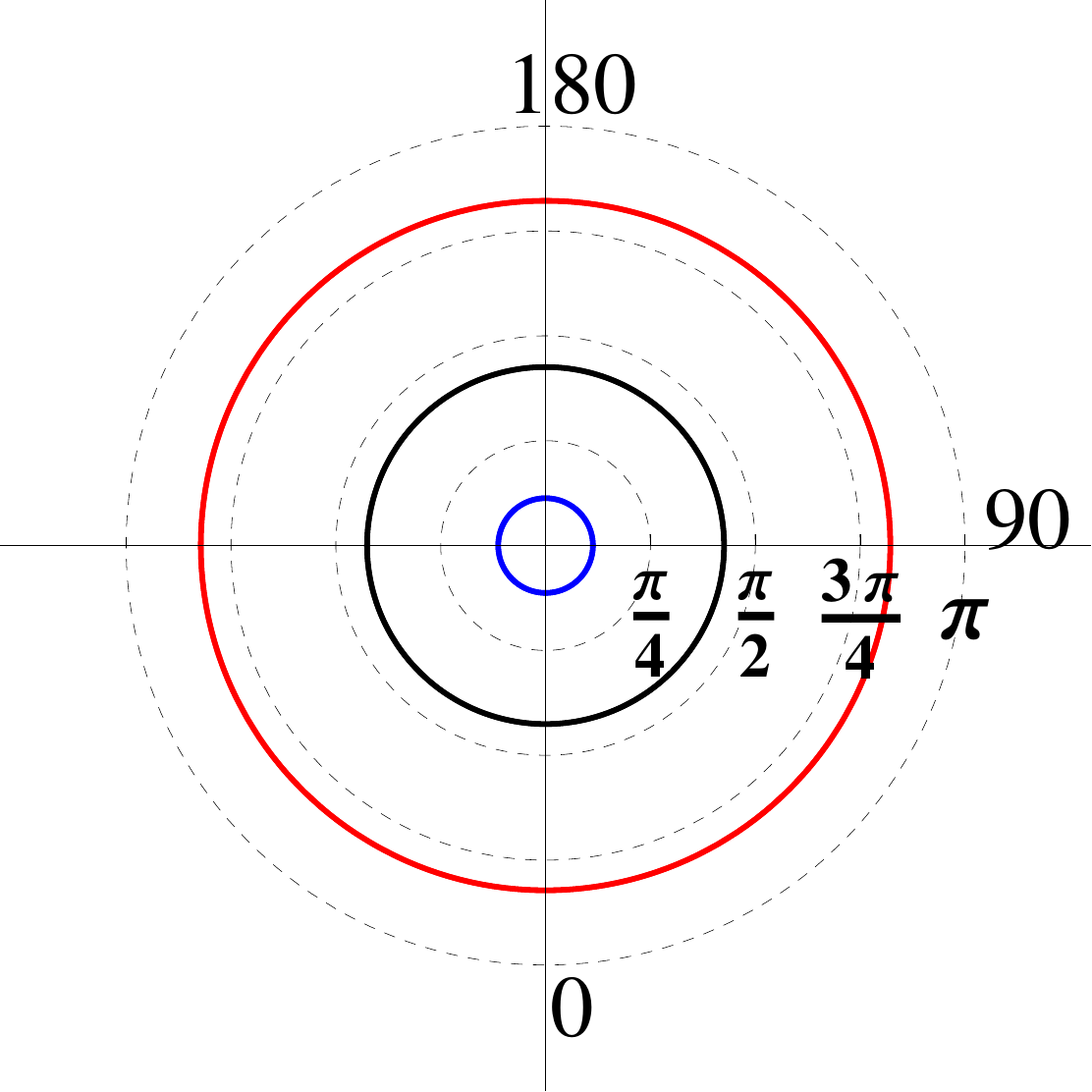}\\
			\hline
		\end{tabular}
			\center (\textit{Figure continued}) 
	\end{figure}
	
	\begin{figure}[H]
		
		\flushleft Class I: $y=0.01,\quad a^2=0.1$\\
		\bet{lccr}
		\hline
		\bet{l}$r_{o}=2.0376$\\$r_{c}=8.7908$\ent & \bet{l}$r_{ph+}=2.6524$\\$r_{ph-}=3.2838$ \ent &\bet{l} $r_{d1}=2.0939$\\$r_{d2}=8.7822$\ent & \bet{l} $r_{pol}=2.9517$ \ent
		\ent
		
		\begin{tabular}{|cc|}
						
			\hline
			\multicolumn{2}{|c|}{}\\
			\multicolumn{2}{|c|}{$r_{e}=r_{pol}$}\\
			\includegraphics[width=5cm]{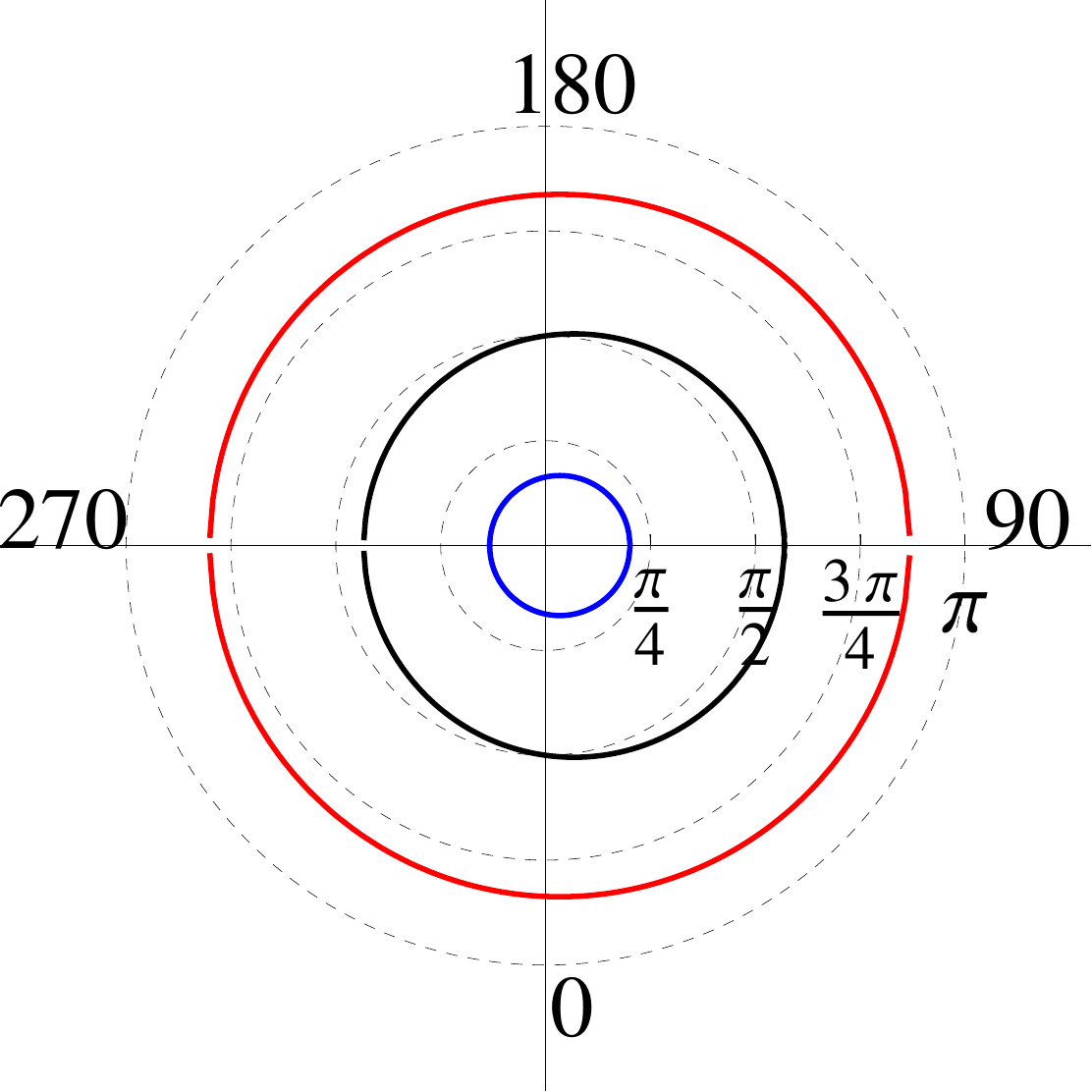}&\includegraphics[width=5cm]{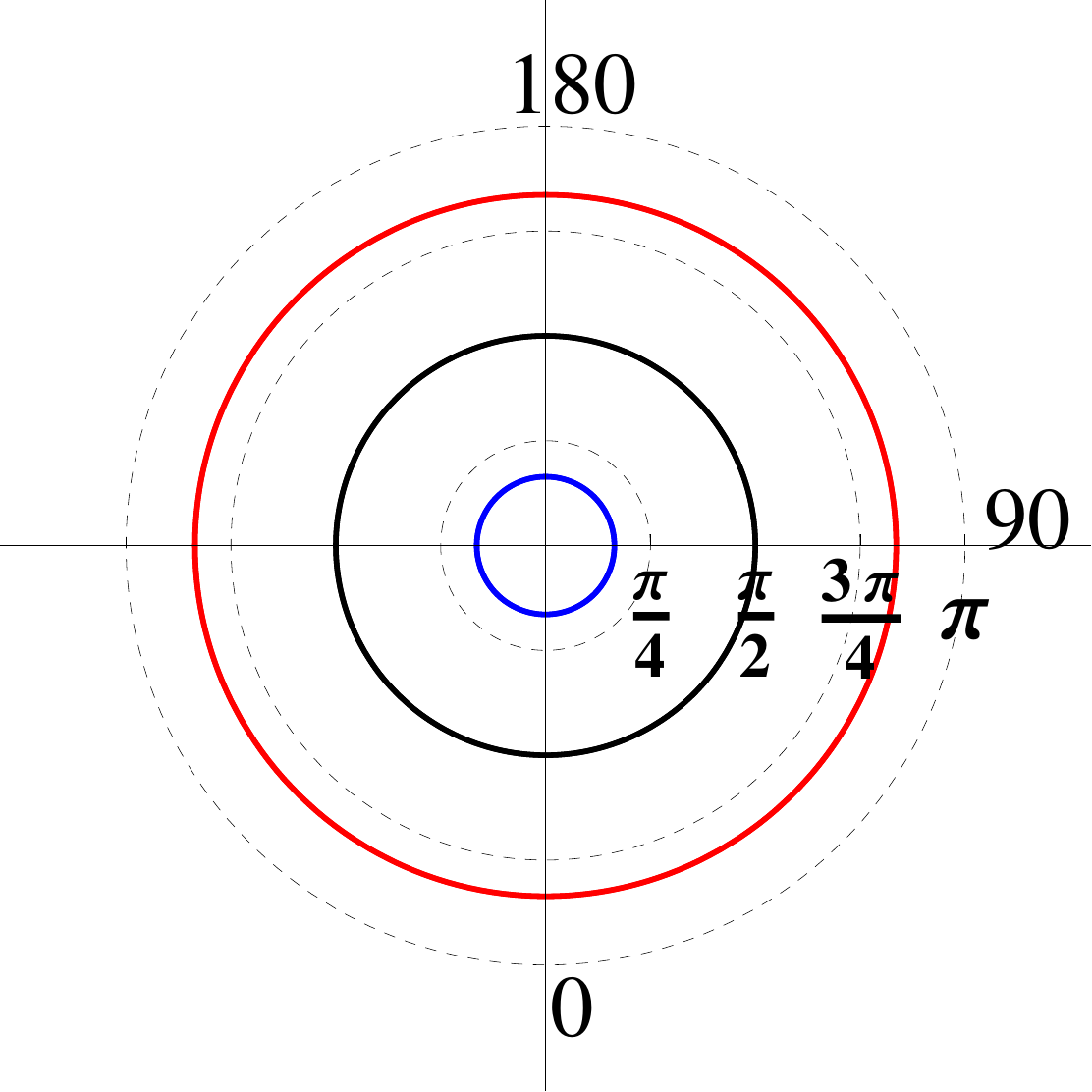}\\	   		
			\hline
			\multicolumn{2}{|c|}{}\\
			\multicolumn{2}{|c|}{$r_{e}=r_{ph-}$}\\
			\includegraphics[width=5cm]{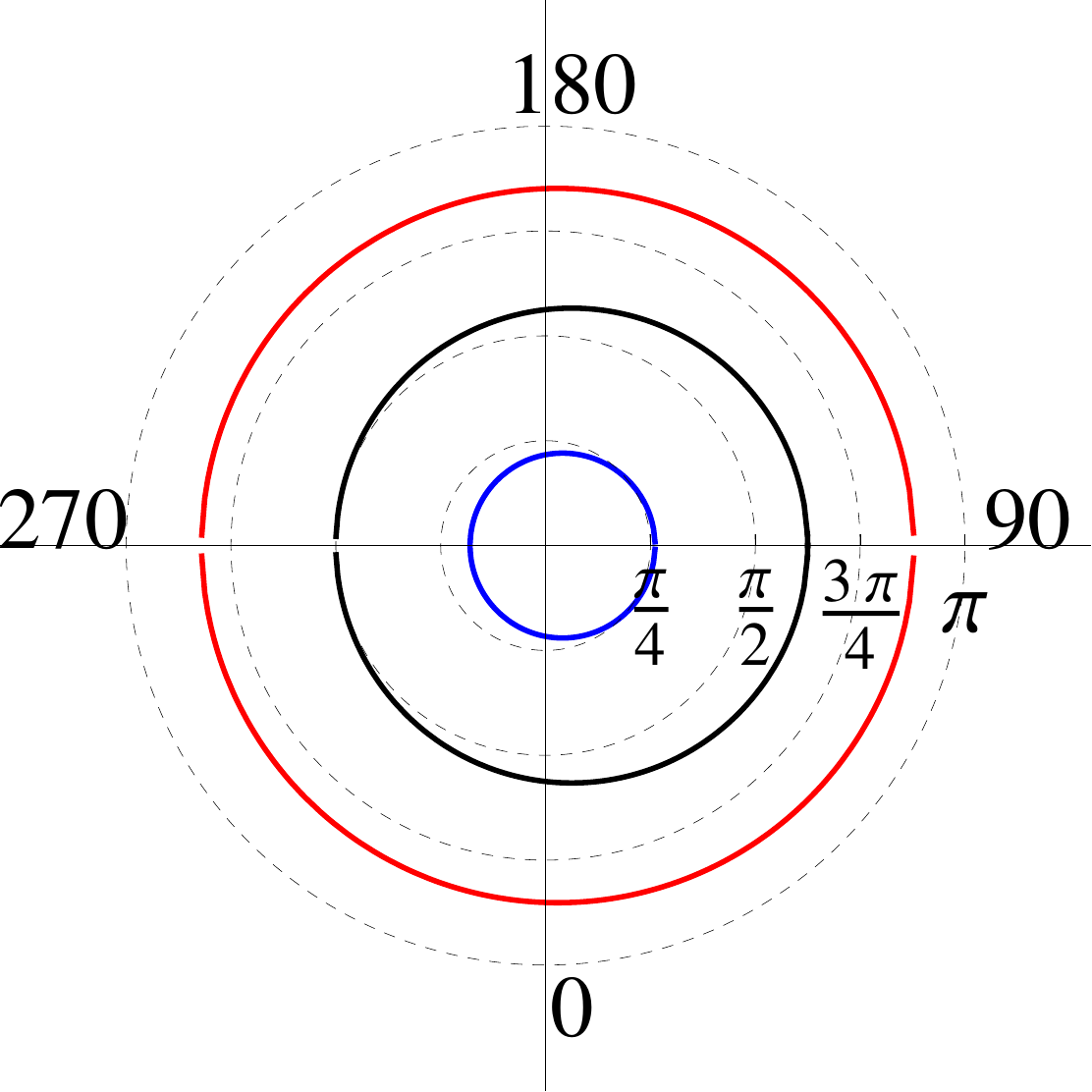}&\includegraphics[width=5cm]{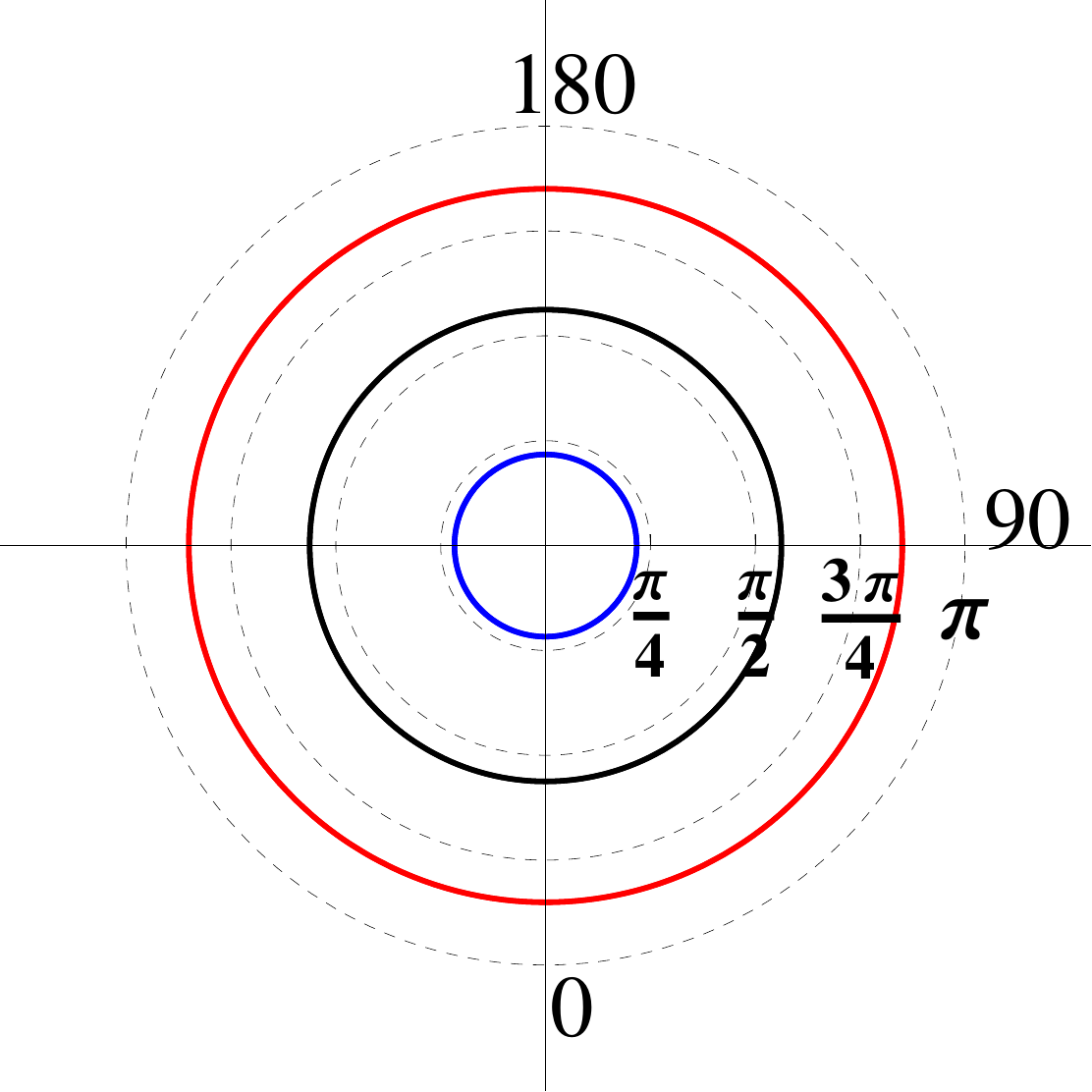}\\
			\hline
			\multicolumn{2}{|c|}{}\\
			\multicolumn{2}{|c|}{$r_{e}=5$}\\
			\includegraphics[width=5cm]{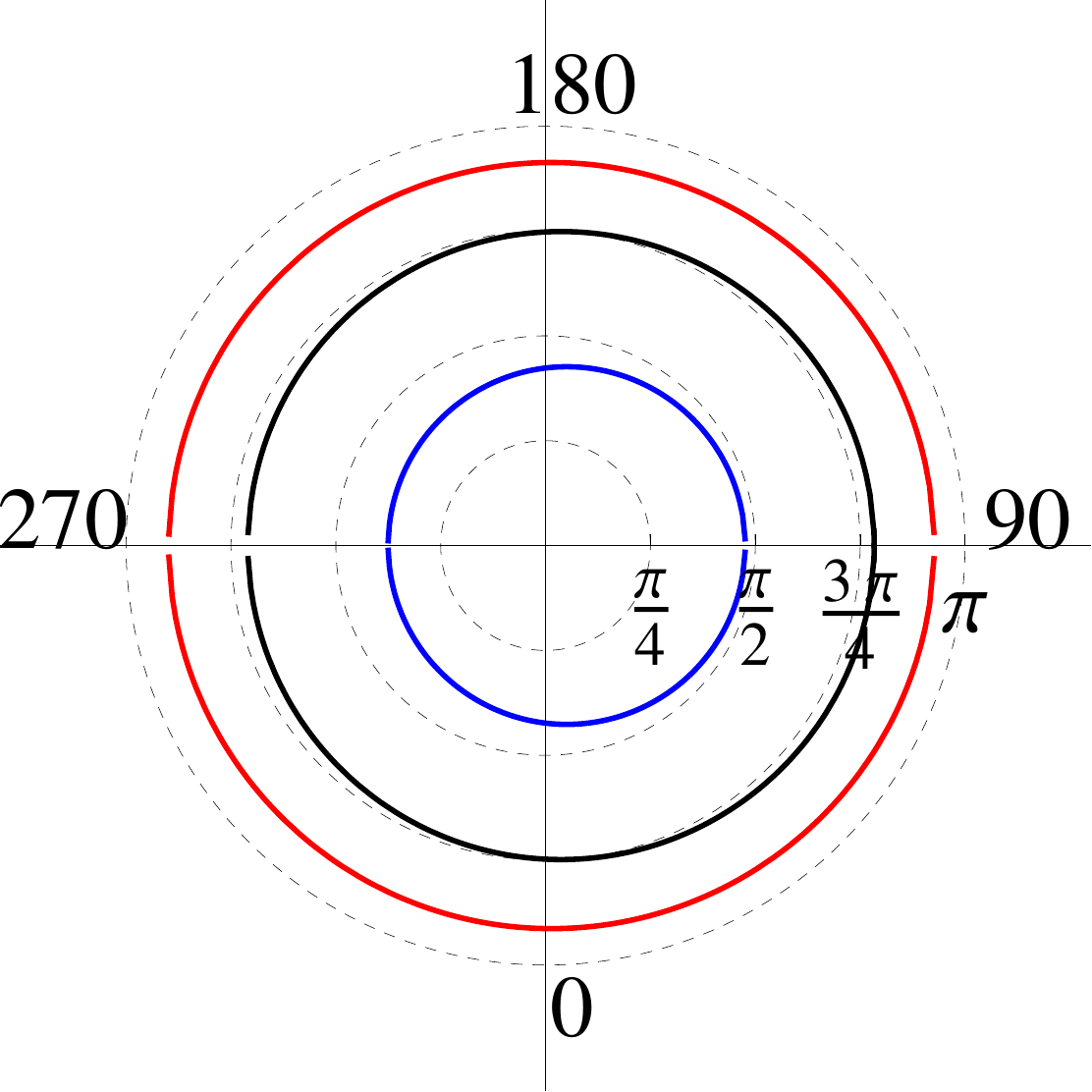}&\includegraphics[width=5cm]{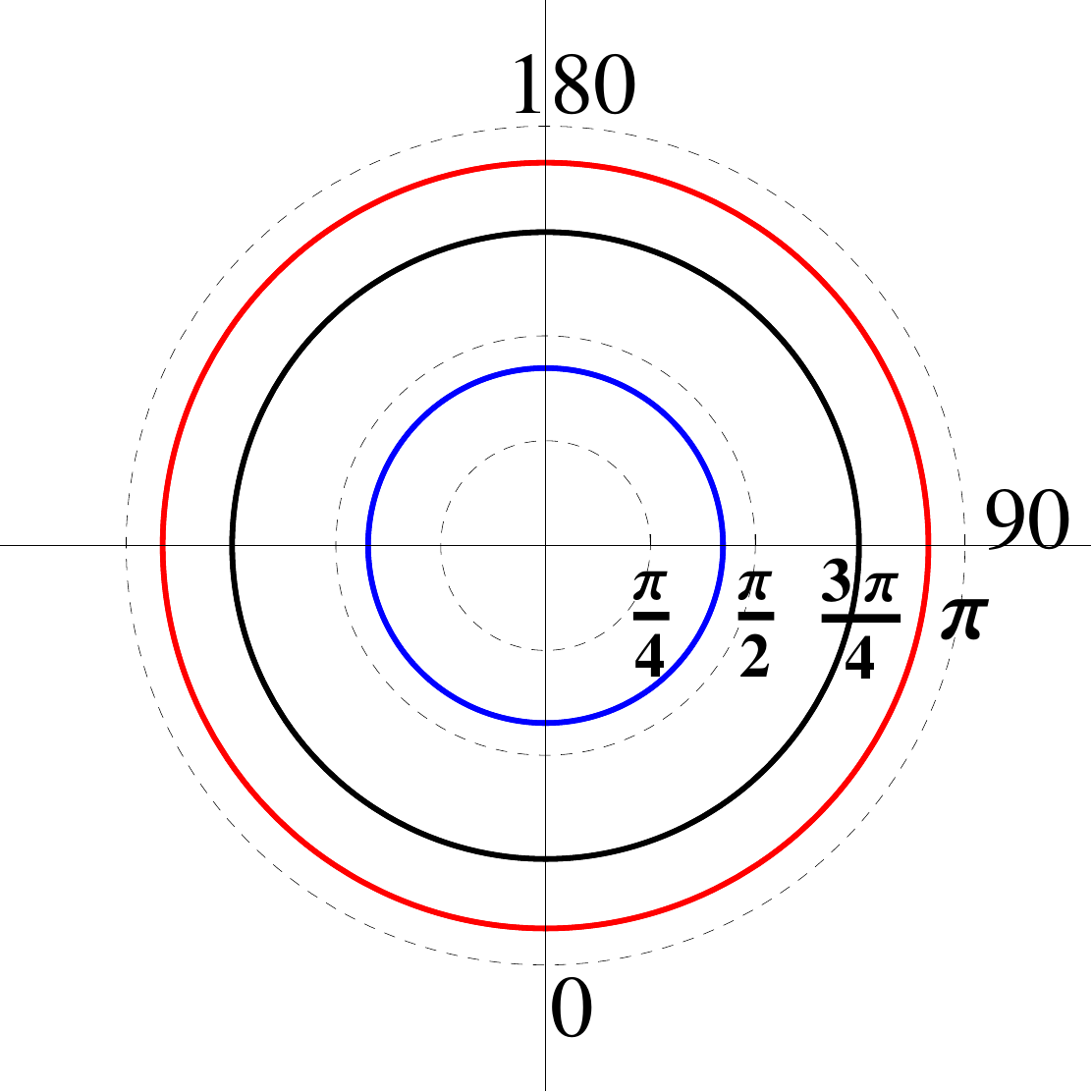}\\
			\hline

		\end{tabular}
		\center (\textit{Figure continued})	   		
	\end{figure}

\begin{figure}[H]
		\flushleft Class I: $y=0.01,\quad a^2=0.1$\\
		\bet{lccr}
		\hline
		\bet{l}$r_{o}=2.0376$\\$r_{c}=8.7908$\ent & \bet{l}$r_{ph+}=2.6524$\\$r_{ph-}=3.2838$ \ent &\bet{l} $r_{d1}=2.0939$\\$r_{d2}=8.7822$\ent & \bet{l} $r_{pol}=2.9517$ \ent
		\ent
		
		\begin{tabular}{|cc|}
			
			\hline
			\multicolumn{2}{|c|}{}\\
			\multicolumn{2}{|c|}{$r_{e}=8.785$}\\
			\includegraphics[width=5cm]{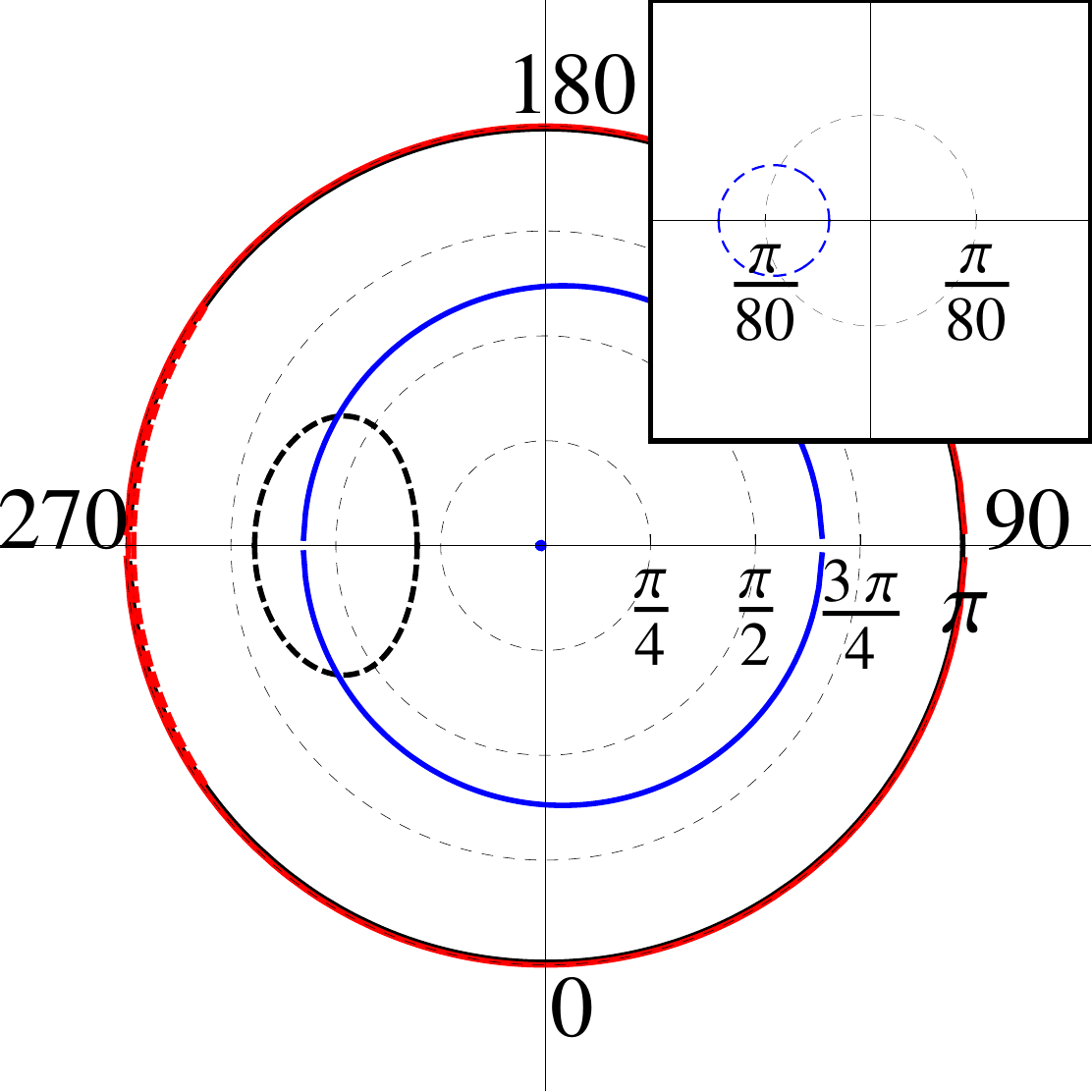}&\includegraphics[width=5cm]{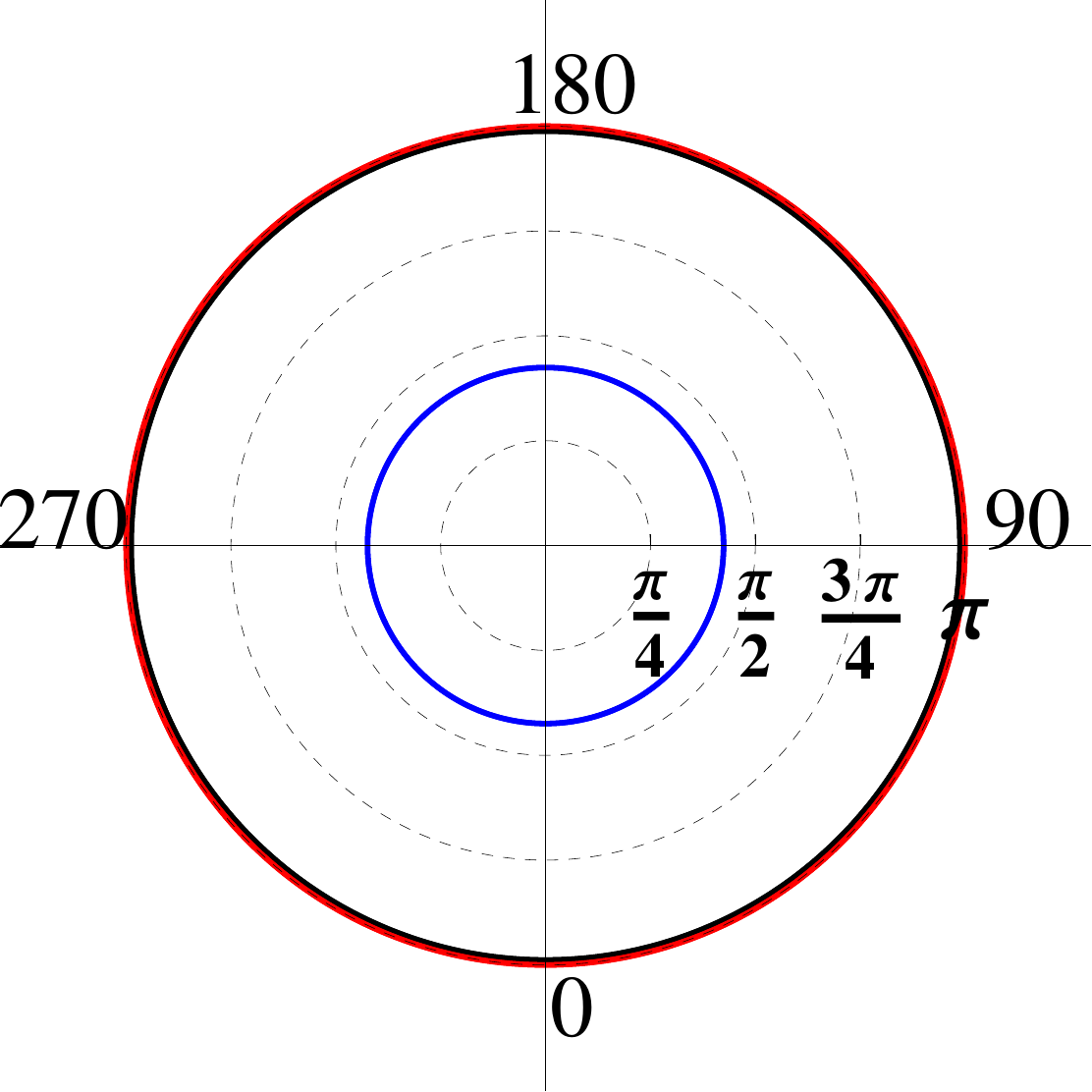}\\
			\hline
			
		\end{tabular}
	\vspace{2cm}
	\flushleft \textbf{Class II: $y=0.02,\quad a^2=0.9$}\\
	
	\bet{lccr}
	
	\hline
	\bet{l}$r_{o}=1.4876$\\$r_{c}=5.7508$\ent & \bet{l}$r_{ph+}=1.5794$\\$r_{ph-}=3.4728$ \ent&\bet{l} $r_{d1}=2.2772$\\$r_{d2}=5.5853$\ent &\bet{l} $r_{pol}=2.4257$\ent
	\ent
	
	\begin{tabular}{|cc|}
		\hline
		\multicolumn{2}{|c|}{}\\
		\multicolumn{2}{|c|}{$r_{e}=1.5$}\\
		\includegraphics[width=5cm]{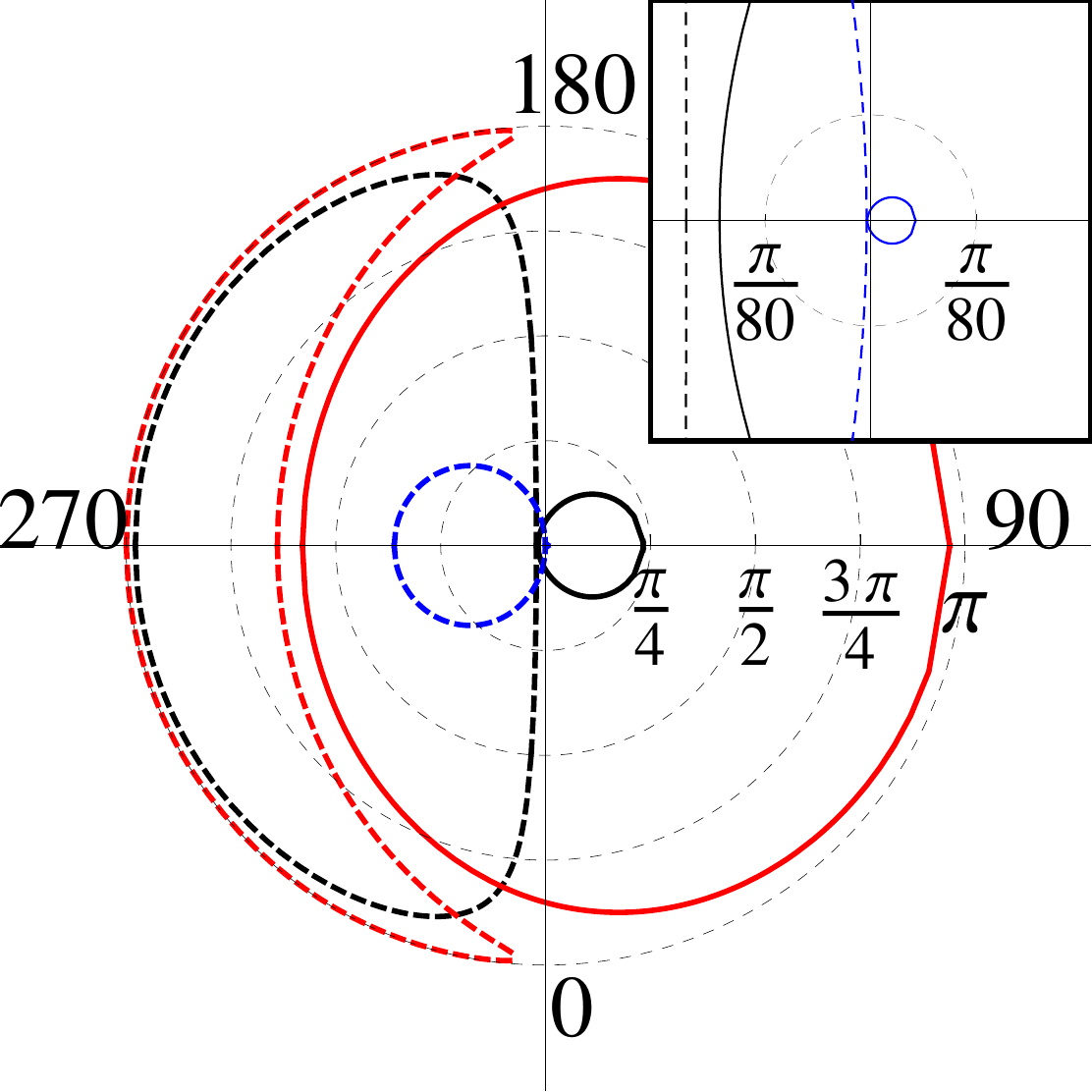}&\includegraphics[width=5cm]{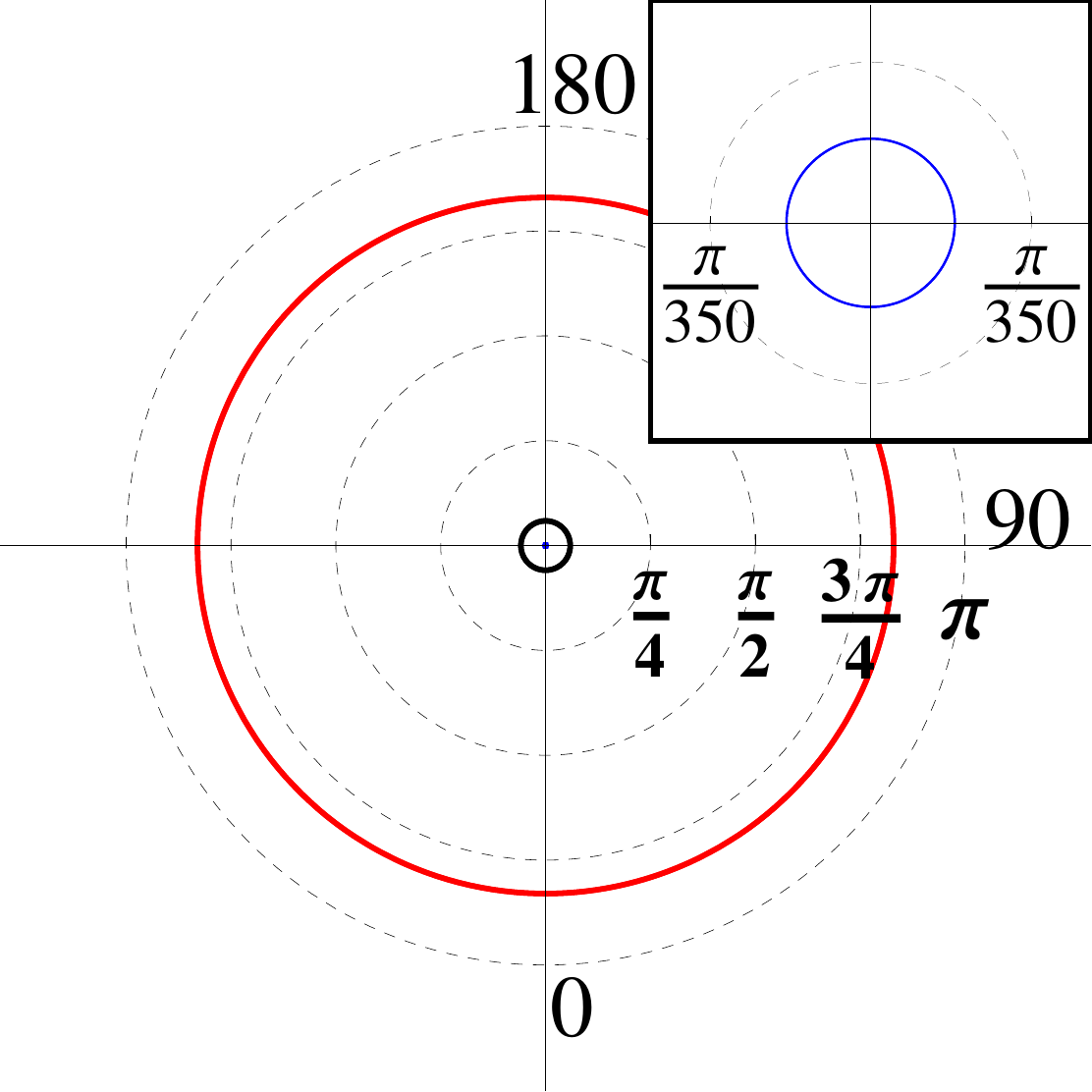}\\
		\hline
	\end{tabular}
	\center (\textit{Figure continued})	 
\end{figure}

\begin{figure}[H]
	\flushleft Class II: $y=0.02,\quad a^2=0.9$\\
	\bet{lccr}
	\hline
	\bet{l}$r_{o}=1.4876$\\$r_{c}=5.7508$\ent & \bet{l}$r_{ph+}=1.5794$\\$r_{ph-}=3.4728$ \ent&\bet{l} $r_{d1}=2.2772$\\$r_{d2}=5.5853$\ent &\bet{l} $r_{pol}=2.4257$\ent
	\ent
	
	\begin{tabular}{|cc|}
		\hline		
		\multicolumn{2}{|c|}{}\\
		\multicolumn{2}{|c|}{$r_{e}=r_{ph+}$}\\
		\includegraphics[width=5cm]{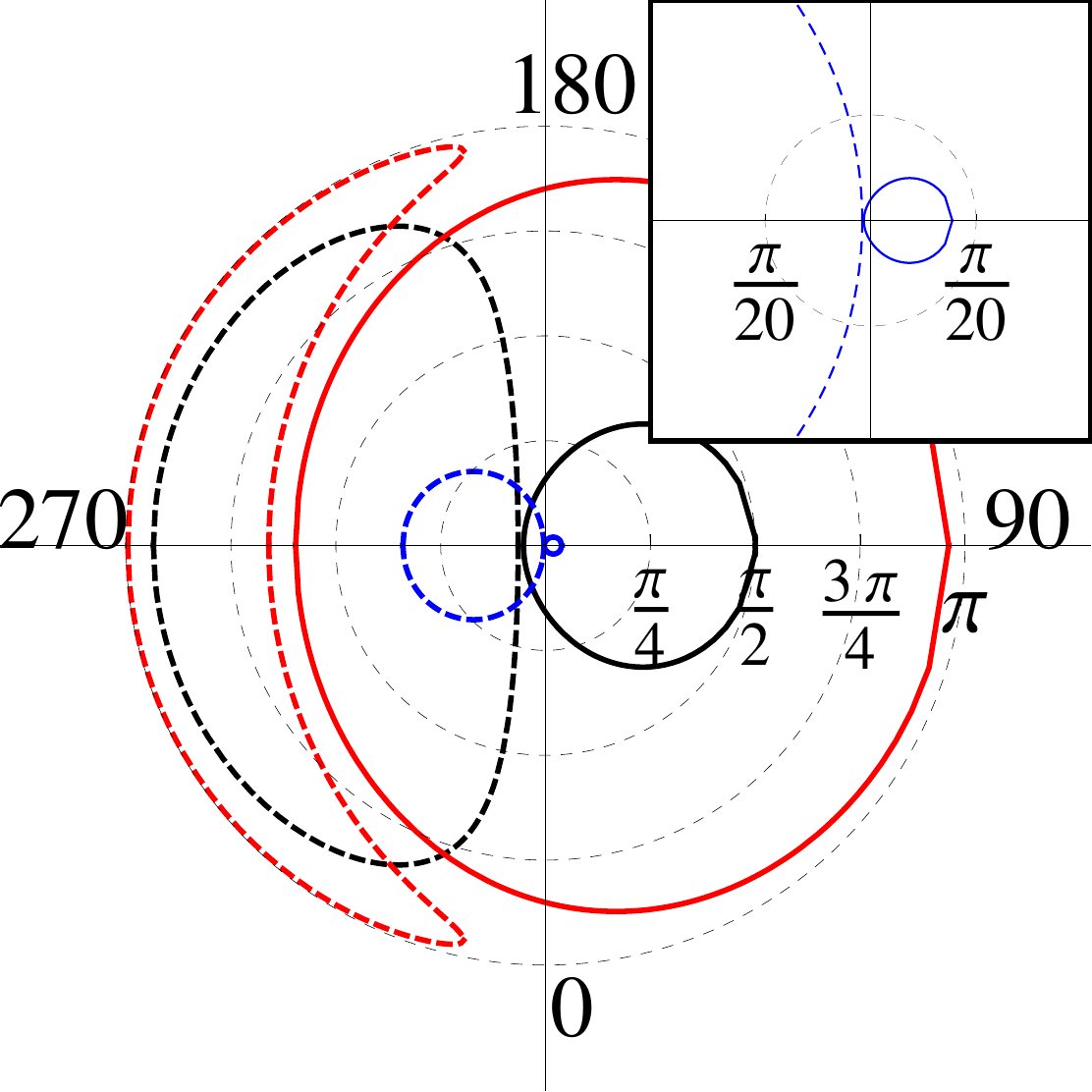}&\includegraphics[width=5cm]{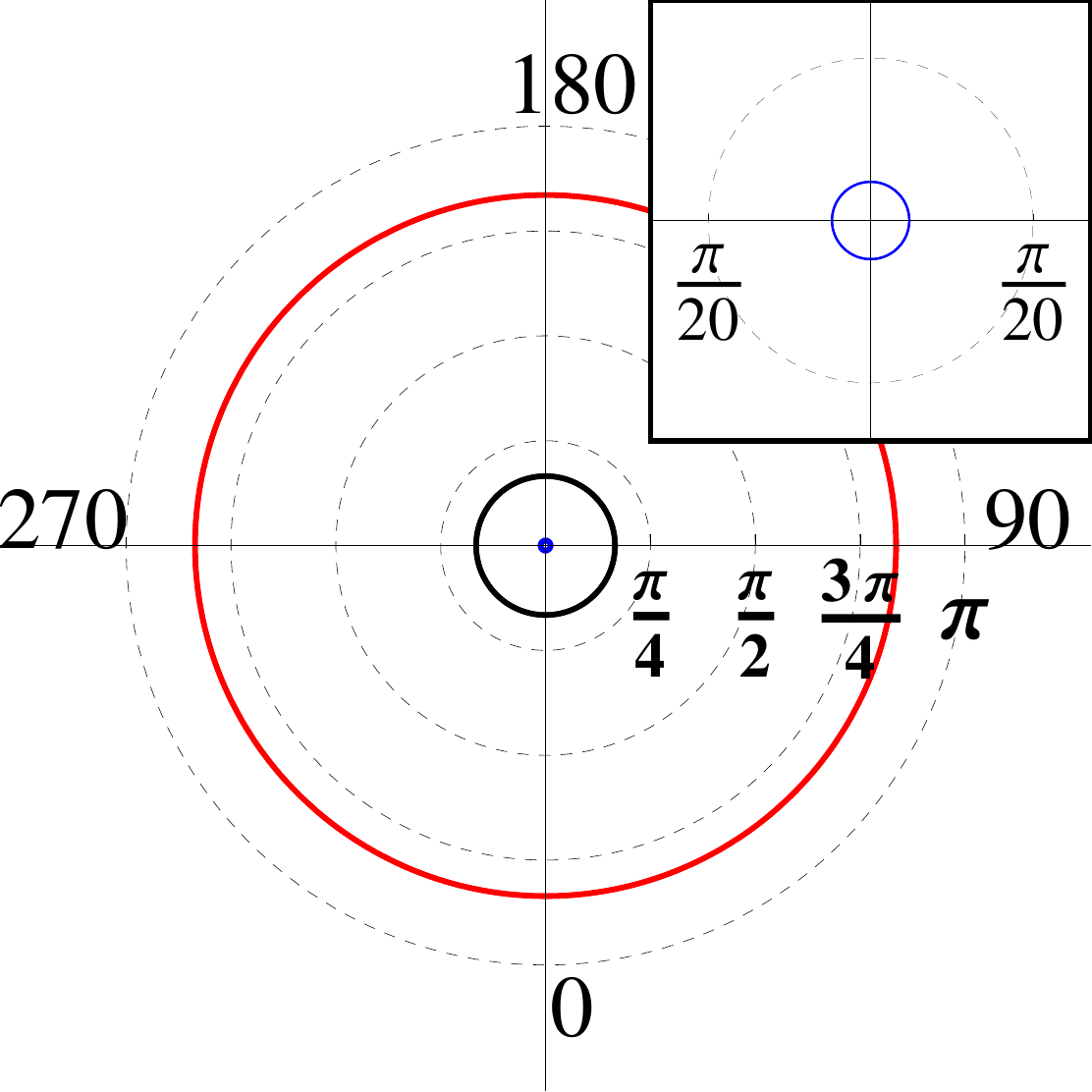}\\
		
		\hline
		\multicolumn{2}{|c|}{}\\
		\multicolumn{2}{|c|}{$r_{e}=1.8$}\\
		\includegraphics[width=5cm]{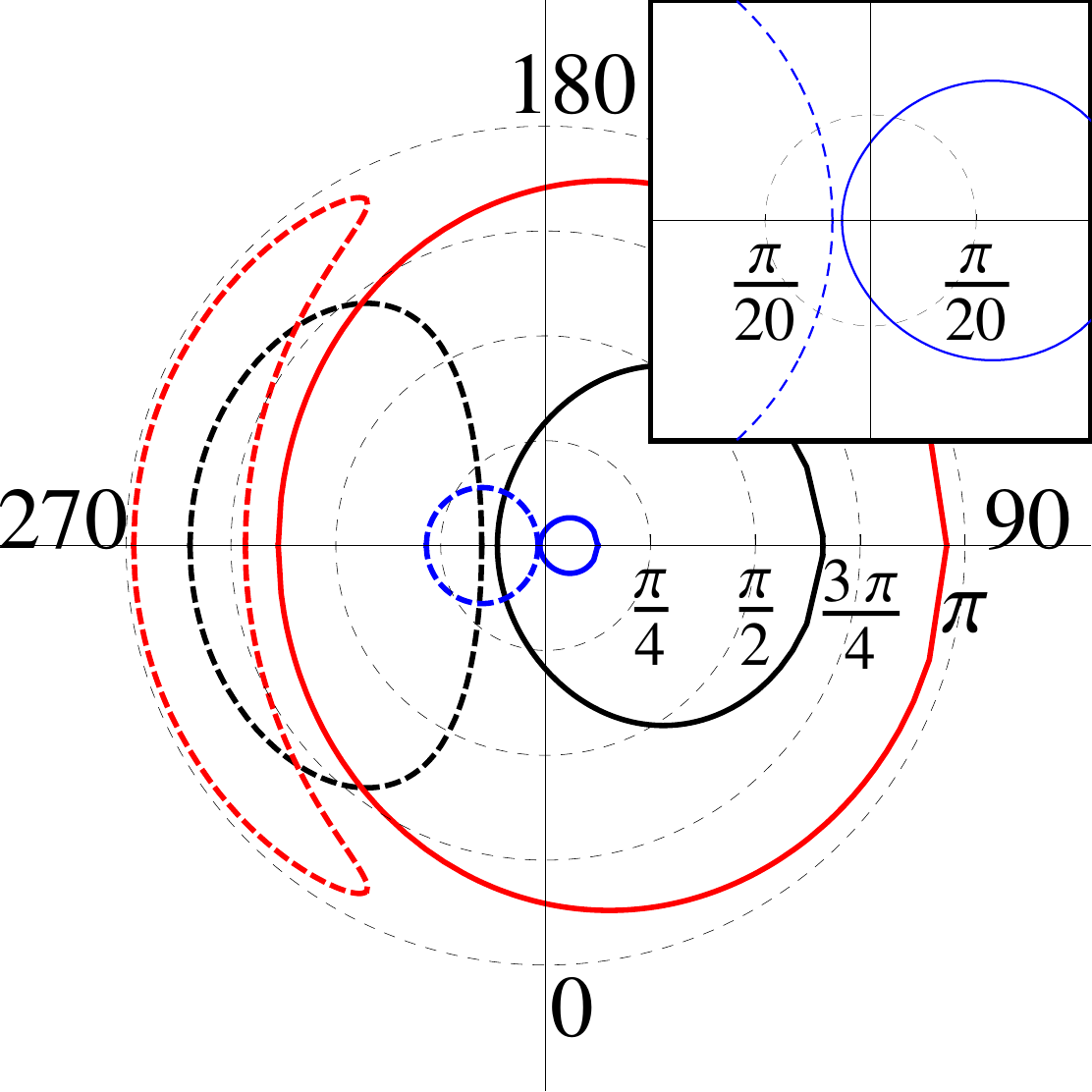}&\includegraphics[width=5cm]{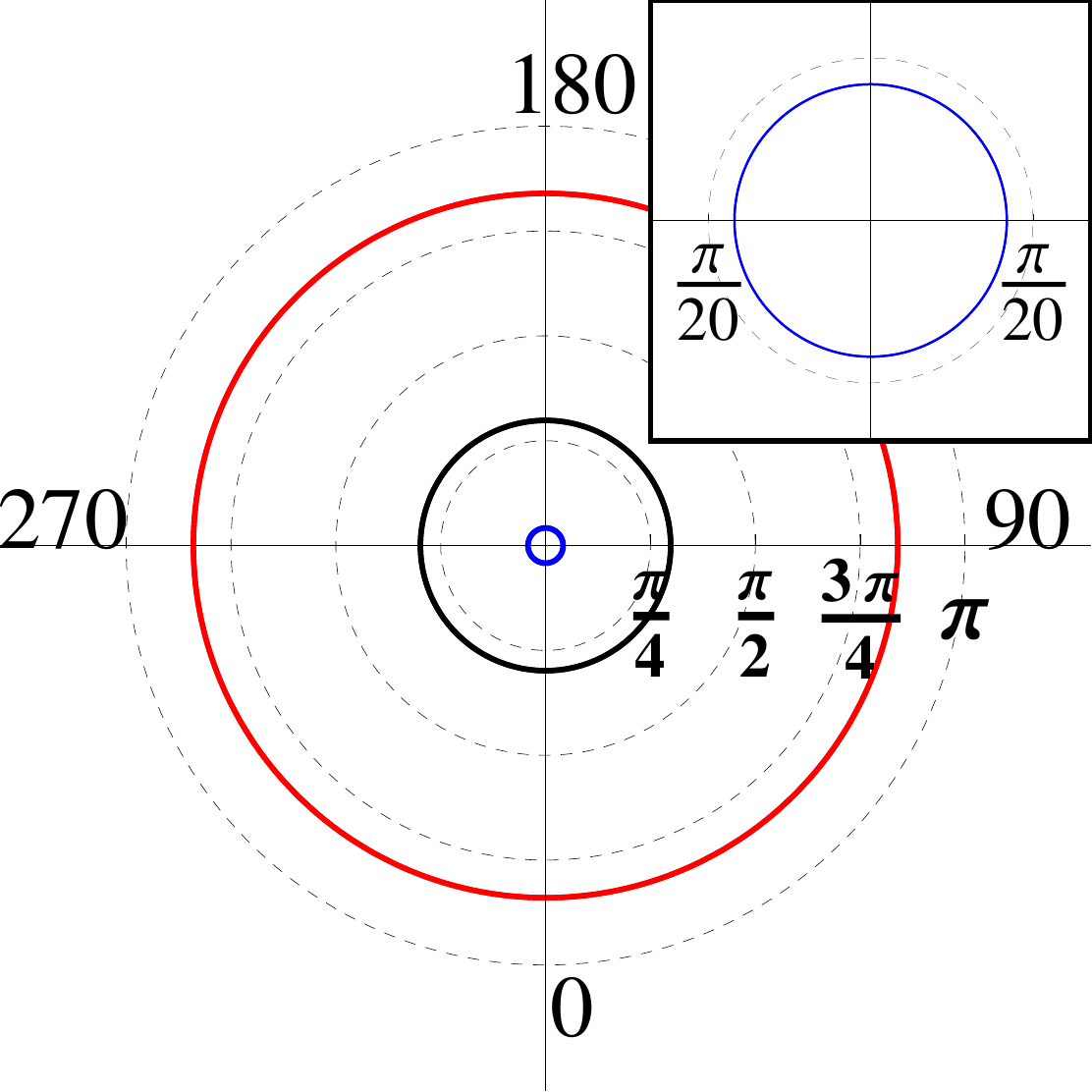}\\
		\hline
		\multicolumn{2}{|c|}{$r_{e}=r_{pol}$}\\
		\includegraphics[width=5cm]{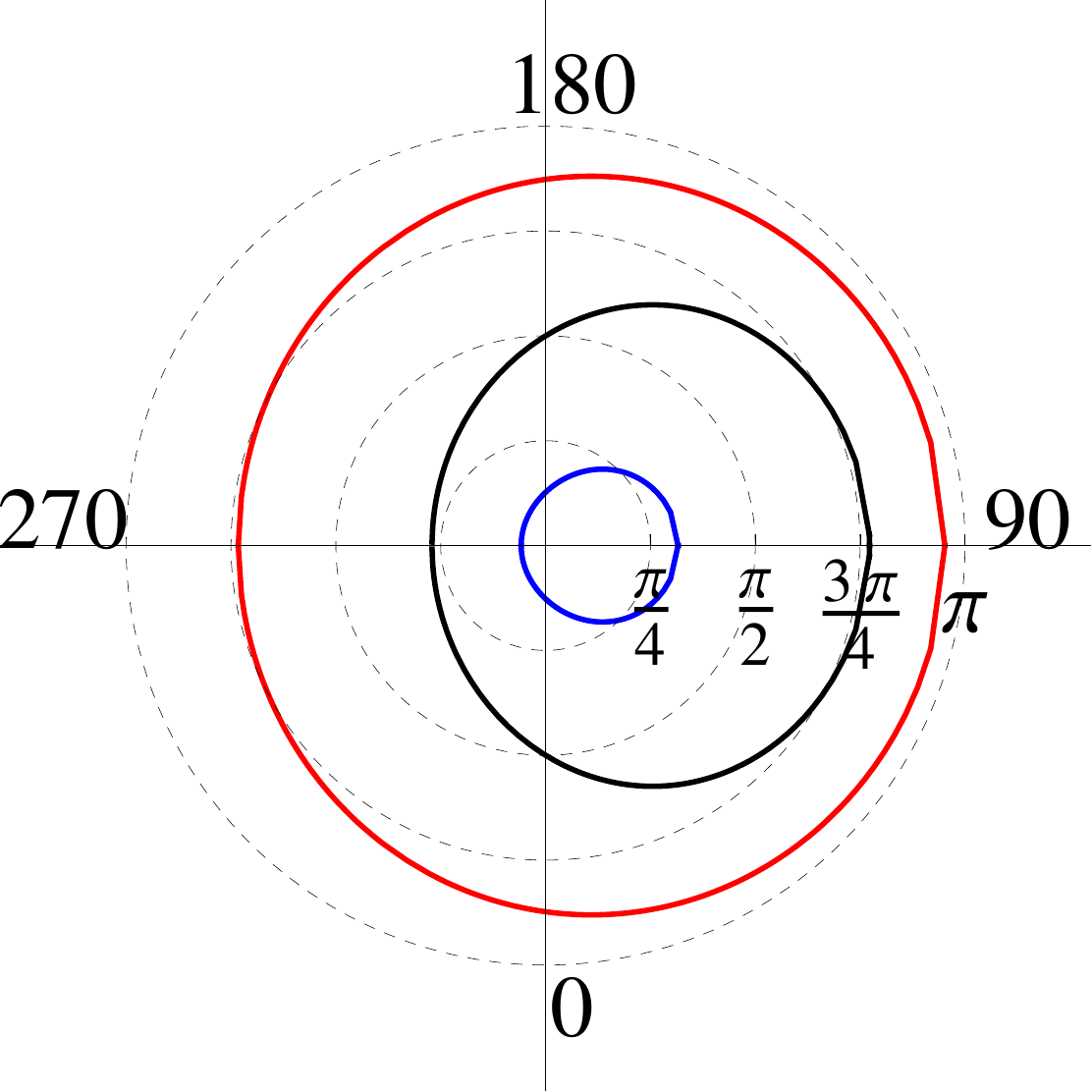}&\includegraphics[width=5cm]{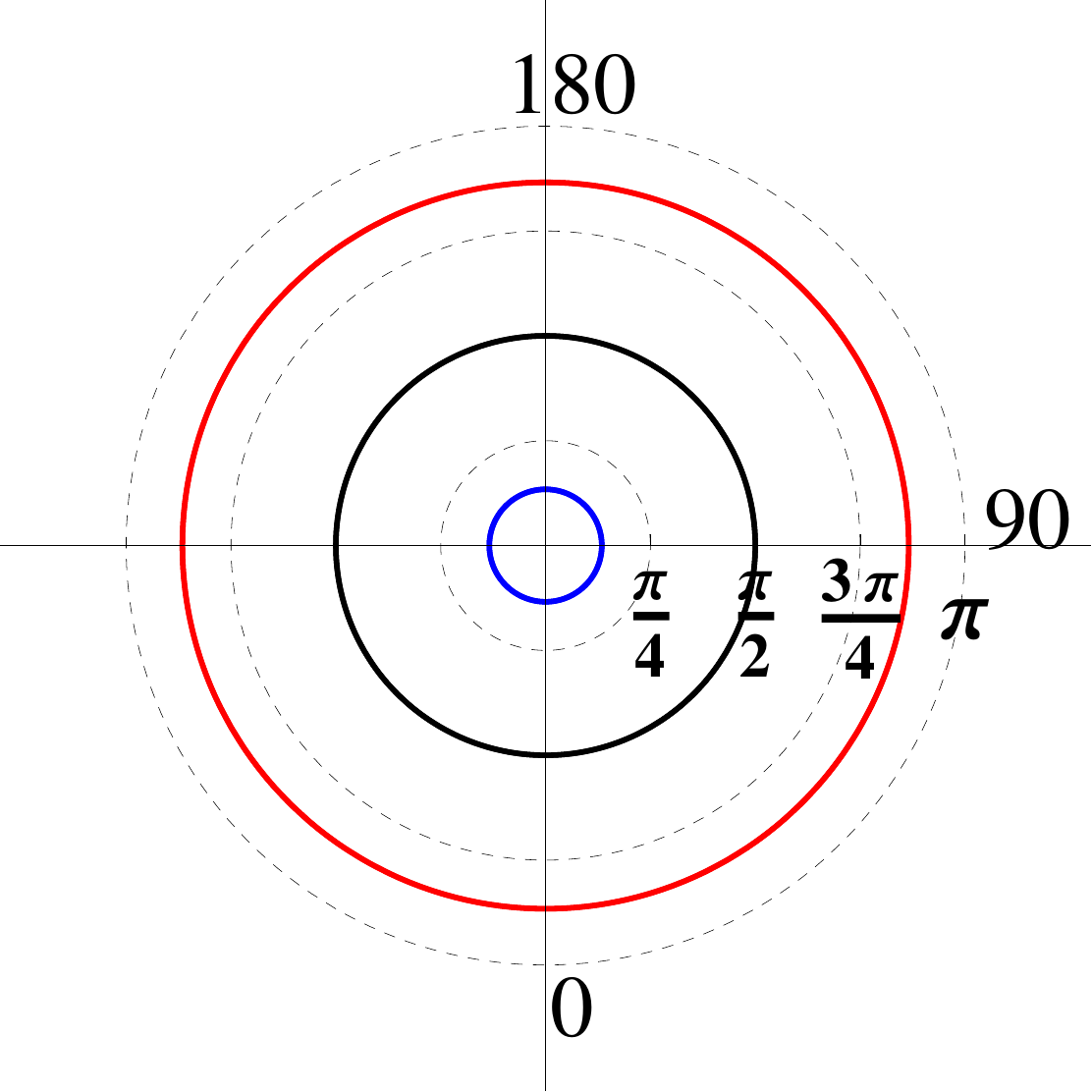}\\
		\hline
	\end{tabular}
	\center (\textit{Figure continued}) 
\end{figure}

\begin{figure}[H]
	\flushleft Class II: $y=0.02,\quad a^2=0.9$\\
	\bet{lccr}
	\hline
	\bet{l}$r_{o}=1.4876$\\$r_{c}=5.7508$\ent & \bet{l}$r_{ph+}=1.5794$\\$r_{ph-}=3.4728$ \ent&\bet{l} $r_{d1}=2.2772$\\$r_{d2}=5.5853$\ent &\bet{l} $r_{pol}=2.4257$\ent
	\ent
	
	\begin{tabular}{|cc|}
		\hline
		\multicolumn{2}{|c|}{$r_{e}=r_{ph-}$}\\
		\includegraphics[width=5cm]{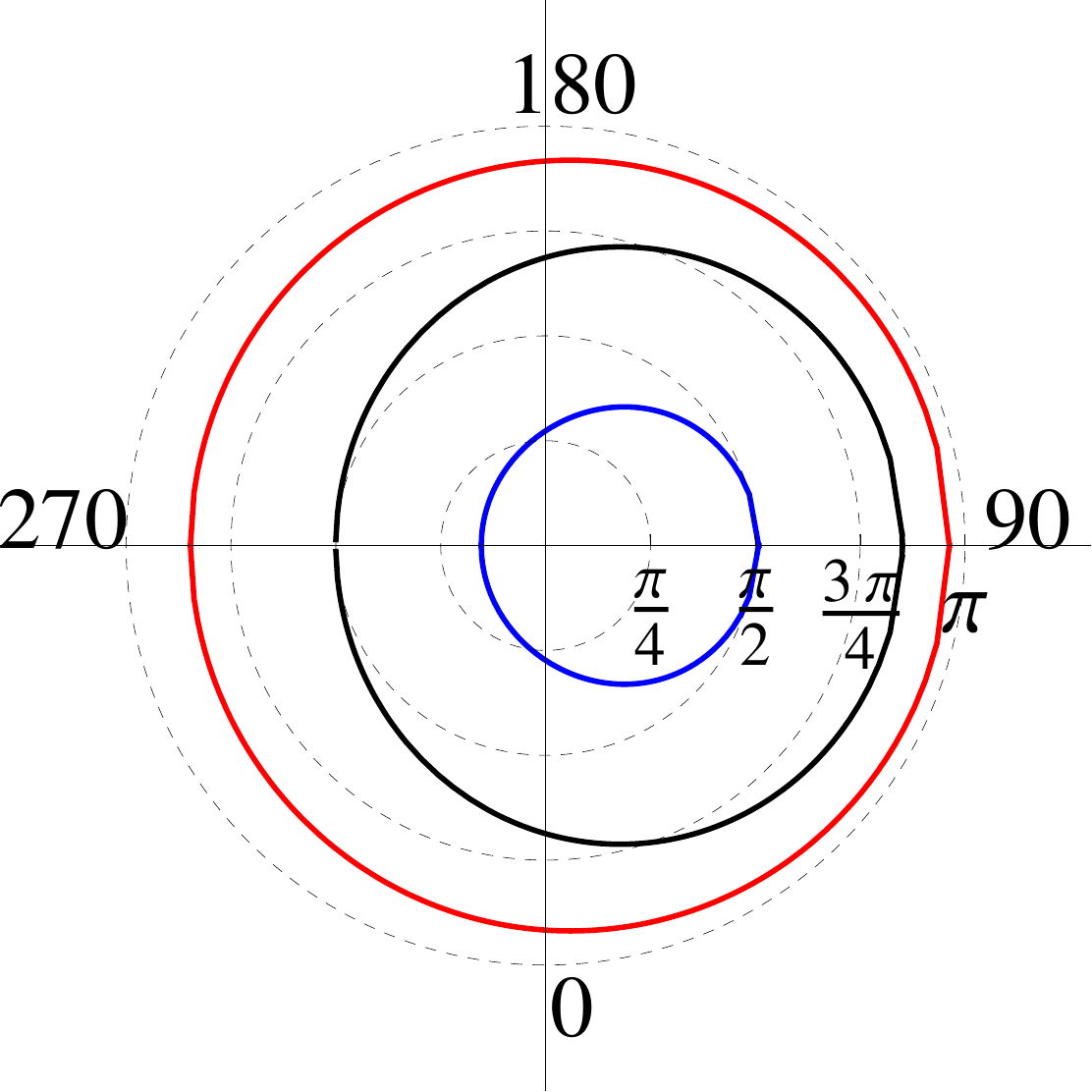}&\includegraphics[width=5cm]{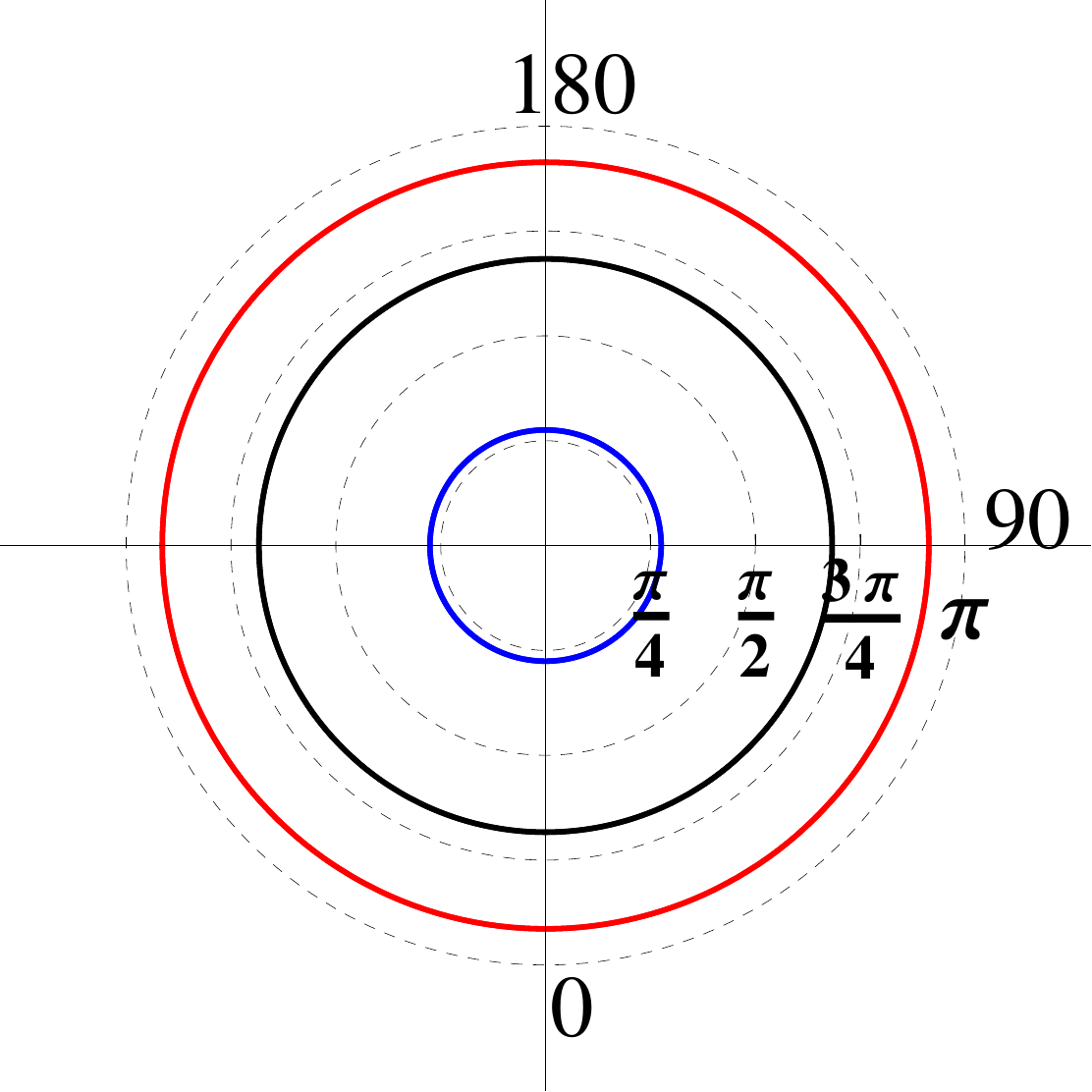}\\
		\hline
		\multicolumn{2}{|c|}{$r_{e}=4$}\\
		\includegraphics[width=5cm]{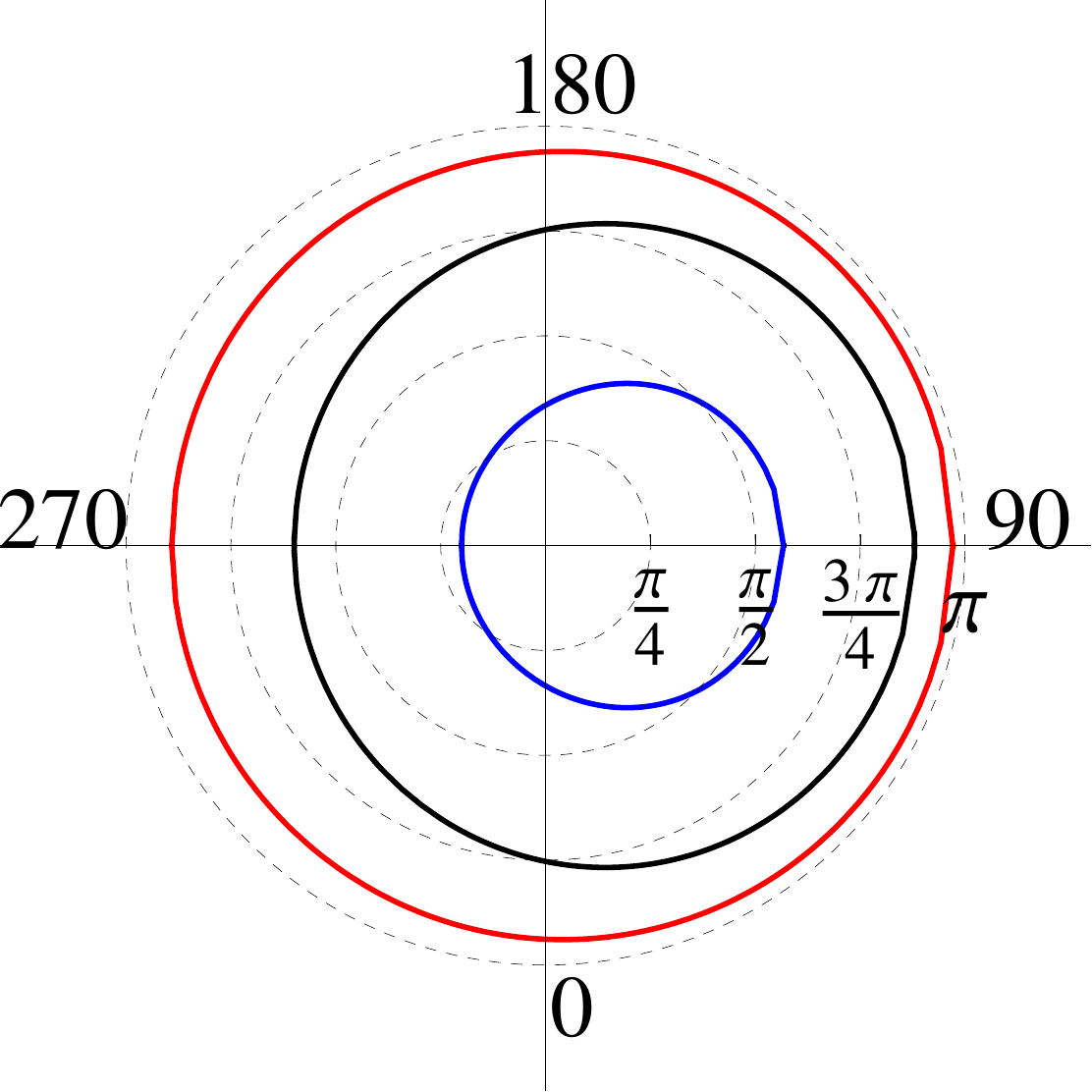}&\includegraphics[width=5cm]{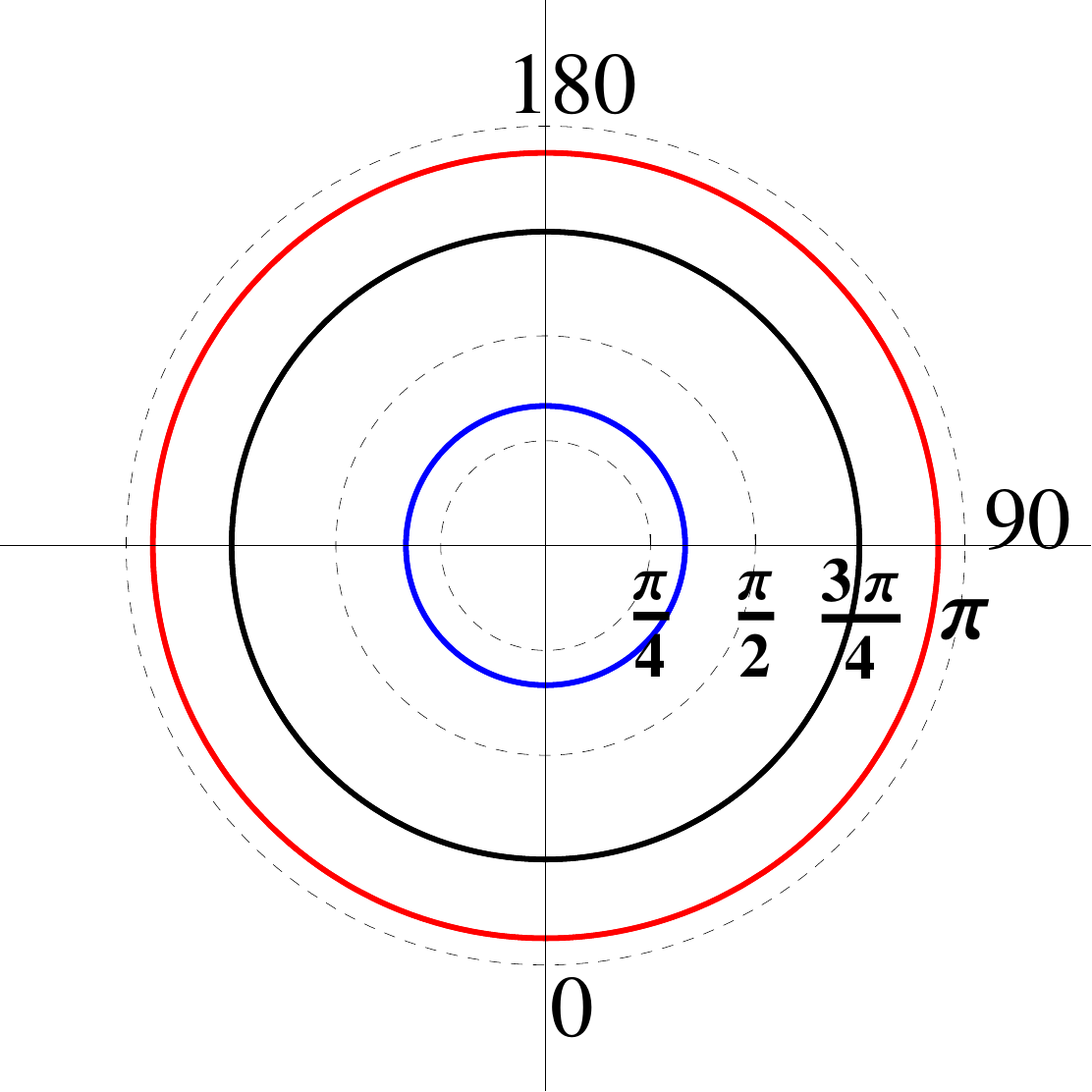}\\
		\hline
		\multicolumn{2}{|c|}{}\\
		\multicolumn{2}{|c|}{$r_{e}=5.6$}\\
		\includegraphics[width=5cm]{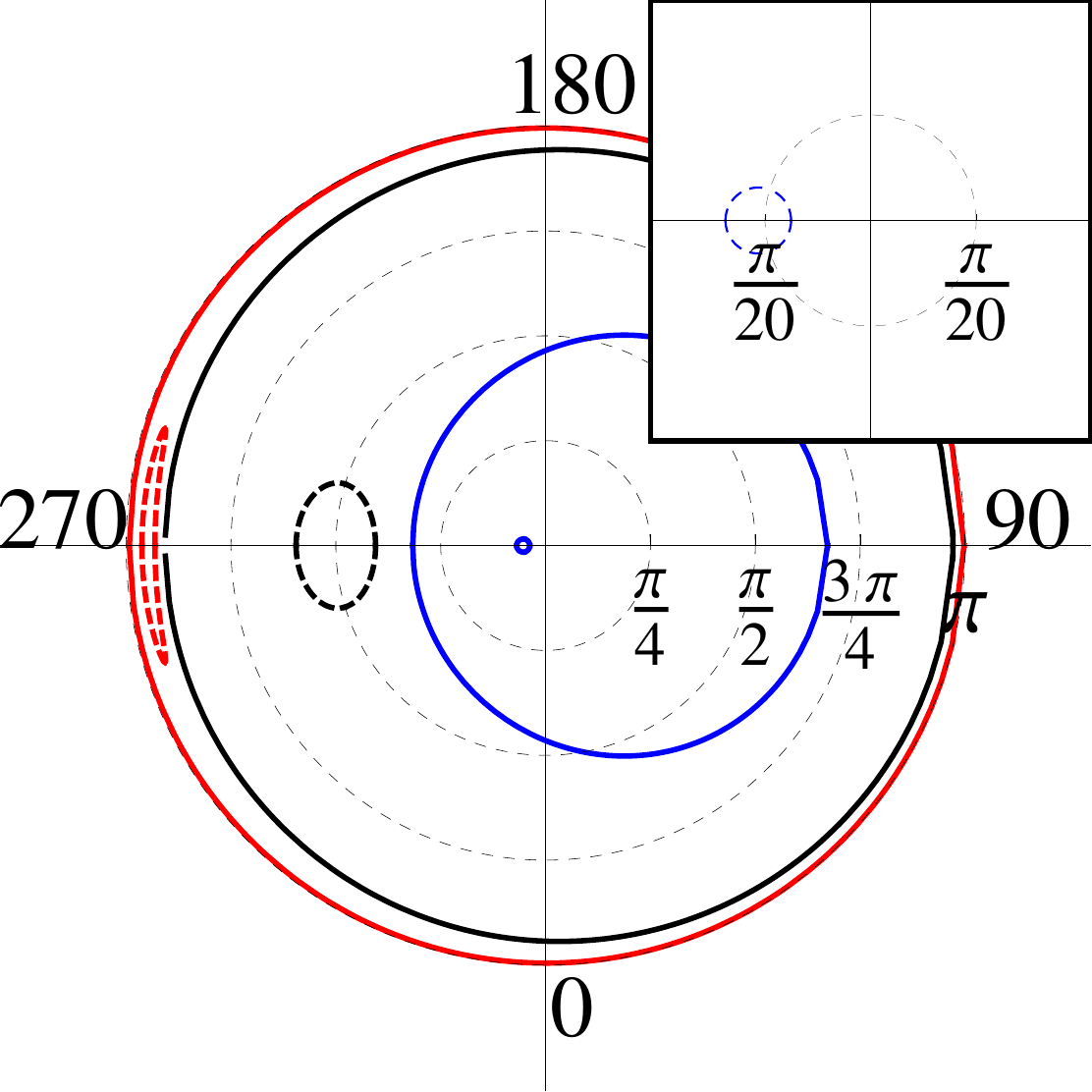}&\includegraphics[width=5cm]{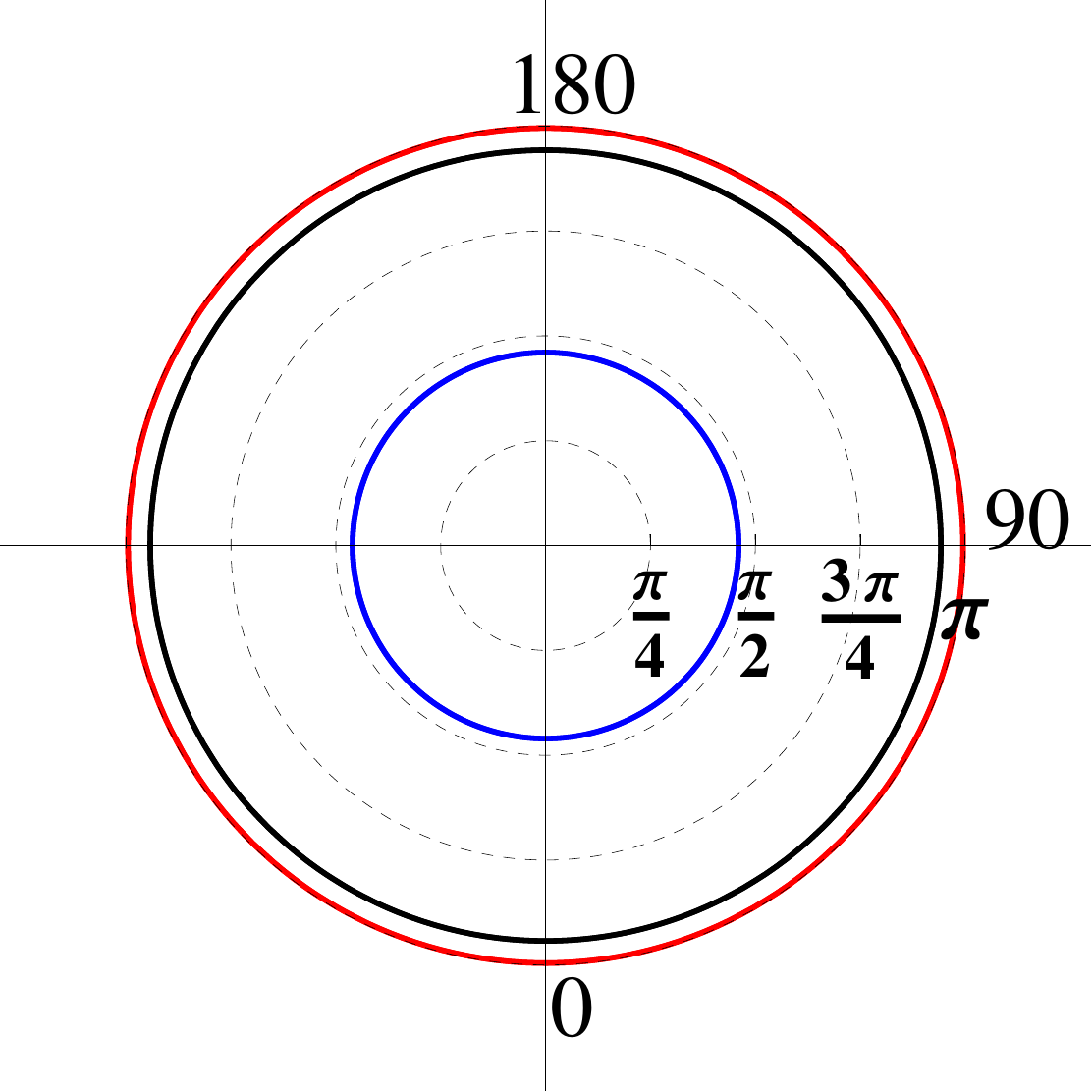}\\
		\hline
	\end{tabular}
	\center (\textit{Figure continued}) 
\end{figure}

\begin{figure}[H]
	\flushleft \textbf{Class III: $y=0.04,\quad a^2=0.9$}\\
	
	\begin{tabular}{lcr}
		\hline
		\bet{l}$r_{o}=1.7936$\\$r_{c}=3.3374$\ent & \bet{c}$r_{ph+}=1.8864$\\$r_{ph-}=2.8867$ \ent&\bet{r} $r_{d(ex)}=2.7340$\\ $r_{pol}=2.3578$ \ent 
	\end{tabular}

	\begin{tabular}{|cc|}
		\hline
		\multicolumn{2}{|c|}{}\\
		\multicolumn{2}{|c|}{$r_{e}=1.83$}\\
		\includegraphics[width=5cm]{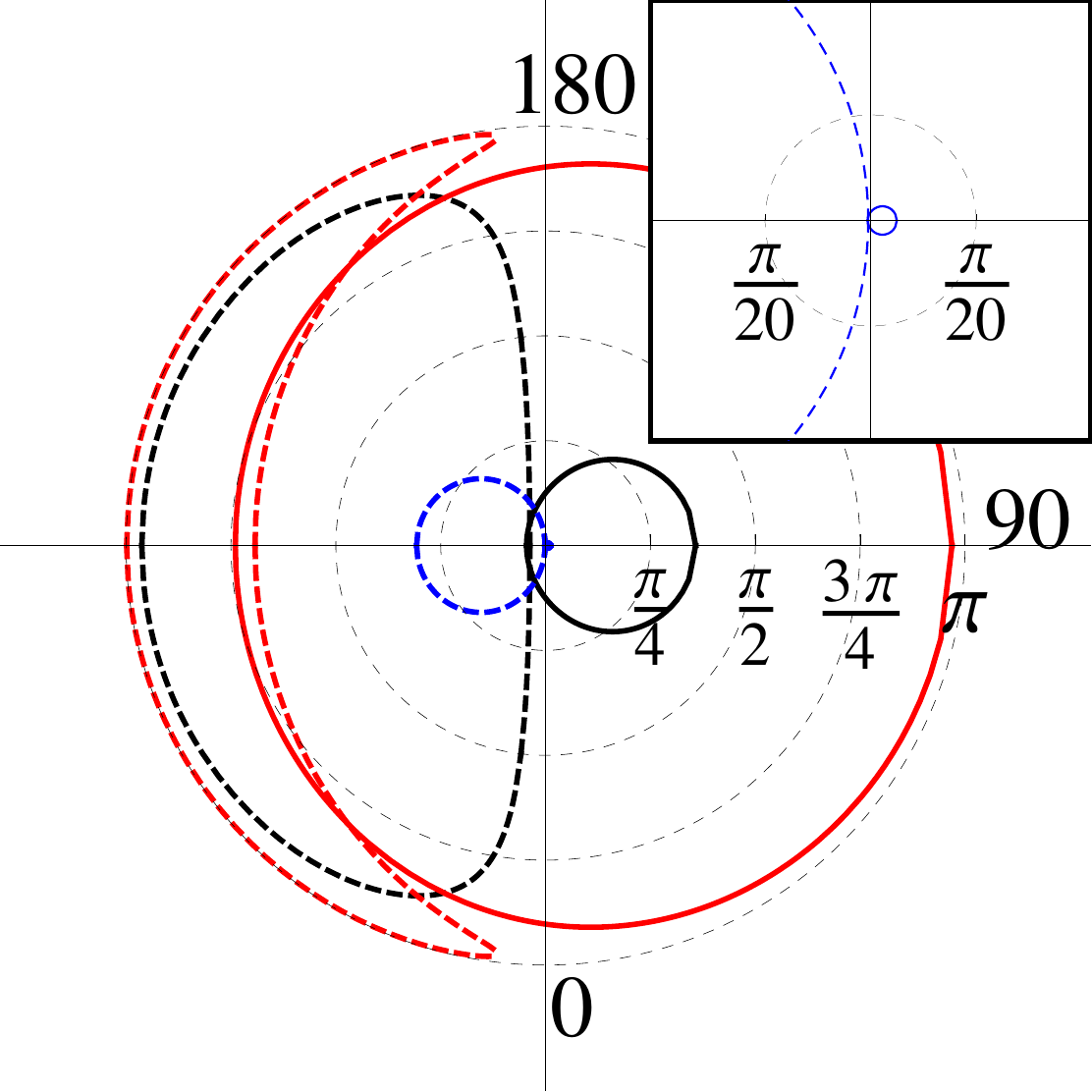} & \includegraphics[width=5cm]{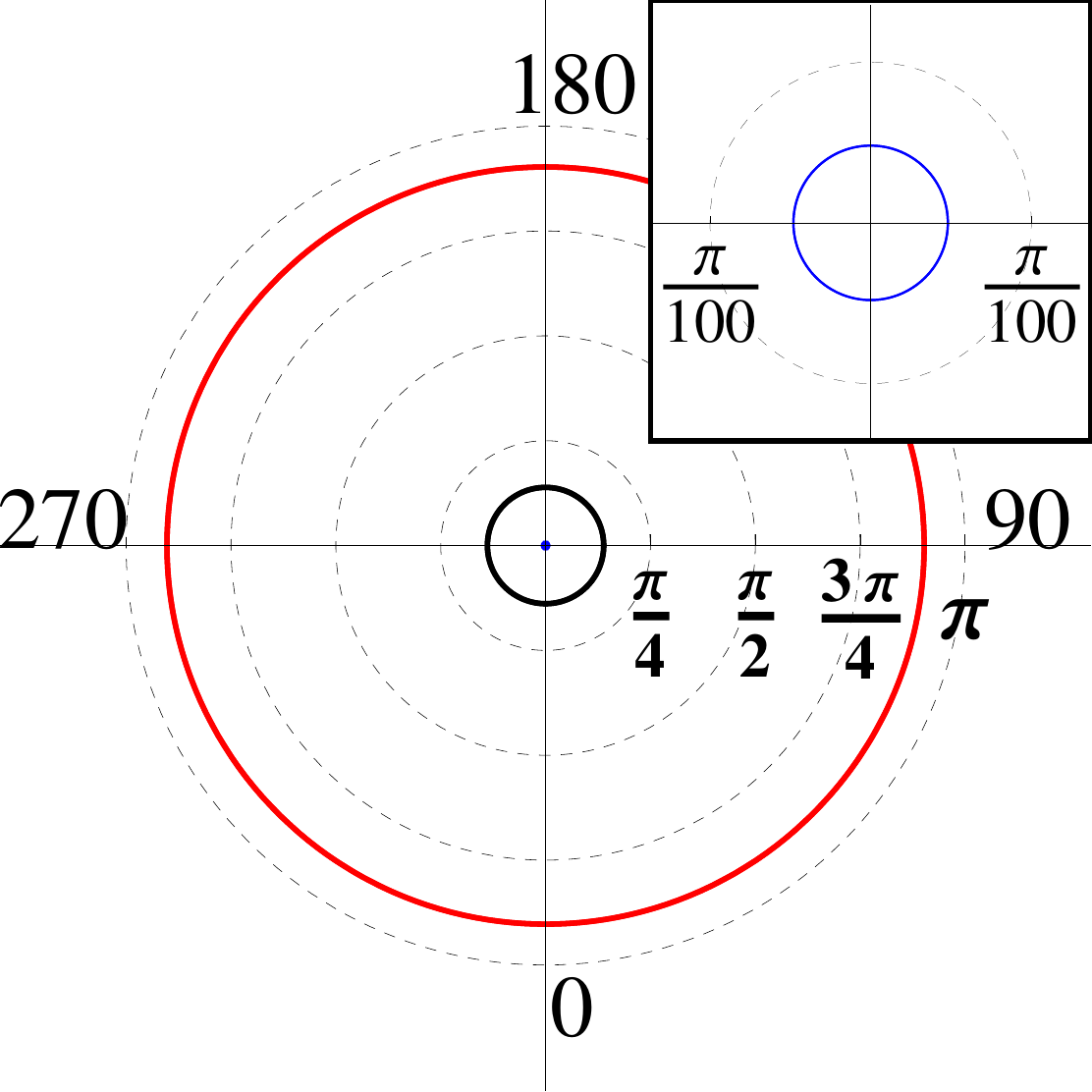}\\
		\hline
		\multicolumn{2}{|c|}{}\\
		\multicolumn{2}{|c|}{$r_{e}=r_{ph+}$}\\
		\includegraphics[width=5cm]{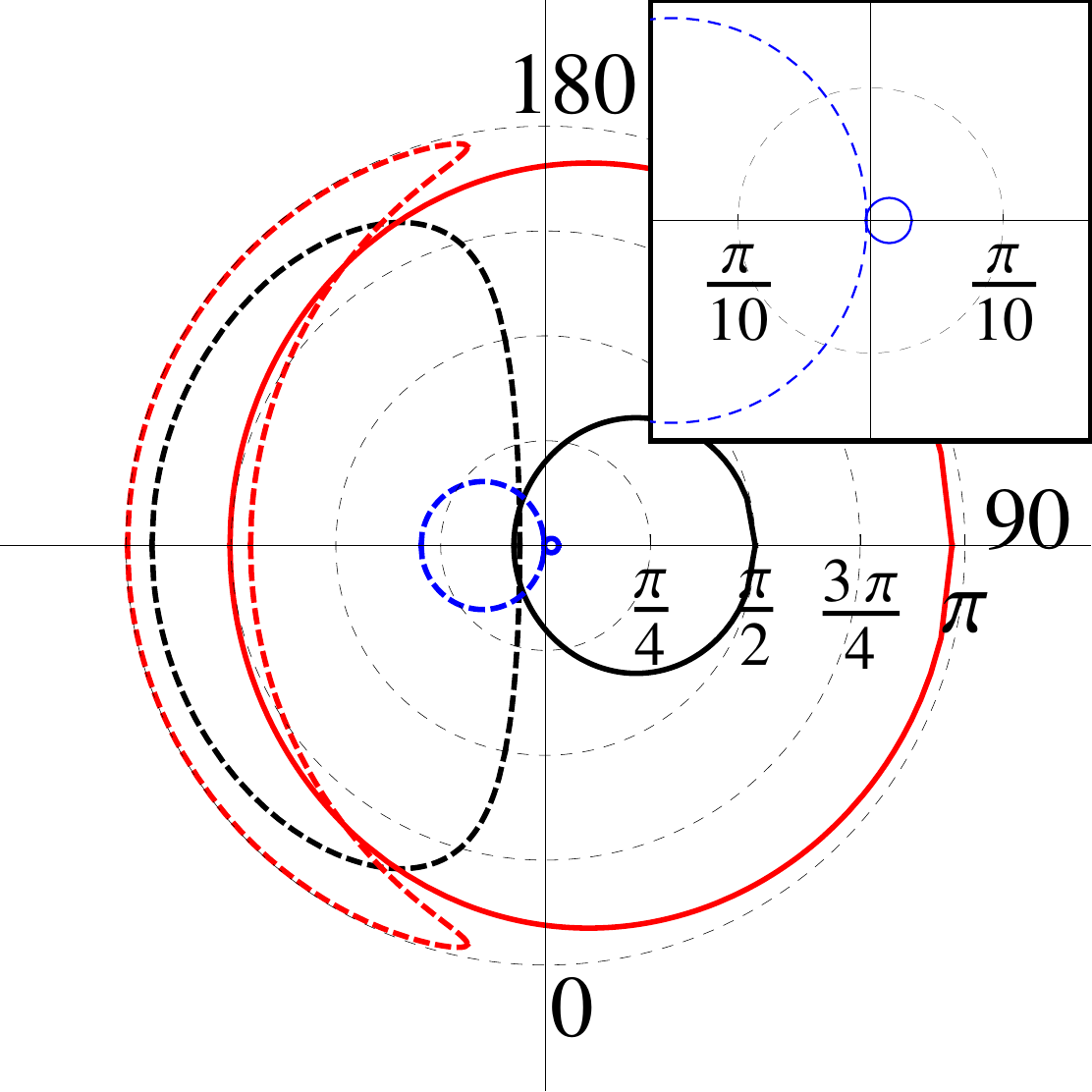}&\includegraphics[width=5cm]{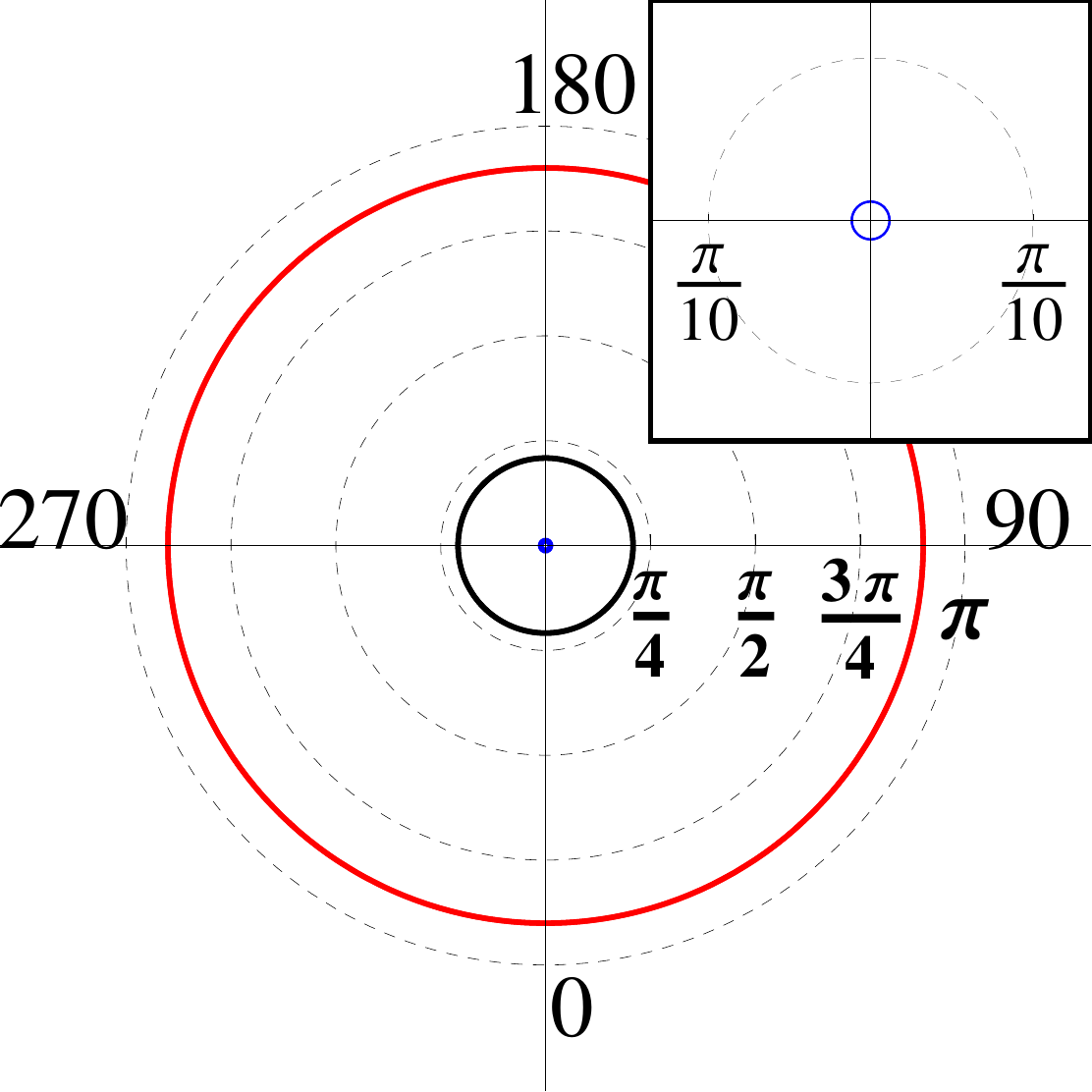}\\
		\hline
		\multicolumn{2}{|c|}{}\\
		\multicolumn{2}{|c|}{$r_{e}=r_{pol}$}\\
		\includegraphics[width=5cm]{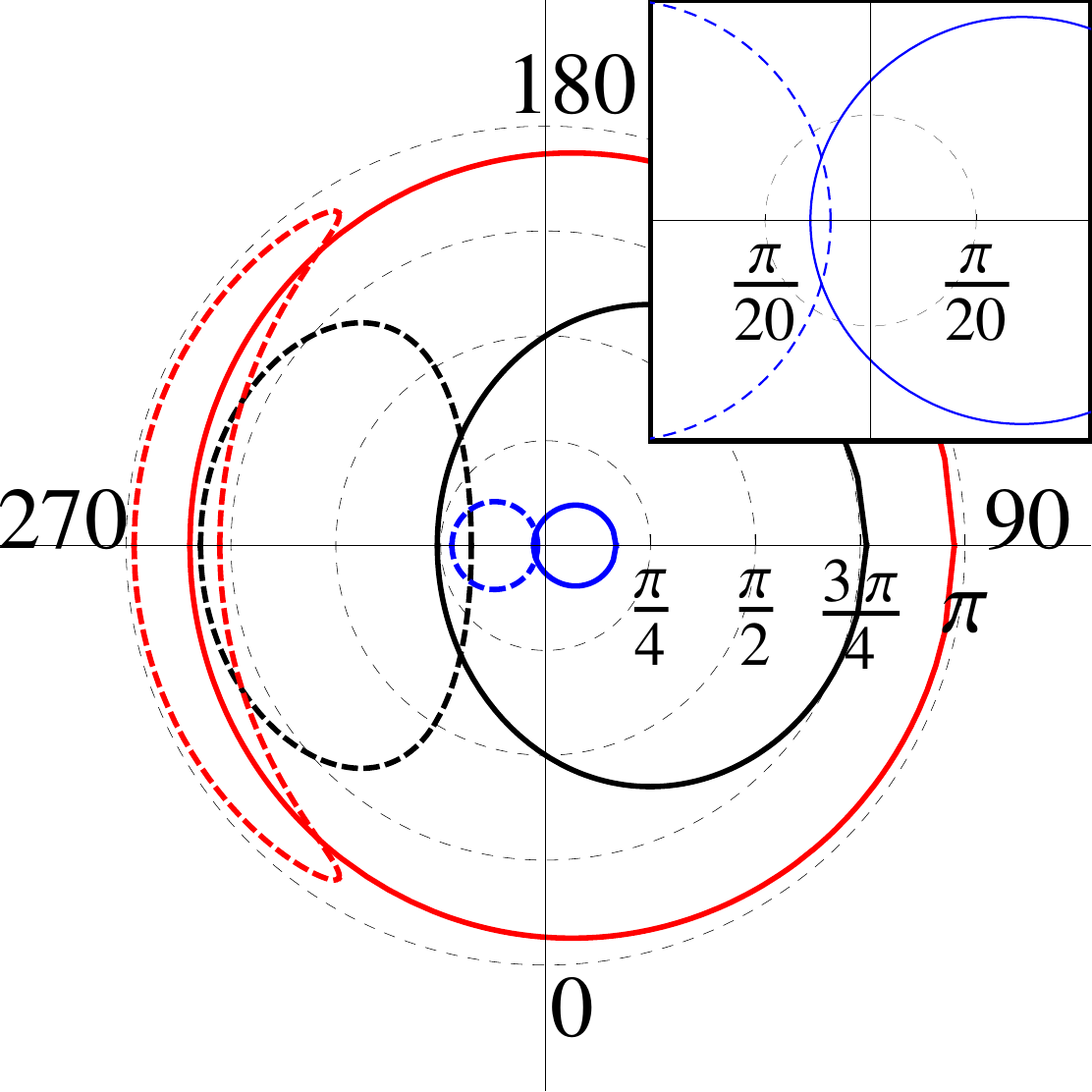}&\includegraphics[width=5cm]{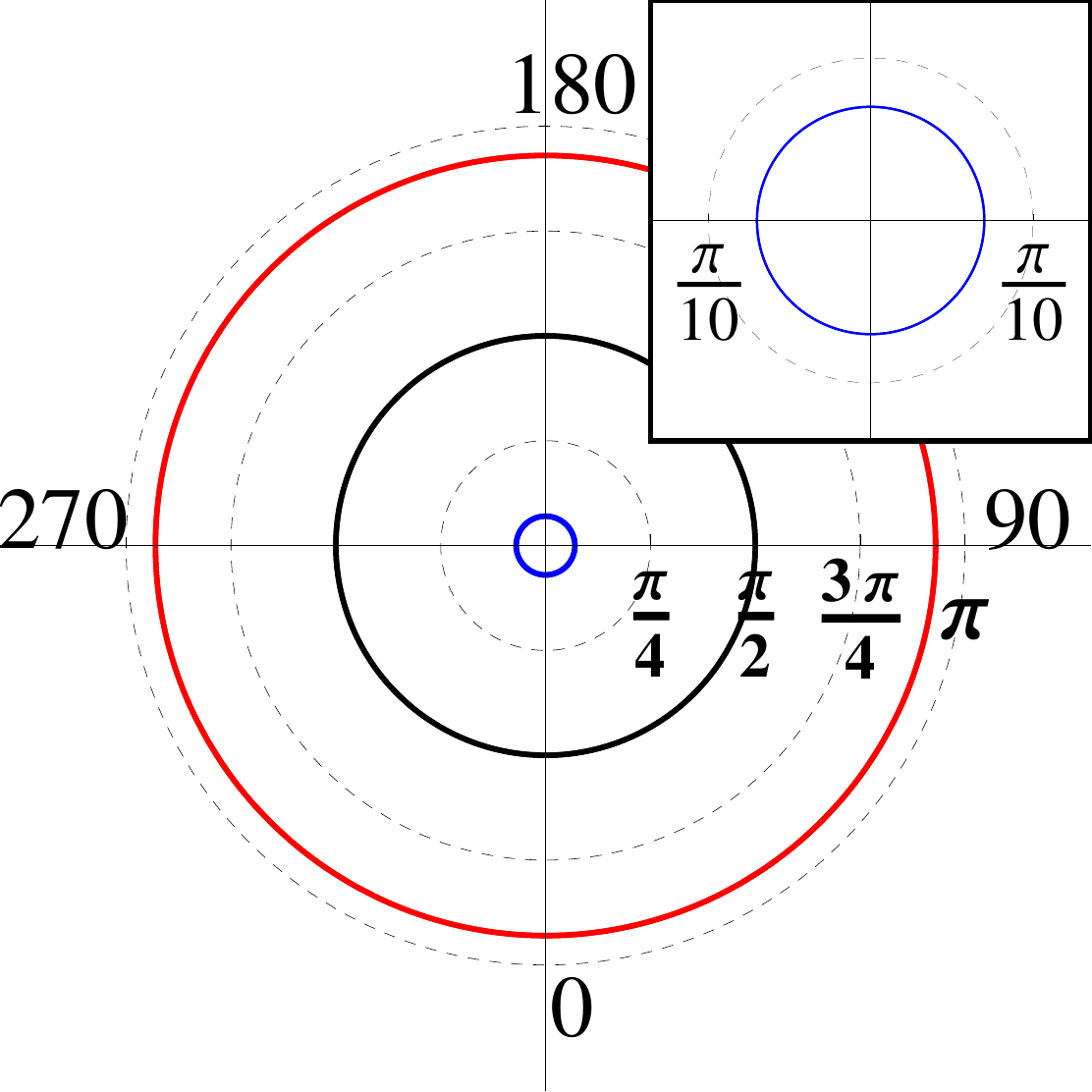}\\
		\hline
	\end{tabular}
	\center (\textit{Figure continued})
\end{figure}
\begin{figure}[H]
	\flushleft Class III: $y=0.04,\quad a^2=0.9$\\
	
	\begin{tabular}{lcr}
		\hline
		\bet{l}$r_{o}=1.7936$\\$r_{c}=3.3374$\ent & \bet{c}$r_{ph+}=1.8864$\\$r_{ph-}=2.8867$ \ent&\bet{r} $r_{d(ex)}=2.7340$\\ $r_{pol}=2.3578$ \ent 
	\end{tabular}
	
	\begin{tabular}{|cc|}
		\hline
		\multicolumn{2}{|c|}{}\\
		\multicolumn{2}{|c|}{$r_{e}=2.6$}\\
		\includegraphics[width=5cm]{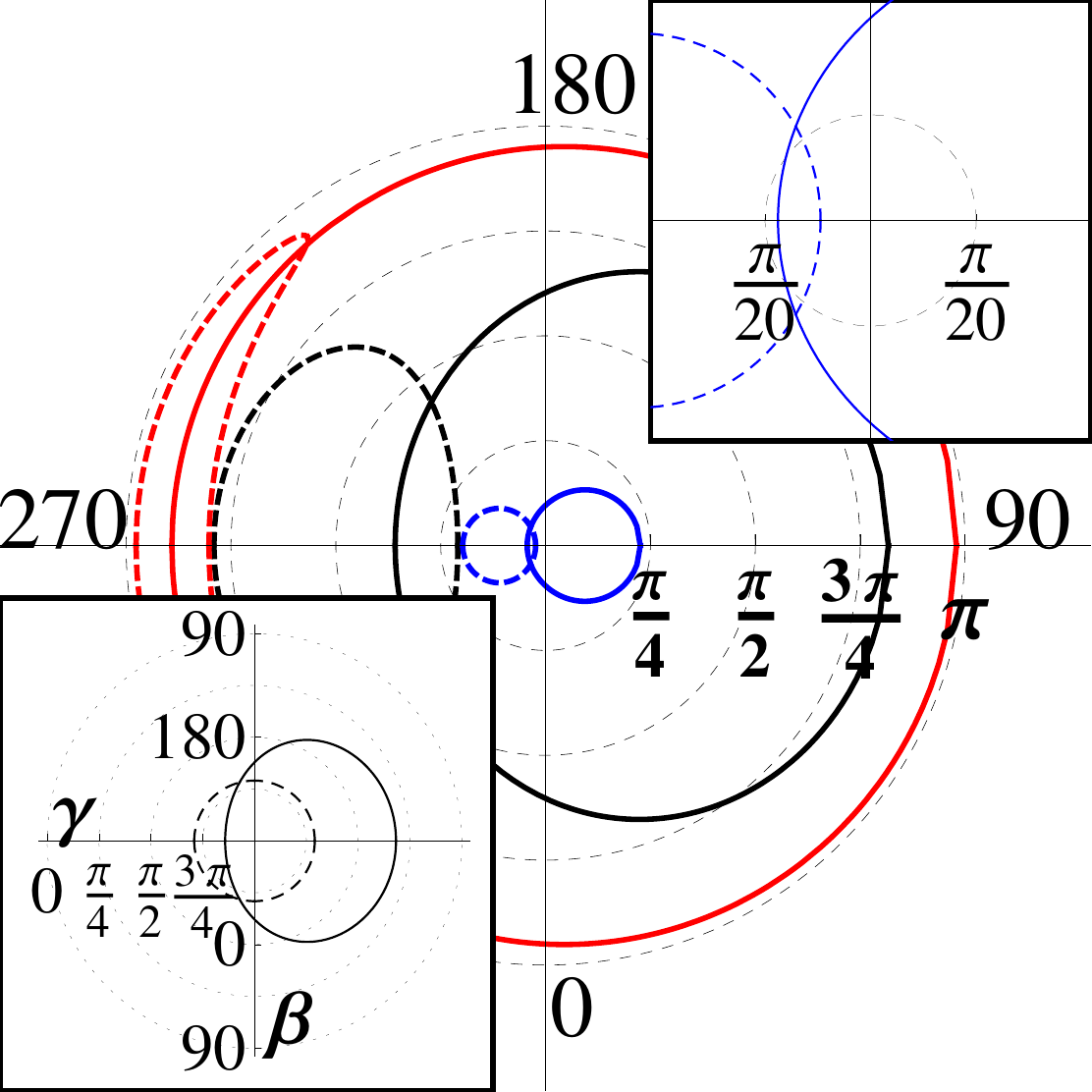} & \includegraphics[width=5cm]{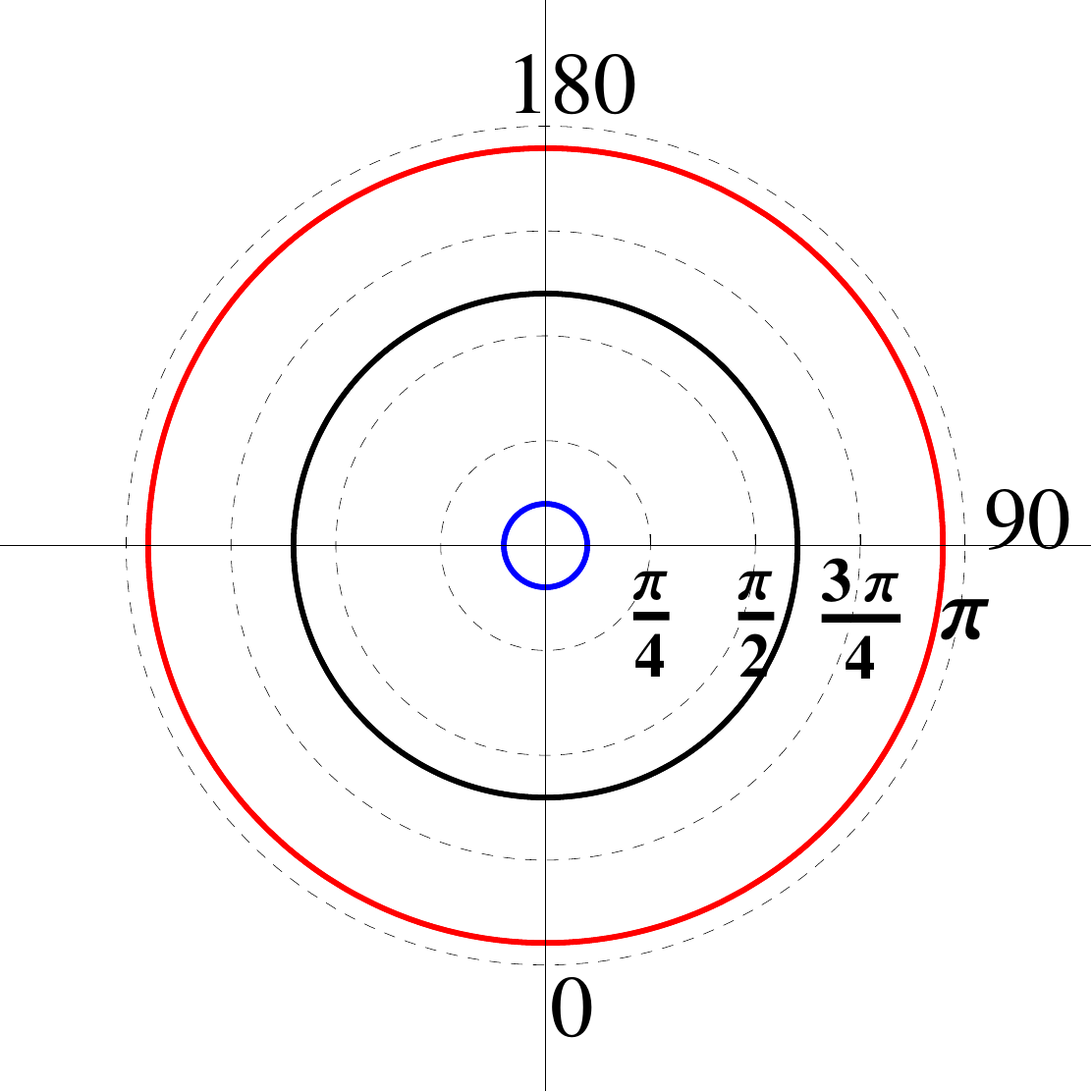}\\
		\hline
		\multicolumn{2}{|c|}{}\\
		\multicolumn{2}{|c|}{$r_{e}=r_{d(ex)}$}\\
		\includegraphics[width=5cm]{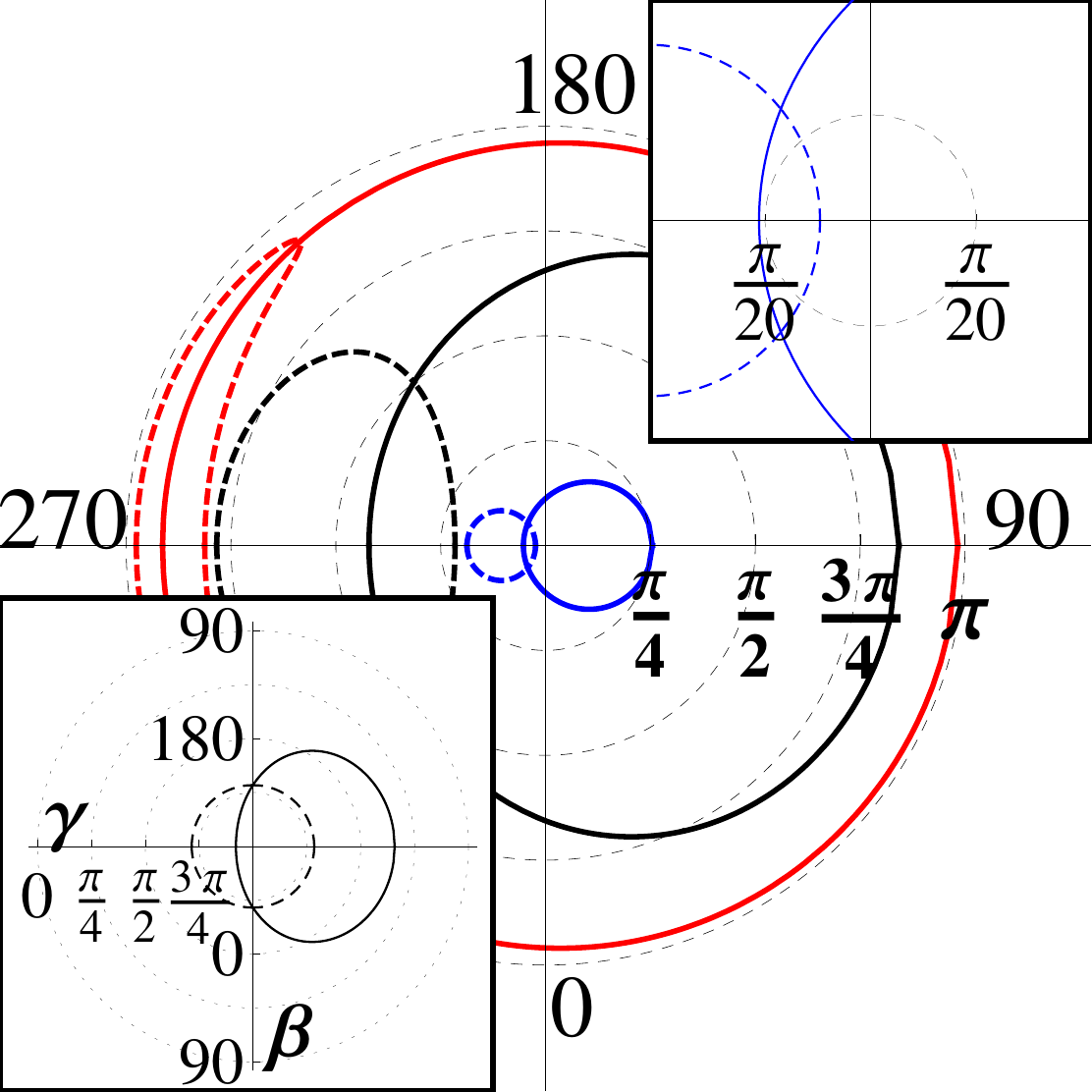} & \includegraphics[width=5cm]{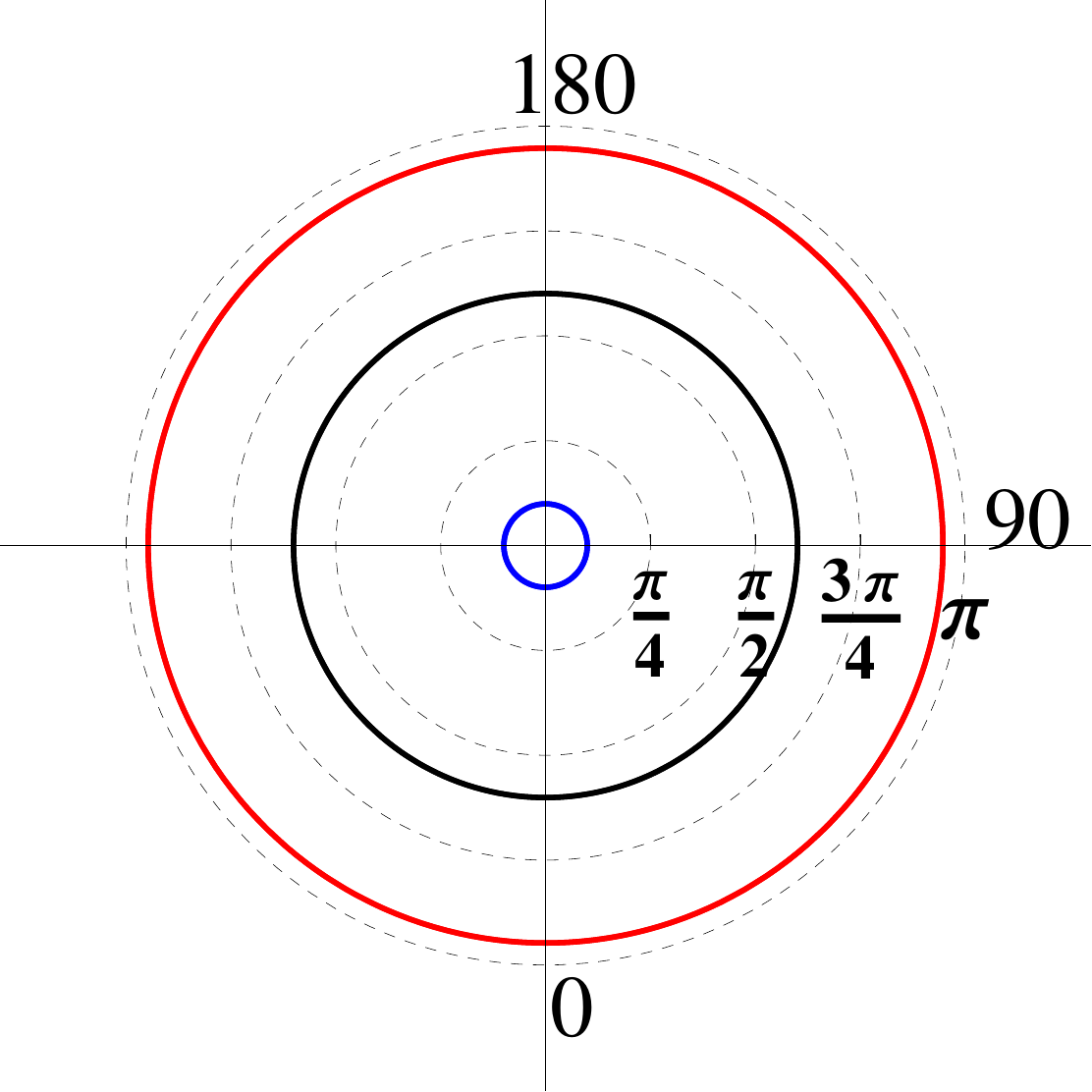}\\ 
		\hline
		\multicolumn{2}{|c|}{}\\
		\multicolumn{2}{|c|}{$r_{e}=r_{ph-}$}\\
		\includegraphics[width=5cm]{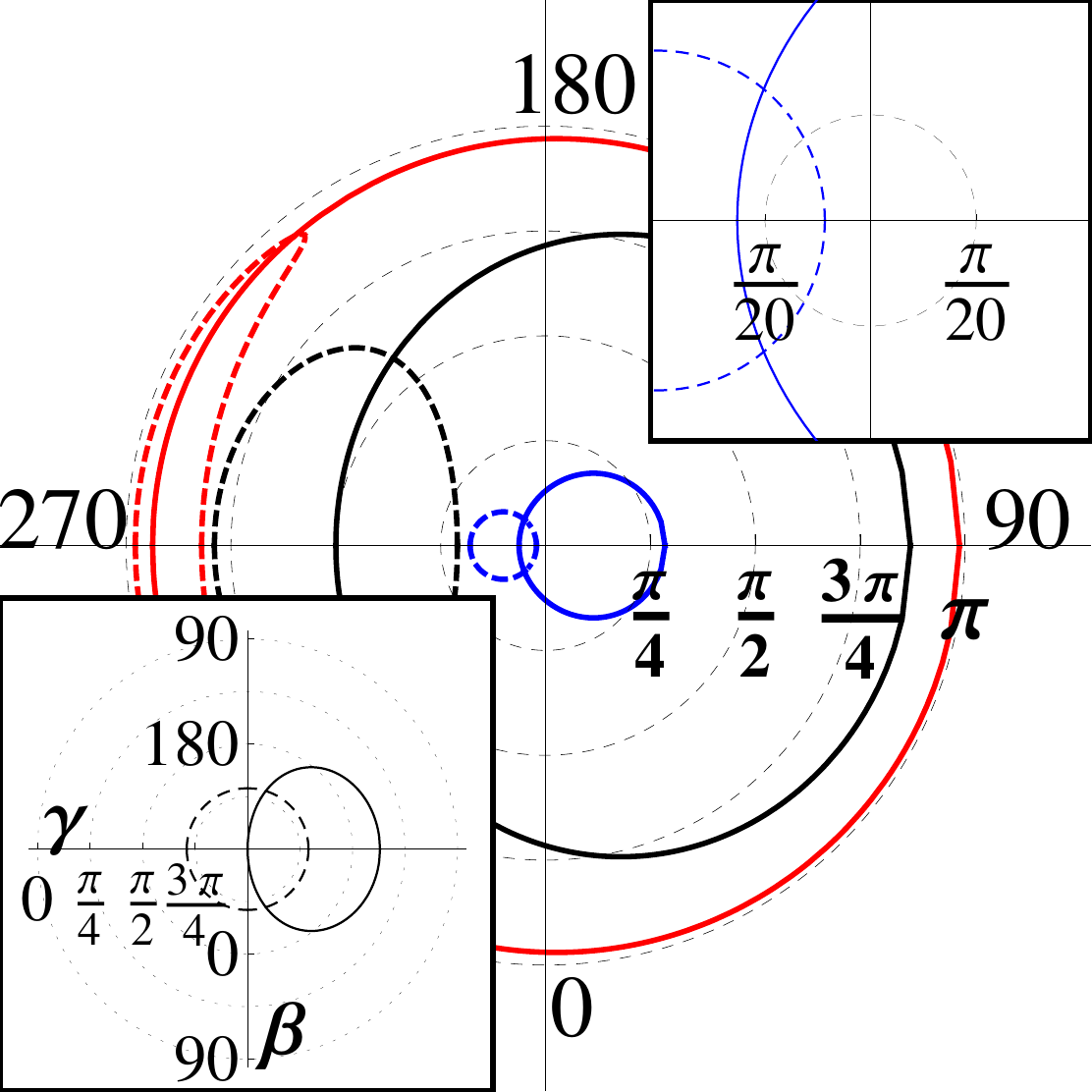} & \includegraphics[width=5cm]{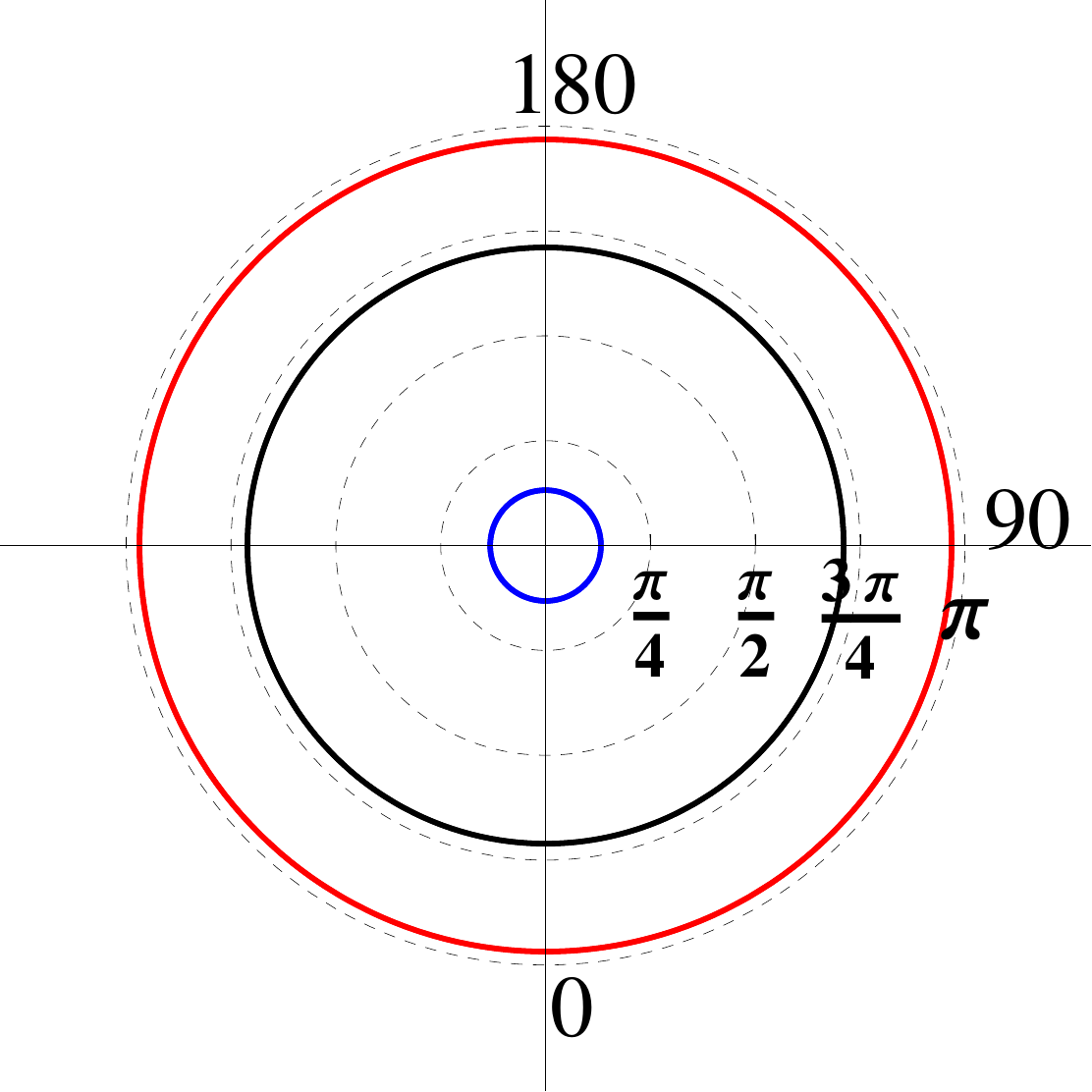}\\
		\hline
	\end{tabular}
	\center (\textit{Figure continued}) \label{cones_rph-}
\end{figure}

 \label{cones_r>rph-}


\section{Light escape cones in circular geodesic frames}

Now we consider the astrophysically most interesting and significant case of local light escaping cones related to the Keplerian accretion disks governed by the circular geodesic motion. 

\subsection{Observers on the equatorial circular geodesics}
	
	Let us introduce the acronym CGF for the local reference frame connected with a massive test particle following a circular geodesic in the equatorial plane of the \KdS\ background. The constants of motion defined by the relations (\ref{parE}) - (\ref{parQ}) for such particle (observer) read (see \cite{Stu-Sla:2004:PHYSR4:}) 
	\be E=E_{\pm}(r;\:a,\:y)\equiv \frac{1-2/r-(r^2+a^2)y\pm a\sqrt{1/r^3-y}}{1-3/r-a^2y\pm2a\sqrt{1/r3-y}}, \label{Eco} \ee
	\be L=L_{\pm}(r;\:a,\:y)\equiv -\frac{2a+ar(r^2+a^2)y\mp r(r^2+a^2)\sqrt{1/r^3-y}}{r\sqrt{1-3/r-a^2y\pm2a\sqrt{1/r^3-y}}} \label{Lco} \ee
	and
	\be Q=0. \label{Qco} \ee
	The CGF moves in $\phi$ direction relatively to the LNRF with velocity $v^{(\phi)},$ which according to (\ref{3-velocity}) reads
	\be
	v^{(\phi)}_{\pm}=\frac{A}{r^2\sqrt{\Delta_{r}}}(\Omega_{\pm}-\Omega_{LNRF}). \label{vphi}
	\ee
	Here
	\be
	\Omega_{\pm}=U^{\phi}/U^{t}=\frac{\pm \sqrt{1-yr^3}}{r\sqrt{r}\pm a\sqrt{1-yr^3}} \label{Oco}
	\ee
	is the angular velocity of the observer. For $y=0$ this relation goes over into $\Omega_{\pm}=\frac{\pm \sqrt{r}}{r^2\pm a\sqrt{r}}$ (c. f. the appropriate relation in \cite{2005ragt.meet..143S}). The $\pm$ signs in (\ref{vphi}), (\ref{Oco}) correspond to plus-family or minus-family orbits introduced in \cite{Stu-Sla:2004:PHYSR4:}, where both cases $v_{+}>0$ or $v_{+}<0$ are possible, but for $v_{-}$ it must be $v_{-}<0.$ Here and in the following we again omit the superscript $(\phi)$ for brevity. The value $v_{+}=1$ corresponds to the co-rotating circular photon orbit located at $r=r_{ph+},$ while $v_{-}=-1$ holds for the counter-rotating photon orbit located at $r=r_{ph-},$ where the orientation is determined relative to LNRF. According to (\ref{vphi}), (\ref{Oco}), there is $v_{+}=v_{-}<0$ at the static radius $r_{s}=y^{-1/3},$ which is also the limiting value for both functions. For the case of astrophysically most important stable orbits it holds $r_{ph+}<r_{(ms+)i}<r_{(ms-)i}<r_{(ms-)o}<r_{(ms+)o}<r_{s},$ where $r_{(ms\pm)i}, r_{(ms\pm)o}$ denote the marginally stable inner and outer plus/minus family orbits, respectively. A detailed analysis of the stability/instability of the equatorial circular orbits can be found in \cite{Stu-Sla:2004:PHYSR4:}.   \par
	
	\subsection{Photons in circular geodesic frames}
	The components of the photon four-momentum $k^{(\tilde{a})}$ locally measured in the CGF and  corresponding quantities $k^{(b)}$ measured in the LNRF are mutually tied by the special Lorentz transformation   
	\be
	k^{(\tilde{a})}=\Lambda^{(\tilde{a})}_{(b)} k^{(b)}, \label{LTCGF}
	\ee
	where the transformation matrix reads 
	\be
	\Lambda^{(\tilde{a})}_{(b)}=\left( \begin{array}{cccc}
									\gamma & 0 & 0 & -\gamma v_{\pm}\\
									0 & 1 & 0 & 0\\
									0 & 0 & 1 & 0\\
									-\gamma v_{\pm} & 0 & 0 & \gamma
									\end{array}	\right) 
	\ee
	and 
	\be
	\gamma=(1-v_{\pm}^2)^{-1/2}
	\ee
	is the Lorentz factor. By the same way, as in the previous section, we find that the relations between directional angles in the CGFs and LNRFs read 
	\bea
	\cos \tilde{\alpha}&=&\frac{\cos \alpha}{\gamma(1-v_{\pm}\sin \alpha \sin \beta)},\\	
	\sin \tilde{\alpha} \sin \tilde{\beta}&=&\frac{\sin \alpha \sin \beta-v_{\pm}}{1-v_{\pm}\sin \alpha \sin \beta},\\	
	\sin \tilde{\alpha} \cos \tilde{\beta}&=&\frac{\sin \alpha \sin \beta}{\gamma(1-v_{\pm}\sin \alpha \sin \beta)},
	\eea
	by which we construct the light escape cones in the CGFs.\par
\begin{figure}[H]
	 	\includegraphics[scale=1.2]{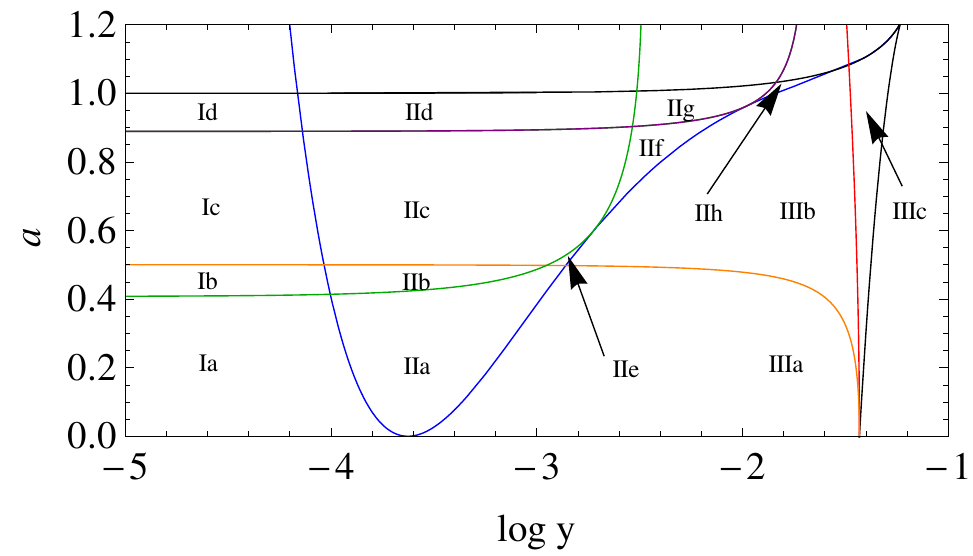} \caption{Separation of parameter space $y-a^2$ according to existence of stable circular orbits and mutual order of radii of marginally stable orbits, photon circular orbits and divergent points of effective potential governing the photon motion. Region bordered by the black curves correspond to black hole spacetimes, the outer part represents naked singularities. Blue curves induce basic separation into three regions as introduced in \cite{Stu-Sla:2004:PHYSR4:}. Red curve separate black hole spacetimes with DRB and RRB; crossing the orange curve upwards indicates spacetimes where the co-rotating photon circular orbit enters the region $0 < \Delta_{r} < a^2$ (ergosphere); area above the green curve corresponds to spacetimes where the stable plus-family orbits spread somehow between co-rotating and counter-rotating photon circular orbits; in a similar manner the purple curve demarcate spacetimes where the plus family orbits reach the ergosphere (see the table for details). } \label{fig_y-a2_plane}
	 \end{figure}

	 The stable circular orbits play important role in the study of accretion processes related to thin Keplerian discs in the vicinity of black holes, therefore we shall focus on construction of the light escape cones at the marginally stable orbits. In the KdS spacetimes, where the stable circular orbits are not allowed, it could be inspiring to deal with such construction at radii differing by the behaviour of $\ell(\tilde{\alpha})$ as described above. Since the shape of the light escape cones is influenced by the radial position of the source with respect to the above mentioned outstanding radial coordinates, we shall divide the KdS spacetime parameter plane $y-a^2$ according to the succession of these radii relative to loci of marginally stable orbits in the individual spacetimes. In the following list we present all possible sequences of that radii, which results in separation of the KdS black hole spacetimes giving their classification according to mixed criteria for the circular geodesics and general photon motion, as illustrated in Fig. \ref{fig_y-a2_plane}:
	 \begin{description}
	 	\item[Case Ia:] $r_o<r_{d1}<r_{ph+}<r_{ph-}<r_{(ms+)i}<r_{(ms-)i}<r_{(ms-)o}<r_{(ms+)o}<r_s<r_{d2}<r_c$
	 	\item[Case Ib:] $r_o<r_{d1}<r_{ph+}<r_{(ms+)i}<r_{ph-}<r_{(ms-)i}<r_{(ms-)o}<r_{(ms+)o}<r_s<r_{d2}<r_c$
	 	\item[Case Ic:] $r_o<r_{ph+}<r_{d1}<r_{(ms+)i}<r_{ph-}<r_{(ms-)i}<r_{(ms-)o}<r_{(ms+)o}<r_s<r_{d2}<r_c$
	 	\item[Case Id:] $r_o<r_{ph+}<r_{(ms+)i}<r_{d1}<r_{ph-}<r_{(ms-)i}<r_{(ms-)o}<r_{(ms+)o}<r_s<r_{d2}<r_c$
	 	\item[Case IIa:] $r_o<r_{d1}<r_{ph+}<r_{ph-}<r_{(ms+)i}<r_{(ms+)o}<r_s<r_{d2}<r_c$
	 	\item[Case IIb:] $r_o<r_{d1}<r_{ph+}<r_{(ms+)i}<r_{ph-}<r_{(ms+)o}<r_s<r_{d2}<r_c$
	 	\item[Case IIc:] $r_o<r_{ph+}<r_{d1}<r_{(ms+)i}<r_{ph-}<r_{(ms+)o}<r_s<r_{d2}<r_c$
	 	\item[Case IId:] $r_o<r_{ph+}<r_{(ms+)i}<r_{d1}<r_{ph-}<r_{(ms+)o}<r_s<r_{d2}<r_c$
	 	\item[Case IIe:] $r_o<r_{ph+}<r_{d1}<r_{ph-}<r_{(ms+)i}<r_{(ms+)o}<r_s<r_{d2}<r_c$
	 	\item[Case IIf:] $r_o<r_{ph+}<r_{d1}<r_{(ms+)i}<r_{(ms+)o}<r_{ph-}<r_s<r_{d2}<r_c$
	 	\item[Case IIg:] $r_o<r_{ph+}<r_{(ms+)i}<r_{d1}<r_{(ms+)o}<r_{ph-}<r_s<r_{d2}<r_c$
	 	\item[Case IIh:] $r_o<r_{ph+}<r_{(ms+)i}<r_{(ms+)o}<r_{d1}<r_{ph-}<r_s<r_{d2}<r_c$
	 	\item[Case IIIa:] $r_o<r_{d1}<r_{ph+}<r_{ph-}<r_s<r_{d2}<r_c$
	 	\item[Case IIIb:] $r_o<r_{ph+}<r_{d1}<r_{ph-}<r_s<r_{d2}<r_c$
	 	\item[Case IIIc:] $r_o<r_{ph+}<r_{ph-}<r_s<r_c$
	 \end{description}

	\subsection{Construction of the light escape cones} 
\begin{figure}[H]
		\begin{tabular}{cc}
			\includegraphics[scale=0.6]{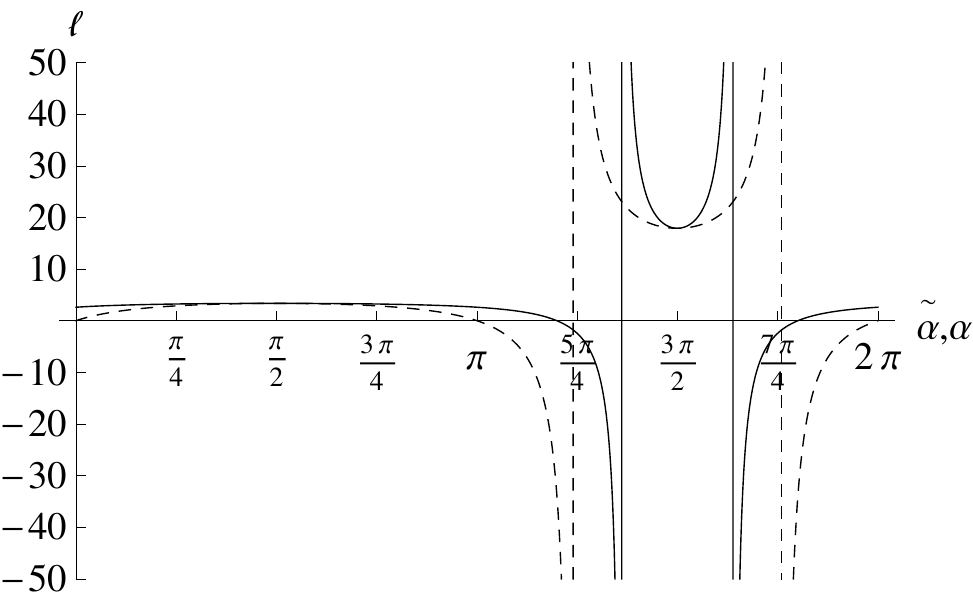}&\includegraphics[scale=0.6]{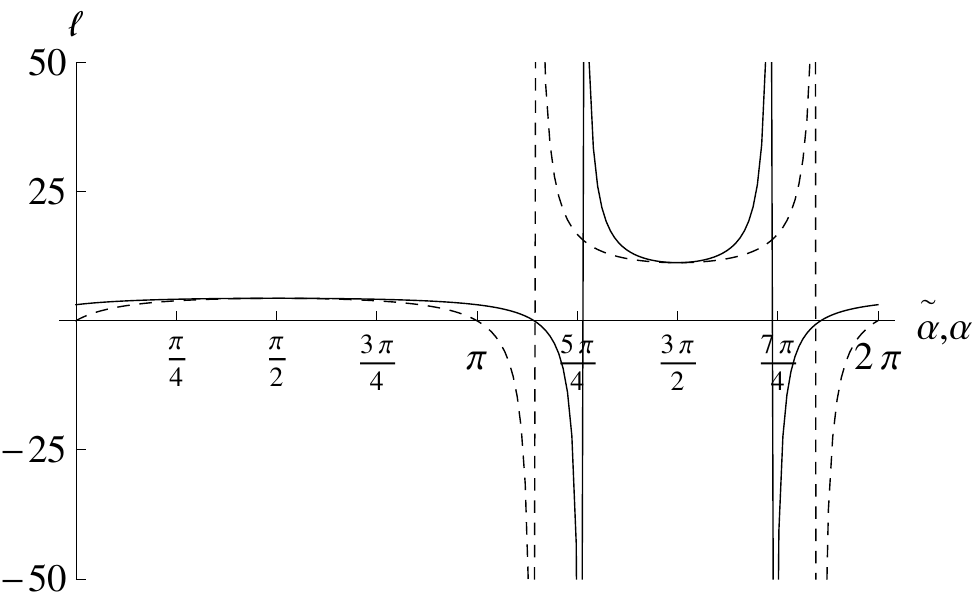}\\
			(a)&(b)\\
			\includegraphics[scale=0.6]{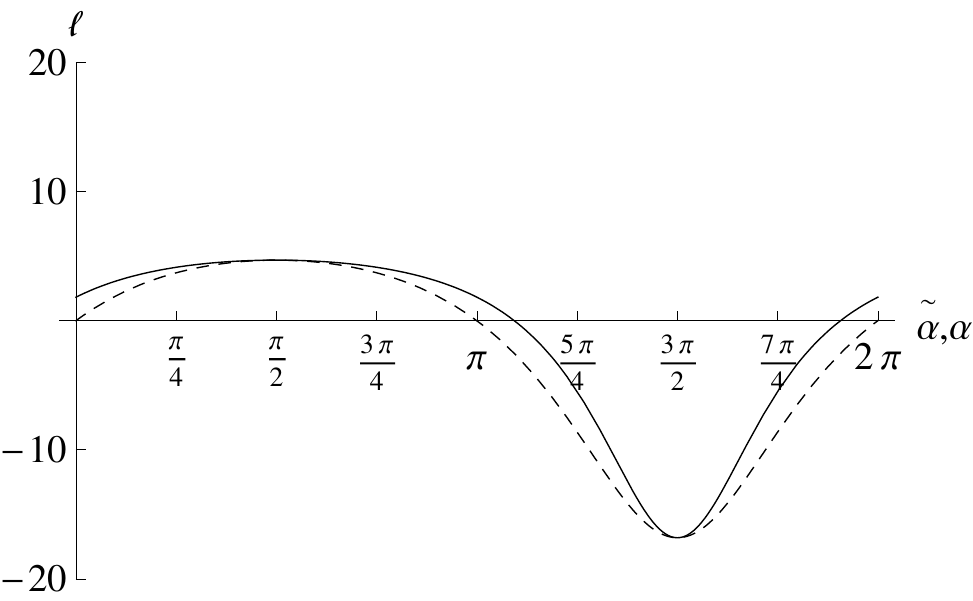}&\includegraphics[scale=0.6]{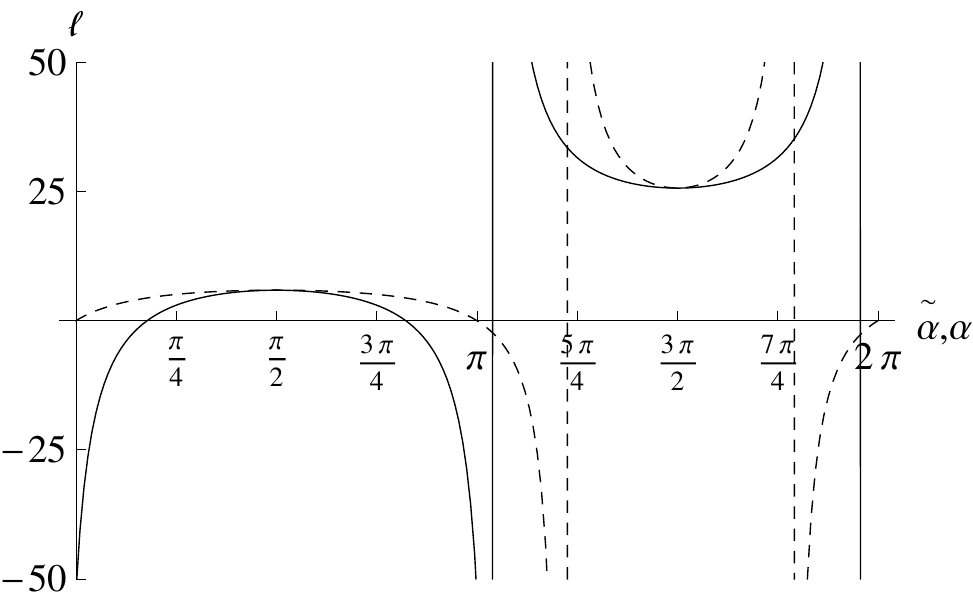}\\
			(c)&(d)\\
			\includegraphics[scale=0.6]{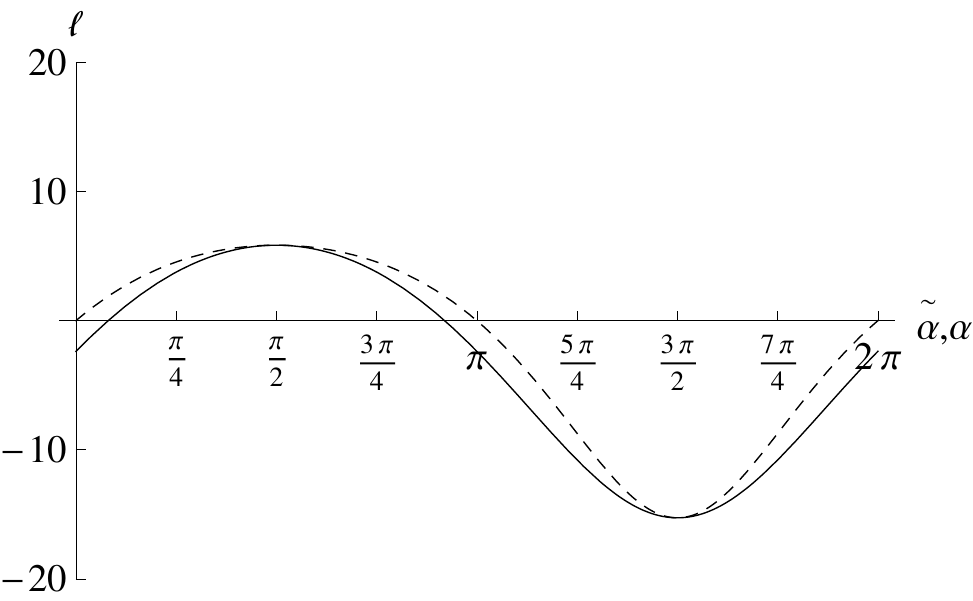}&\includegraphics[scale=0.6]{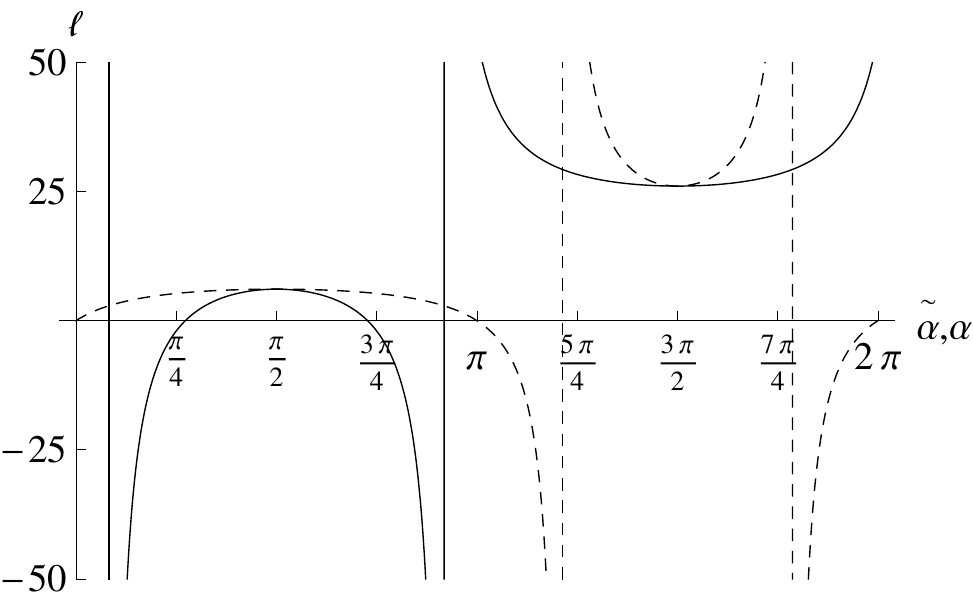}\\
			(e)&(f)
		\end{tabular} \caption{Different types of behaviour of functions $\ell(\tilde{\alpha})$ (full curve) for $v=v_{+}$ and comparison with $\ell(\alpha)$ (dashed curve) for $\tilde{\beta},\beta=90\,^{\circ}$ when the local extrema, located for both functions at $\tilde{\alpha},\alpha=\pi/2,3\pi/2,$ have the same value. The KdS spacetime parameters are $a^2=0.9$ combined with $y=0.02$ (left column) and $y=0.04$ (right column) representing thus both types of spacetimes with DRB and RRB, respectively. The radii of the source are chosen so that they cover all significant cases relevant for spacetimes of both types-DRB: \textbf{(a)} $r_o<r_e=2<r_{d1},$ \textbf{(c)} $r_{d1}<r_e=3<r_{(+)z},$ \textbf{(e)} $r_{(+)z}<r_e=3.65<r_s;$ RRB: \textbf{(b)} $r_o<r_e=2.2<r_{(+)z},$ \textbf{(d)} $r_{(+)z}<r_e=2.8<r_{crit},$ \textbf{(f)} $r_{crit}<r_e=2.85<r_s.$ The behaviour of $\ell(\tilde{\alpha})$ for $v=v_{-}$ is qualitatively the same as in the case \textbf{(c)}. }  \label{fig_ell(alpha)CGF}
	\end{figure}

	The dependence of the impact parameter $\ell$ of a photon on its directional angles in the CGF is given by
	\be
	\ell=\frac{A(\sin \tilde{\alpha} \sin \tilde{\beta}+v_{\pm})}{r^2\sqrt{\Delta_{r}}+v_{\pm}A\Omega+(v_{\pm}r^2\sqrt{\Delta_{r}}+A\Omega_{LNRF})\sin \tilde{\alpha} \sin \tilde{\beta}}, \label{ell(alpha)CGF} 
	\ee
	which, in accordance with (\ref{ell(alpha)LNRF}), takes the form valid in the LNRF for $v_{\pm}=0$. In Fig. \ref{fig_ell(alpha)CGF} we illustrate all qualitatively possible types of behaviour of the function $\ell(\tilde{\alpha}).$ It is compared with  $\ell(\alpha)$ in common graph for $\tilde{\beta},\beta=90\,^{\circ},$ since their local extrema located at $\tilde{\alpha}=\pi/2,3\pi/2$ then have the same value. In order to clearly display the behaviour of the divergency points of $\ell(\tilde{\alpha})$, these functions are plotted for $\tilde{\beta},\beta=90\,^{\circ}$ in the range $0\leq\alpha \leq 2\pi,$ although the natural interval is $0\leq\alpha \leq \pi.$ However, this can be done because any direction determined by double $(\alpha,\beta),$ where $\pi\leq \alpha\leq2\pi,$ $0\,^{\circ}\leq\beta\leq180\,^{\circ}$ can be identified with that described by double $(\acute{\alpha}, \acute{\beta}),$ $0\leq\acute{\alpha}\leq\pi,$ where $\acute{\alpha}=2\pi-\alpha,$ $\acute{\beta}=\beta+180\,^{\circ}.$ With this note in mind we have to understand the graphs bellow. \par
	The divergent points $\tilde{\alpha}_{d}$ of $\ell(\tilde{\alpha})$ again exist in a regions where $0\leq \Delta_{r} \leq a^2,$ i.e., for $r_{o}\leq r \leq r_{d1}$ (Fig. \ref{fig_ell(alpha)CGF}a) or $r_{d2} \leq r \leq r_{c}$ in spacetimes with a divergent barrier of a photon motion, or anywhere in the stationary region, i.e., for $r_{o}\leq r \leq r_{c}$ in spacetimes with a restricted barrier of a photon motion (Figs. \ref{fig_ell(alpha)CGF}b, d, e). The 'most common' is the behaviour shown in Figs. \ref{fig_ell(alpha)CGF}a, b, c when $v=v_{+}>0,$ or $v=v_{-},$ but for $v_{+}<0$ some qualitative changes arise (see Figs. \ref{fig_ell(alpha)CGF}d, e, f). 
	The radii of orbits with  $v_{+}<0$ satisfy the inequality $r_{(+)z} \leq r \leq r_{s},$ where $r_{(+)z}$ is a solution between the outer black hole and cosmological horizon of the equation
		\be
		y=\frac{-\sqrt{r}[4a^2+r(r^2+a^2)]+\sqrt{(r^2+a^2)[r^3(r^2+a^2)+4a^2(3r^2+a^2) ] } }{2a^2\sqrt[3]{r^2}(r^2+a^2)}. \label{yZAM}
		\ee
	The function in (\ref{yZAM}) is in \cite{Stu-Sla:2004:PHYSR4:} denoted $y_{(L=0)}(r; a),$ since it determines loci of circular orbits with zero angular momentum. However, this solution does not satisfy conditions necessary for existence of the stable circular orbits, see  \cite{Stu-Sla:2004:PHYSR4:}. Hence, orbits with $v_{+}<0$ must be unstable. Note that the stable orbits with $L=0$ exist in the KdS naked singularity spacetimes \cite{Stu:1980:BULAI:,Stu-Sla:2004:PHYSR4:} that are not considered in the present paper. \par 
	Further, it can be shown that in spacetimes with RRB there exist radii $r_{(+)z} < r_{crit} < r_{s}$ of the plus-family orbits with velocities
	\be
	 -1< v_{+} <-\frac{r^2\sqrt{\Delta_{r}}}{A\Omega_{LNRF}}\equiv v_{(+)crit} \label{vlemez},
	 \ee
	 for which the divergent points $\tilde{\alpha}_{d}$ of $\ell(\tilde{\alpha})$ occur even when $0\leq \tilde{\beta}\leq90\,^{\circ}.$ This case is illustrated in Fig.\ref{fig_ell(alpha)CGF}f. \par

\begin{figure}[H]
			\flushleft \textbf{Case IIIc: $y=0.04,\quad a^2=0.9$}\\
			
			\bet{lcr}
			\hline
			\bet{l}$r_{o}=1.79355$\\$r_{c}=2.92402$\ent & \bet{c}$r_{ph+}=1.88639$\\$r_{ph-}=2.88665$ \ent&\bet{l} $r_{s}=3.33743$\\ $r_{crit}=2.82314$ \ent 
			\ent
			
			\begin{tabular}{|lcr|}
				\hline
				\multicolumn{3}{|c|}{}\\
				\includegraphics[width=3cm]{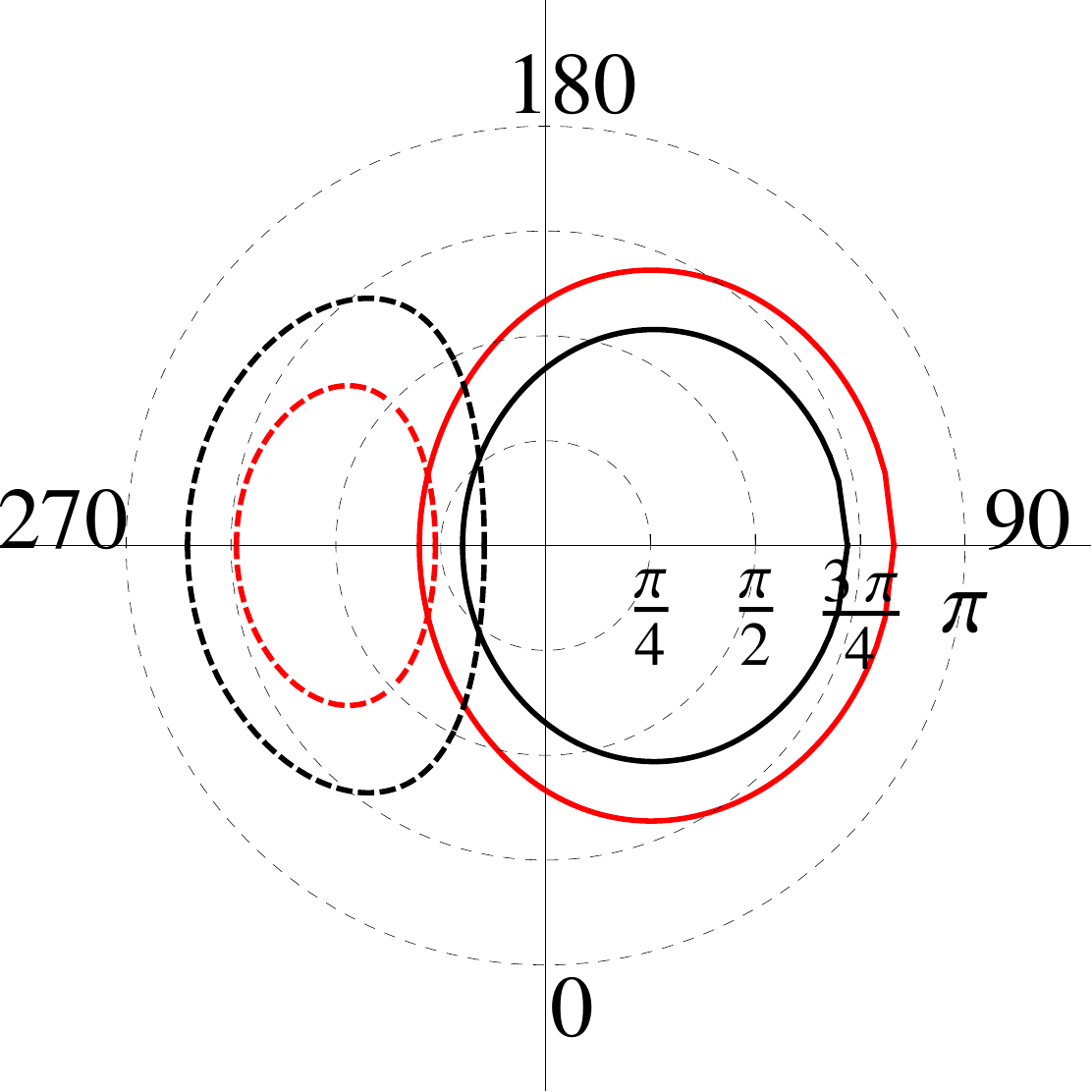}&\includegraphics[width=3cm]{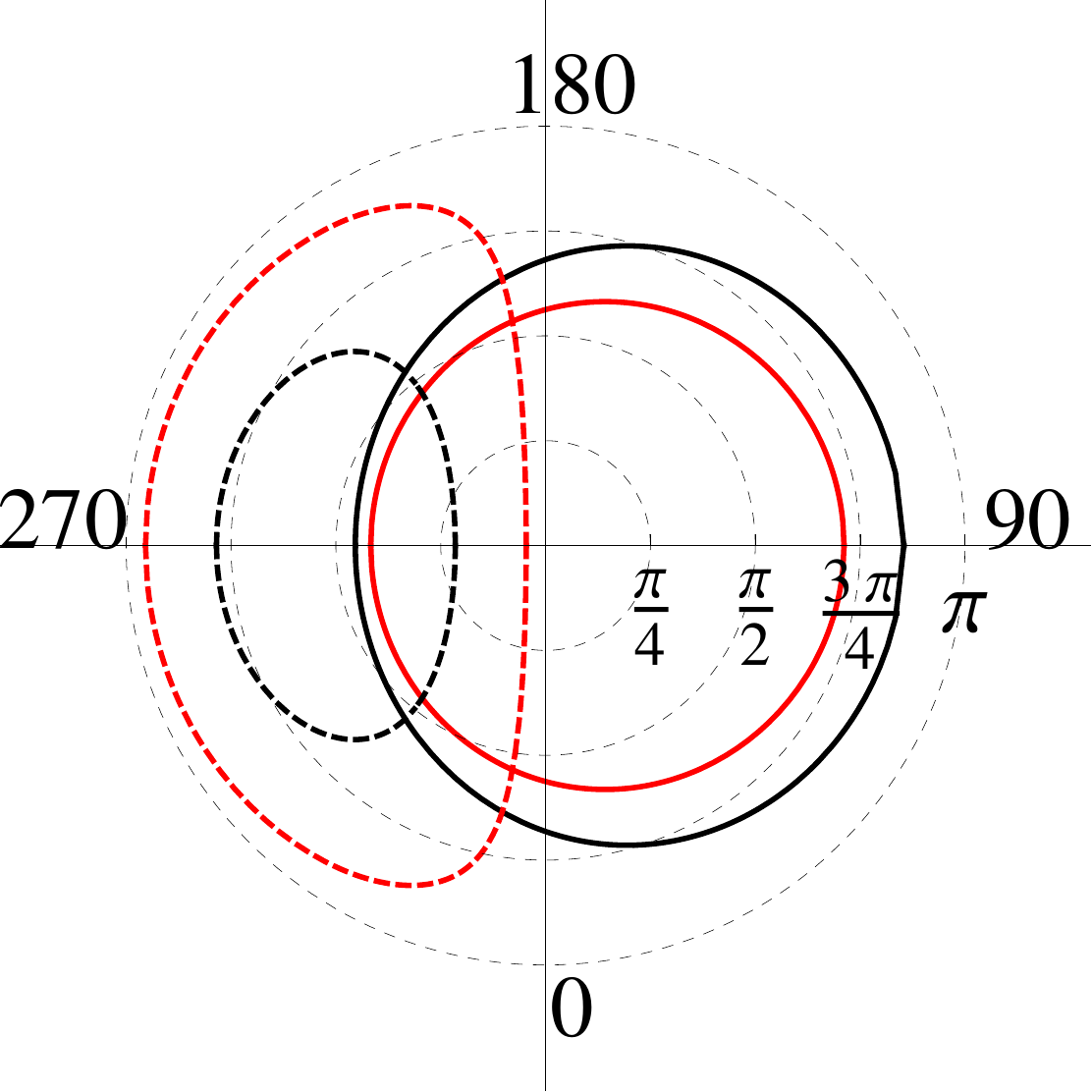}&\includegraphics[width=3cm]{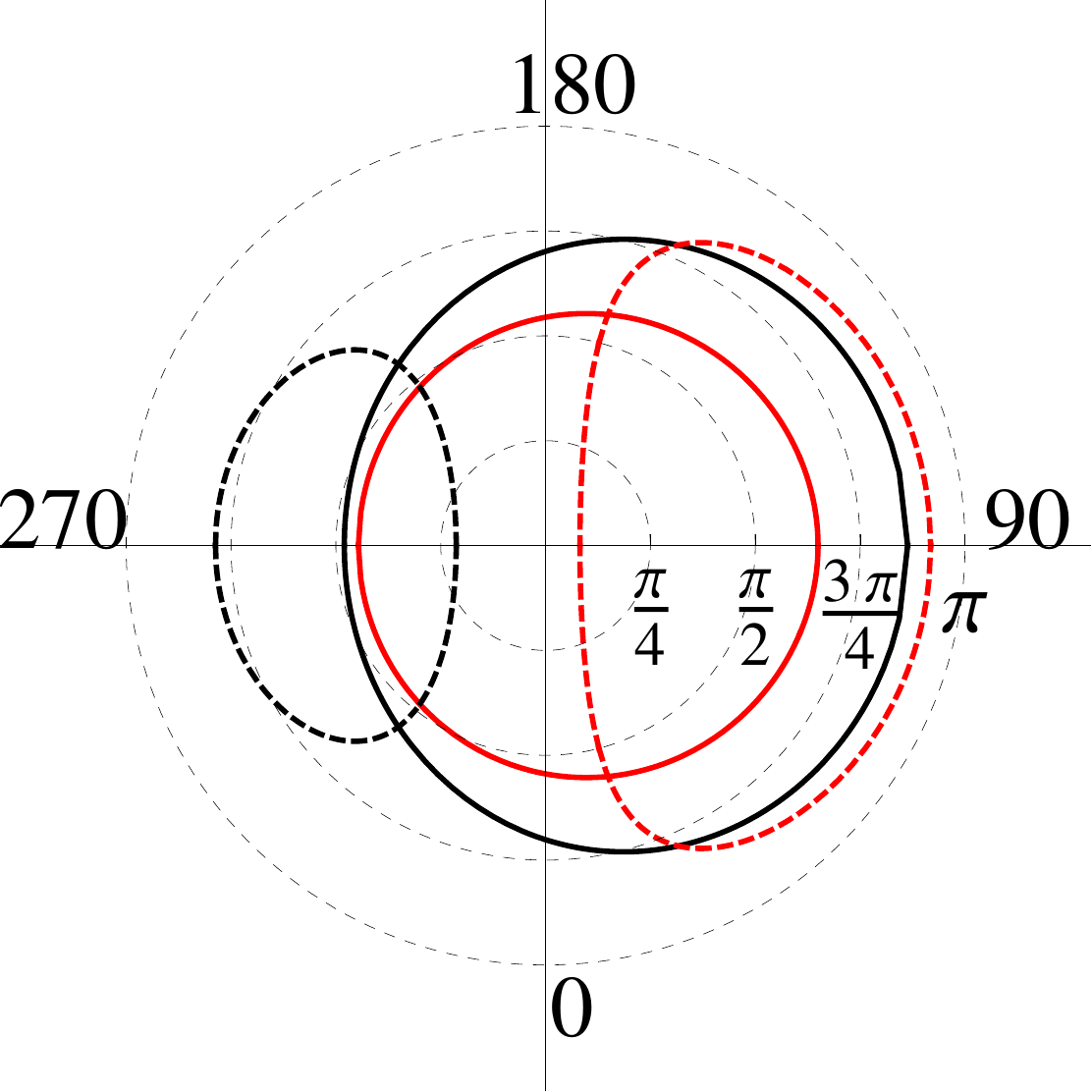}\\
				$r_{1}=2.2$ & $r_{2}=2.8$& $r_{3}=2.85$ \\
				\hline
				
			\end{tabular}
			\caption{Light escape cones are constructed for the CGF sources orbiting at the KdS black hole spacetimes along the innermost and outermost marginally stable equatorial circular orbits of both plus and minus family orbits, if they exist (red curves), and compared with local escape cones constructed for the LNRF sources located at the same radii (black curves). The local escape cones are constructed successively for all types of black hole spacetimes differing by order of outstanding radii. The curves determining local escape cones represent a boundary between two areas of observer's sky, which he recognizes as photons captured by the black hole or escaping to cosmological horizon. The area of escaping photons can be easily identified as the one containing the centre of considered figure ('zenith'). The dashed curves, if present, delimit flux transmitted by counter-rotating photons with positive impact parameter $X$ and negative energy $E.$ Directional angle $\alpha$ is measured in radians, directional angle $\beta$ in degrees. In the case III.c there are no stable circular orbits, the radii are chosen so that they successively correspond to unstable plus-family circular orbits with velocities $v_+>0,$ $v_{(+)crit}<v_+<0,$ $-1<v_+<v_{(+)crit}.$ In the last case, some of the locally counter-rotating photons with positive impact parameter appears in CGF as being co-rotating. } \label{LEC_CGF}
		\end{figure}

For all of the \KdS black hole spacetimes we give for typical radii of circular geodesics in Fig. \ref{LEC_CGF} the light escape cones in the related CGFs in the 2D representation. We again use the same procedure as in the case of the LNRFs and RGFs. The resulting CGF cones are compared to the related cones constructed in the corresponding LNRFs. 

	\newpage
		\begin{figure}[H]
			\flushleft \textbf{Case Ia: $y=3.10^{-5},\quad a^2=0.2$}\\
			
			\bet{lccr}
			\hline
			\bet{l}$r_{o}=1.89466$\\$r_{c}=181.566$\ent & \bet{c}$r_{ph+}=2.42587$\\$r_{ph-}=3.47923$ \ent&\bet{l} $r_{d1}=2.00025$\\$r_{d2}=181.565$\ent & $r_{s}=32.183$
			\ent
			
			\begin{tabular}{|lccr|}
				\hline
				\multicolumn{4}{|c|}{}\\
				\includegraphics[width=3cm]{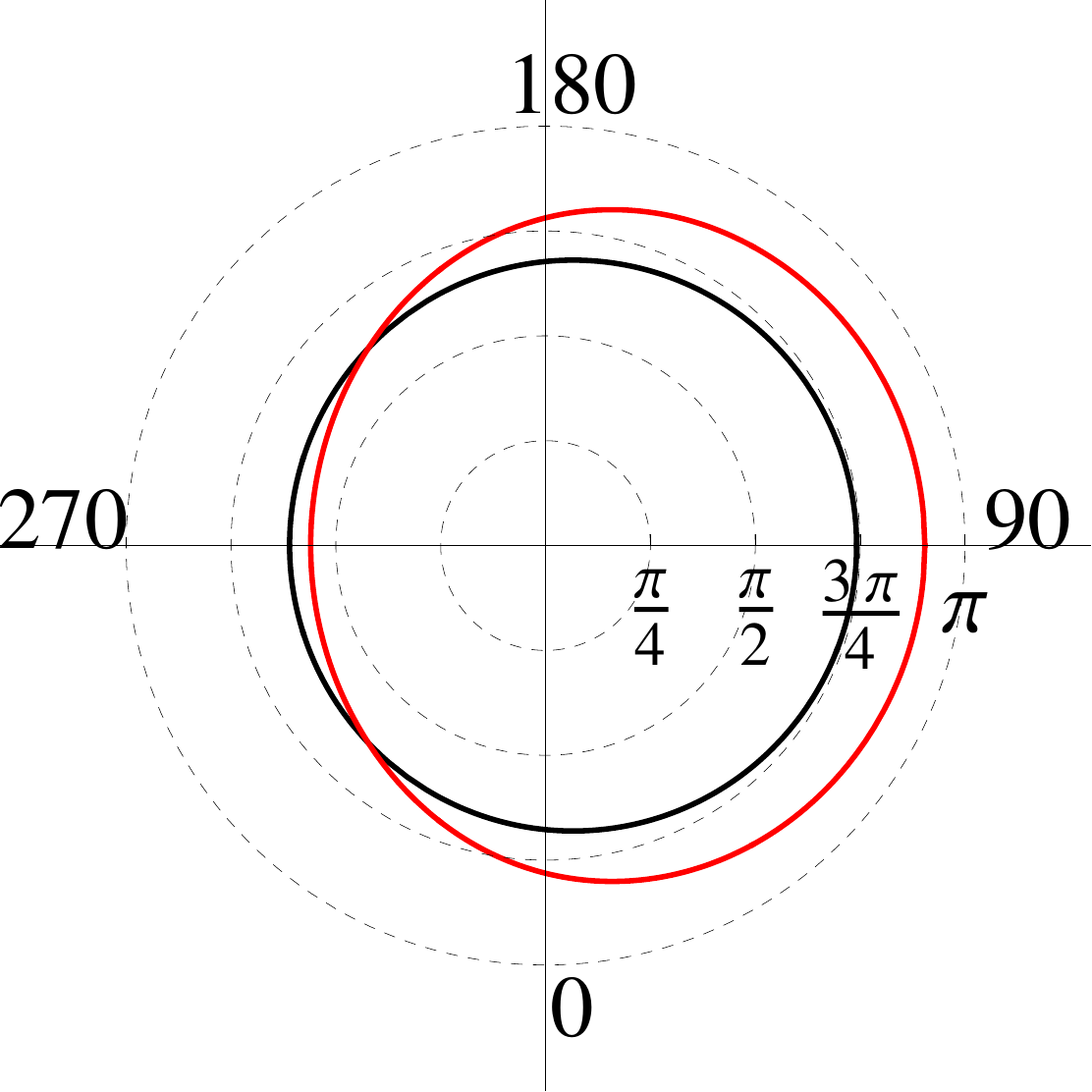}&\includegraphics[width=3cm]{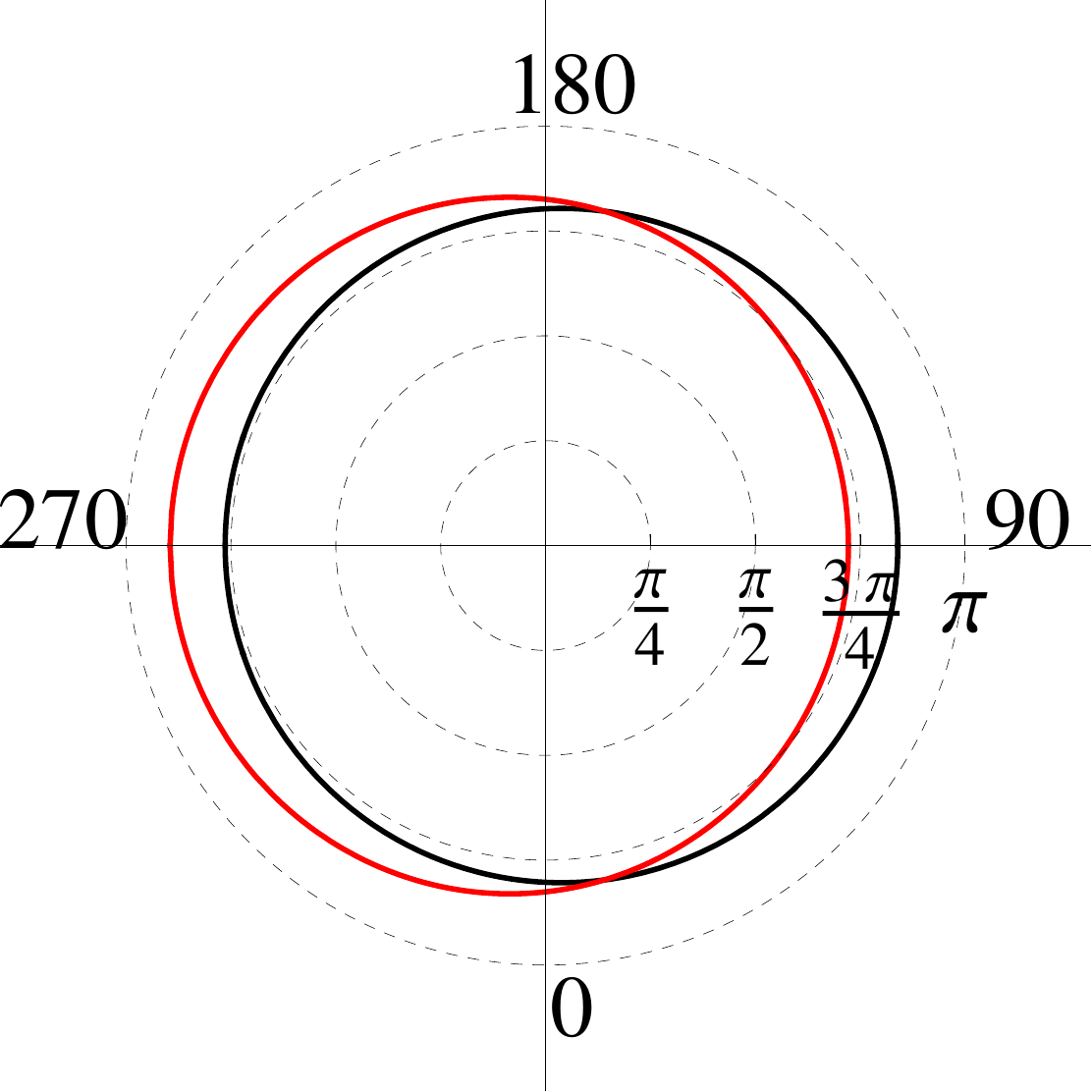}&\includegraphics[width=3cm]{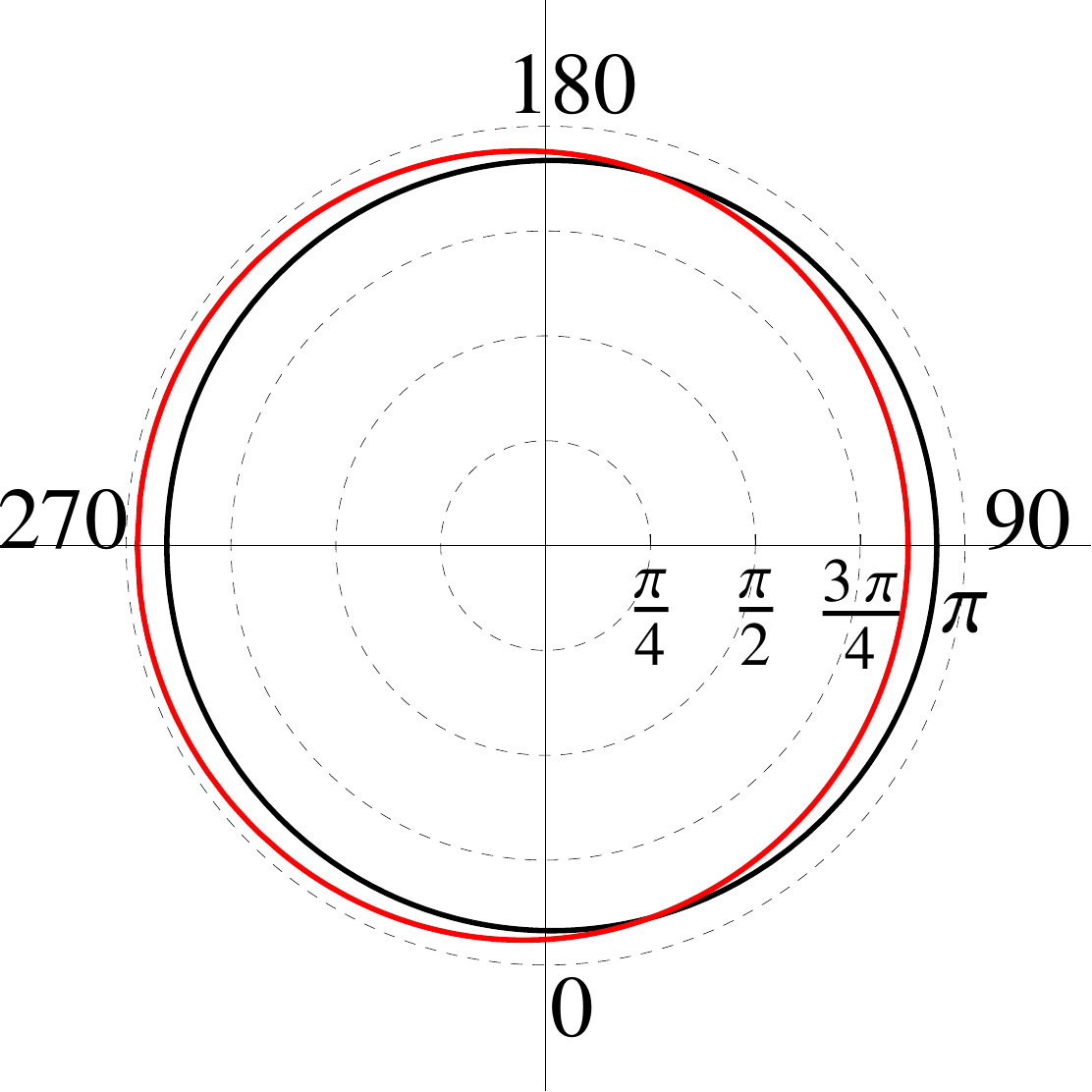}&\includegraphics[width=3cm]{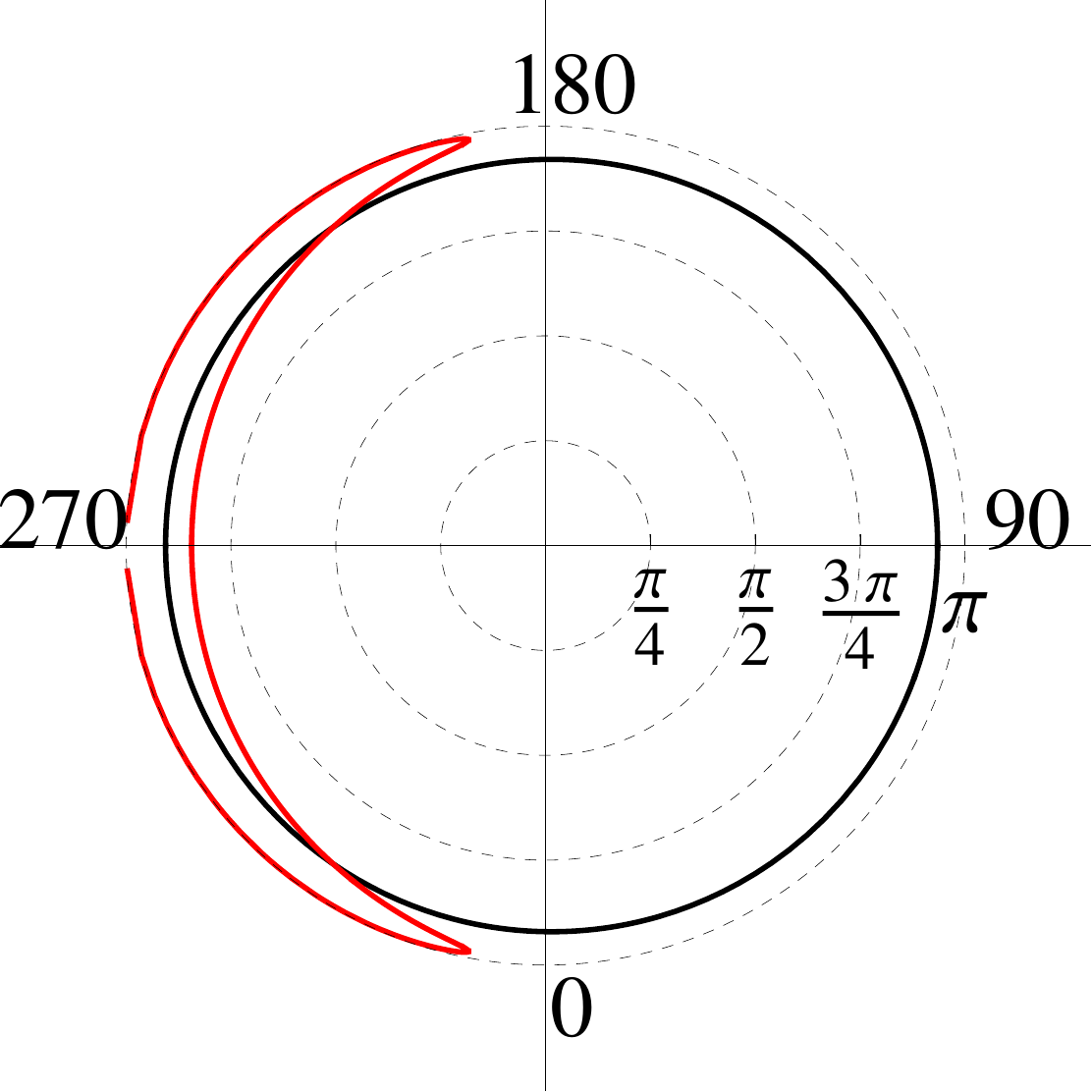}\\
				$r_{(ms+)i}=4.45364$ & $r_{(ms-)i}=7.55276$&$r_{(ms-)o}=18.9049$ & $r_{(ms+)o}=19.5051$ \\		
			  	\hline
				
			\end{tabular}
				 
		\end{figure}
		
		\begin{figure}[H]
			\flushleft \textbf{Case Ib: $y=3.10^{-5},\quad a^2=0.45$}\\
			
			\bet{lccr}
			\hline
			\bet{l}$r_{o}=1.74183$\\$r_{c}=181.566$\ent & \bet{c}$r_{ph+}=2.06696$\\$r_{ph-}=3.69726$ \ent&\bet{l} $r_{d1}=2.00027$\\$r_{d2}=181.565$\ent & $r_{s}=32.183$
			\ent
			
			\begin{tabular}{|lccr|}
				\hline
				\multicolumn{4}{|c|}{}\\
				\includegraphics[width=3cm]{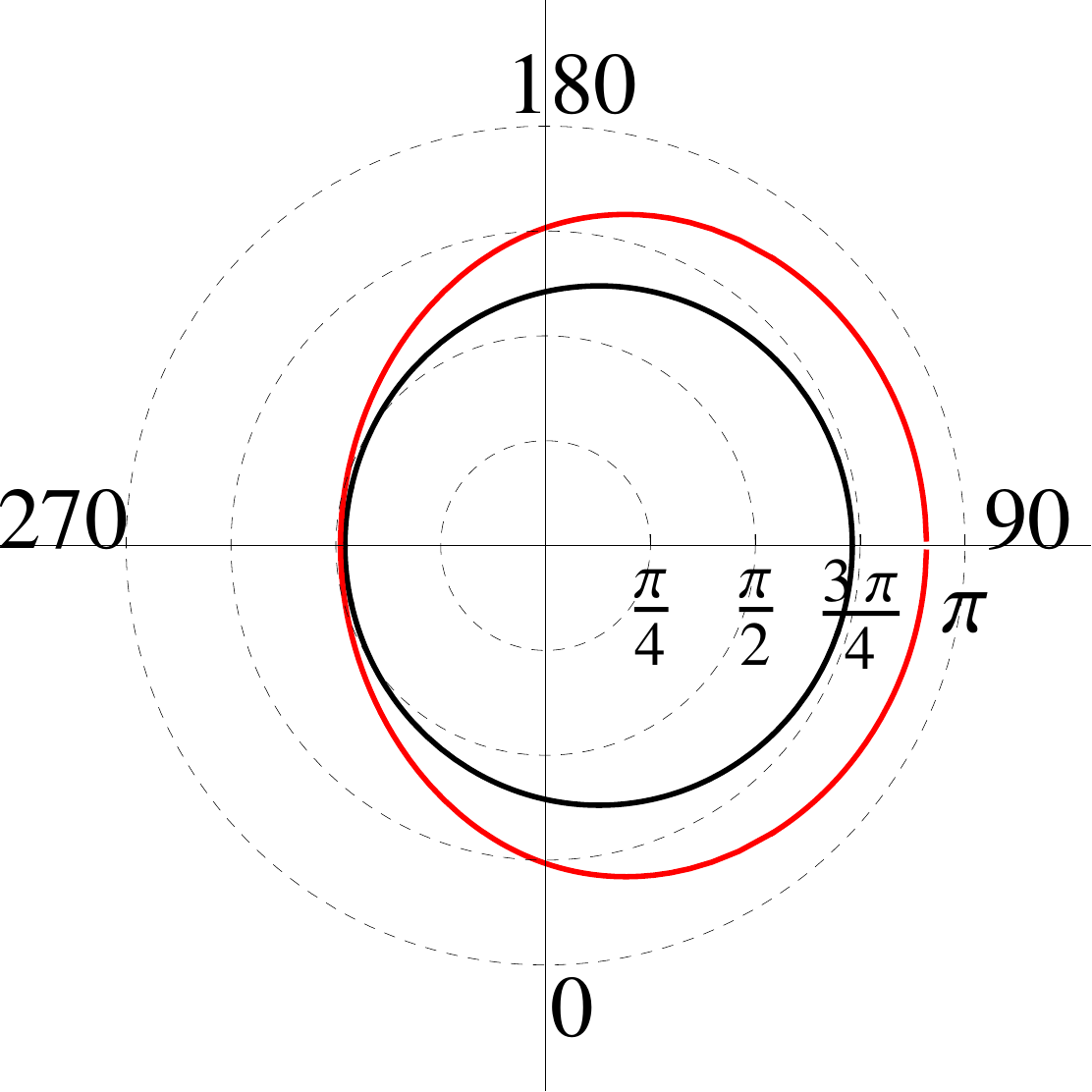}&\includegraphics[width=3cm]{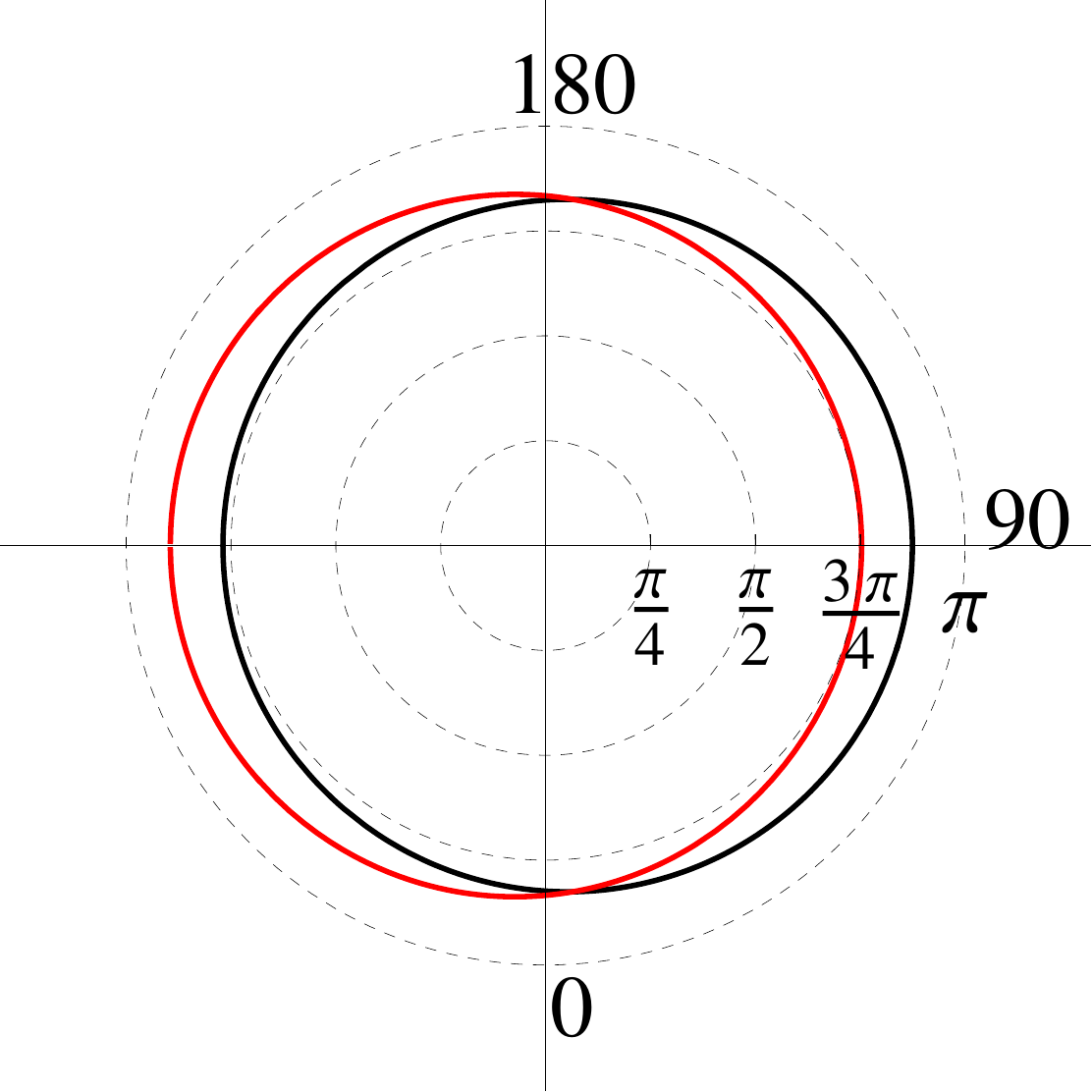}&\includegraphics[width=3cm]{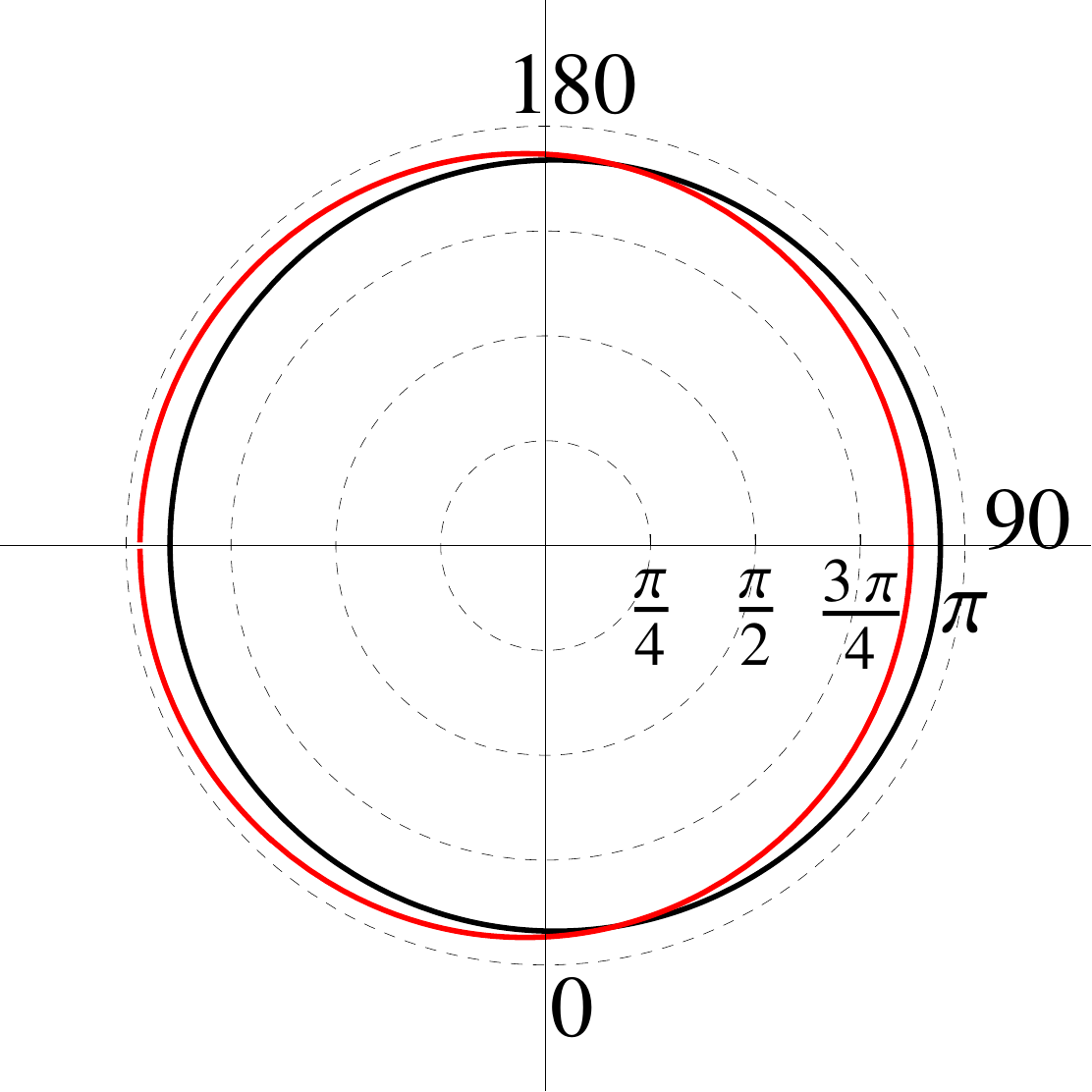}&\includegraphics[width=3cm]{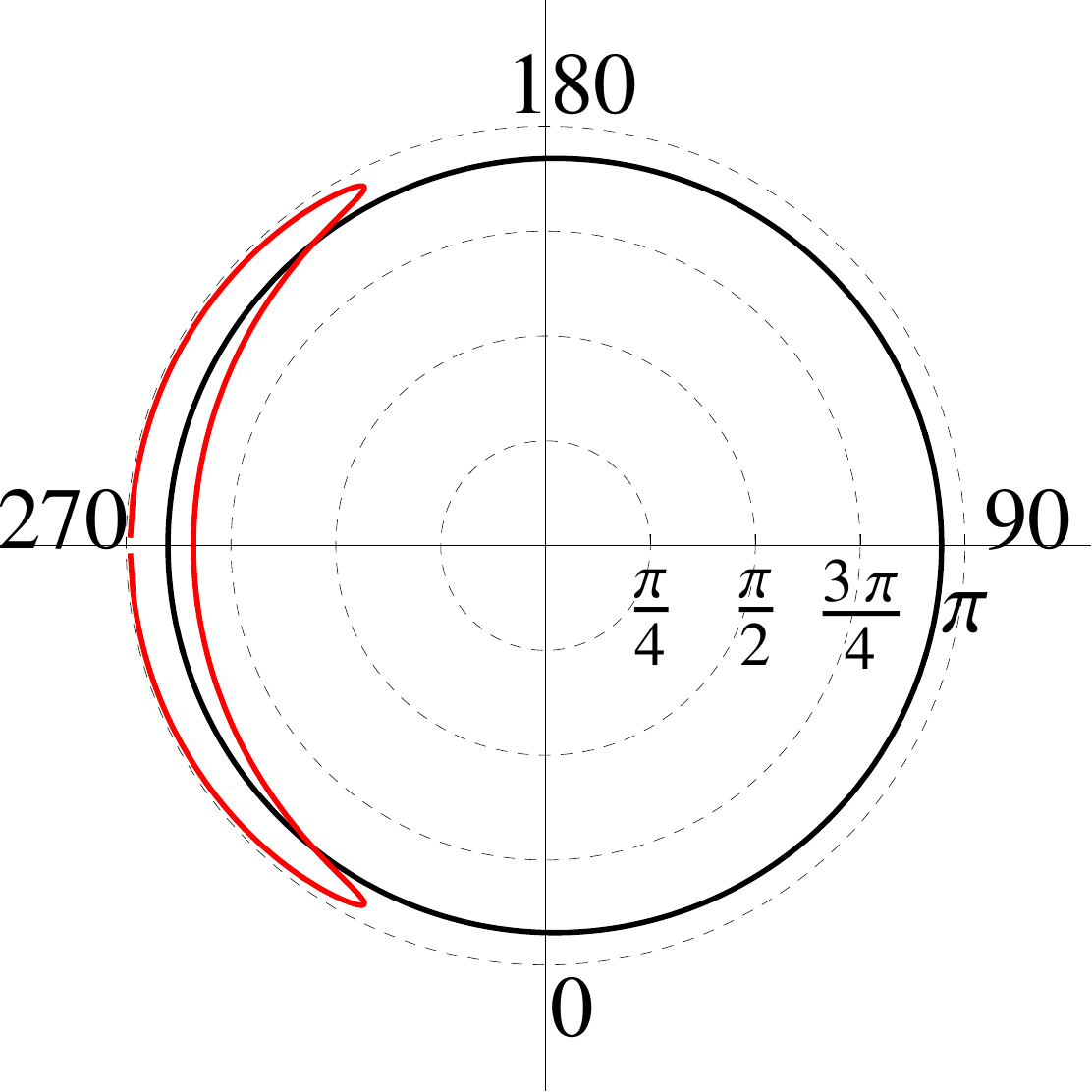}\\
				$r_{(ms+)i}=3.53092$ & $r_{(ms-)i}=8.28963$& $r_{(ms-)o}=18.7013$ & $r_{(ms+)o}=19.6164$ \\
				\hline
				
			\end{tabular}
			 
		\end{figure}
		
		\begin{figure}[H]
			\flushleft \textbf{Case Ic: $y=3.10^{-5},\quad a^2=0.6$}\\
			
			\bet{lccr}
			\hline
			\bet{l}$r_{o}=1.63266$\\ $r_{c}=181.566$\ent & \bet{c}$r_{ph+}=1.86607$\\$r_{ph-}=3.79469$ \ent&\bet{l} $r_{d1}=2.00028$\\$r_{d2}=181.564$\ent & $r_{s}=32.183$
			\ent
			
			\begin{tabular}{|llll|}
				\hline
				\multicolumn{4}{|c|}{}\\
				\includegraphics[width=3cm]{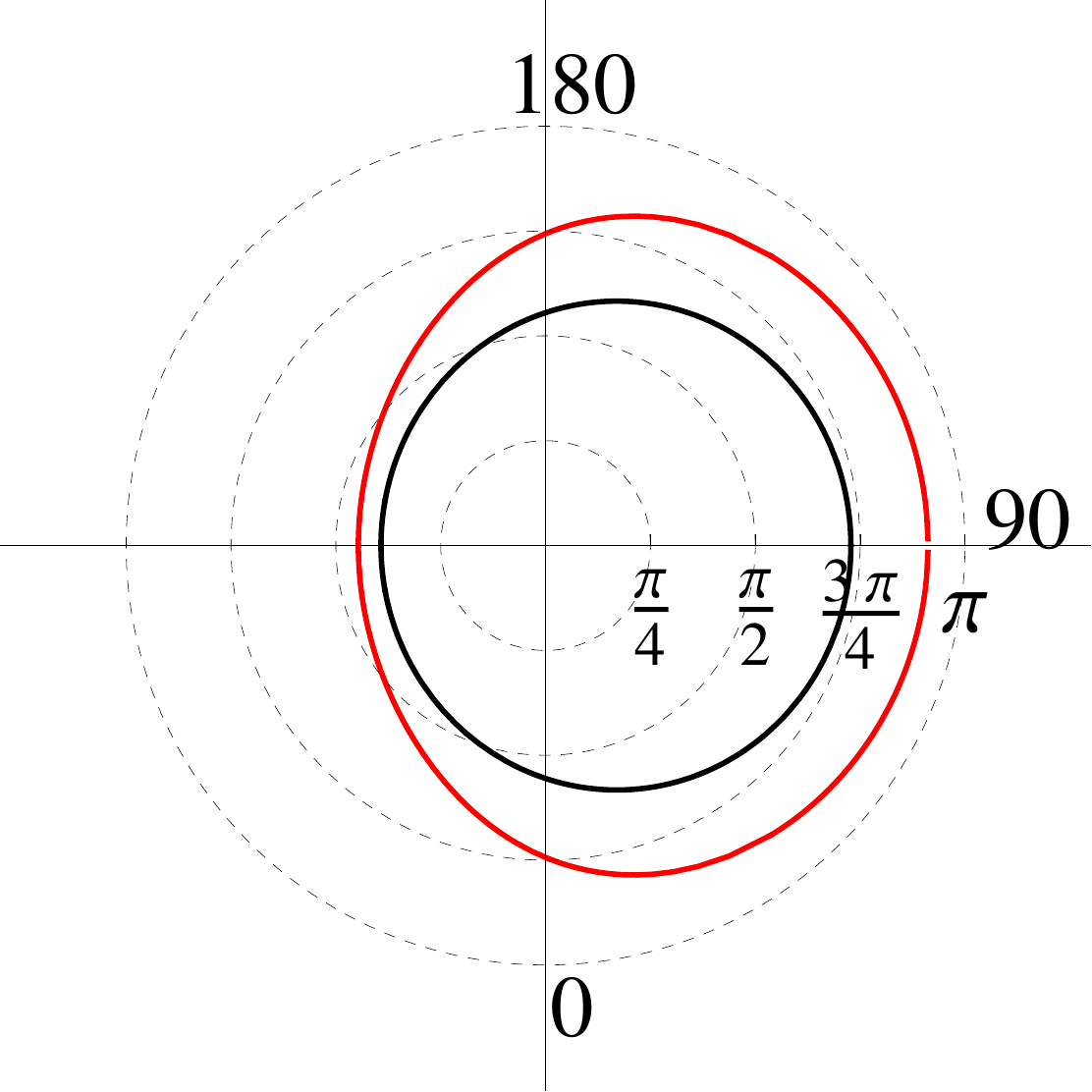}&\includegraphics[width=3cm]{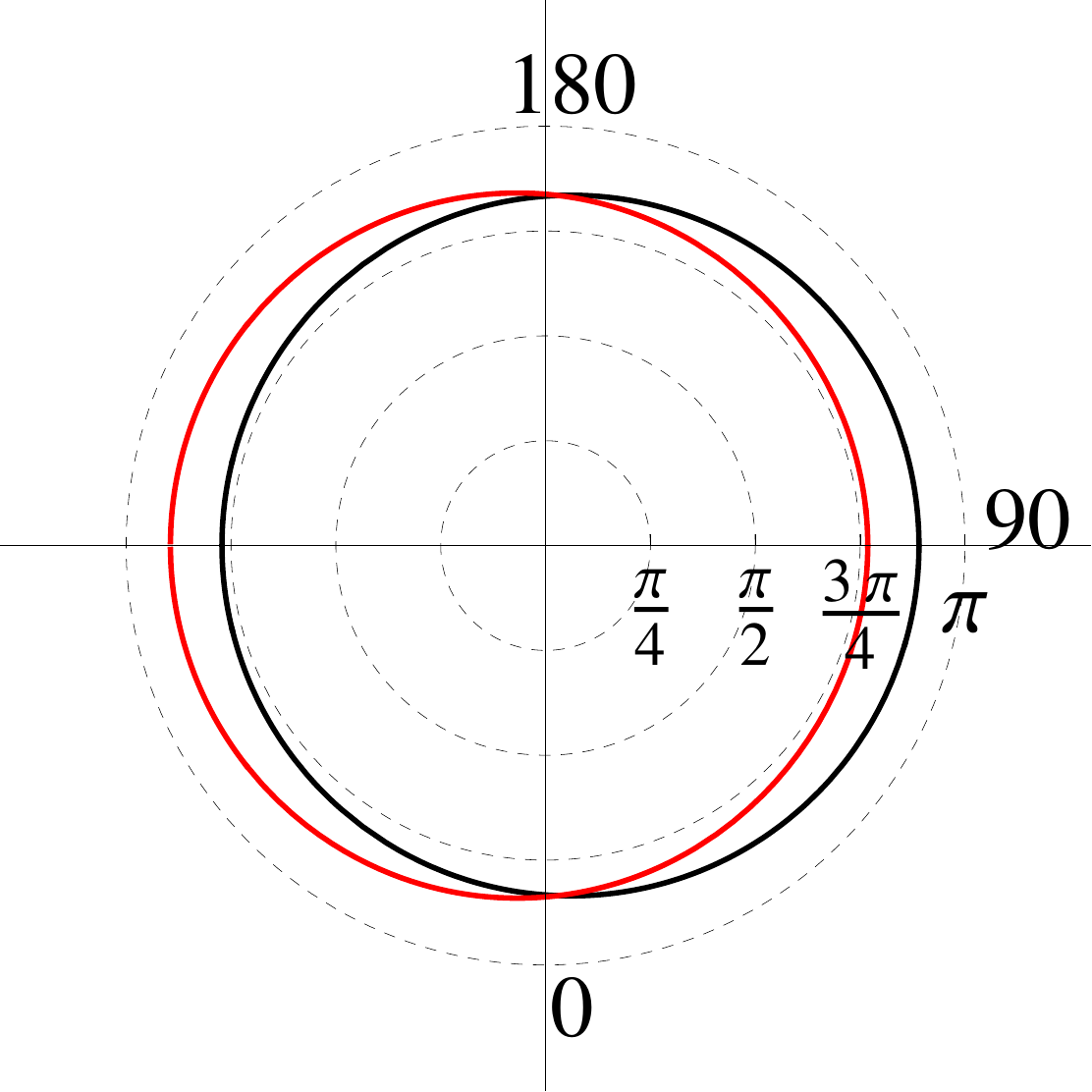}&\includegraphics[width=3cm]{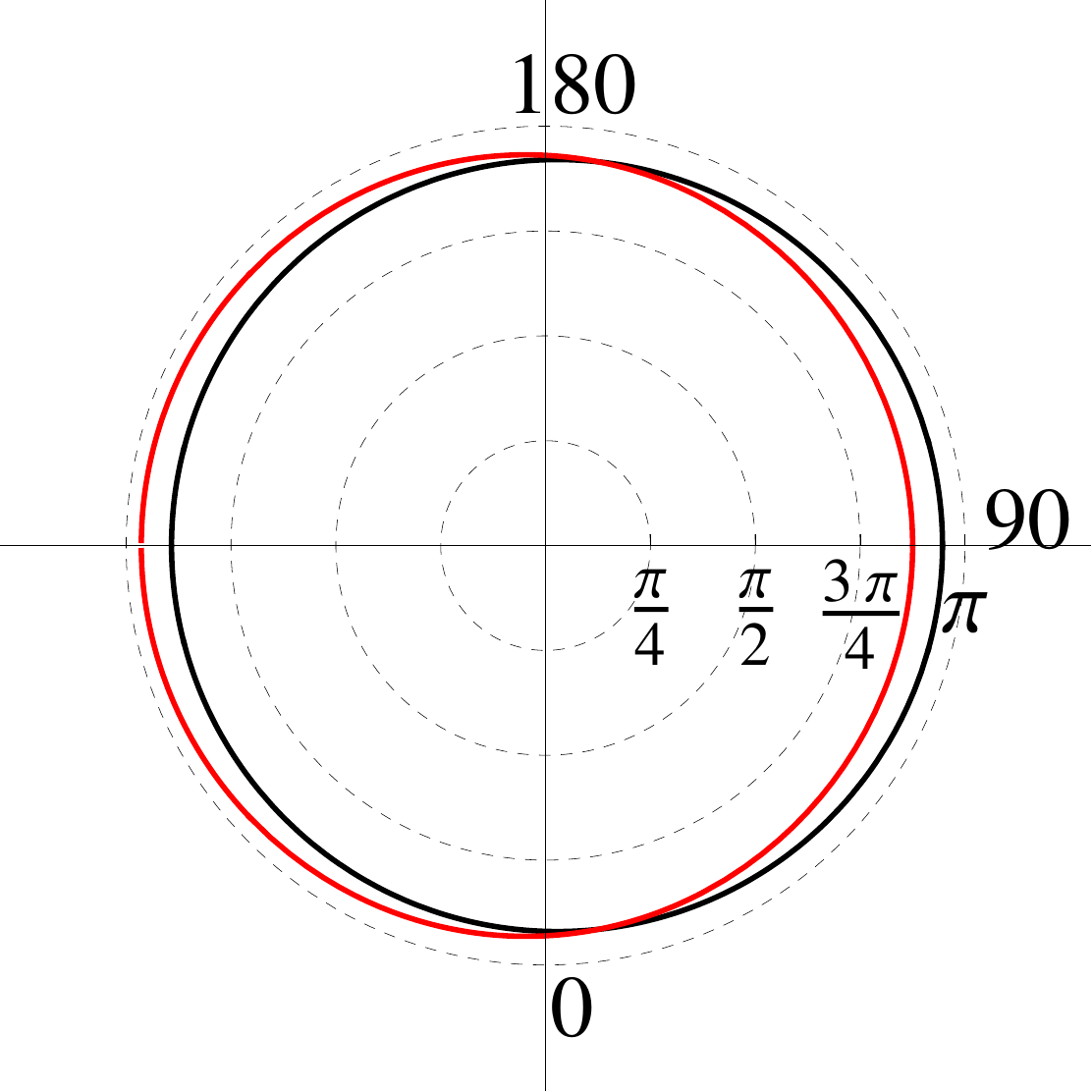}&\includegraphics[width=3cm]{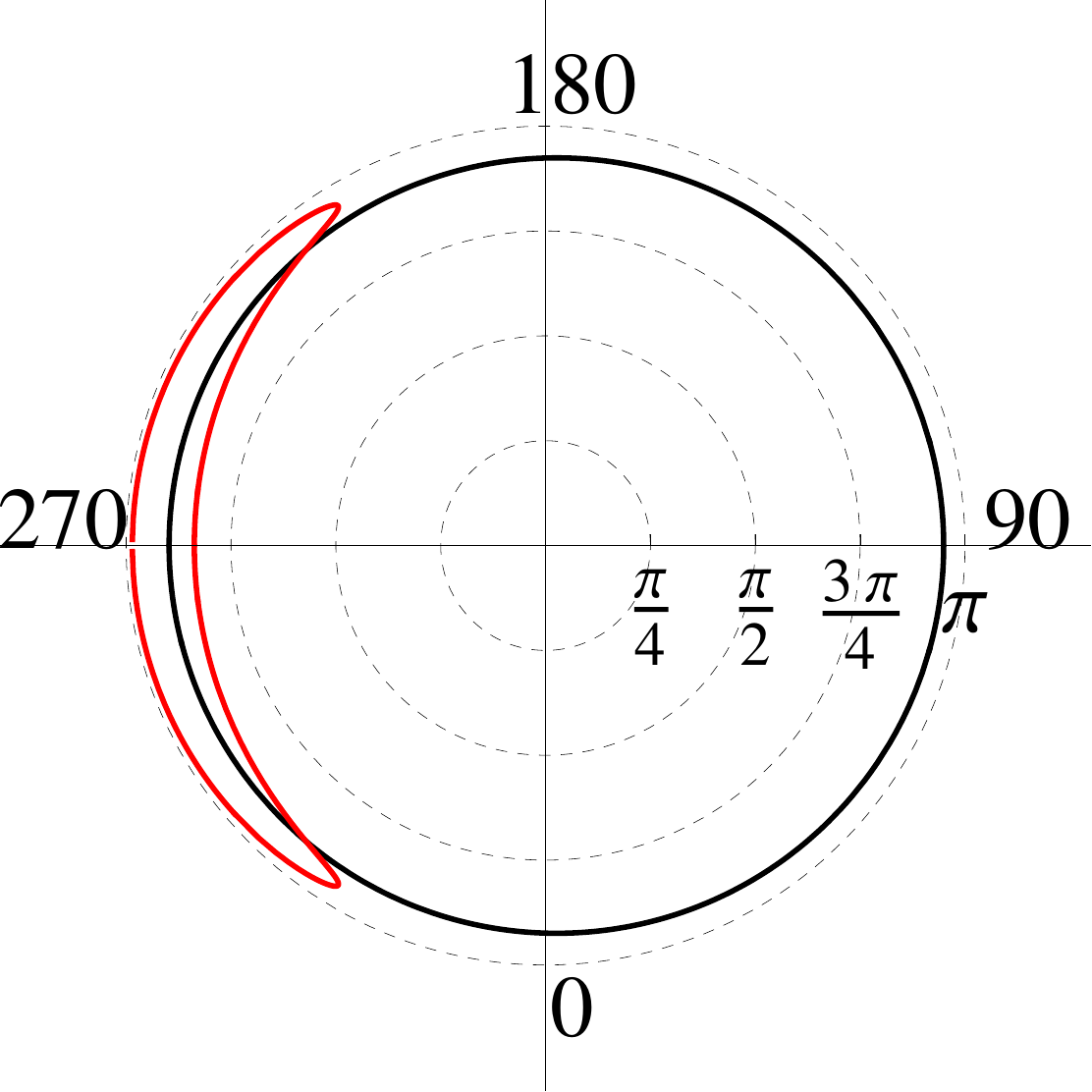}\\
				$r_{(ms+)i}=3.04027$ & $r_{(ms-)i}=8.63397$ & $r_{(ms-)o}=18.5965$ & $r_{(ms+)o}=19.6639$ \\
				\hline
				
			\end{tabular}
		\end{figure}
		\newpage
		\begin{figure}[H]
			\flushleft \textbf{Case Id: $y=3.10^{-5},\quad a^2=0.95$}\\
			
			\bet{lccr}
			\hline
			\bet{l}$r_{o}=1.22385$\\$r_{c}=181.563$\ent & \bet{c}$r_{ph+}=1.27098$\\$r_{ph-}=3.97672$ \ent&\bet{l} $r_{d1}=2.0003$\\$r_{d2}=181.563$\ent & $r_{s}=32.183$
			\ent
			
			\begin{tabular}{|llll|}
				\hline
				\includegraphics[width=3cm]{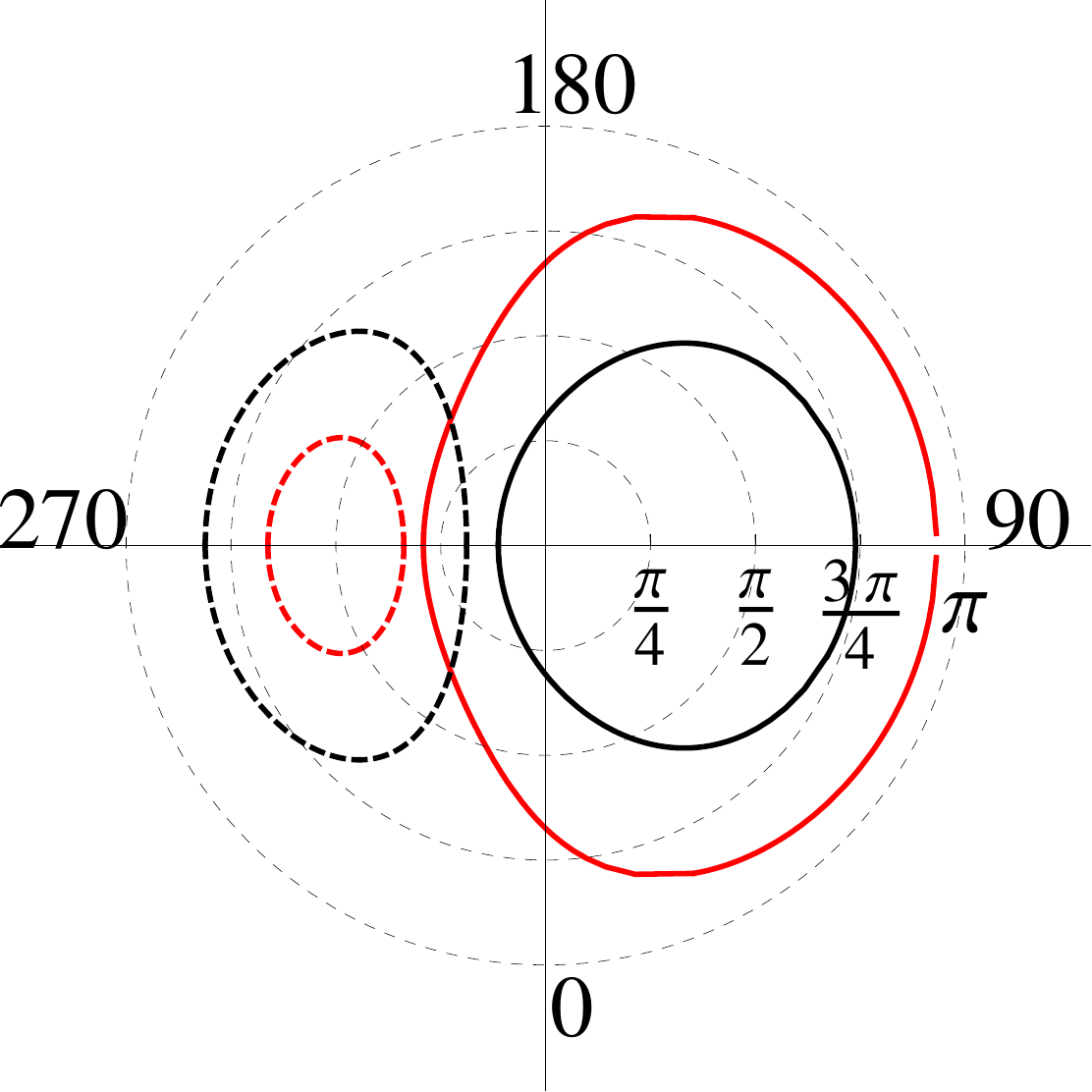}&\includegraphics[width=3cm]{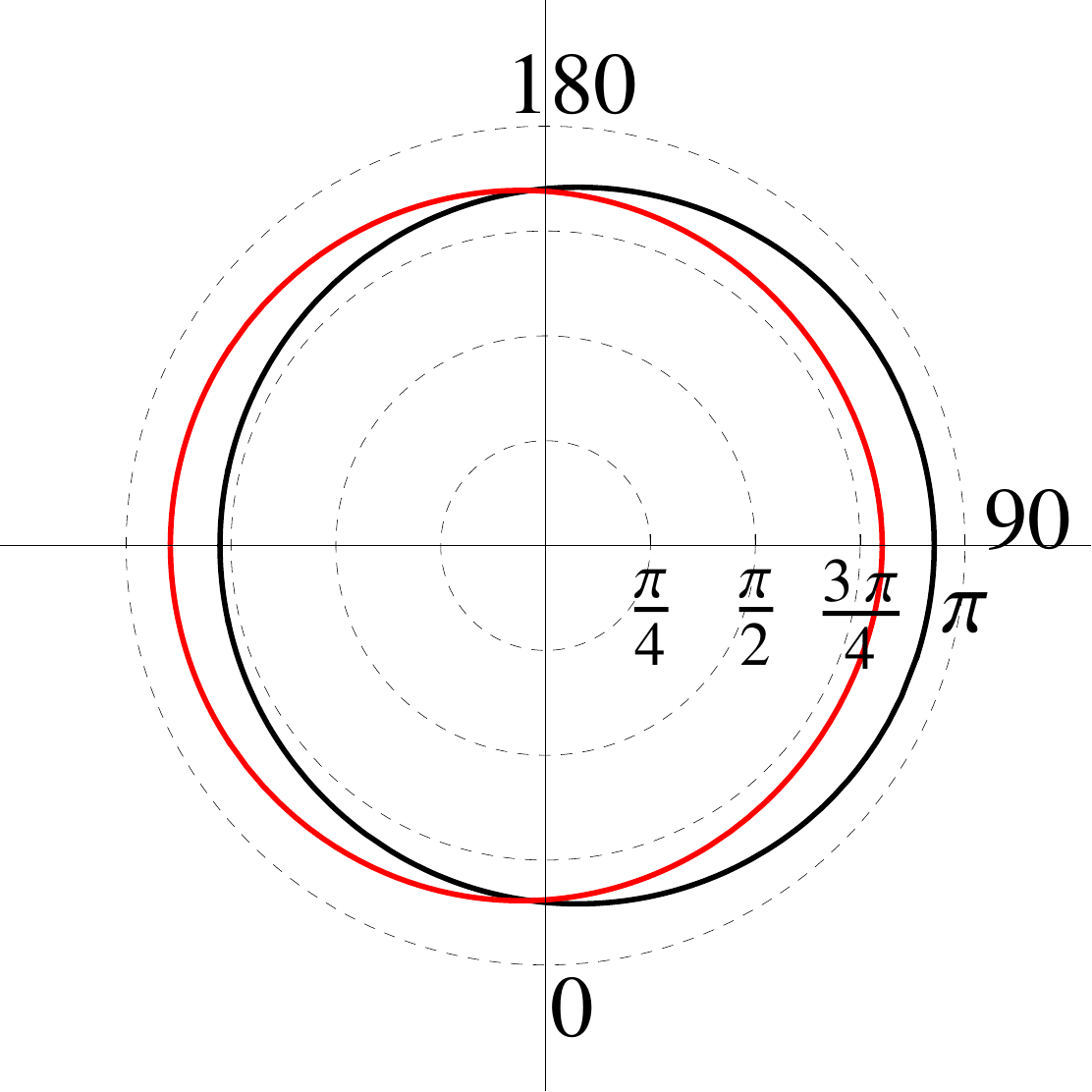}&\includegraphics[width=3cm]{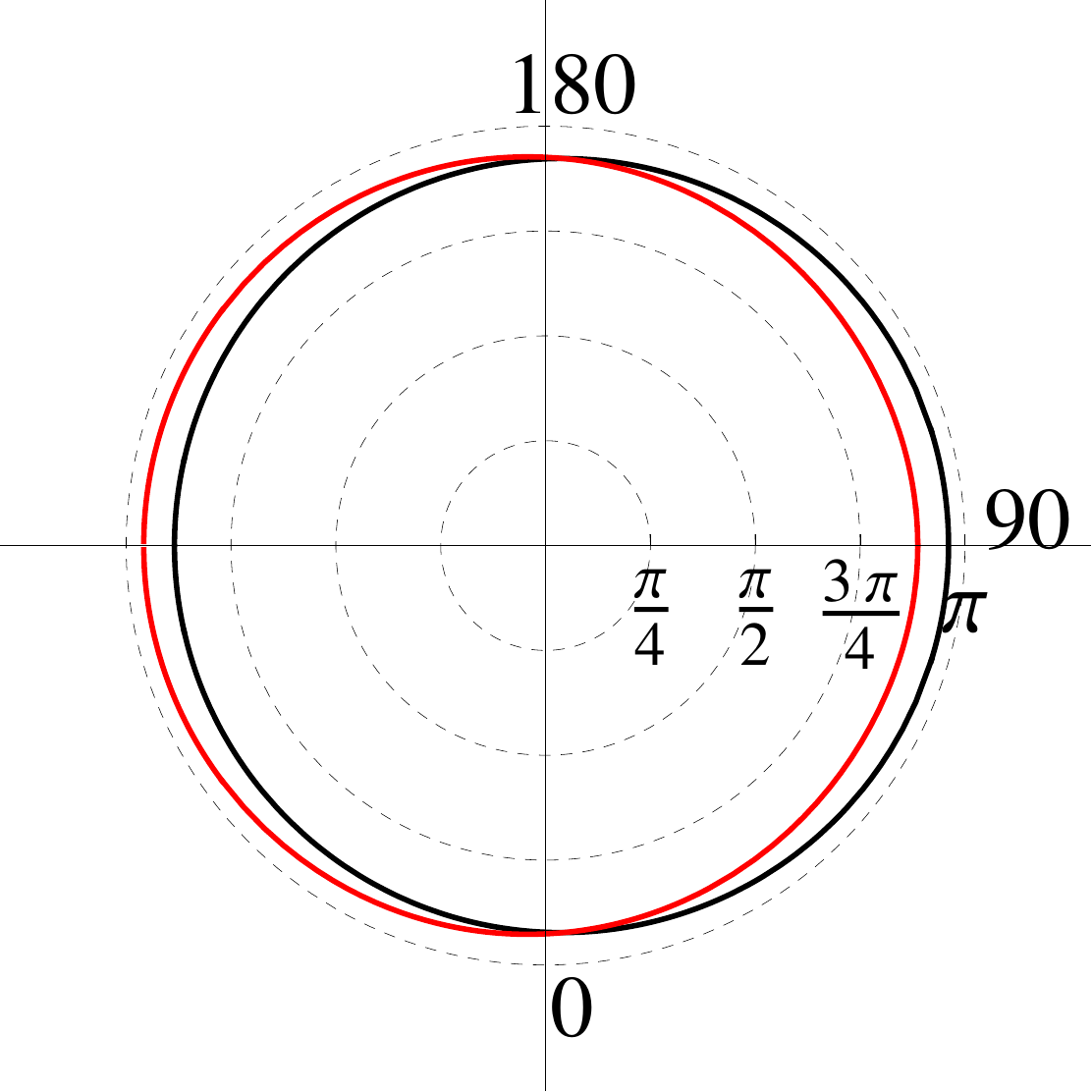}&\includegraphics[width=3cm]{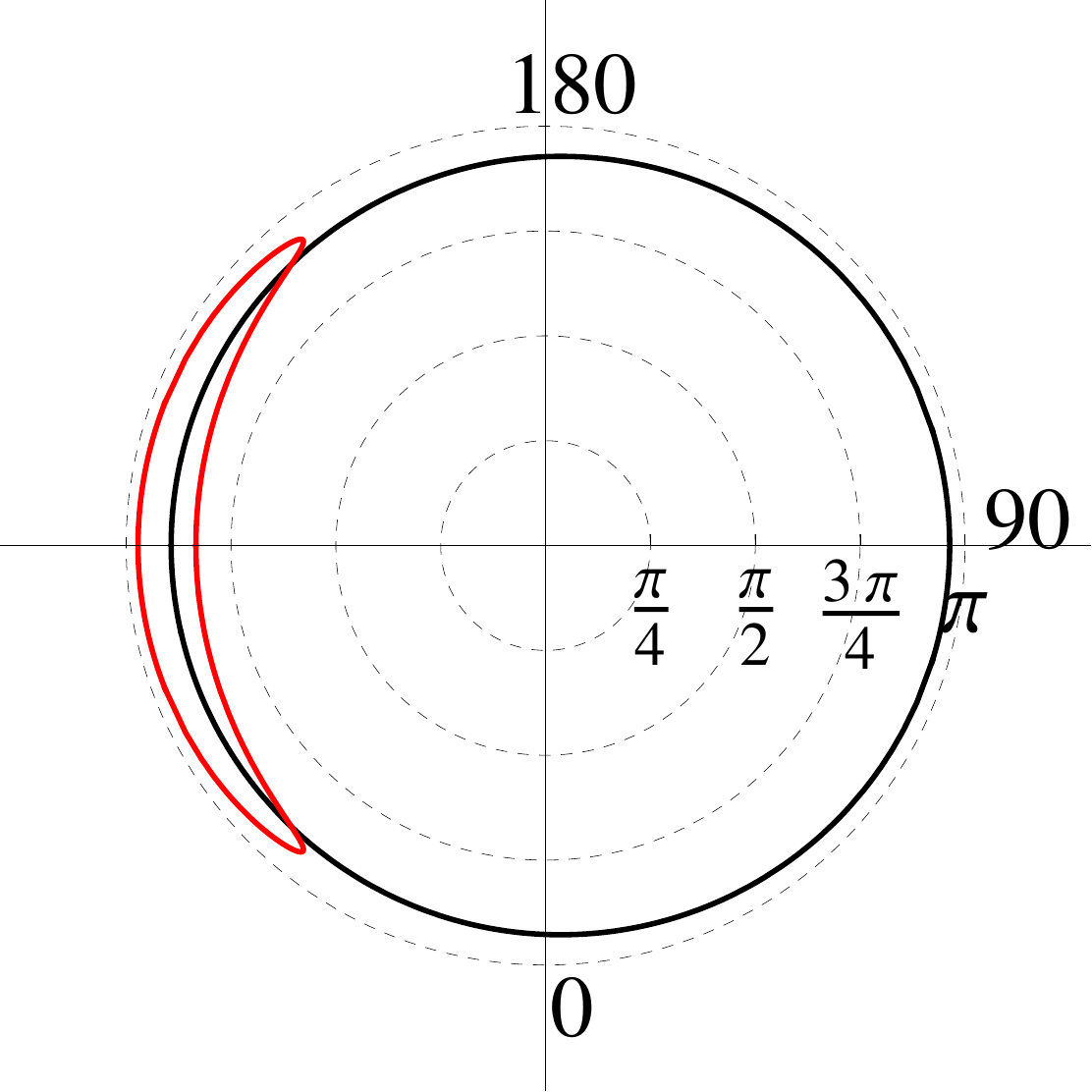}\\
				$r_{(ms+)i}=1.68332$ & $r_{(ms-)i}=9.30787$ & $r_{(ms-)o}=18.3721$ & $r_{(ms+)o}=19.7489$ \\
				\hline
				
			\end{tabular}
			
		\end{figure}
		
		\begin{figure}[H]
			\flushleft \textbf{Case IIa: $y=3.10^{-4},\, a^2=0.2$}\\
			
			\begin{tabular}{|lcr|}
				\hline
				
				\bet{l}$r_{o}=1.89672$\\$r_{c}=56.7078$\\$r_{ph+}=2.42728$\\$r_{ph-}=3.47685$ \\ $r_{d1}=2.00253$\\$r_{d2}=56.706$\\$r_{s}=14.938$\ent
				&
				\bet{c} \includegraphics[width=3cm]{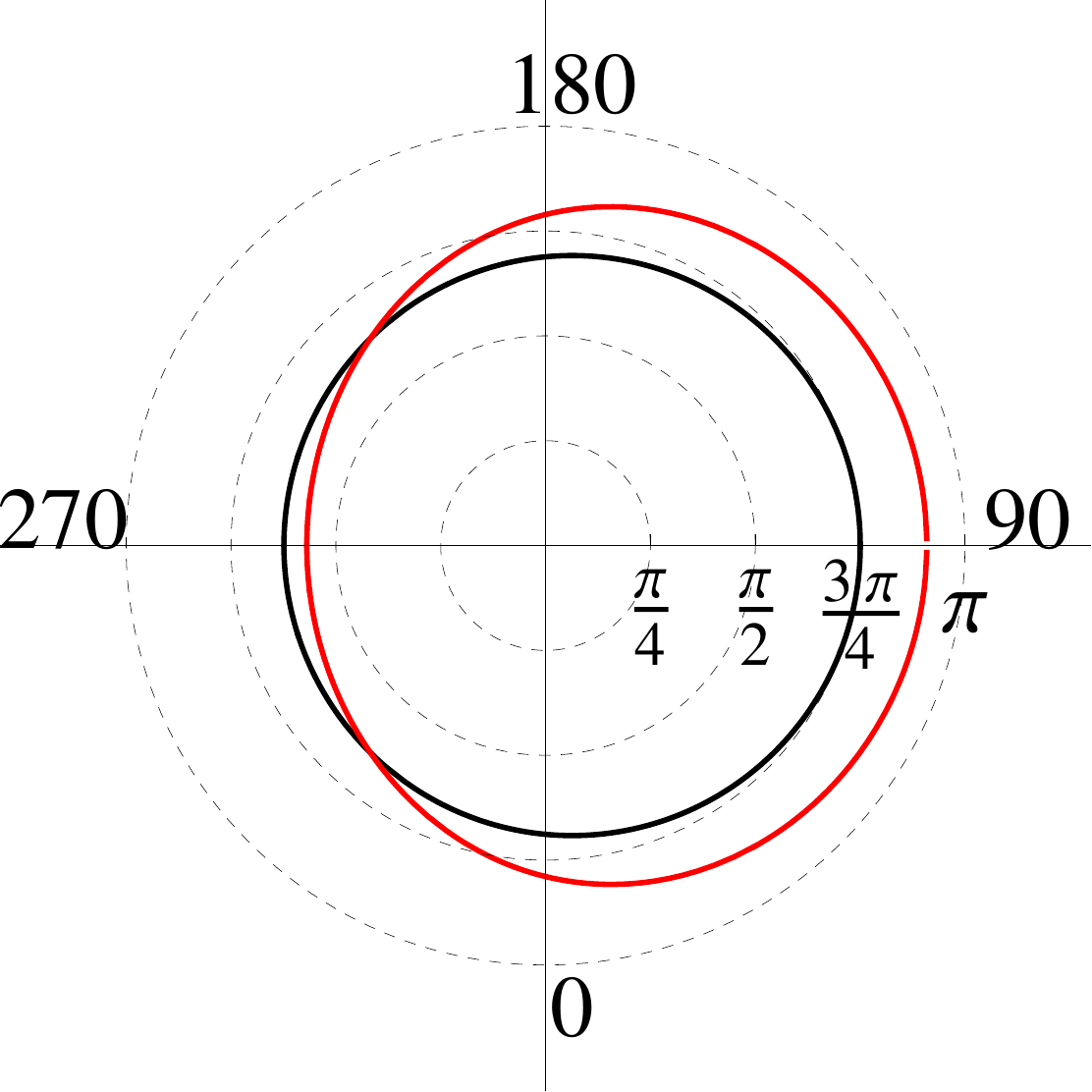}\\$r_{(ms+)i}=4.65524$ \ent & 
				\bet{c} \includegraphics[width=3cm]{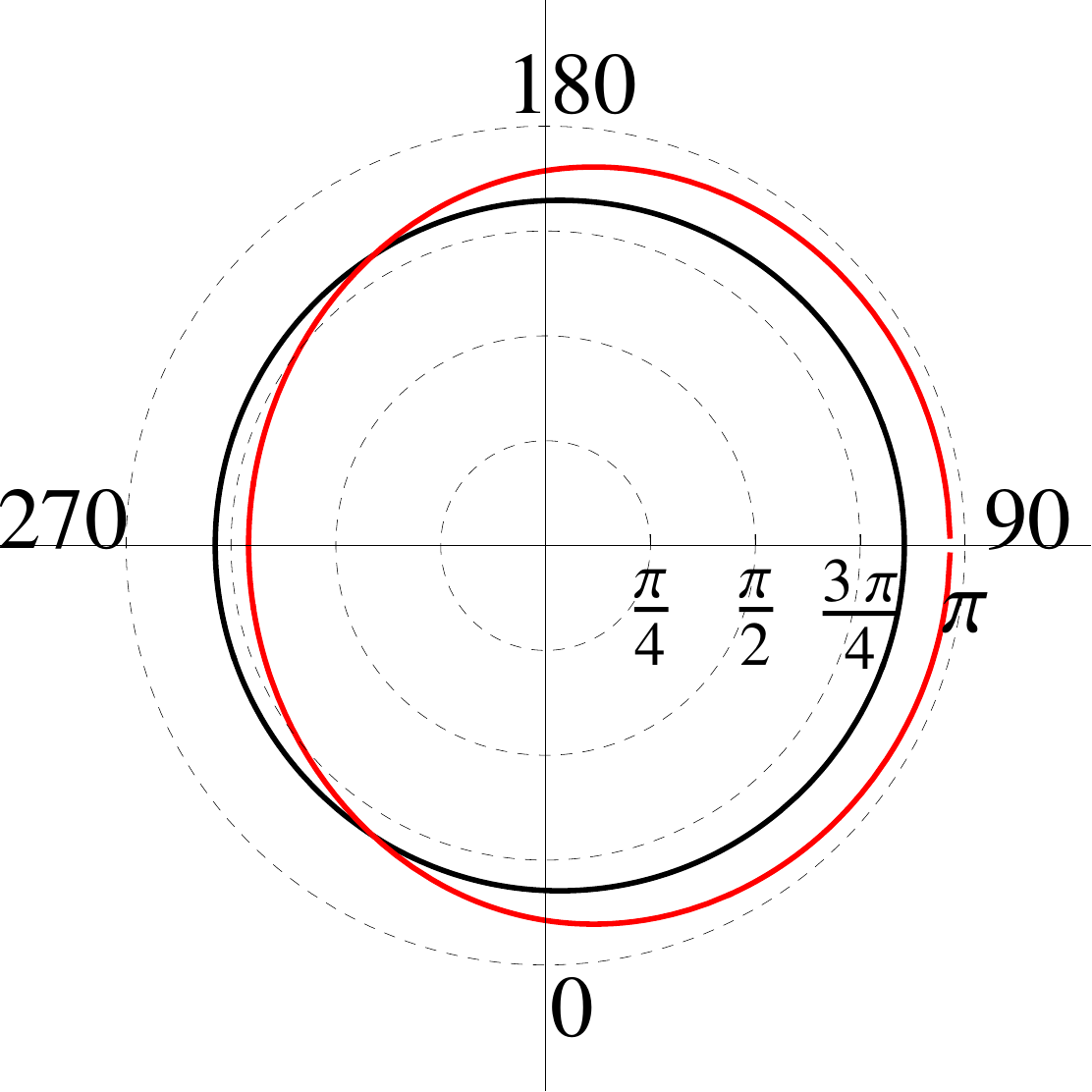}\\$r_{(ms+)o}=8.34272$ \ent \\
				\hline			
				
			\end{tabular}
			
		\end{figure}

		\begin{figure}[H]
			\flushleft \textbf{Case IIb: $y=3.10^{-4},\, a^2=0.45$}\\
			
			\begin{tabular}{|lcr|}
				\hline
				
				\bet{l}$r_{o}=1.74376$\\$r_{c}=56.7079$\\$r_{ph+}=2.06873$\\$r_{ph-}=3.69332$ \\ $r_{d1}=2.00268$\\$r_{d2}=56.7037$\\$r_{s}=14.938$\ent
				&
				\bet{c} \includegraphics[width=3cm]{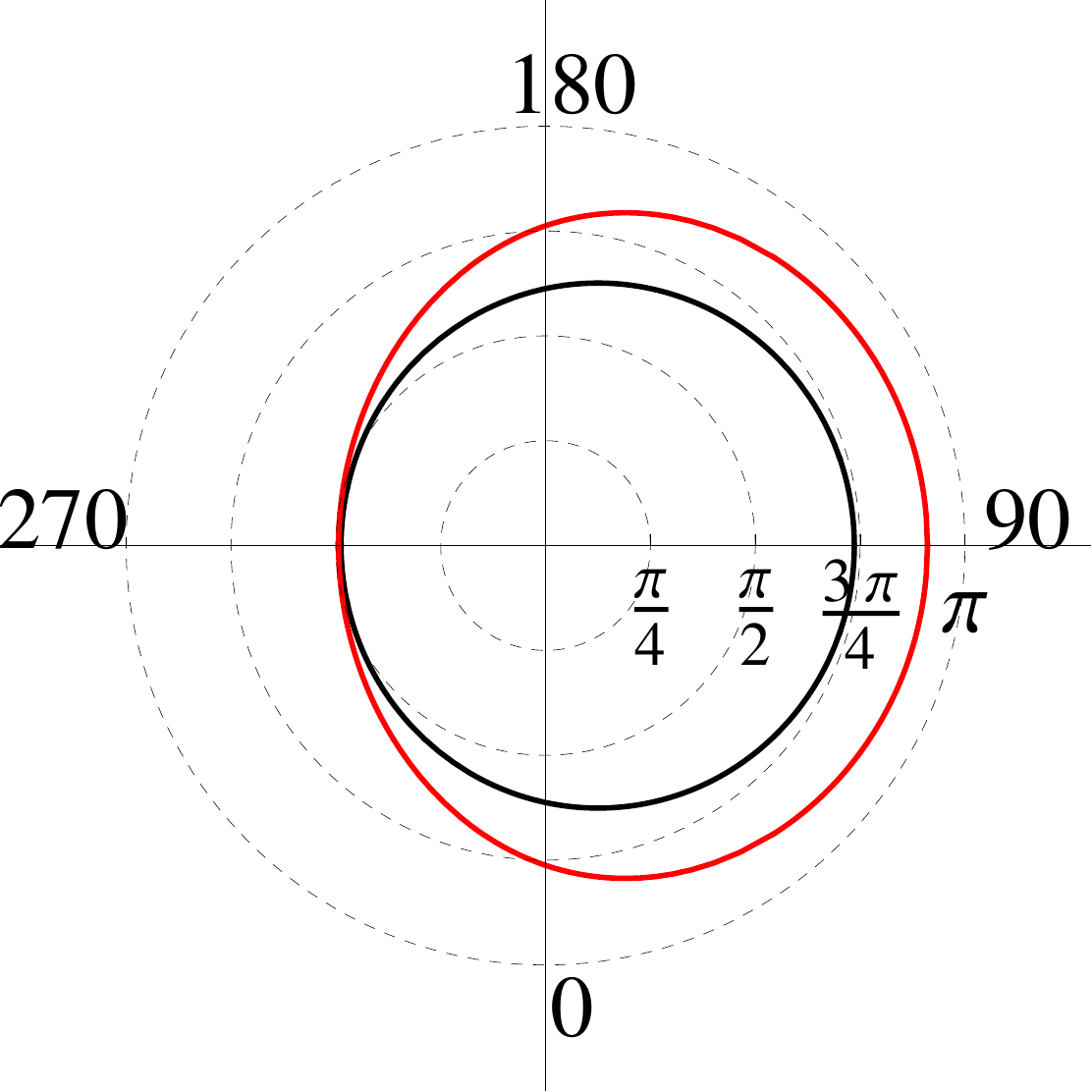}\\$r_{(ms+)i}=3.59647$ \ent & 
				\bet{c} \includegraphics[width=3cm]{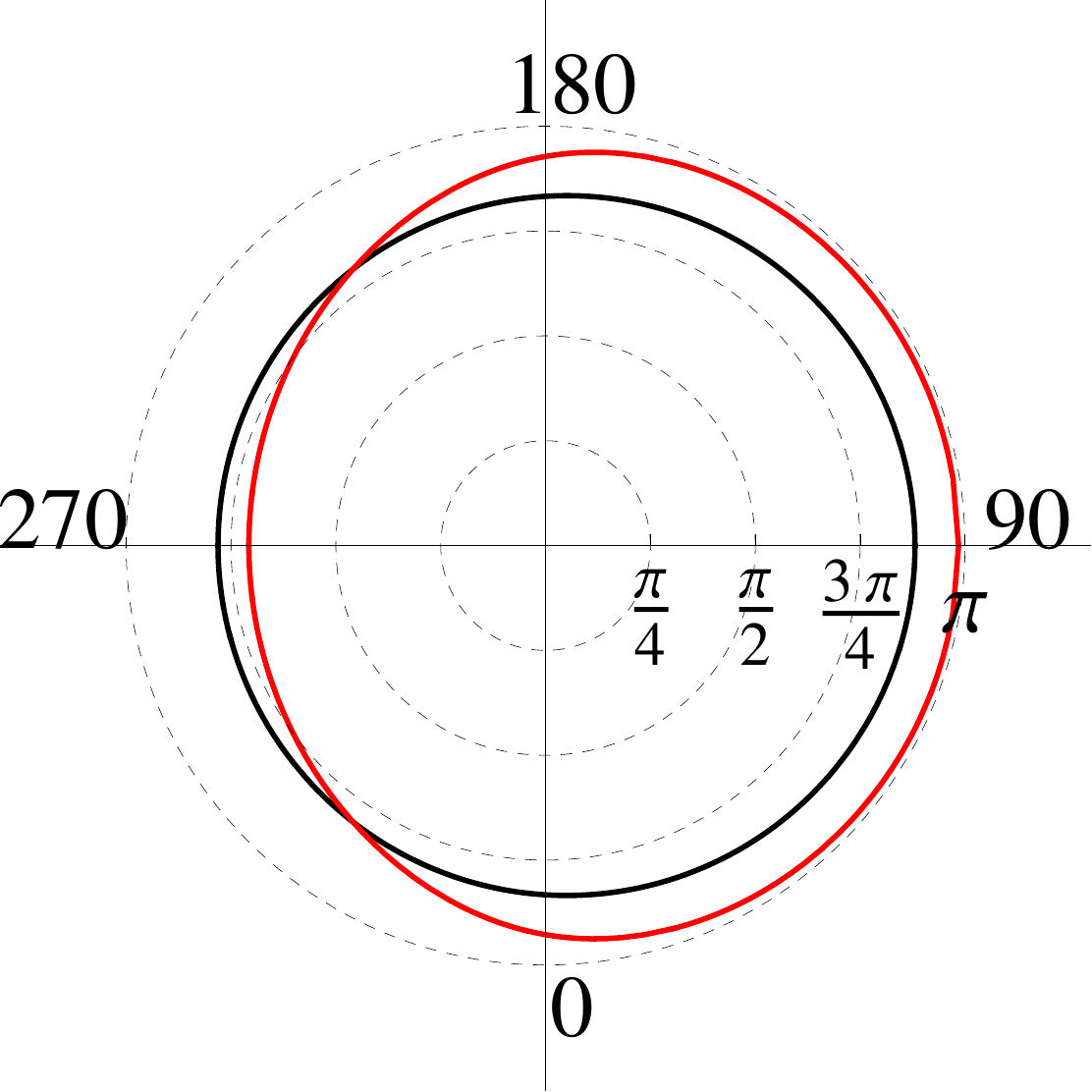}\\$r_{(ms+)o}=8.65605$ \ent \\
				\hline			
				
			\end{tabular}
			
		\end{figure}
		
		\begin{figure}[H]
			\flushleft \textbf{Case IIc: $y=3.10^{-4},\, a^2=0.6$}\\
			
			\begin{tabular}{|lcr|}
				\hline
				
				\bet{l}$r_{o}=1.63453$\\$r_{c}=56.708$\\$r_{ph+}=1.86794$\\$r_{ph-}=3.78993$ \\ $r_{d1}=2.00277$\\$r_{d2}=56.708$\\$r_{s}=14.938$\ent
				&
				\bet{c} \includegraphics[width=3cm]{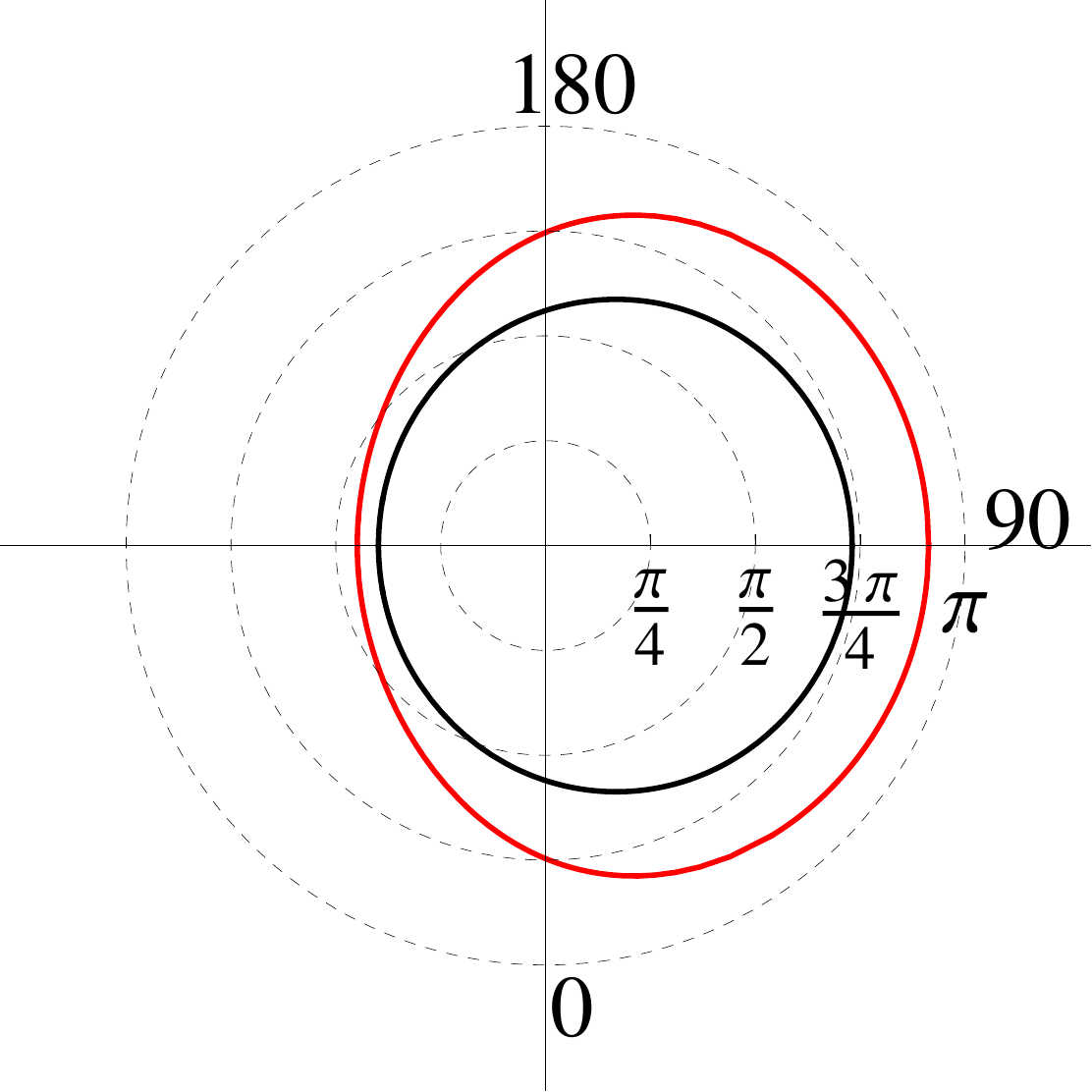}\\$r_{(ms+)i}=3.07443$ \ent & 
				\bet{c} \includegraphics[width=3cm]{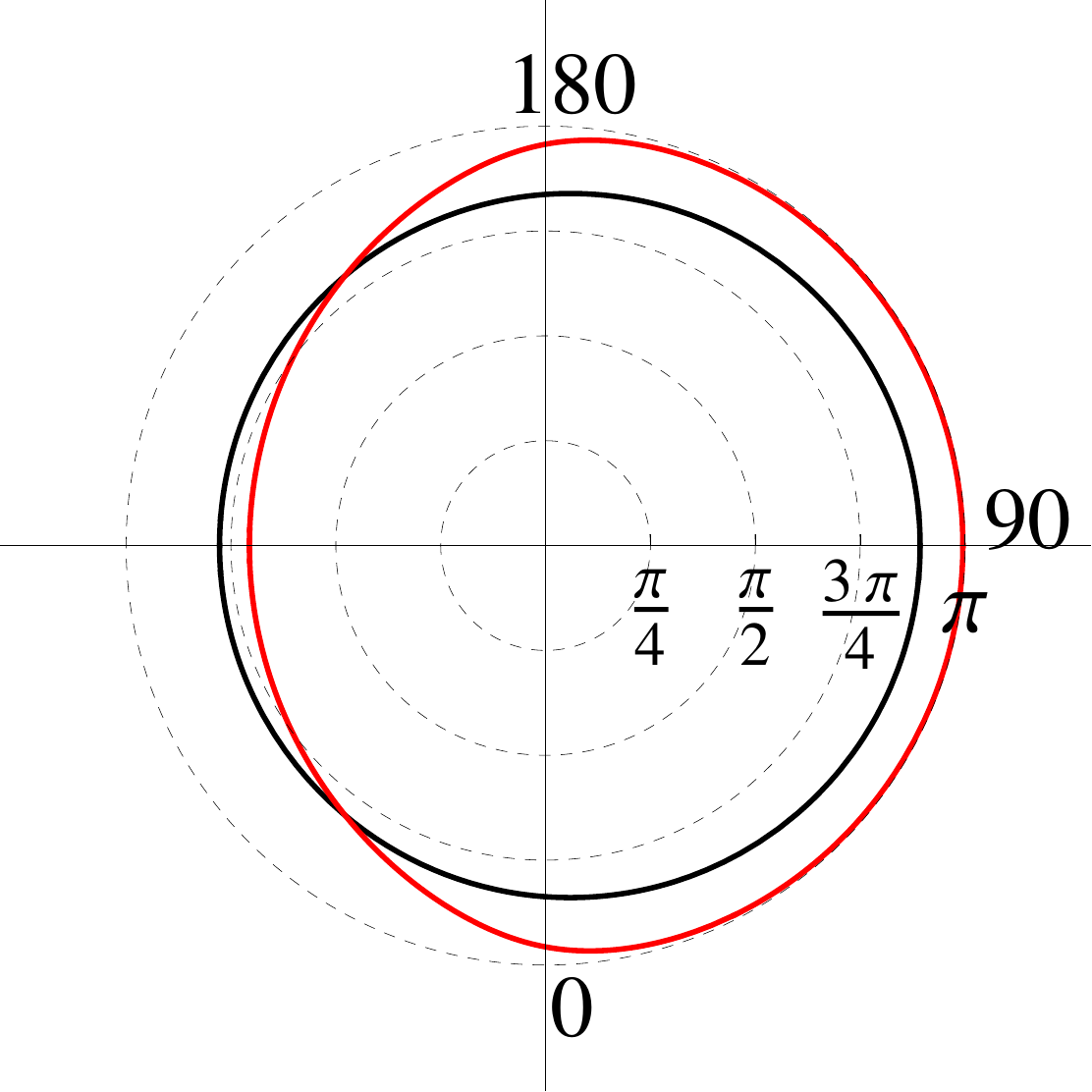}\\$r_{(ms+)o}=8.76433$ \ent \\
				\hline			
				
			\end{tabular}
			
		\end{figure}
		
		\begin{figure}[H]
			\flushleft \textbf{Case IId: $y=3.10^{-4},\, a^2=0.95$}\\
			
			\begin{tabular}{|lcr|}
				\hline
				
				\bet{l}$r_{o}=1.22607$\\$r_{c}=56.7081$\\$r_{ph+}=1.2735$\\$r_{ph-}=3.97023$ \\ $r_{d1}=2.00298$\\$r_{d2}=56.6992$\\$r_{s}=14.938$\ent
				&
				\bet{c} \includegraphics[width=3cm]{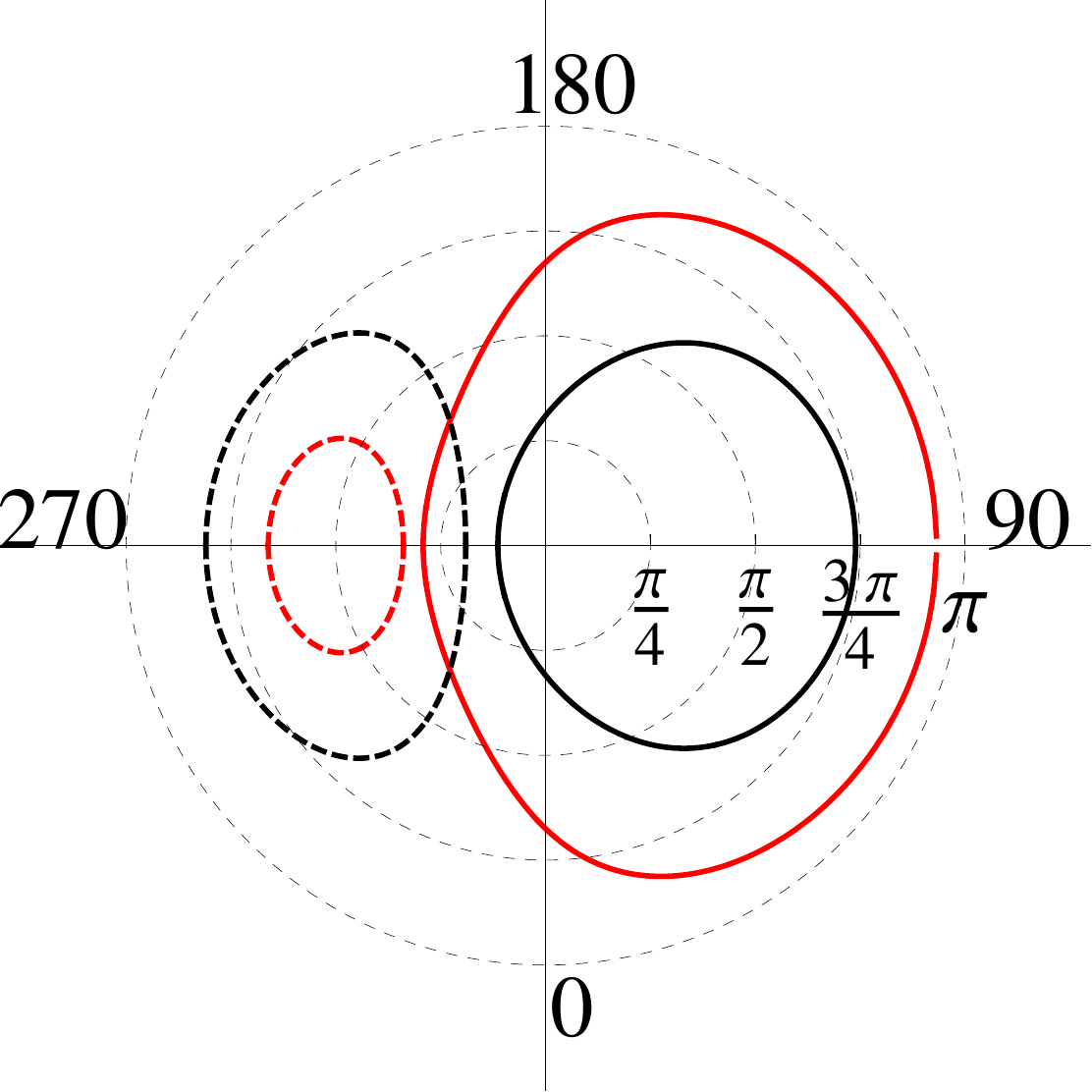}\\$r_{(ms+)i}=1.6897$ \ent & 
				\bet{c} \includegraphics[width=3cm]{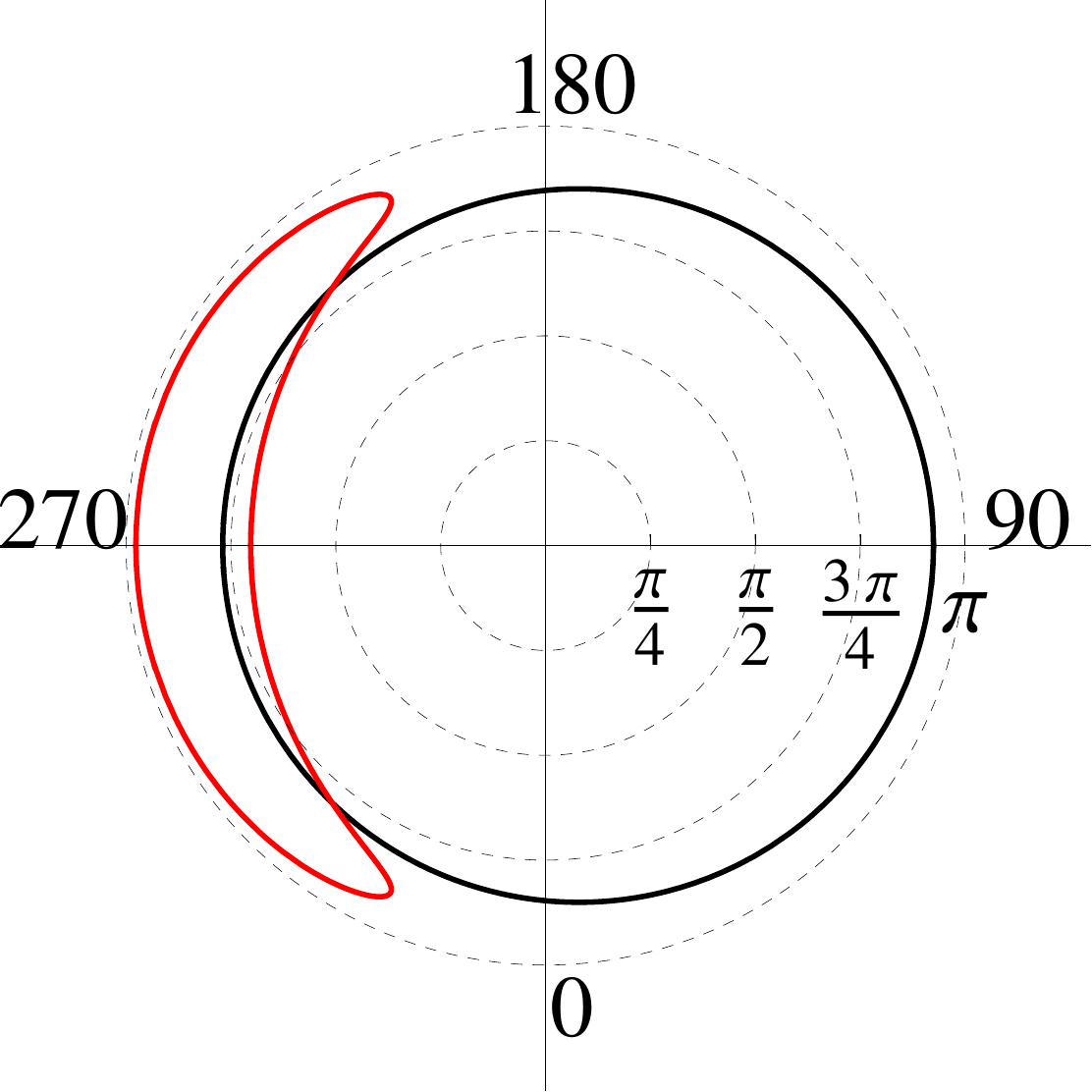}\\$r_{(ms+)o}=8.93209$ \ent \\
				\hline			
				
			\end{tabular}
			
		\end{figure}
		
		\begin{figure}[H]
			\flushleft \textbf{Case IIe: $y=0.0013,\, a^2=0.51$}\\
			
			\begin{tabular}{|lcr|}
				\hline
				
				\bet{l}$r_{o}=1.70925$\\$r_{c}=26.6758$\\$r_{ph+}=1.99546$\\$r_{ph-}=3.71802$ \\ $r_{d1}=2.01192$\\$r_{d2}=26.6758$\\$r_{s}=9.1626$\ent
				&
				\bet{c} \includegraphics[width=3cm]{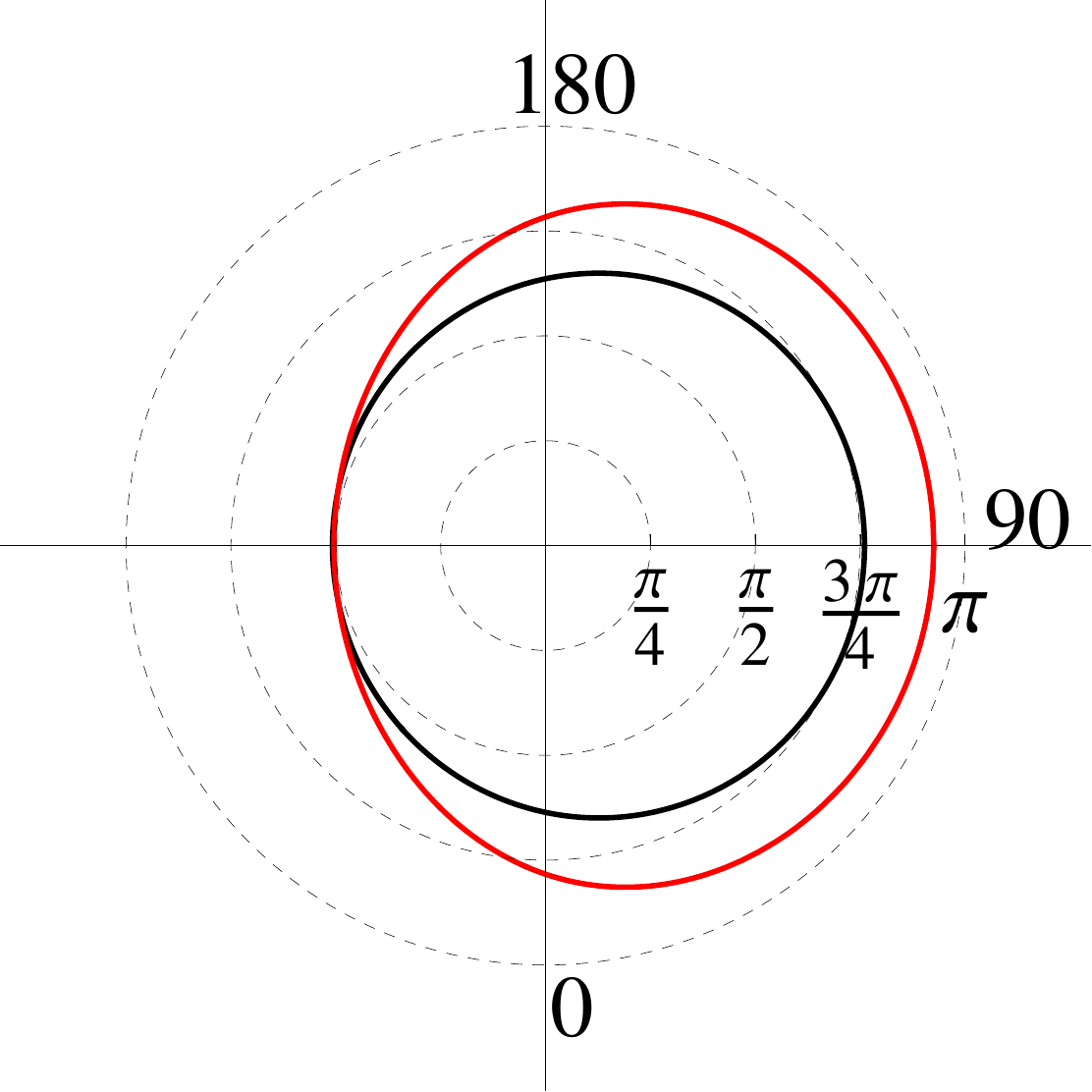}\\$r_{(ms+)i}=3.78194$ \ent & 
				\bet{c} \includegraphics[width=3cm]{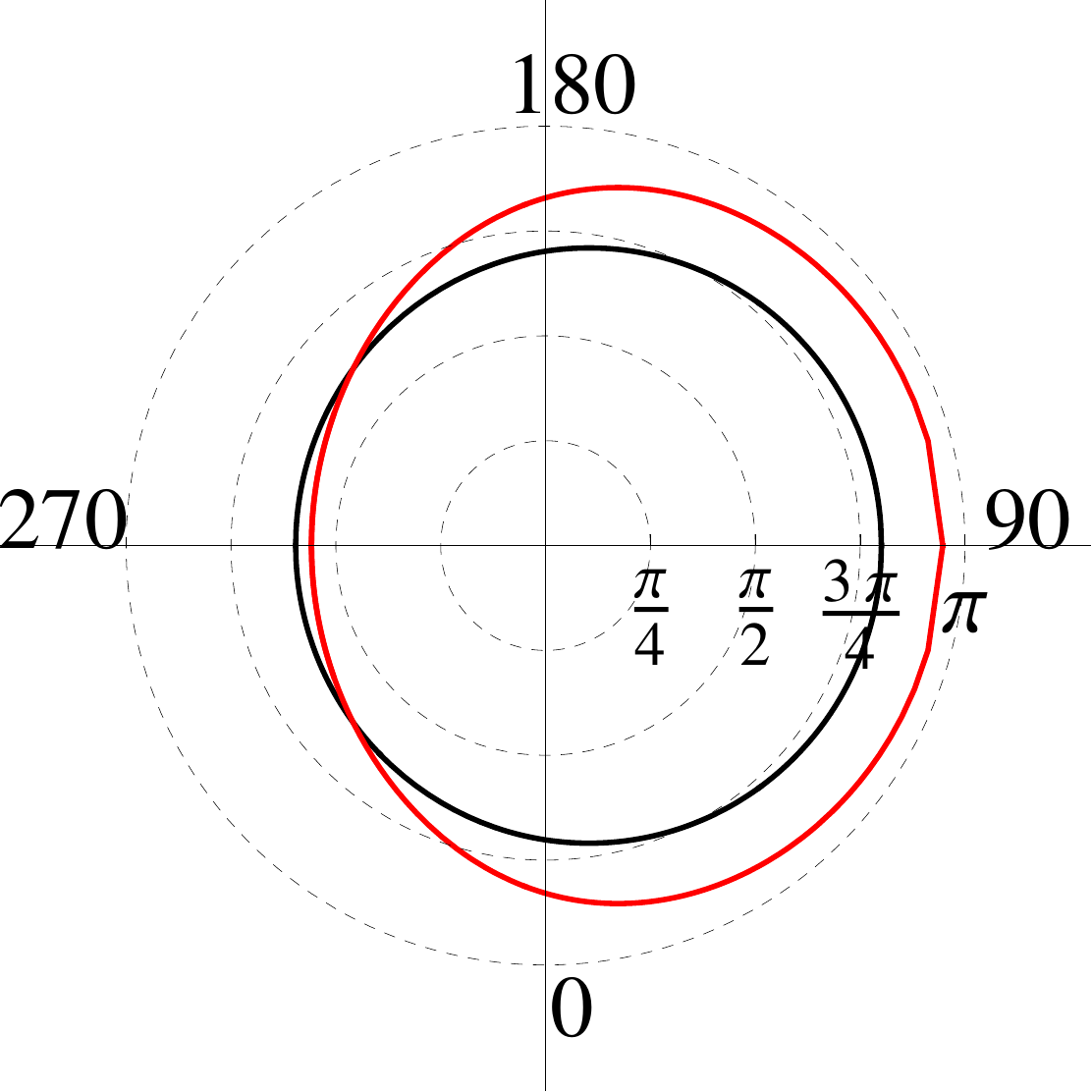}\\$r_{(ms+)o}=4.63727$ \ent \\
				\hline			
				
			\end{tabular}
			
		\end{figure}
		
		\begin{figure}[H]
			\flushleft \textbf{Case IIf: $y=0.0035,\, a^2=0.85$}\\
			
			\begin{tabular}{|lcr|}
				\hline
				
				\bet{l}$r_{o}=1.41213$\\$r_{c}=15.801$\\$r_{ph+}=1.51618$\\$r_{ph-}=3.85179$ \\ $r_{d1}=2.03558$\\$r_{d2}=15.7678$\\$r_{s}=6.58634$\ent
				&
				\bet{c} \includegraphics[width=3cm]{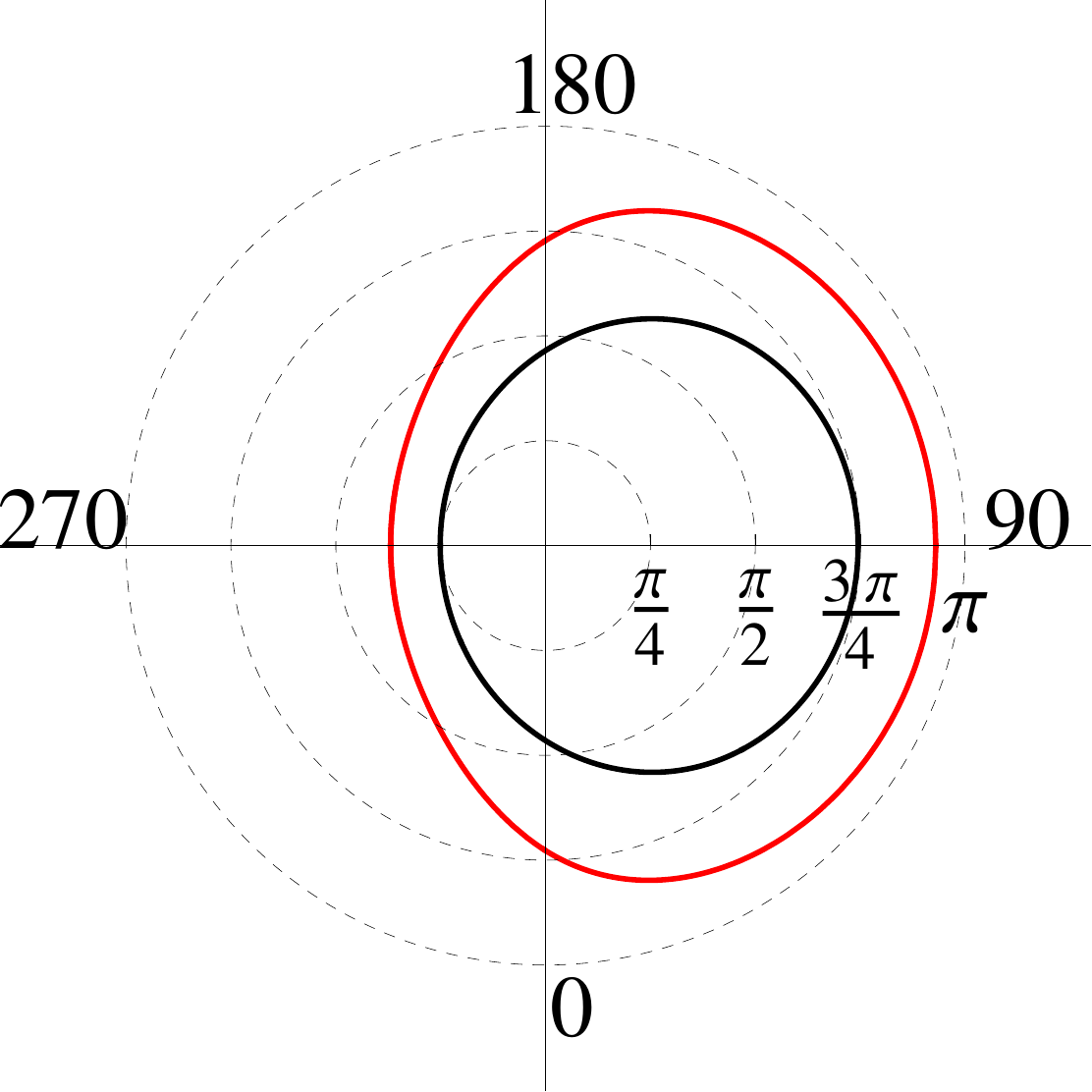}\\$r_{(ms+)i}=2.34663$ \ent & 
				\bet{c} \includegraphics[width=3cm]{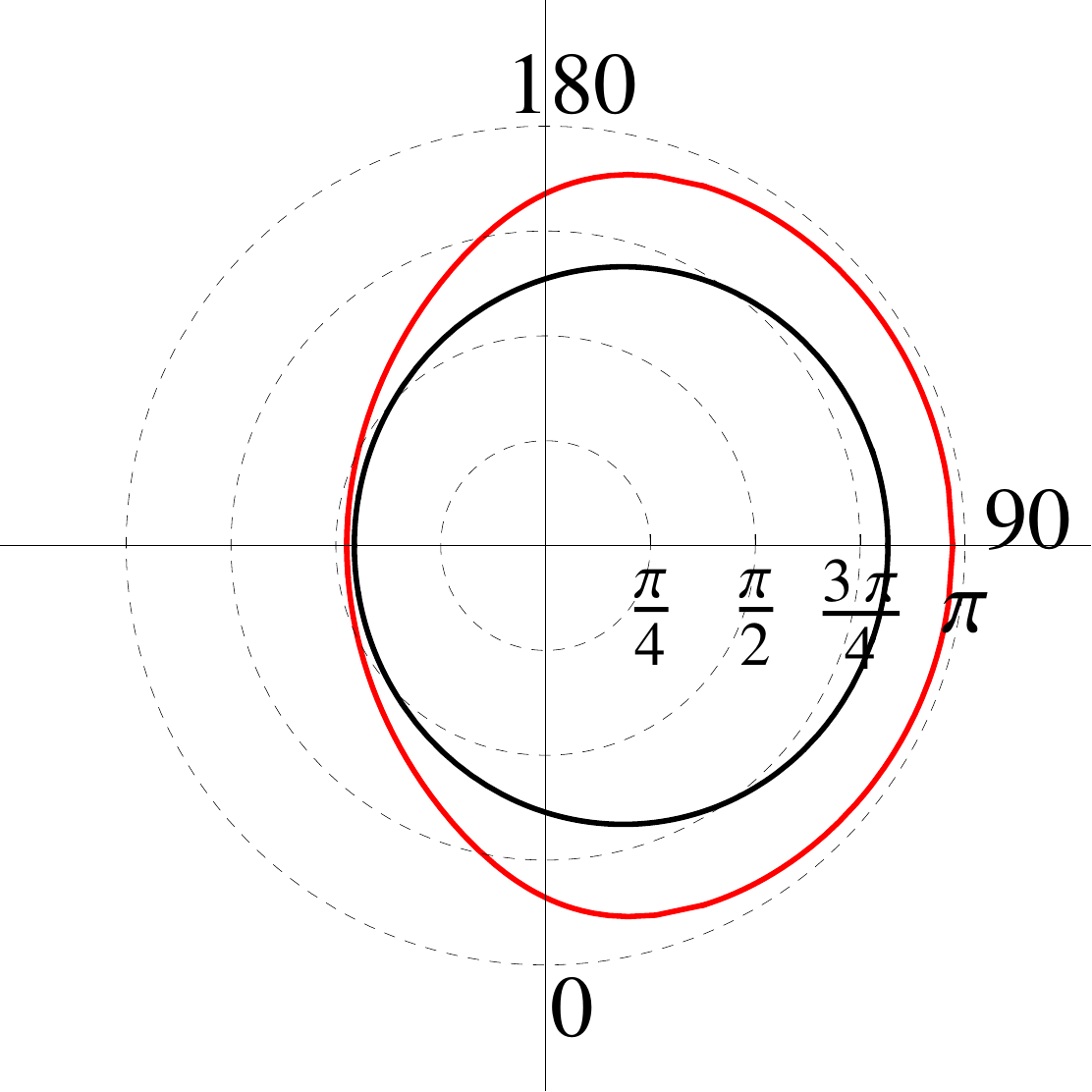}\\$r_{(ms+)o}=3.51592$ \ent \\
				\hline			
				
			\end{tabular}
			
		\end{figure}
		
		\begin{figure}[H]
			\flushleft \textbf{Case IIg: $y=0.005,\, a^2=0.95$}\\
			
			\begin{tabular}{|lcr|}
				\hline
				
				\bet{l}$r_{o}=1.26538$\\$r_{c}=13.017$\\$r_{ph+}=1.31806$\\$r_{ph-}=3.19443$ \\ $r_{d1}=2.05302$\\$r_{d2}=12.9695$\\$r_{s}=5.84804$\ent
				&
				\bet{c} \includegraphics[width=3cm]{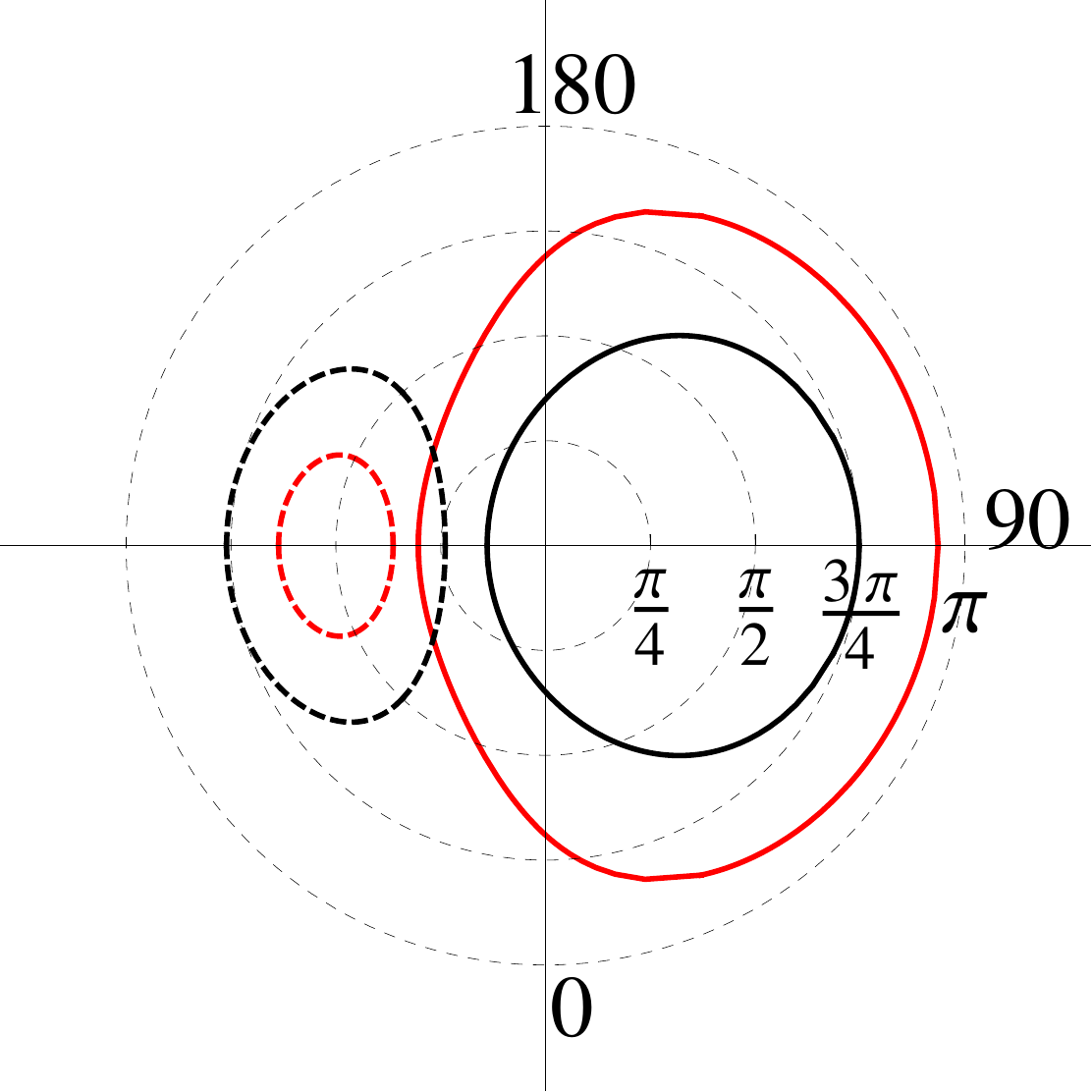}\\$r_{(ms+)i}=1.8219$ \ent & 
				\bet{c} \includegraphics[width=3cm]{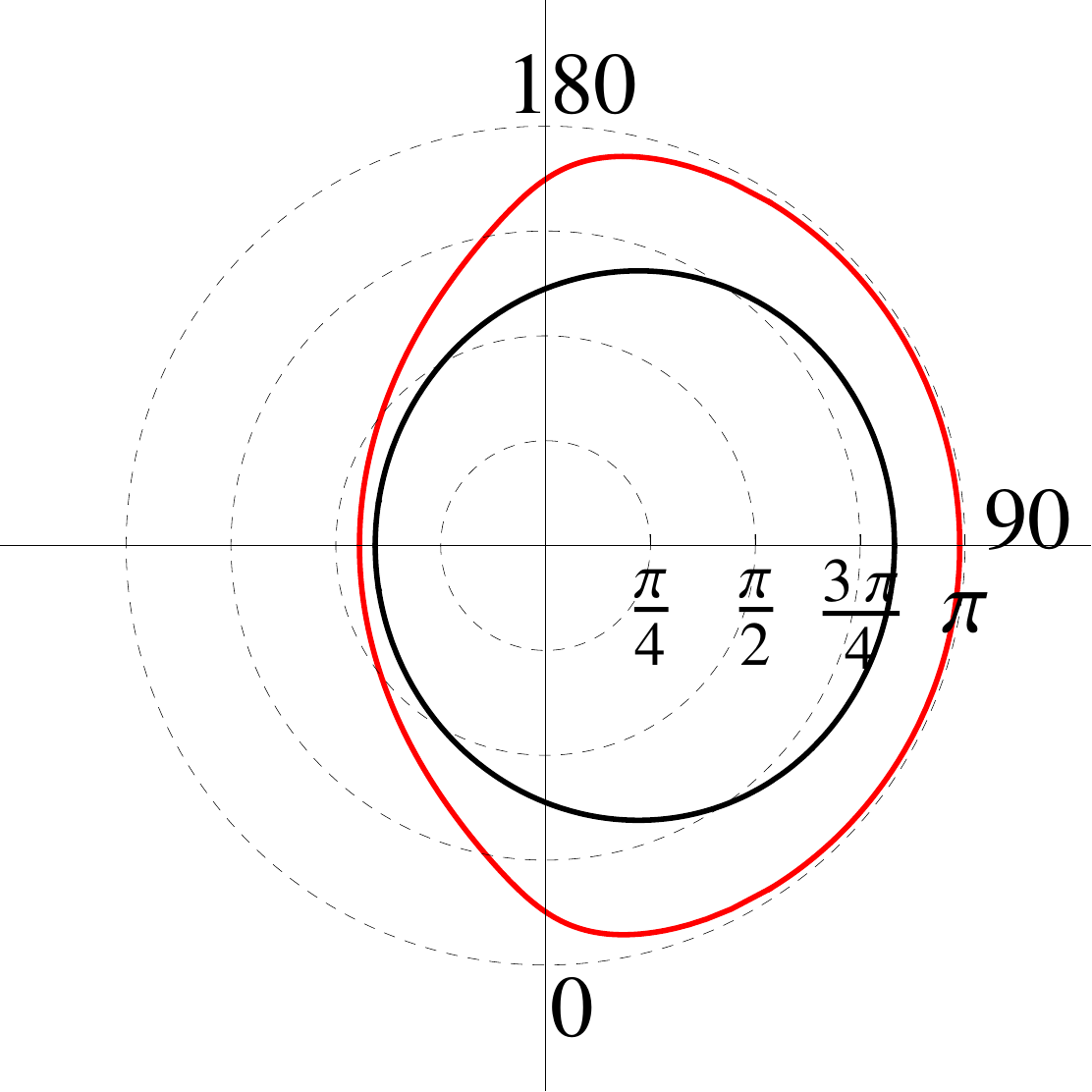}\\$r_{(ms+)o}=3.19443$ \ent \\
				\hline			
				
			\end{tabular}
			
		\end{figure}
		
		\begin{figure}[H]
			\flushleft \textbf{Case IIh: $y=0.016,\, a^2=1.02$}\\
			
			\begin{tabular}{|lcr|}
				\hline
				
				\bet{l}$r_{o}=1.18607$\\$r_{c}=13.017$\\$r_{ph+}=1.20814$\\$r_{ph-}=3.19443$ \\ $r_{d1}=2.05302$\\$r_{d2}=12.9695$\\$r_{s}=5.84804$\ent
				&
				\bet{c} \includegraphics[width=3cm]{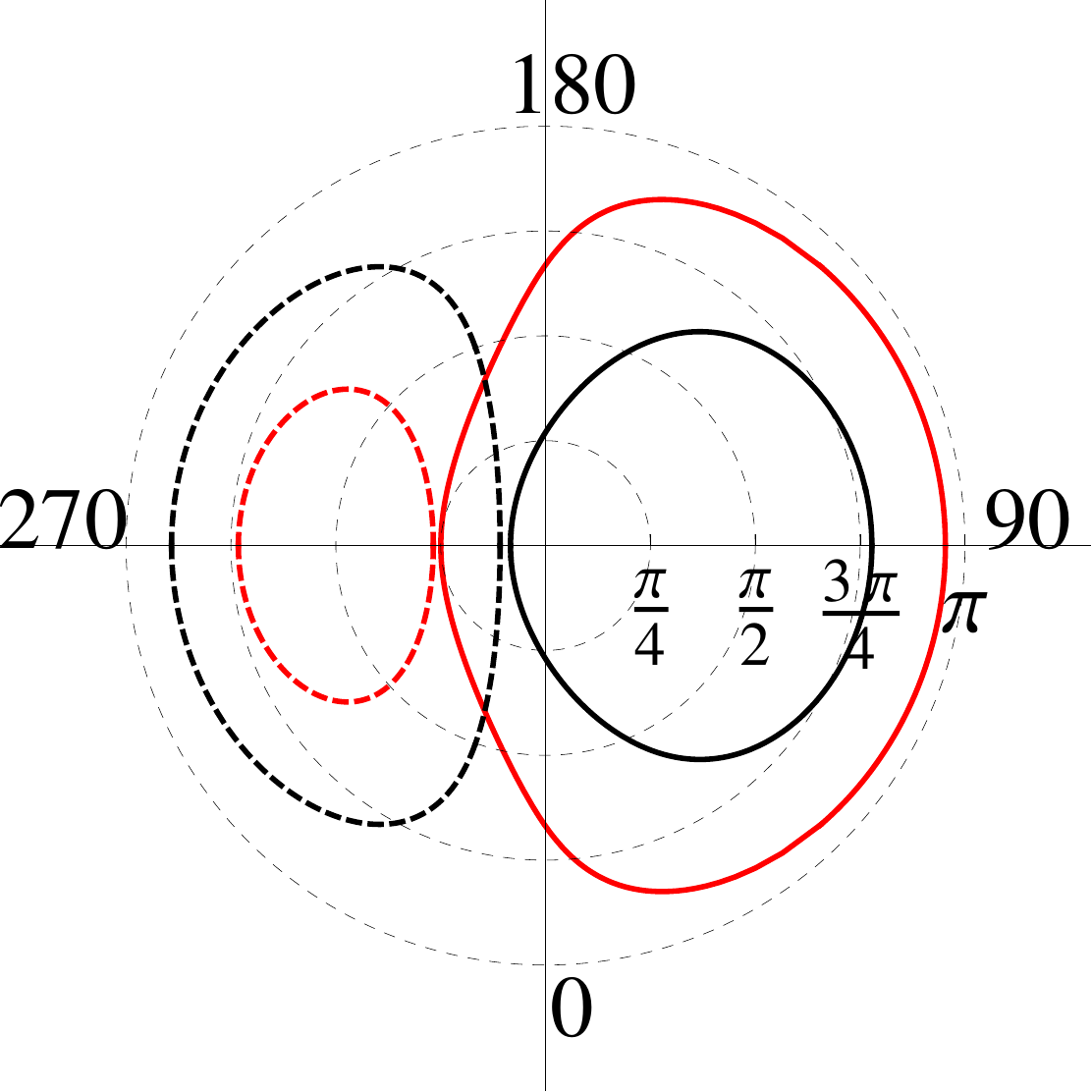}\\$r_{(ms+)i}=1.56402$ \ent & 
				\bet{c} \includegraphics[width=3cm]{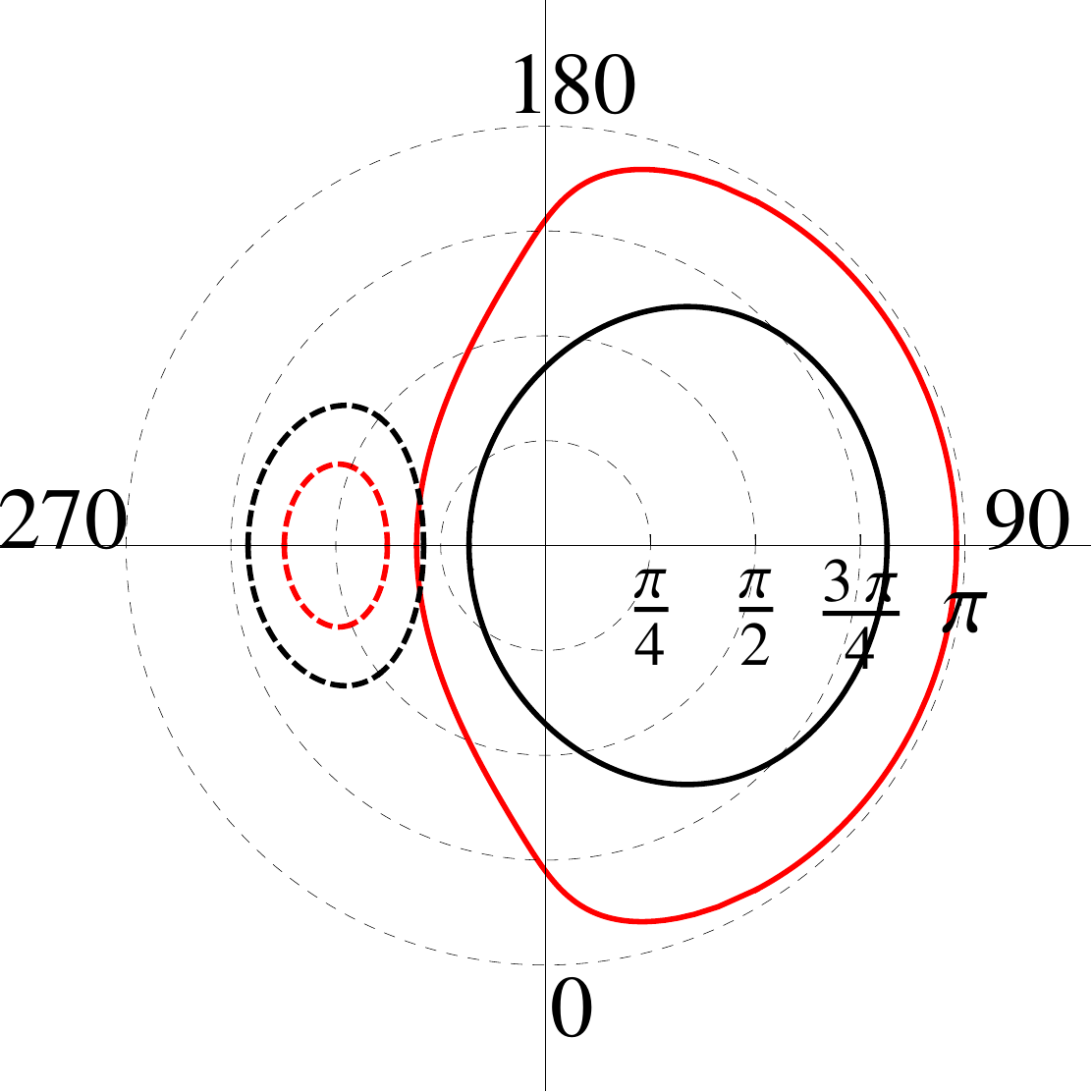}\\$r_{(ms+)o}=2.02345$ \ent \\
				\hline			
				
			\end{tabular}
			
		\end{figure}
		
		\newpage
	\section{Conclusions}
	Inspecting the behaviour of the light escape cones we can summarize following results:
	
	\begin{description}
		\item[LNRF]
		\begin{enumerate}
			\item The rotational effects of the \KdS\ geometry are strongest in the equatorial plane, where the light escape cones are most distorted. This distortion decreases towards the polar axis, where the cones are symmetric.
			
			\item The range of the spherical photon orbits is widest in the equatorial plane, where the limits are radii of the inner co-rotating $r_{ph+}$  and outer counter-rotating $r_{ph-}$ equatorial circular photon orbits. Towards the polar axis these marginal orbits approach some value $r_{pol}$ at which they coalesce on the polar axis. 
			
			\item At radii between the outer black hole horizon $r_{o}$ and $r_{ph+}$, all escaping photons are outwards directed. Approaching the horizon, the cone shrinks to infinitesimal extension.
			
			\item At the radius $r_{ph+}$, and in the equatorial plane, photons emitted in the direction of motion of LNRF follow circular orbit of this radius, another escaping photons are outwards directed.
			
			\item At radial distances between $r_{ph+}$ and radius $r_{pol}$, and sufficiently close to the equatorial plane, some escaping photons with positive azimuthal motion component inwards directed exist. Approaching the polar axis all escaping photons become outwards directed. On the polar axis and at the radius $r_{pol}$, the escaping and captured photons are separated by the plane perpendicular to the polar axis.
			
			\item There exist circular photon orbits with radius $r_{pol}$ crossing the polar axis as seen by locally non-rotating observers. Photons on such orbits have impact parameter $X=-a$  ($\ell=0$).
			
			\item Above the radius $r_{pol}$ and under $r_{ph-}$, some inwards directed escaping photons enter the region of negative azimuthal component of motion.  Approaching the polar axis all captured photons become inwards directed.
			
			\item At radii $r_{ph-}$ and in the equatorial plane, photons emitted in the opposite direction of the LNRF revolution follow circular orbit of this radius, another captured photons are all inwards directed.   
			
			\item At radii greater than $r_{ph-}$, all captured photons are inward directed at any latitude. Close to the cosmological horizon $r_{c}$, the captured cone again reduces to infinitesimal extension.
			
			\item In any kind of the \KdS\ black hole spacetimes, there exist photons with positive impact parameter $X$ that are counter-rotating relatively to the locally non-rotating observers. Such phenomenon exists in the ergosphere. In the KdS spacetimes with the DRB, the ergosphere is confined between the radii $r_{o}$  and $r_{d1},$ and between $r_{d2}$ and $r_{c}$ in the equatorial plane, where $r_{d1},\,r_{d2}$ are the points where the potential $X_{+}(r;\:q,\:y,\;a)$ diverges. For each radius $r_e$ between these limits, there exists a marginal latitude $\theta_{m}(r_e)$ as a corresponding latitudinal coordinate of the static limit surface, such that this phenomenon appears between the latitudes $\theta_{m}$ and $\pi-\theta_{m}.$ In the KdS spacetimes with the RRB, the ergosphere spreads anywhere between the radii $r_{o}$ and $r_{c}$ in the equatorial plane. 
			
			\item The marginal latitude increases with increasing radius from $\theta_{m}=0$ at $r=r_{o}$ to $\theta_{m}=\pi/2$ at $r=r_{d1}$, and decreases with increasing radius from $\theta_{m}=\pi/2$ at $r=r_{d2}$ to $\theta_{m}=0$ at $r=r_{c}$ in the KdS spacetimes with the DRB, otherwise it is not defined (see Fig. \ref{fig_ergos}a). Here we restrict ourselves on the latitudes 'above' the equatorial plane, since the situation is symmetric with respect to the equatorial plane. In the KdS spacetimes with the RRB, the function $\theta_{m}(r_{e})$ is defined between both the horizons where it has zeros again, and for $r=r_{d(ex)}$ it has a local maximum (Figs. \ref{fig_ergos}c,d, \ref{fig_RGF_cones}). 
			
			\item In the KdS spacetimes with the DRB, all the locally counter-rotating photons with positive impact parameter in the 'inner' ergosphere are captured by the black hole, the photons in the 'outer' ergosphere escape through the cosmological horizon. In the KdS black hole spacetimes with the RRB, and sufficiently close to the equatorial plane, some of these photons are captured and others escape; at radii $r_e\geq r_{d(ex)}$ some inwards directed escaping photons emerge, for $r_e\geq r_{ph-}$ all captured photons are inwards directed. Approaching the marginal latitude $\theta_{m}(r_e),$ all photons are trapped for $r_e<r_{d(ex)}$ and all escape for $r_{d(ex)}<r_e.$    
		\end{enumerate}
		\item[RGF]
		\begin{enumerate}
			\item Similarly to the case of the LNRF, the distortion of the light escape cones for both radially escaping and radially falling observers are most apparent in the equatorial plane, while on the polar axis it disappears.
			\item The light escape cones appear to be shifted in the opposite direction to the movement of the source. In the same manner this shift affects the photons with high positive impact parameter $X,$ moving in negative azimuthal direction. Such photons exist in the ergosphere.  
			\item In any kind of the \KdS\ black hole spacetimes, there exist inwards directed escaping photons at any radii in the stationary region, as seen by the radially escaping observers. For radially falling observers, such inwards directed escaping photons can occur in the vicinity of the cosmological horizon and sufficiently close to the equatorial plane.
			\item Due to the time symmetry of the \KdS\ geometry, one can reverse the flow of the time and commute the falling/escaping source emitting the light through the light escape cone by the escaping/falling observer receiving the light incoming from the bright celestial background. The light escape cone itself then represents the silhouette of the black hole, rotating, unlike normal, due to the time reversion clockwise. The shift of the silhouette is the result of dragging of the spacetime by the rotation of the black hole and aberration due to the observer motion. Note that near the so called static radius, the radially falling or escaping observers are closely related to the corresponding LNRFs, and can be considered as nearly static distant observers. At distances larger than the static radius of the KdS black hole spacetime, the escaping radial observers have to be considered as a proper generalization of the static distant observers related to asymptotically flat Kerr black hole spacetimes. 
			 
		\end{enumerate}
		\item[CGF]
		\begin{enumerate}
			\item In all cases of the KdS black hole spacetimes, at all the stable circular orbits, there is $v_{(+)}>0,$ and $v_{(-)}<0,$ hence the symmetry axis of the light escape cones of the source orbiting along the plus family orbits, located between the inner and outer marginally stable orbits, is shifted in positive azimuthal direction, while in the case of the minus family orbits, it is shifted in negative azimuthal direction. 
			\item The light escape cones of the sources related to the whole range of stable co-rotating circular orbits, limited by the inner and outer marginally stable orbits, appear to be more opened (wider) than in the case of the LNRF sources. This is also the case of all stable counter-rotating orbits in the spacetimes with the rotational parameter $a$ low enough; however, in the KdS spacetimes of the type Id, the cones are narrower in comparison to the LNRF cones. Such effect can be assigned to the dragging by rotation of the spacetime.  
			\item In the ergosphere, the part of the local escape cones related to the locally non-rotating photons with positive impact parameter $X$ and negative energy $E$ appear to be narrower in CGFs in comparison with LNRFs. This is the effect of the relative motion of the source with respect to the LNRF. Note that only co-rotating stable circular orbits can reside in the ergosphere. 
			\item As the inner marginally stable plus family orbit approaches the outer event black hole horizon due to modification of the spacetime parameters $a,y$, the local escape cone related to the CGF at this orbit becomes more and more opened in comparison with the related LNRF escape cone. This effect can be assigned to an increase of the orbital velocity of the source with approaching to the outer black hole event horizon. 
			\item As in the case of radially moving frames, from the construction of the light escape cones one can get an immediate idea of the shape of the black hole silhouette, observed on the circular orbits, by simultaneous change of the orientation of the black hole spin and orbital velocity of the source/observer.           
		\end{enumerate}
		
	\end{description}
		
		We can summarize that we have constructed local escape cones for the sources related to the LNRFs, RGFs, and CGFs in all types of the KdS black hole spacetimes. Such cones in a direct complementary way determine shadow, or silhouette, of a KdS black hole located in front of a radiating screen (or whole radiating sky of the observer). Such shadows can be constructed for any orbiting or falling (escaping) observer. However, the shadows related to distant observers, nearly stationary near the static radius of the KdS spacetimes, or those related to radially escaping observers at distances overcoming the static radius, are clearly astrophysically significant, having direct observational relevance. 

Construction of the local photon escape cones enables detailed modelling of the whole variety of the optical observational phenomena (light curves of radiating hot spots, spectral continuum profiles, profiled spectral lines, appearance of the radiating structure) related to the Keplerian accretion disks \cite{Bar:1973:BlaHol:}, including the self-occult effect \cite{Bao-Stu:1992:ApJ:}, and a minor modification of the CGF model enables modelling of the optical effects connected with the radiating toroids, or complex toroidal configurations \cite{Pug-Stu:2015:ApJS:,Pug-Stu:2016:ApJS:,Pug-Stu:2017:ApJS:}. 

In addition to the construction of the photon escape cones related to local sources, we have considered in the complementary local photon capture cones their special parts related to the special class of (counter-rotating) photons with positive impact parameters and negative covariant energy. The photons with negatively-valued energy can have a specific role in calculations of the black hole spin and mass evolution due to accretion, with modifications implied by the captured radiation of the accretion disk \cite{Tho:1974:ApJ:}. 
	
In future we plan to extend our study of the local escape cones to the KdS naked singularity spacetimes, where the situation is more complex than in the black hole spacetimes due to non-existence of the black hole event horizon, as only the cosmological horizon exists in such spacetimes. Moreover, along with the escape and captured photons that occur in the KdS black hole spacetimes, also the so called trapped photons exist in the naked singularity spacetimes, as shown in \cite{Stu-Sche:2010:CLAQG:}. 

\section*{Acknowledgement}

The authors acknowledge the Albert Einstein Centre for Gravitation and Astrophysics, supported by the Czech Science Foundation grant No 14-37986G, and the Silesian university at Opava internal grant SGS/14/2016.

\bibliographystyle{abbrv}
\bibliography{bibliography}

\end{document}